\documentclass[
11pt, 
twoside, 
english, 
onehalfspacing,
headsepline, 
]{DoctoralThesis} 

\usepackage[utf8]{inputenc} 
\DeclareUnicodeCharacter{202F}{~}

\usepackage[toc, page]{appendix}
\usepackage[nottoc,notlot,notlof]{tocbibind} 
\usepackage[numbers, compress]{natbib} 

\usepackage{amsmath, amssymb}
\usepackage{mathtools} 
\usepackage{dsfont} 
\usepackage{slashed}

\newcommand{\dd}{\mathrm{d}} 
\newcommand{\dvol}{d{\mathrm{vol}}}
\newcommand{\pd}{\partial}
\newcommand{\cD}{\mathcal{D}}
\newcommand{\e}{\mathbf{e}} 
\newcommand{\pp}{=\kern-0.40em{\vert}} 

\newcommand{\bR}{\mathbb{R}}
 
\newcommand{\bC}{\mathbb{C}}

\newcommand{\defeq}{\vcentcolon=} 
\newcommand{\eqdef}{=\vcentcolon} 

\newcommand{\hol}[1]{\textrm{hol}(#1)}
\newcommand{\hodge}[1]{(\star #1)} 

\newcommand{\Dirac}[2]{\left \langle #1, #2 \right \rangle_D} 
\newcommand{\Herm}[2]{\left \langle #1, #2 \right \rangle} 
\newcommand{\Spin}[2]{\left \langle #1, #2 \right \rangle_s} 

\thesistitle{Geometric Aspects of Type IIA Supersymmetric Backgrounds and Heterotic Anomalies}
\supervisor{Prof. Papadopoulos} 
\degree{Doctor of Philosophy} 
\author{Jake \textsc{Phillips}} 
\university{\href{https://www.kcl.ac.uk}{King's College London}} 
\department{\href{https://www.kcl.ac.uk/mathematics}{Department of Mathematics}}
\group{\href{https://www.kcl.ac.uk/research/theoretical-physics}{Theoretical Physics Group}}
\faculty{\href{https://www.kcl.ac.uk/nmes}{Faculty of Natural, Mathematical \& Engineering Sciences}}

\AtBeginDocument{
\hypersetup{pdftitle=\ttitle} 
\hypersetup{pdfauthor=\authorname} 
\hypersetup{urlcolor=black}
\hypersetup{linkcolor=black}
\hypersetup{citecolor=black}
}

\begin{document}

\frontmatter 

\pagestyle{plain} 


\begin{titlepage}
	\begin{center}

		\vspace*{-.05\textheight}
		{\scshape\LARGE \univname\par}\vspace{1cm} 
		\textsc{\Large Doctoral Thesis}\\[0.5cm] 

		\HRule \\[0.4cm] 
		{\huge \bfseries \ttitle\par}\vspace{0.4cm} 
		\HRule \\[1.5cm] 

		\vspace{-0.5cm} 
		\href{https://www.kcl.ac.uk/people/jake-phillips}{\huge\bfseries Jake Phillips} 
		\vspace{1cm} 

		\includegraphics[scale=2.6]{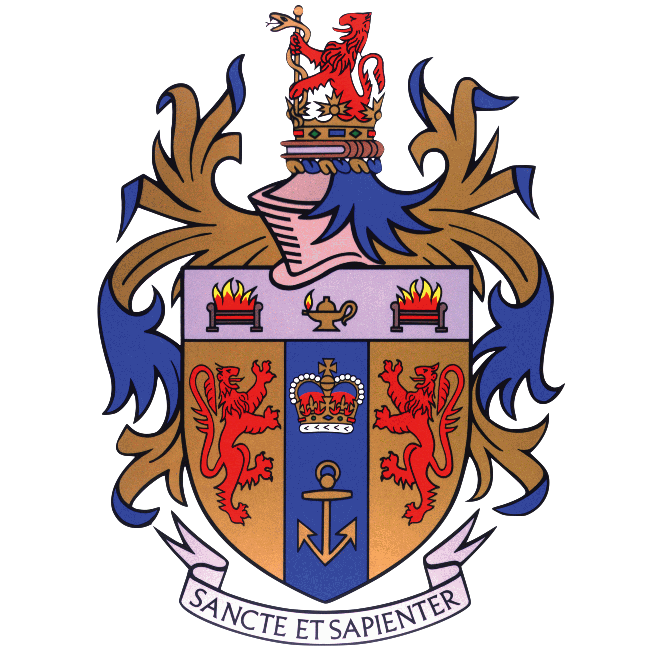}\\[1cm] 
		\vfill

		\large \textit{A thesis submitted in partial fulfilment of the requirements\\ for the degree of \degreename}\\[0.3cm] 
		\textit{in the}\\[0.4cm]
		\textit{\groupname\\\deptname}\\[1cm] 

		\vfill
	\end{center}
\end{titlepage}
\cleardoublepage


\begin{abstract}
\addchaptertocentry{\abstractname} 
We construct the twisted covariant form hierarchies (TCFH) of (massive) type IIA supergravity for common sector, D-brane and warped product AdS supersymmetric backgrounds and show that the Killing spinor bilinears satisfy a generalisation of the conformal Killing-Yano equation with respect to the TCFH connections. The Killing-Stäckel, Killing-Yano and closed conformal Killing-Yano tensors of all spherically symmetric (massive) type IIA brane backgrounds are computed and one demonstrates that the geodesic flow on these solutions is completely integrable by giving all independent charges in involution. The Killing spinor form bilinears that generate hidden symmetries for spinning particle and string probe actions on such backgrounds are identified. The interplay between TCFHs and hidden symmetries of probes propagating on these backgrounds is investigated and used to explore the question of whether charges constructed from these bilinears are sufficient to prove the integrability of such probes on this class of backgrounds. Additionally, some of the properties of TCFHs, such as the reduced holonomy of the minimal TCFH connections for generic backgrounds, are investigated. After this, the algebra of holonomy symmetries of sigma models propagating on supersymmetric heterotic backgrounds with a non-compact holonomy group is determined. One demonstrates that these close as a W-algebra that is specified by a Lie algebra structure on the space of covariantly constant forms that generate the holonomy symmetries. In addition, the chiral anomalies associated with these symmetries are identified. Finally, it is argued that these anomalies are consistent and can be cancelled up to two loops in the sigma model perturbation theory.
\end{abstract}


\begin{acknowledgements}
\addchaptertocentry{\acknowledgementname} 

First and foremost, I would like to thank my supervisor, Professor Papadopoulos. His analytical approach, unwavering work ethic, and mathematical rigour have not only pushed me to work to the best of my ability but also profoundly influenced my development as a researcher. Certainly, without him, this thesis would not have been possible. 

I am grateful to have shared this PhD journey with my academic brothers Edgar and Loukas. I vividly recall our numerous discussions on the intricacies of our research, as well as the times when we would laugh about our joint plight. These day-to-day interactions remain a highlight of my experience.

Further, I wish to thank my peers in the theoretical physics group at King's; I have enjoyed our frequent coffee and lunch breaks, where we would have a much-needed reprieve from work. Frequently informal discussions arising from these breaks led to epiphanies in my research. I would also like to thank Sebastian for the many enjoyable conversations centred around our shared experiences and for being a standard to which to aspire.

I want to thank my friends for indulging me in my (many!) ramblings on supergravity, string theory, black holes and geometry. In this regard, I want to particularly thank Joseph for allowing me to soundboard a myriad of ideas, equations, and concepts on an almost weekly basis throughout the PhD and Jack for the many late-evening debates on the philosophical and foundational aspects of the subject. 

To Mayya, Goronwy and Joseph, I am eternally grateful for the emotional support you have given me over the past years (and more). Our frequent conversations and excursions have kept me sane through the tough times and have provided me with great relief and joy. I also want to say thanks for your detailed feedback on my manuscript, especially when given the density of equations! To Charlie, thank you for your endless belief and faith in me. Knowing this has helped calm me down when I have felt overwhelmed and anxious.

Most of all, I would like to thank my mother. The way you have kindled my inquisitiveness from a young age and encouraged me to think deeply about matters has shaped who I am today. Without your unwavering support in all things, I could not have made it this far. I feel blessed and am truly grateful to have had such a supportive figure throughout my life. I know Grandad would have loved to have read this thesis.

This work was supported by an EPSRC studentship grant. 

\end{acknowledgements}


\tableofcontents 


\dedicatory{For mum.} 


\mainmatter 

\hypersetup{allcolors=blue}

\pagestyle{thesis} 

\chapter*{Introduction}
\addcontentsline{toc}{chapter}{Introduction}

From the early days of classical mechanics to modern theories of spacetime and the fundamental interactions, symmetries have consistently acted as a driving force for the development of physics. For instance, Galilean relativity is central to the formulation of Newtonian mechanics, taking the laws of physics to be the same in all inertial frames, with two such frames being related by a Galilean transformation\footnote{This group of symmetry transformations between inertial frames is given by the Galilean group $G_3$.}. While Galilean relativity laid the groundwork for the concept of relative motion, the discovery of the invariance of the speed of light suggested that there is a different symmetry between inertial frames, namely Poincaré symmetry. From this extension of the rotation group to transformations involving both the spatial and temporal coordinates, one arrives at the notion of spacetime based on symmetry considerations. The realisation that one should consider \emph{spacetime} rather than space and time separately led to the theory of Special Relativity and a fundamental change in the way in which we formulate physical laws. Building on the Poincaré symmetry between inertial frames, Einstein later realised that when generalising to non-inertial frames, the laws of physics must be invariant under arbitrary diffeomorphisms. This discovery gave rise to the theory of General Relativity and the reinterpretation of gravity as the geometry of spacetime.

One of the first solutions of the field equations of General Relativity was the Schwarzschild solution, which describes a static, spherically symmetric black hole: this was found by imposing a $\bR \times SO(3)$ symmetry on the spacetime. Symmetries often play a central role in guiding the search for solutions to theories of spacetime; by starting with a symmetry ansatz, one can significantly simplify the form of the equations of motion, readily giving way to closed-form solutions. Beyond this, symmetries give rise to powerful tools such as Noether's theorem and Liouville's theorem --- with Noether's theorem establishing a deep link between conserved charges and symmetries and Liouville's theorem providing conditions for the integrability of a dynamical system in terms of the number of independent, mutually Poisson commuting conserved charges.

It is clear that symmetry plays an integral role in physics, and from this end, unification has been an ongoing theme of research. One only needs to look back towards the electromagnetic theory constructed by Maxwell\footnote{A modern description of Maxwell theory is the gauge theory of $U(1)$.} to see this. Maxwell's unification of the electromagnetic interaction was built upon by the electroweak theory of Glashow, Weinberg and Salam in 1967 \cite{Glashow:1961tr, Weinberg:1967tq, Salam:1968rm}, which led to the formulation of the $SU(3) \times SU(2) \times U(1)$ Standard Model, providing a quantum description of the strong and electroweak interactions. The goal of further unification, namely to reconcile gravity with the other fundamental forces, is, however, fraught with difficulty. This difficulty arises due to the gravitational coupling constant having a mass dimension of minus two in four dimensions, and so by simple power counting one can see that attempts to quantise General Relativity lead to non-renormalisable ultraviolet divergences.

String Theory emerged as the leading candidate for such a ``theory of everything''. Work first began on String Theory in the late 1960s as a model of the strong nuclear interaction by Veneziano \cite{Veneziano:1968yb}. This description of the strong nuclear force was quickly superseded in favour of quantum chromodynamics. Work on the amplitude proposed by Veneziano continued where it was reinterpreted as a scattering amplitude for relativistic bosonic strings \cite{Nambu:1969se, Susskind:1969ha, Susskind:1970qz}. Schwarz, Scherk and others continued research into String Theory, discovering that the spectrum of these bosonic strings contains a massless spin 2 state that corresponds to the graviton \cite{Scherk:1974mc, Yoneya:1974jg}. Excitingly, while the theory was initially proposed to describe the strong nuclear interaction, gravity was naturally emergent. This work developed into what has become known as ``bosonic'' String Theory, given that its spectrum of states comprises exclusively of bosons.

Despite these promising aspects of bosonic String Theory, there were several unsatisfactory features. For instance, the spectrum of states did not contain any fermionic matter and that the ground state contained a tachyon, indicating instability of the vacuum. Moreover, the bosonic string is critical in 26 dimensions. Ramond et al. conducted research with the goal of introducing fermionic states to the spectrum of the bosonic string, leading to the discovery of supersymmetry\footnote{Supersymmetry was also independently discovered outside of the context of String Theory by Golfand and Likhtman where they were researching spinorial extensions of the Poincaré algebra \cite{Golfand:1971iw}.} \cite{Ramond:1971gb, Neveu:1971rx, Gervais:1971ji}. This gave way to what is now known as \emph{Superstring} Theory\footnote{Though Superstring Theory is frequently referred to as ``String Theory'' for brevity.}, which is critical in 10 dimensions. Strings that possess supersymmetry are known as superstrings. As research continued into the late 1970s, it was realised that the spectrum of states of the superstring admits a consistent truncation where the tachyon is removed, and the remaining states can be arranged into supermultiplets, resulting in the first consistent superstring theories \cite{Gliozzi:1976qd}. 

The ten year period between 1984 and 1994 was host to several important discoveries in String Theory\footnote{The period between 1984-1985 is known as the first superstring revolution, and the mid-1990s is referred to as the second superstring revolution.}: the Green-Schwarz anomaly cancellation of the type I superstring \cite{mgjs} and the discovery of the heterotic string by Gross et al. \cite{Gross:1984dd}. During this time, Candelas, Horowitz, Strominger and Witten showed that by compactifying on Calabi-Yau manifolds, one can construct $\mathcal{N} = 1$ supersymmetric theories in four dimensions \cite{Candelas:1985en}. By 1985, it was identified that there exist only five consistent superstring theories: type I, type IIA, type IIB, heterotic $SO(32)$ and heterotic $E_8 \times E_8$. In the mid 1990s, it was realised that these five theories are related to one another by dualities, and each can be constructed by taking different limits of a single 11 dimensional theory of branes, known as M-theory \cite{Witten:1995ex}.

On the heels of the second superstring revolution, the proposal of the AdS/CFT correspondence by Maldacena \cite{Maldacena:1997re} in the late 1990s marked a pivotal moment for String Theory. The correspondence states that superstring theories on $(n + 1)$-dimensional Anti-de Sitter (AdS$_{n + 1}$) geometries are dual to $n$-dimensional conformal field theories (CFT), providing a holographic duality between gravitational theories and traditional quantum field theories. As the AdS/CFT correspondence is a strong-weak coupling duality, it provides new directions for studying strongly coupled quantum field theories through the framework of String Theory that are otherwise inaccessible through traditional perturbative approaches. This discovery has ignited a surge of interest in studying superstring and supergravity theories on AdS geometries over the past years. The most well-understood examples of this duality are those with a high degree of supersymmetry, with the most celebrated example being the duality between type IIB String Theory on AdS$_5 \times S^5$ and $\mathcal{N} = 4$ super Yang-Mills theory. This has prompted great interest in the study of supersymmetric backgrounds.

We have frequently discussed symmetries of the space(time) geometry: this class of symmetries\footnote{Referring to a symmetry of the metric tensor.} are known as isometries and are generated by Killing vector fields. In the context of geodesic flow of dynamical systems, isometries are also known as explicit symmetries. In this context, another class of symmetries, known as \emph{hidden} symmetries, play a crucial and subtle role in studying such systems. A hidden symmetry is a symmetry of the phase space of a dynamical system that does not have a well-defined projection to the configuration space. Phrased more geometrically, if the dynamical system is propagating on a manifold, $M$, then the phase space will be naturally identified with the cotangent bundle, $T^*M$: thus a hidden symmetry is a symmetry of $T^*M$ that does not have a well-defined projection to $M$. Physically, one can think of such symmetries as mapping solutions of the equations of motion to other solutions. Hidden symmetries are generated by generalisations of Killing vectors, known as Killing-Stäckel and Killing-Yano tensors. 

While hidden symmetries are not physical symmetries of the spacetime geometry but rather symmetries of the space of solutions, and through Noether's theorem, they nonetheless give rise to conserved charges. Such charges contribute towards the conditions required for integrability of the dynamical system as set by Liouville's theorem. The Kerr black hole is one of the most prodigious examples of the application of hidden symmetries to the integrability of dynamical systems on gravitational backgrounds. Here Carter discovered an integral of geodesic motion that was quadratic in the particle's momenta \cite{carter-b, Carter:1968ks}, leading to the complete separability of the Hamilton-Jacobi equation in the Kerr geometry. Penrose and Walker later showed that this integral of motion was in one-to-one correspondence with a rank two Killing-Stäckel tensor \cite{Walker:1970un}. 

Beyond the Kerr geometry, Killing-Stäckel and Killing-Yano tensors have been instrumental in the integrability of geodesic flows of several black hole spacetimes, whilst also playing a significant role in the separability and integrability of other classical field equations on curved backgrounds \cite{penrose, floyd, chandrasekhar, carter-a, carter-c, page, sfetsos, lun}. For additional applications of these tensors see e.g. \cite{kt, Ggp, howe2, ls1, ls2}.

It has been shown in \cite{gptcfh} that the conditions imposed by the gravitino Killing spinor equation on the (Killing spinor) form bilinears can be arranged as a twisted covariant form hierarchy (TCFH) \cite{jggp}. This means that there is a connection, $\mathcal{D}^\mathcal{F}$, on the space of spacetime forms which depends on the fluxes, $\mathcal{F}$, of the theory such that the highest weight representation of $\mathcal{D}^\mathcal{F}\Omega$ vanishes, where $\Omega$ is a collection of forms of various degrees and $\mathcal{D}^\mathcal{F}$ may not be form degree preserving. Equivalently, this condition can be written as
\begin{equation}\label{tcfheqn}
	\mathcal{D}_X^\mathcal{F}\Omega= i_X \mathcal{P}+ \alpha_X \wedge \mathcal{Q}~,
\end{equation}
for every spacetime vector field $X$, where $\mathcal{P}$ and $\mathcal{Q}$ are appropriate multi-forms and $\alpha_X$ denotes the associated 1-form constructed from the vector field $X$ after using the spacetime metric $g$, $\alpha_X(Y) = g(X,Y)$. The proof of this result is rather general and includes supergravities on spacetimes of any signature as well as the effective theories of strings which include higher order curvature corrections. It also puts the conditions imposed by the Killing spinor equations on the form bilinears on a firm geometric basis.

One consequence of the TCFH is that the form bilinears satisfy a generalisation of the conformal Killing-Yano equation with respect to the connection $\mathcal{D}^\mathcal{F}$. This can be easily seen after taking the skew-symmetric part and contraction with respect to the metric $g$ of \eqref{tcfheqn}, and so one identifies $\mathcal{P}$ with an exterior derivative constructed from $\mathcal{D}^\mathcal{F}$ and $\mathcal{Q}$ with a formal adjoint of $\mathcal{D}^\mathcal{F}$ acting on $\Omega$. As it has been demonstrated in \cite{gibbons} that spinning particle probes \cite{bvh} propagating on backgrounds equipped with a Killing-Yano form admit (hidden) symmetries generated by the form, it raises the possibility that, as a consequence of the TCFH, the form bilinears of supersymmetric backgrounds may be associated with the (hidden) symmetries of certain probes whose actions may include couplings associated with the supergravity fields. Thus, there may be an interplay between TCFHs and probe conservation laws.

This thesis is organised as follows: in chapter \ref{background} we aim to provide a review of the necessary theoretical background to provide context for the work of the thesis. In chapter \ref{TCFH-General-Theory-Backgrounds} we present the TCFH of (massive) type IIA supergravity and show that the Killing spinor bilinears satisfy a generalisation of the conformal Killing-Yano equation. We also compute the Killing-Yano tensors for all spherically symmetric IIA brane backgrounds and show that they are integrable. We then identify those Killing spinor bilinears corresponding to hidden symmetries and investigate whether the charges constructed from these bilinears are sufficient to prove the integrability of the relevant probes on this class of backgrounds. This work was published in \cite{lggpjp}. Chapter \ref{TCFH-AdS-Backgrounds} is the natural extension of chapter \ref{TCFH-General-Theory-Backgrounds} where we refine our focus to warped-product AdS backgrounds, which has been published in \cite{Papadopoulos:2023chf}. In chapter \ref{W-symmetries-heterotic-backgrounds} we determine the algebra of holonomy symmetries of sigma models propagating on heterotic backgrounds with a non-compact holonomy group. We demonstrate that they close as a W-algebra specified by a Lie algebra structure on the space of covariantly constant forms. We then identify the chiral anomalies associated with these symmetries and argue that they are consistent and can be cancelled up to two loops in the sigma model perturbation theory. This chapter has been published in \cite{Papadopoulos:2023exe}. In appendices \ref{Conventions}, \ref{sec:Warped-AdS-Backgrounds} and \ref{IIA-bilinears} we state our conventions, provide a more thorough treatment of warped product AdS backgrounds than is presented in chapter \ref{TCFH-AdS-Backgrounds} and present the Killing spinor bilinears of type IIA supergravity backgrounds, respectively.

\chapter{Symmetries, Supergravity and TCFHs} \label{background}

In this chapter, we provide a lightning review of the essential theoretical background to provide context for the work outlined in this thesis. We begin by examining the classical symmetries of geodesic flow, discussing hidden symmetries and their relation to Killing-Stäckel and Killing-Yano tensors and their application to the integrability and separability of the geodesic and Klein-Gordon equations. Next, we shift our focus to the introduction of the maximal supergravities in 10 dimensions. Here, we examine the key aspects of supersymmetric backgrounds, such as the Killing spinor equations and their associated integrability conditions. We then explore non-linear sigma models and their supersymmetric extensions, relating the symmetries of these actions to the previously discussed Killing-Stäckel and Killing-Yano tensors. Following this, we review anomalies and present the Fujikawa construction of the chiral anomaly. Finally, we introduce the notion of a twisted covariant form hierarchy and demonstrate how it gives rise to a generalisation of the conformal Killing-Yano equations. 

\section{Geometric Formulation of Symmetries}
\subsection{Explicit Symmetries}
Consider the motion of a free particle propagating on an $n$-dimensional spacetime manifold, $M$, with metric $g$. Such a system has an action
\begin{equation} \label{Free-Particle-Action}
	S = \frac{1}{2} \int \dd \tau\, g_{MN} \dot{x}^M \dot{x}^N \,,
\end{equation}
which describes the geodesic flow\footnote{When viewing the geodesic flow as a dynamical system, $M$ is identified with its configuration space.} on $M$, where $\dot x$ denotes the derivative of the coordinate $x$ with respect to the affine parameter $\tau$. Isometries are diffeomorphisms that preserve the spacetime geometry and are generated by Killing vector fields, $K \in \Gamma(TM)$, which satisfy the Killing equation:
\begin{equation}
	\mathcal{L}_K g = 0 \,,
\end{equation}
where $\mathcal{L}_K$ is the Lie derivative with respect to the vector field $K$. This can be equivalently written in a coordinate basis as:
\begin{equation}
	\nabla^{(M}K^{N)} = 0 \,,
\end{equation}
where $\nabla$ is the Levi-Civita connection. It is well-known that Killing vector fields generate symmetries of geodesic motion,
\begin{equation} \label{Isometry}
	\delta x^M = \epsilon \, K^M(x) \,,
\end{equation}
where $\epsilon$ is a constant that parametrises the transformation. The transformation \eqref{Isometry} is a symmetry of \eqref{Free-Particle-Action} and is known as an \emph{explicit} symmetry of the action. In general, explicit symmetries of particle dynamics are symmetries generated by Killing vector fields. Through Noether's theorem, these symmetries give rise to an associated conserved charge
\begin{equation} \label{Isometry-Charge}
	Q(K) = g(K, \dot{x}) \,,
\end{equation}
which is conserved along time-like geodesics.

There is an extension of Killing vector fields that generate conformal transformations of the spacetime geometry, known as \emph{conformal} Killing vector fields. These give rise to conserved charges for the geodesic flow of massless particles. A vector field $\xi \in \Gamma(TM)$ is known as a conformal Killing vector field if it satisfies the conformal Killing equation:
\begin{equation}
	\mathcal{L}_\xi g = \lambda \, g \,,
\end{equation}
where $\lambda \in C^\infty(M)$ is a smooth function. Equivalently, this can be written in a coordinate basis:
\begin{equation}
	\nabla^{(M} \xi^{N)} = \lambda(x) \, g^{MN} \, .
\end{equation}
The infinitesimal diffeomorphisms generated by $\xi$ are of the same form as \eqref{Isometry} and they give rise to explicit symmetries of the massless version of the action \eqref{Free-Particle-Action}.

\subsection{Hidden Symmetries and Killing-Stäckel Tensors}

Other than explicit symmetries, there also exist less evident symmetries of geodesic flow that are determined by the spacetime geometry. Such symmetries are generated by generalisations of Killing vector fields. Considering a transformation that is monomial in the particle's velocity:
\begin{equation} \label{Hidden-Symmetry}
	\delta x^M = \epsilon \, d^M{}_{N_1 \dots N_{k-1}}(x) \, \dot{x}^{N_1} \dots \dot{x}^{N_{k-1}} \, ,
\end{equation} 
and requiring this to be a symmetry of the action \eqref{Free-Particle-Action}, it constrains the tensor $d$ to satisfy the equation
\begin{equation} \label{Killing-Tensor-Equation}
	\nabla_{(M} d_{N_1 \dots N_k)} = 0 \, .
\end{equation}
This is known as the Killing tensor equation, and the tensors that solve it are known as Killing-Stäckel tensors. To be precise, a rank $k$ Killing-Stäckel tensor, $d$, is a totally symmetric $(0, k)$-tensor that satisfies the Killing tensor equation \eqref{Killing-Tensor-Equation}. From this definition, it is clear that a Killing-Stäckel tensor of rank 1 is a Killing vector field. As Killing-Stäckel tensors generate symmetries of the action, by Noether's theorem, they give rise to charges, $Q$, that are conserved on geodesic motion
\begin{equation} \label{Hidden-Symmetry-Charge}
	Q(d) = d_{N_1 \dots N_k} \dot{x}^{N_1} \dots \dot{x}^{N_k} \, .
\end{equation}
Symmetries of the type \eqref{Hidden-Symmetry} are known as \emph{hidden} symmetries of the free particle action \eqref{Free-Particle-Action}. It is important to note that Killing-Stäckel tensors do not generate spacetime diffeomorphisms; therefore, unlike explicit symmetries, hidden symmetries are not symmetries of the geometry itself. Instead, they correspond to symmetries of the phase space, $T^*M$, governing the particle dynamics. 

Moreover, given two Killing-Stäckel tensors $d$ and $e$, of rank $k$ and $l$ respectively, one can construct a new Killing-Stäckel of rank $(k + l)$ by taking their symmetrised product:
\begin{equation}
	(d \otimes_s e)_{N_1 \dots N_{k + l}} \defeq d_{(N_1 \dots N_k} e_{N_{k + 1} \dots N_{k + l})} \, .
\end{equation} 
The metric tensor, $g$, is a (trivial) example of a Killing-Stäckel tensor: given that the metric is covariantly constant with respect to the Levi-Civita connection $\nabla_X g = 0$, one can see that it satisfies \eqref{Killing-Tensor-Equation}. The associated conserved charge is simply the norm of the velocity of the particle $Q(g) = g_{MN} \dot{x}^M \dot{x}^N \equiv -1$.

Just as there is a conformal extension of Killing vectors, there is also a conformal generalisation of Killing-Stäckel tensors. A conformal Killing-Stäckel tensor is a symmetric $(0, k)$-tensor, $d$, that satisfies the conformal Killing tensor equation:
\begin{equation} \label{Conformal-Killing-Tensor-Equation}
	\nabla_{(M} d_{N_1 \dots N_k)} = g_{(M N_1} q_{N_2 \dots N_k)} \,,
\end{equation}
where $q$ is a symmetric $(0, k-1)$-tensor. It is apparent that a conformal Killing-Stäckel tensor of rank 1 is a conformal Killing vector. In the same manner that Killing-Stäckel tensors generate hidden symmetries of the action \eqref{Free-Particle-Action}, conformal Killing-Stäckel tensors generate hidden symmetries of the geodesic flow for massless particles and give rise to conserved charges along null geodesics. Further, one can take the symmetrised product to two conformal Killing-Stäckel tensors to generate a new conformal Killing-Stäckel tensor. 

\subsection{Killing-Yano Forms}

Killing-Yano forms are closely related to hidden symmetries and Killing-Stäckel tensors and can, in some respects, be regarded as a more fundamental object. Such objects can be defined by studying the decomposition of the covariant derivative of a general $p$-form into its irreducible terms. This covariant derivative takes values in the space of tensors $T^* M \otimes \Lambda^p T^* M$ that are totally skew-symmetric in all but the first index; the space decomposes as \cite{Semmelmann:2002fra, frolov}
\begin{equation}
	T^* M \otimes \Lambda^p T^* M \cong \Lambda^{p-1} T^* M \oplus \Lambda^{p+1} T^* M \oplus \Lambda^{p, 1} T^* M \,.
\end{equation}
One can project the covariant derivative onto these subspaces by the projectors $\mathcal{A}$, $\mathcal{C}$ and $\mathcal{T}$ as follows:
\begin{align}
	(\mathcal{A}\, \Psi)_{M N_{1} \dots N_{p}} &= \Psi_{[M N_{1} \dots N_{p}]} \, , \\
	(\mathcal{C}\, \Psi)_{M N_{1} \dots N_{p}} &= \frac{p}{n-p+1} g_{M[N_{1}} \Psi^{P}{}_{|P| N_{2} \dots N_{p}]} \, , \\
	(\mathcal{T}\, \Psi)_{M N_{1} \dots N_{p}} &= \Psi_{M N_{1} \dots N_{p}} - \Psi_{[M N_{1} \dots N_{p}]}-\frac{p}{n-p+1} g_{M[N_{1}} \Psi^{P}{}_{|P| N_{2} \dots N_{p}]} \, ,
\end{align}
where $\Psi \in T^* M \otimes \Lambda^p T^* M$, i.e. $\Psi_{M N_{1} \dots N_{p}} = \Psi_{M [N_{1} \dots N_{p}]}$. The covariant derivative of a general $p$-form therefore decomposes as:
\begin{equation} \label{Covariant-Deriv-Decomp}
	\nabla_X \omega = \mathcal{A}\, \nabla_X \omega + \mathcal{C}\, \nabla_X \omega + \mathcal{T}\, \nabla_X \omega \, .
\end{equation}
The first term, known as the antisymmetric part, is constructed from the exterior derivative $\dd \omega$. The second term, referred to as the divergence part, is built from the adjoint of the exterior derivative\footnote{The adjoint of the exterior derivative is also referred to as the co-derivative.}, $\delta \omega \equiv - \nabla \cdot \omega$. The third term is related to the twistor operator's action, which is not relevant to the discussion here.

\subsubsection{Conformal Killing-Yano Forms}

Forms where the twistor term vanishes are known as \emph{conformal Killing-Yano} forms and therefore satisfy: 
\begin{equation} \label{CKY1}
	\nabla_{M} \omega_{N_{1} \dots N_{p}} = \nabla_{[M} \omega_{N_{1} \dots N_{p}]} + \frac{p}{n-p+1} g_{M[N_{1}} \nabla^{P} \omega_{|P| N_{2} \dots N_{p}]} \, .
\end{equation}
This condition can be equivalently formulated as a $p$-form, $\omega$, is a conformal Killing-Yano form if and only if there exists a $(p+1)$-form, $\kappa$, and a $(p-1)$-form, $\xi$, such that:
\begin{equation} \label{CKY2}
	\nabla_M \omega_{N_1 \dots N_p} = \kappa_{MN_1 \dots N_p} + p g_{M[N_1}\xi_{N_2 \dots N_p]} \,.
\end{equation} 
From \eqref{CKY2} the forms $\kappa$ and $\xi$ are uniquely determined as:
\begin{equation}
	\kappa_{MN_1 \dots N_p} = \nabla_{[M} \omega_{N_{1} \dots N_{p}]}, \quad \xi_{N_2 \dots N_p} = \frac{1}{n-p+1} \nabla^{P} \omega_{P N_{2} \dots N_{p}} \,.
\end{equation}
We can contract \eqref{CKY2} with a vector field $X \in \Gamma(TM)$ to write the conformal Killing-Yano condition in a coordinate-free form:
\begin{equation} \label{cky}
	\nabla_X \omega = \frac{1}{p + 1} i_X \dd \omega - \frac{1}{n-p+1} \alpha_X \wedge \delta \omega \,,
\end{equation}
where $i_X$ denotes the interior derivative with respect to $X$, $\dd \omega$ is the exterior derivative of $\omega$ and $\delta \omega$ is the co-derivative of $\omega$. $\alpha_X$ represents the associated one-form constructed from the vector field $X$ after using the spacetime metric $g$, $\alpha_X(Y) = g(X, Y)$.

\subsubsection{Killing-Yano Forms and Closed Conformal Killing-Yano Forms}

There are two important subclasses of conformal Killing-Yano forms: Killing-Yano forms and closed conformal Killing-Yano forms. If $\omega$ is co-closed, $\delta \omega = 0$,
\begin{equation}
	\nabla_X \omega = \frac{1}{p + 1} \, i_X \dd \omega \,,
\end{equation}
the form $\omega$ is known as a rank $p$ Killing-Yano form\footnote{Equivalently, if $\nabla_X \omega$ is given by the antisymmetric term of \eqref{Covariant-Deriv-Decomp}.}. On the other hand, if $\omega$ is closed, $\dd \omega = 0$,
\begin{equation}
	\nabla_X \omega = - \frac{1}{n - p + 1} \, \alpha_X \wedge \delta \omega \, ,
\end{equation}
then $\omega$ is known as a rank $p$ closed conformal Killing-Yano form\footnote{Equivalently, if $\nabla_X \omega$ is given by the divergence term of \eqref{Covariant-Deriv-Decomp}.}. 

One property of such forms is that the Hodge dual of a Killing-Yano form yields a closed conformal Killing-Yano form and vice versa. The Hodge dual of a conformal Killing-Yano form is also a conformal Killing-Yano form.

An important property of closed conformal Killing-Yano forms is that the exterior product of two such forms generates a new closed conformal Killing-Yano form\footnote{This can be viewed as an ``antisymmetric parallel'' to how one can generate (conformal) Killing-Stäckel tensors from the symmetrised product of two (conformal) Killing-Stäckel tensors.}. Namely, if $h_1$ and $h_2$ are closed conformal Killing-Yano forms of rank $p$ and $q$, then their exterior product:
\begin{equation}
	h = h_1 \wedge h_2 \, ,
\end{equation} 
yields a closed conformal Killing-Yano $(p+q)$-form.

Moreover, Killing-Yano forms can be thought of as the ``square root'' of Killing-Stäckel tensors in the following sense: taking the symmetrised product of two rank $k$ (conformal) Killing-Yano forms, $\omega$, $\eta$
\begin{equation}
	k^{MN} = \omega^{(M}{}_{P_1 \dots P_{k-1}} \eta^{N)P_1 \dots P_{k-1}} \, ,
\end{equation}
yields a rank 2 (conformal) Killing-Stäckel tensor, $k$. This generalises to a greater number of (increasing rank) forms. Due to this, Killing-Yano forms are associated with hidden symmetries of the action \eqref{Free-Particle-Action}, while the conformal Killing-Yano forms are associated with hidden symmetries of geodesic flow of massless particles. 

In addition to giving rise to hidden symmetries through the above ``squaring'' to Killing-Stäckel tensors, Killing-Yano forms directly generate symmetries for spinning particle actions, which we discuss in section \ref{SUSY-Sigma-Models}.

\subsection{The Principal Tensor and the Killing Tower}

Although we do not make use of a principal tensor in the work of this thesis, we will present it here for completeness of the discussion as the existence of a principal tensor has been instrumental for the complete integrability of the Kerr geometry and its generalisations.

A tensor, $h$, is known as a principal tensor if it is a non-degenerate\footnote{The non-degeneracy condition imposes that the principal tensor is of maximal rank and possesses the maximum number of functionally independent eigenvalues.} closed conformal Killing-Yano 2-form. Thus $h$ must satisfy:
\begin{equation} \label{principalTensor}
	\nabla_X h = \alpha_X \wedge \Phi, \quad \Phi = - \frac{1}{n - 1} \delta h \, .
\end{equation} 
It can be shown that the associated vector field to $\Phi$ is Killing and preserves $h$ \cite{Krtous:2008tb,Houri:2008ng,Yasui:2011pr}:
\begin{equation}
	\mathcal{L}_\phi g = 0, \quad \mathcal{L}_\phi h = 0 \,,
\end{equation} 
where $\phi = \alpha_\Phi$ is determined by \eqref{principalTensor}. This vector field $\phi$ is known as the \emph{primary Killing vector} field. From these properties, one can show that a principal tensor generates a rich symmetry structure known as the Killing tower. From the principal tensor, $h$, one can construct \cite{page, Frolov:2007cb, Frolov:2008jr}:
\begin{enumerate}
	\item Closed conformal Killing-Yano forms, $h_{(j)}$, of rank $2j$: $h_{(j)} = \frac{1}{j!}h^{\wedge j}$.
	\item Killing-Yano forms, $\omega_{(j)}$, of rank $(n - 2j)$: $\omega_{(j)} = \star h_{(j)}$.
	\item Rank 2 Killing-Stäckel tensors $k_{(j)}$: $k_{(j)}^{M N}=\frac{1}{(n-2 j-1) !} \omega_{(j)}^{(M}{}_{P_{1} \dots P_{n-2j-1}} \omega_{(j)}^{N)P_{1} \dots P_{n-2j-1}}$.
	\item Rank 2 conformal Killing-Stäckel tensors $Q_{(j)}$: $Q_{(j)}^{M N}=\frac{1}{(2 j-1) !} h_{(j)}^{(M}{}_{P_{1} \ldots P_{2 j-1}} h_{(j)}^{N) P_{1} \ldots P_{2 j-1}}$.
	\item Killing vectors $l_{(j)}$: $l_{(j)} = i_\phi k_{(j)}$.
\end{enumerate}
Where $\phi$ is the primary Killing vector field. The existence of a principal tensor places strong conditions on the geometry, with the most general background that admits a principal tensor given by the Kerr-NUT-(A)dS spacetime. Such geometries have many applications to higher-dimensional black holes \cite{frolov}. The Killing tower generates sufficiently many conserved charges to yield the geodesic flow of dynamical systems to be completely integrable, giving rise to the separability of the Hamilton-Jacobi, Klein-Gordon and Dirac equations. Liouville's theorem gives the conditions for the complete integrability of a dynamical system, which we will now discuss.

\subsection{Integrability and Separability}

Of particular importance are the class of dynamical systems known as \emph{integrable} systems. Such systems are exceptionally rare and are characterised as those with sufficiently many conserved charges that lead to the exact solubility of the dynamics. Their applications in modern theoretical physics are vast, ranging from General Relativity, to Condensed Matter Physics, String Theory and the AdS/CFT correspondence; see \cite{frolov, Retore:2021wwh, Arutyunov:2009ga, Bena:2003wd} for some selected reviews.

\subsubsection{Liouville's Theorem}

A Hamiltonian system propagating on a $n$-dimensional manifold $M$ has a $2n$-dimensional phase space, $P$, which is identified with the cotangent bundle of $M$, $P = T^* M$. Such a system is said to be \emph{Liouville integrable} if it admits $n$ functionally independent constants of motion, $Q^r$, including the Hamiltonian, $H$, that are in involution. Functional independence of the constants of motion refers to the linear independence of the set of associated one-forms $\dd Q^r$, meaning that the gradients span an $n$-dimensional space. The constants of motion are said to be in involution if their Poisson bracket algebra vanishes:
\begin{equation}
	\{Q^r, Q^s\}_{PB} = 0 \,, \quad \forall r, s \in \{1, \dots, n\} \, .
\end{equation} 
By performing a canonical transformation of the phase space coordinates, the equations of motion to a Liouville integrable system can be solved in a finite number of integration steps. One takes the conserved charges, $Q^r$, as the momentum coordinates, with new position coordinates, $\theta^r$, constructed as their conjugates. These coordinates satisfy the canonical Poisson bracket relation:
\begin{equation}
	\{\theta^r, Q^s\}_{PB} = \delta^{rs} \, .
\end{equation}
In these coordinates, the equations of motion take on a particularly simple form. The solution of the canonical position coordinates are linear in time:
\begin{equation}
	\theta^r(\tau) = \omega^r \tau + \theta^r_0 \,,
\end{equation}
where $\theta^r_0 = \theta^r(\tau = 0)$ and the ``frequencies'', $\omega^r$, are functions of the conserved charges, $Q^r$, and given by the derivative of the Hamiltonian,
\begin{equation}
	\omega^r = \frac{\pd H}{\pd Q^r} \,.
\end{equation}
The frequencies are constant along the dynamical trajectories, making the analysis of the system trivial.

\subsubsection{The Hamilton-Jacobi Equation}

The Hamilton-Jacobi equation provides an alternative method to study geodesic flow by reformulating the system in terms of Hamilton's principal function\footnote{While in the literature it is known as Hamilton's principal function it simply is equal to the action of a system, up to an additive constant.}, $\bar{S}(x, \tau)$. The Hamilton-Jacobi equation is a non-linear first-order partial differential equation for $\bar{S}(x, \tau)$ and is given by:
\begin{equation} \label{Hamilton-Jacobi}
	\frac{\pd \bar{S}}{\pd \tau} + H\left(x, \frac{\pd \bar{S}}{\pd x}, \tau\right) = 0 \,.
\end{equation}
A solution to the Hamilton-Jacobi equation, $\bar{S}(x, \tau)$, generates a canonical transformation of phase space. In this transformation, the old phase space coordinates $(x^M, p_M)$ are transformed into new phase space coordinates $(q^M, \tilde{p}_M)$, where the new momenta are defined as:
\begin{equation}
	\tilde{p}_M = \frac{\pd \bar{S}}{\pd q^M} \,,
\end{equation}
and $q^M$ are the new coordinates of the configuration space. This generates a corresponding transformation of the Hamiltonian
\begin{equation}
	\mathcal{H} = H + \frac{\pd \bar{S}}{\pd \tau} \,,
\end{equation}
which by \eqref{Hamilton-Jacobi} implies $\mathcal{H} = 0$, thereby trivialising the dynamics of the system.

For a time-independent Hamiltonian, the time dependence of Hamilton's principal function can be solved by taking $\bar{S}(x, \tau) = S(x) - E\tau$, where $E$ is a constant. This gives rise to the time-independent Hamilton-Jacobi equation:
\begin{equation}
	H\left(x, \frac{\pd S}{\pd x}\right) = E \,.
\end{equation}
The function $S$ is known in the literature as Hamilton's characteristic function\footnote{A solution to the time-independent Hamilton-Jacobi equation is also known as a complete integral.} and the constant $E$ is identified as the energy. 

The relevant case here is when the time-independent Hamilton-Jacobi equation can be solved by a complete additive separation of variables. In this case, the characteristic function can be decomposed as:
\begin{equation}
	S(x) = \sum\limits_{i=1}^n S_i(q^i) \,,
\end{equation}
where the Hamiltonian must now also separate into a sum of independent functions for each $q^i$, leading the time-independent Hamilton-Jacobi equation to take the form
\begin{equation}
	\sum\limits_{i=1}^n H_i\left(q^i, \frac{\pd S_i}{\pd q^i}\right) = E \,,
\end{equation}
implying that each term $H_i\left(q^i, \frac{\pd S_i}{\pd q^i}\right) = E_i$, where $E=\sum\limits_i E_i$. As the solution to the Hamilton-Jacobi equation generates a canonical transformation of phase space, the Poisson bracket structure is preserved, and so one has
\begin{equation}
	\{E_i, E_j\}_{PB} = 0 \,, \quad \forall i, j \in \{1, \dots, n\} \, .
\end{equation}
Therefore, given a geometry where the Hamilton-Jacobi equation can be solved by complete additive separation of variables, one has that the dynamical system is Liouville integrable. While we have shown that this condition is sufficient, it can be shown that it is also necessary \cite{frolov}.

Further, suppose that the manifold is an Einstein space. In this case, the complete additive separability of the Hamilton-Jacobi equation implies that the Klein-Gordon equation, $g^{MN} \nabla_M \nabla_N \phi = m^2 \phi$, can be solved by a multiplicative separation of variables \cite{Benenti:1979erw}.

\section{Supergravity}

The low-energy effective theory of string theory is uniquely given by supergravity. In particular, the effective theories of type IIA and IIB string theories are type IIA and type IIB supergravity, respectively, which are collectively known as type II supergravity. One can also view supergravity as the theory of gauge supersymmetry. The first supergravity theories were discovered in 1976 by Freedman et al. \cite{Freedman:1976xh}. This was minimal ($\mathcal{N}=1$) supergravity in four dimensions. Shortly after, generalisations to various dimensions and a greater number of supersymmetries were discovered. A supergravity theory can have at most 32 supersymmetries and any such supergravity is known as \emph{maximal}. Of the maximal supergravities we will be most interested in type II supergravity as it constitutes the effective theory of strings.

\subsection{Type IIA Supergravity}

The field content of type IIA supergravity is given by the massless sector of type IIA string theory; the bosonic field content consists of the metric, $g$, the dilaton, $\Phi \in C^\infty(M)$, the NS-NS $B$-field, $B \in \Omega^2(M)$, and the $p$-form RR fields $A^{(1)} \in \Omega^1(M)$ and $A^{(3)} \in \Omega^3(M)$. The fermionic content consists of the gravitino, $\psi_M$, and the dilatino, $\lambda$ which are both in the 32-dimensional\footnote{The fermionic content of the massless sector of the type IIA superstring consists of two gravitini and two dilatini of \emph{opposite} chirality in the 16-dimensional Majorana-Weyl representation; these are typically assembled into single 32-dimensional Majorana spinors.} Majorana representation of $\mathfrak{spin}(9,1)$. Type IIA supergravity is a non-chiral theory.

The action of the bosonic sector of type IIA supergravity is \cite{Campbell:1984zc, Giani:1984wc, Huq:1983im}:
\begin{equation}
	S_B = S_{NSNS} + S_{RR} + S_{CS} \, ,
\end{equation}
where we have split the action into the NS-NS (common) sector, RR sector and Chern-Simons term. These are given by:
\begin{equation}
	S_{NSNS} = \frac{1}{2\kappa_{10}^2} \int \dd^{10}x\, \sqrt{-g}\, e^{-2\Phi}\left(R + 4\,\pd_M \Phi \pd^M \Phi -\frac{1}{3}H_{MNR}H^{MNR}\right) \,,
\end{equation}
\begin{equation}
	S_{RR} = -\frac{1}{4\kappa_{10}^2} \int F_2 \wedge \star{F_2} + \tilde{F}_4 \wedge \star{\tilde{F}_4} \,,
\end{equation}
\begin{equation}
	S_{CS} = -\frac{1}{4\kappa_{10}^2}\int B \wedge F_4 \wedge F_4 \,,
\end{equation}
where $H = \dd B$ is the field strength of the NS-NS $B$-field, $F_2 = \dd A^{(1)}$, $F_4 = \dd A^{(3)}$ are the field strengths of the RR fields and $\tilde{F}_4 = F_4 - A^{(1)} \wedge H$. $\kappa_{10}$ is the 10-dimensional gravitational constant and is related to the string scale via $2 \kappa_{10}^2 = (2 \pi)^7 \alpha^{\prime 4}$, where $\alpha^\prime = l_s^2 / (2\pi)^2$ is the usual string scale. The Chern-Simons part of the action is notable as it is a purely topological term.

Type IIA supergravity can also be constructed by means of dimensional reduction from 11-dimensional supergravity along $S^1$. 11-dimensional supergravity is special as 11 is the maximum number of spacetime dimensions whereby the massless multiplets contain spins no greater than 2 \cite{Nahm:1977tg}. Furthermore, the theory is also unique and can be considered as the low-energy limit of M-theory \cite{Cremmer:1978km, Witten:1995ex, Townsend:1995kk}.

\subsubsection{Massive IIA}
There exists a generalisation of type IIA supergravity, first discovered by Romans \cite{romans}, known as \emph{massive} type IIA that includes an additional scalar field $S = e^\Phi m$; the constant $m$ is the mass parameter of massive IIA supergravity and is related to the cosmological constant of the theory \cite{bgpiia}. The presence of this scalar field introduces new D8 brane solutions \cite{d8}.

\subsection{Type IIB Supergravity}

In much the same way as type IIA supergravity, the field content of type IIB supergravity is given by the massless sector of type IIB string theory. The bosonic field content consists of the metric, $g$, the dilaton, $\Phi \in C^\infty(M)$ and the NS-NS $B$-field, $B \in \Omega^2(M)$. The RR fields are a complex scalar $\varphi \in C^\infty(M, \bC)$, a complex two-form $A^{(2)} \in \Omega^2(M, \bC)$ and a complex four-form $A^{(4)} \in \Omega^4(M, \bC)$. The fermionic sector consists of the gravitino, $\psi_M$, and the dilatino, $\lambda$, which are both in the 32-dimensional\footnote{In the same way as the type IIA case, the 16-dimensional spinors of type IIB string theory are assembled into 32-dimensional spinors.} Majorana-Weyl representation of $\mathfrak{spin}(9,1)$. Type IIB supergravity is a chiral theory.

Given that type IIB supergravity is the effective theory of the type IIB superstring, which only differs from the type IIA case by means of differing G.S.O. projections, the bosonic action takes a similar form:
\begin{equation}
	S_B = S_{NSNS} + S_{RR} + S_{CS} \, ,
\end{equation}
where the action has been split into terms arising from the NS-NS sector, RR sector and the Chern-Simons term. The action of the NS-NS is identical to that of the IIA case, whereas the action for the RR sector and Chern-Simons terms are different and are given by \cite{Howe:1983sra, js, Schwarz:1983wa}:
\begin{equation}
	S_{NSNS} = \frac{1}{2\kappa_{10}^2} \int \dd^{10}x\, \sqrt{-g}\, e^{-2\Phi}\left(R + 4\,\pd_M \Phi \pd^M \Phi -\frac{1}{3}H_{MNR}H^{MNR}\right) \,,
\end{equation}
\begin{equation}
	S_{RR} = -\frac{1}{4\kappa_{10}^2} \int F_1 \wedge \star{F_1} + \tilde{F}_3 \wedge \star{\tilde{F}_3} + \tilde{F}_5 \wedge \star{\tilde{F}_5}\,,
\end{equation}
\begin{equation}
	S_{CS} = -\frac{1}{4\kappa_{10}^2}\int A^{(4)} \wedge H \wedge F_3 \,,
\end{equation}
where $H = \dd B$ is the field strength of the NS-NS $B$-field, $F_3 = \dd A^{(2)}$ and $F_5 = \dd A^{(4)}$ are the field strengths of the RR fields and $\tilde{F}_3 = F_3 - \varphi \wedge H$ and $\tilde{F}_5 = F_5 - \frac{1}{2} A^{(2)} \wedge H + \frac{1}{2} B \wedge F_3$. 

Strictly speaking, the above action is a `pseudo-action' as it does not encode all the properties of the fields. This is due to the fact that the 5-form field strength of type IIB string theory/supergravity is self-dual and as such the kinetic term for this field will vanish identically. One must impose the self-duality of the 5-form field strength by hand as an added constraint on the solutions as it cannot follow from the (pseudo) action itself.

\subsection{Supersymmetric Backgrounds}

A supergravity background \cite{Polchinski:1998rr, Becker:2006dvp, Green:2012pqa} is a collection, $(M, g, \Phi, \Psi, \cal{F})$, that is a solution to the classical equations of motion of the theory, where $M$ is a spacetime manifold, $g$ is the metric, $\Phi$ are the scalar fields, $\Psi$ are the spinor fields\footnote{In this thesis we set the spinor fields to vanish, which is typical of the literature in this context.} and $\cal{F}$ are the fluxes. Such solutions are referred to as ``backgrounds'' as they are taken to be a fixed configuration of the geometry and fields to which one can later consider perturbations around this classical solution.

Just as a solution to a classical field theory may not exhibit the full symmetry group of the action, a supergravity background will not, in general, preserve the complete symmetry of the underlying supergravity theory. In particular, while every supergravity theory is supersymmetric, a given solution may only be invariant under a subset of the supersymmetry transformations. If a solution preserves some residual supersymmetry, then the solution is known as a \emph{supersymmetric background}.

As a supergravity background is evaluated on the locus that all fermionic fields vanish, the supersymmetry variations of the bosonic fields are identically zero. Consequently, for the solution to be supersymmetric, one must only impose the vanishing of the fermionic supersymmetry variations.

The supersymmetry variations for the gravitino, $\psi$, and the dilatino, $\lambda$, are associated with a parallel transport equation for the supercovariant connection, $\mathcal{D}$, and an algebraic equation, respectively:
\begin{equation}
	\left. \delta \psi_M \right|_{\psi, \lambda = 0}= \mathcal{D}_M \epsilon, \quad \left. \delta \lambda \right|_{\psi, \lambda = 0} = \mathcal{A}\epsilon \, ,
\end{equation}
where $\epsilon$ is the (fermionic) supersymmetry parameter, $\mathcal{D}$ is the supercovariant connection and $\mathcal{A}$ is a Clifford algebra element depending on the fields. The condition for a background to be supersymmetric is therefore:
\begin{equation} \label{KSE}
	\mathcal{D}_M \epsilon = 0, \quad \mathcal{A}\epsilon = 0 \, .
\end{equation}
These equations are known as the \emph{Killing spinor equations} and the spinor, $\epsilon$, is known as a Killing spinor\footnote{This name originates due to the property that the ``square'' of a Killing spinor gives rise to a Killing vector field in the following sense: given a Killing spinor, $\epsilon$, the one-form bilinear $k = \Spin{\epsilon}{\Gamma_A \epsilon}\e^A$ is dual to a Killing vector field by means of the metric $g$, where $\Spin{\cdot}{\cdot}$ is any spin-invariant inner product.}. As the gravitino Killing spinor equation is a parallel transport equation for the supercovariant connection, the condition for supersymmetric solutions can be formulated as the requirement that the background admits (super)parallel spinors.

In general, the supercovariant connection for a supergravity theory will be of the form:
\begin{equation}
	\mathcal{D}_M \defeq \nabla_M + \sigma_M(\e, \mathcal{F}) \,,
\end{equation}
where $\nabla$ is the Levi-Civita spin connection of the spacetime acting on the spinors:
\begin{equation}
	\nabla_M \defeq \pd_M + \frac{1}{4} \Omega_{M, AB}\Gamma^{AB} \,,
\end{equation}
and $\sigma_M(\e, \mathcal{F})$ is a theory dependent Clifford algebra element that is a function of the spacetime co-frame, $\e$, and the fluxes of the theory, $\mathcal{F}$. The spacetime co-frame is adapted to the metric tensor, $g = \eta_{AB}\e^A \e^B$, and $\Gamma$ represents the standard Dirac gamma matrices. 

The number of supersymmetries preserved by a supersymmetric background, $N$, is equal to the number of linearly independent solutions, $\epsilon$, to the Killing spinor equations. Frequently one will express the preserved supersymmetry of a background as a fraction of the original supersymmetry:
\begin{equation}
	f = \frac{N}{N_{\textrm{total}}} \,,
\end{equation}
where $N_{\textrm{total}}$ is the total number of supersymmetry generators of the theory and is given by:
\begin{equation}
	N_{\textrm{total}} = \mathcal{N} \dim{S} \, .
\end{equation}
Here $\mathcal{N}$ is the number of supercharges and $\dim{S}$ is the dimension of the minimal spinor representation of the theory. A solution is known as \emph{maximally supersymmetric} if $N = N_{\textrm{total}}$. For example, in 10-dimensional type II supergravity one has $\mathcal{N} = 2$ supercharges and $\dim{S} = 16$ as the (minimal) spinors are Majorana-Weyl; the theory is maximally supersymmetric with $N_{\textrm{total}} = 32$ supersymmetry generators.

Supersymmetric backgrounds are also known as BPS solutions as they may saturate certain Bogomol'nyi-Prasad-Sommerfield bounds \cite{Bogomolny:1975de, Prasad:1975kr}. A background preserving a fraction $f$ of the original supersymmetry is called $f$-BPS.

The integrability conditions of the Killing spinor equations assist in the study of the field equations and the holonomy of the supercovariant connection in supersymmetric backgrounds. The integrability conditions of the Killing spinor equations are given by \cite{review}:
\begin{equation}
	\mathcal{R}_{MN} \epsilon \defeq [\mathcal{D}_M, \mathcal{D}_N] \epsilon = 0 \,, \quad
	[\mathcal{D}_M, \mathcal{A}] \epsilon = 0 \,, \quad
	[\mathcal{A}, \mathcal{A}] \epsilon = 0 \, ,
\end{equation} 
where $\mathcal{R}$ is the curvature of the supercovariant connection.

As the gravitino Killing spinor equation is a parallel transport equation for the supercovariant connection, it is natural to consider the holonomy of this connection. First consider the vacuum case, $\mathcal{F} = 0$; here the supercovariant connection coincides with the spin connection, $\nabla$. In this scenario one has:
\begin{equation}
	\mathcal{R}_{MN} \epsilon \equiv [\nabla_M, \nabla_N] \epsilon = \frac{1}{4} R_{MN}{}^{AB} \Gamma_{AB} \epsilon \, .
\end{equation}
On an $n$-dimensional background, $\Gamma_{AB}$ generates $Spin(n - 1, 1)$ and therefore, for vacuum solutions to the equations of motion, $\mathcal{F} = 0$, the (reduced) holonomy group of the supercovariant connection satisfies $\hol{\mathcal{D}} \equiv \hol{\mathcal{\nabla}} \subseteq Spin(n -1, 1)$.

In general one will have $\mathcal{F} \neq 0$; the supercovariant connection will contain additional Clifford algebra elements on a generic background, leading $\hol{\mathcal{D}}$ to be contained within a $SL$ group rather than a $Spin$ group. To determine the Lie algebra of $\hol{\mathcal{D}}$ one computes the span of the action of the supercovariant curvature $\mathcal{R}$ and its derivatives $\mathcal{D}^k \mathcal{R}$ on vector fields. This will typically yield an expression in terms of all possible skew-symmetric products of Dirac gamma matrices. The holonomy of the supercovariant connection has been computed for generic 11-dimensional \cite{hull, duff, gpdtx} and type II supergravity backgrounds \cite{gpdt} showing that for these cases $\hol{\mathcal{D}} \subseteq SL(32, \bR)$. See \cite{Batrachenko:2004su} for a list of holonomies of lower dimensional supergravity theories.

The gauge symmetry of the Killing spinor equations have been an essential aid for the analysis of supersymmetric backgrounds. These gauge transformations are local transformations of the spinors, $\epsilon$, spacetime co-frame, $\e$, and the fluxes, $\mathcal{F}$ that leave the Killing spinor equations covariant:
\begin{equation}
	\ell \mathcal{D}(\e, \mathcal{F}) \ell^{-1} = \mathcal{D}(\e^{\ell}, \mathcal{F}^{\ell}), \quad \ell \mathcal{A}(\e, \mathcal{F}) \ell^{-1} = \mathcal{A}(\e^{\ell}, \mathcal{F}^{\ell}) \, .
\end{equation}
The gauge group, $G$, always contains $Spin(n - 1, 1)$ as a subgroup, and for most supergravity theories, will be smaller than $\hol{\mathcal{D}}$.

\section{Non-Linear Sigma Models, Probes and Anomalies}

\subsection{Non-Linear Sigma Models}

A non-linear sigma model is a scalar field theory in which the scalar fields take values in a Riemannian manifold, which is typically referred to as the ``target space''. Non-linear sigma models were first introduced by Gell-Mann and Lévy in 1960 as a description for the low energy interactions of the spinless $\sigma$ meson \cite{Gell-Mann:1960mvl} and since their inception they have found widespread applications in both string and supergravity theories.

Consider an $n$-dimensional spacetime $\Sigma$ with coordinates $\{x^\alpha\}$ and metric $\gamma(x)$. An action for a set of $d$ free scalar fields $\phi^M$, $M = 1, \dots, d$, is
\begin{equation} \label{sigma-model}
	S[\phi] = \frac{1}{2} \mu^{n - 2} \int \dd^n x \sqrt{\gamma} \gamma^{\alpha\beta}\, g_{MN}(\phi) \pd_\alpha \phi^M \pd_\beta \phi^N \,.
\end{equation}
In a non-linear sigma model\footnote{From the perspective of the quantum theory, a sigma model can be considered to be a field theory with an infinite set of coupling constants, $g^{(n)}$, given by the Taylor expansion of the target space metric:
\begin{equation*}
	g_{MN}(\phi) = g^{(0)}_{MN} + g^{(1)}_{MN,P} \phi^P + \frac{1}{2} g^{(2)}_{MN, PQ} \phi^P \phi^Q + \dots \, .
\end{equation*}
Moreover, when written in this way the non-linear interactions of the sigma model are manifest.},
the scalar fields, $\phi^M$, take values in a $d$-dimensional Riemannian manifold, $M$, with a metric tensor $g(\phi)$. The manifold $(M, g)$ is known as the target space. As such, one can think of $\phi^M$ in two ways: from the perspective of the spacetime $\Sigma$, they are a collection of $d$ free scalar fields, whereas from the perspective of the target space $M$, they are coordinates of the manifold. As the fields $\phi^M$ are taken to be dimensionless, it is necessary to introduce the mass scale $\mu$ to ensure that the action is dimensionless. Given that the coupling constant is dimensionful, such theories are non-renormalisable for $n \geq 3$. Typically, units are chosen such that $\mu = 1$. 

To illustrate this, consider the following examples. When $n = 1$, the sigma model describes the action of a free relativistic particle with $\Sigma$ being identified as the worldline of the particle. For the case where $n = 2$, the sigma model corresponds to a description of the bosonic string, where $\Sigma$ is the worldsheet, and \eqref{sigma-model} coincides with the Polyakov action. 

A sigma model is independent of the choice of coordinates on $M$ and $\Sigma$; under diffeomorphisms of the target space:
\begin{equation}
	\phi^M \mapsto \phi^{\prime M}(\phi) \,, \quad g_{M N}(\phi) \mapsto g_{M N}^{\prime}\left(\phi^{\prime}\right)=\frac{\partial \phi^P}{\partial \phi^{\prime M}} \frac{\partial \phi^Q}{\partial \phi^{\prime N}} g_{P Q}(\phi) \,,
\end{equation}
one has that the action transforms as
\begin{equation}
	S^\prime[\phi^{\prime}] = \frac{1}{2} \mu^{n - 2} \int \dd^n x \sqrt{\gamma} \gamma^{\alpha\beta} \, g_{M N}^{\prime}\left(\phi^{\prime}\right) \partial_\alpha \phi^{\prime M} \partial_\beta \phi^{\prime N} \, .
\end{equation}
While the action is not strictly invariant under a reparametrisation of the target space, it retains the same functional form. Consequently, the equations of motion for $\phi^{\prime M}$ take the same structure with the metric expressed in the new coordinates. As this reparametrisation describes the same physics written in a different coordinate system, we regard the transformed sigma model as equivalent to the original. Due to this, the sigma model is determined by the equivalence class of metrics related by diffeomorphisms.

Moreover, for theories where $n \leq d$, one can introduce the Wess-Zumino term
\begin{equation} \label{sigma-model-WZ-term}
	S_{WZ} = \frac{q}{n!} \int \dd^n x \sqrt{\gamma} \, \epsilon^{\alpha_1 \dots \alpha_n} b_{M_1 \dots M_n}(\phi) \pd_{\alpha_1} \phi^{M_1} \dots \pd_{\alpha_n} \phi^{M_n} \, ,
\end{equation}
where $q$ is a constant and $b \in \Omega^n(M)$. Up to surface terms, \eqref{sigma-model-WZ-term} is invariant under the gauge transformation $b \mapsto b + \dd \lambda$, where $\lambda \in \Omega^{n-1}(M)$.

Furthermore, one can couple spacetime fermions $\psi^a(x)$ to the sigma model by the term
\begin{equation}
	S_F = i \int \dd^n x \sqrt{\gamma}\, G_{ab}(\phi) \bar{\psi}^a \sigma^\alpha D_\alpha \psi^b \, ,
\end{equation}
where $\sigma^\alpha$ denotes a spacetime Dirac gamma matrix, that satisfies the Clifford algebra on $M$, $\sigma^\alpha \sigma^\beta + \sigma^\beta \sigma^\alpha = 2 \gamma^{\alpha\beta}$. The covariant derivative $D_\alpha$ is
\begin{equation}
	D_\alpha \psi^a = \nabla_\alpha \psi^a + \Upsilon_M{}^a{}_b(\phi) \pd_\alpha \phi^M \lambda^b \, , 
\end{equation}
where $\nabla_\alpha$ is the covariant derivative, $\Upsilon_M{}^a{}_b(\phi)$ is a connection on the vector bundle\footnote{A common case is where $V = TM$.} $V$ over $M$ and $G_{ab}$ is a fibre metric.

\subsection{Supersymmetric Non-Linear Sigma Models} \label{SUSY-Sigma-Models}

One can extend non-linear sigma models to describe supersymmetric field theories; this is done by taking $\Sigma$ to be a $\mathbb{Z}_2$-graded manifold\footnote{In the literature, it is also referred to as a \emph{supermanifold}.} $\Xi^{p|q}$, where $p$ are the commuting dimensions and $q$ are the anti-commuting dimensions. The dimensions $p$ and $q$ are identified with the spacetime dimension $n$ and the number of supersymmetry generators $N$, respectively. Thus, a $N = 1$ supersymmetric non-linear sigma model in $d$-dimensions would be a field theory on $\Xi^{d|1}$. An action for a one-dimensional $N = 1$ supersymmetric non-linear sigma model is \cite{bvh}
\begin{equation} \label{SUSY-Free-Particle-Action}
	S = -\frac{i}{2} \int \dd \tau \dd \theta \, g_{MN}(x) D x^M \dot{x}^N \,,
\end{equation}
where $x(\tau, \theta)$ is a $N = 1$ superfield and $(\tau, \theta)$ are the commuting and anti-commuting coordinates of the worldline superspace $\Xi^{1|1}$, respectively. $D$ is the superspace covariant derivative which satisfies $D^2 \equiv i \pd_\tau$ and $\dot{x}$ denotes the derivative with respect to $\tau$. This action describes a spinning particle propagating on the target space $M$ and is a supersymmetric extension of the free particle action \eqref{Free-Particle-Action}. By varying the action, one arrives at the equation of motion for the superfield $x(\tau, \theta)$
\begin{equation}
	\nabla_\tau D x^M \equiv \frac{\dd}{\dd \tau} \left(D x^M\right) + \Gamma^M_{PQ} \dot{x}^P Dx^Q = 0 \,.
\end{equation}
It is known that Killing-Yano forms generate symmetries of spinning particle actions \cite{gibbons}. In particular, the Killing-Yano form $\alpha$ generates the infinitesimal symmetry
\begin{equation}\label{svar}
	\delta x^M = \epsilon\, \alpha^M{}_{N_1\cdots N_{k - 1}} Dx^{N_1}\cdots Dx^{N_{k - 1}}~,
\end{equation}
for the action \eqref{SUSY-Free-Particle-Action}, where $\epsilon$ is an infinitesimal parameter. The associated conserved charge is
\begin{equation}\label{ccalpha}
	\begin{split}
		Q(\alpha) =& (k + 1) \alpha_{N_1N_2\cdots N_k} \partial_\tau x^{N_1} Dx^{N_2}\cdots Dx^{N_k} \\
		&-\frac{i}{k + 1} (\dd \alpha)_{N_1N_2\cdots N_{k + 1}} Dx^{N_1} Dx^{N_2}\cdots Dx^{N_{k + 1}}~.
	\end{split}
\end{equation}
Observe that $Q(\alpha)$ is conserved, $DQ(\alpha) = 0$, subject to the equations of motion of \eqref{SUSY-Free-Particle-Action}.

Note that if the Killing-Yano form $\alpha$ is closed, $\dd \alpha = 0$, and so $\alpha$ is covariantly constant (or equivalently parallel) with respect to the Levi-Civita connection, then
\begin{equation}
	\tilde Q(\alpha) = \alpha_{N_1N_2\cdots N_k} D x^{N_1} Dx^{N_2} \cdots Dx^{N_k}~,
\end{equation}
is also conserved subject to the field equations of \eqref{SUSY-Free-Particle-Action}, $\partial_\tau \tilde Q(\alpha) = 0$. This gives the conservation of two charges $\tilde Q(\alpha)$ and $D \tilde Q(\alpha)$. The latter is proportional to that in \eqref{ccalpha} with $\dd \alpha = 0$.

There are several generalisations of conformal Killing-Yano tensors \cite{gggpks, kubiznak, houri1, houri2, kygp, howe2}. One of the most common ones is to replace the Levi-Civita connection that appears in the definition \eqref{cky} with another connection, for example a connection with skew-symmetric torsion. Some of the properties mentioned above extend to the generalised Killing-Yano tensors. For an application of the Killing-Yano forms to G-structures see \cite{Ggp, sat}.

\subsection{Probes}

Here we shall describe the probe actions that we will be considering in this thesis with a primary focus on string and particle probes. The dynamics of string probes propagating on a spacetime with metric $g$ and a 2-form gauge potential $b$ \cite{zumino, lag, ghull, howesierra} is described by the action
\begin{equation}\label{sact}
	S = \int \dd^2\sigma\, \dd^2\theta\, (g + b)_{MN}\, D_+x^M\, D_-x^N~,
\end{equation}
where $x = x(\sigma, \theta)$ are real superfields that depend on the worldsheet superspace with commuting $(\sigma^0, \sigma^1)$ and anti-commuting $(\theta^+, \theta^-)$ real coordinates. The action above has been given as in \cite{howegp1, howegp2} where one defines the lightcone coordinates, $\sigma^{\pp}= \sigma^0 + \sigma^1$, $\sigma^{ = }= - \sigma^0 + \sigma^1$, and the algebra of superspace derivatives is $D_-^2 = i\partial_{ = }$, $D_+^2 = i\partial_{{\pp}}$ and $D_+ D_- + D_- D_+ = 0$\,. Note that the sign labelling of the worldsheet superspace coordinates denotes $\mathfrak{spin}(1,1)$ chirality.

The infinitesimal symmetries of \eqref{sact} that we shall be considering are given by
\begin{equation}\label{ssym1}
	\delta x^M= \epsilon^{(+)} \beta^M{}_{P_1\dots P_k}\, D_+x^{P_1}\dots D_+x^{P_k}~,
\end{equation}
where $\beta$ is a spacetime ($k + 1$)-form and $\epsilon^{(+)}$ is an infinitesimal parameter; the superscript $(+)$ indicates that the weight of the infinitesimal parameter $\epsilon$ is such that the right-hand side of \eqref{ssym1} is a $\mathfrak{spin}(1,1)$ scalar. The action \eqref{sact} is invariant under such transformations provided that
\begin{equation}\label{conp}
	\nabla^{(+)}_M \beta_{P_1\dots P_{k + 1}} = 0~,
\end{equation}
where
\begin{equation}
	\nabla^{(\pm)} = \nabla\pm \frac{1}{2}C~,
\end{equation}
with $C = \dd b$, i.e. $\nabla^{(\pm)}_M Y^N = \nabla_M Y^N\pm\frac{1}{2} C^N{}_{MR} Y^R$. Therefore, $\beta$ generates a symmetry provided that it is a $\nabla^{(+)}$-covariantly constant form.

One can also consider symmetries of \eqref{sact} generated by the infinitesimal transformation
\begin{equation}\label{ssym2}
	\delta x^M= \epsilon^{(-)} \beta^M{}_{P_1\dots P_k} D_-x^{P_1}\dots D_-x^{P_k}~,
\end{equation}
where $\epsilon^{(-)}$ is an infinitesimal parameter. The condition for invariance of the action in such a case is
\begin{equation}\label{conm}
	\nabla^{(-)}_M \beta_{P_1\dots P_{k + 1}} = 0~,
\end{equation}
i.e. $\beta$ is a $\nabla^{(-)}$-covariantly constant form. In many examples that follow the spacetime will admit several $\nabla^{(\pm)}$-covariantly constant forms which generate symmetries of the string probe action \eqref{sact}. All $\nabla^{(+)}$-covariantly constant forms of the common sector backgrounds coincide with those of heterotic supersymmetric backgrounds. In turn these can be computed using the classification results of \cite{hetgpug1, hetgpug2} for all heterotic background Killing spinors. The $\nabla^{(-)}$-covariantly constant forms of common sector backgrounds can also be read from the classification results of \cite{hetgpug1, hetgpug2}. One can easily investigate the commutators of these symmetries \eqref{ssym1} and \eqref{ssym2}. In general these symmetries are of W-type and have been previously explored in \cite{phgp1, phgp2} both in the context of string compactifications and special geometric structures.

Actions of spinning particle probes are also invariant under the symmetries generated by either $\nabla^{(+)}-$ or $\nabla^{(-)}-$ covariantly constant forms $\beta$. One such worldline probe action is
\begin{equation}\label{pact}
	S = \int\, \dd \tau\, \dd^2\theta\, (g + b)_{MN} D_+x^M D_-x^N~,
\end{equation}
which in addition to the metric exhibits a 2-form coupling $b$, where the superfields $x^M = x^M(\tau, \theta)$ depend on the worldline superspace with commuting $\tau$ and anti-commuting $(\theta^+, \theta^-)$ real coordinates; see \cite{colesgp} for a systematic description of spinning particle actions with form and other couplings. The algebra of the worldline superspace derivatives is $D_+^2 = D_-^2 = i\partial_\tau$ and $D_+D_- + D_-D_+ = 0$. The signs on $\theta^\pm$ are just labels --- there is no chirality in one dimension. The infinitesimal variation of the superfields is as in either \eqref{ssym1} or \eqref{ssym2}, but now the fields are worldline superfields and the superspace derivatives are those of the worldline superspace. The conditions for invariance of the action above are given in either \eqref{conp} or \eqref{conm}, respectively.

Another class of spinning particle probes we shall be considering are described by the action \cite{colesgp}
\begin{equation}\label{1part}
	S = -\frac{1}{2} \int\, \dd \tau\, \dd \theta\, \big(i g_{MN} Dx^M \partial_\tau x^N + \frac{1}{6} C_{MNR} Dx^M Dx^N Dx^R\big)~,
\end{equation}
where $g$ is the spacetime metric and $C$ is a 3-form on the spacetime --- $C$ is not a necessarily closed 3-form. Moreover $x^M$ is a superfield that depends on the worldline superspace coordinates $(\tau, \theta)$ and $D^2 = i\partial_\tau$. Given a ($k + 1$)-form $\beta$
one can construct the infinitesimal transformation
\begin{equation}\label{ssym3}
	\delta x^M= \alpha\,\, \beta^M{}_{P_1\dots P_k} Dx^{P_1}\dots Dx^{P_k}~,
\end{equation}
where $\alpha$ is an infinitesimal parameter. The conditions required for this action to be invariant under the transformation \eqref{ssym3} can be arranged in two different ways. One way is to require, as in previous cases, that $\beta$ is $\nabla^{(+)}$-covariantly constant. An alternative way to arrange the conditions for invariance of \eqref{1part} is
\begin{equation}\label{addcon}
	\begin{gathered}
		\nabla^{(+)}_M \beta_{P_1\dots P_{k + 1}} = \nabla^{(+)}_{[M} \beta_{P_1\dots P_{k + 1}]}~, \\
		\dd i_\beta C + (-1)^k \frac{k + 2}{2} i_\beta \dd C = 0~.
	\end{gathered}
\end{equation}
These conditions and an explanation of the notation can be found in \cite{kygp}. Therefore, this set of conditions implies that $\beta$ is a $\nabla^{(+)}$ Killing-Yano form. For $C = 0$, one obtains that $\beta$ is a Killing-Yano form as for the spinning particles described by the action \eqref{SUSY-Free-Particle-Action}.

\subsection{Anomalies}

Thus far we have discussed symmetries of geodesic flow and those of supersymmetric non-linear sigma models. These are symmetries of classical actions. However, when moving to the quantum theory, symmetries of the classical action may not be preserved at the quantum level. When a symmetry transformation of a classical action does not represent a symmetry of the quantum effective action, the theory is said to have an \emph{anomaly}. 

Anomalies arising from global symmetries are associated with classically conserved currents failing to be conserved in the quantum theory. These quantum corrections frequently lead to interesting physical consequences, with the most historically significant example being the Adler-Bell-Jackiw (ABJ) anomaly \cite{Adler:1969gk, Bell:1969ts}. The ABJ anomaly is also known as the chiral (or axial) anomaly as it stems from a symmetry between the chiral fermions in the theory. As gauge symmetries do not represent physical symmetries in the traditional sense, but rather redundancies of our description of a physical theory, anomalous gauge symmetries indicate a fundamental inconsistency of the theory. Thus, for a theory to be consistent, anomalies associated with gauge symmetries must vanish.

\subsubsection{Chiral Anomalies}

In chapter \ref{W-symmetries-heterotic-backgrounds} we investigate the chiral anomalies of holonomy symmetries of sigma models on supersymmetric heterotic backgrounds. To provide context for this work, we provide a recap for the emergence of the chiral anomaly. For simplicity, we work with the case of quantum electrodynamics in four dimensions with a massless Dirac spinor. The fermionic sector of the action is given by
\begin{equation}
	S = i \int \dd^4 x \, \bar{\Psi} \slashed{D} \Psi \,,
\end{equation}
where $D_\mu = \pd_\mu - i e A_\mu$ is the $U(1)$ gauge covariant derivative. The action is invariant under a $U(1)$ gauge symmetry 
\begin{equation}
	\Psi \mapsto \Psi^{\prime} = e^{i\theta(x)}\Psi \,, \quad \bar{\Psi} \mapsto \bar{\Psi}^{\prime} = e^{-i\theta(x)}\bar{\Psi} \,.
\end{equation}
By Noether's theorem, this gives rise to the conserved current $j^\mu = \bar{\Psi} \gamma^\mu \Psi$. This action is also invariant under the global chiral symmetry
\begin{equation} \label{chiral-symmetry}
	\Psi \mapsto \Psi^{\prime} = e^{i\epsilon \gamma^5}\Psi \,, \quad \bar{\Psi} \mapsto \bar{\Psi}^{\prime} = \bar{\Psi}e^{i\epsilon \gamma^5} \,,
\end{equation}
where $\gamma^5 = i \gamma^0\gamma^1\gamma^2\gamma^3$ is the chirality matrix in four-dimensions. This is associated with the ``axial'' Noether current $j_A^\mu = \bar{\Psi} \gamma^\mu \gamma^5\Psi$. While \eqref{chiral-symmetry} is a symmetry of the classical action, it is not a symmetry of the quantum effective action. To see this, consider the argument posed by Fujikawa \cite{Fujikawa:1980eg}.

The partition function is given by
\begin{equation}
	\mathcal{Z} = \int \cD \Psi \cD \bar{\Psi} \, \exp{\left[i \int \dd^4 x \, \bar{\Psi} (i \slashed{D}) \Psi \right]} \, .
\end{equation}
Under the transformation \eqref{chiral-symmetry}, while the exponent will be invariant, the path integral measure will transform. One can expand $\Psi$ in a basis of orthonormal eigenspinors of the Dirac operator $\slashed{D}$,
\begin{equation}
	\Psi(x) = \sum_m a_m \psi_m(x) \,, \quad \bar{\Psi}(x) = \sum_m \bar{b}_m \bar{\psi}_m(x) \,,
\end{equation}
where $i \slashed{D} \psi_m = \lambda_m \psi_m$. The coefficients of this expansion are Grassmannian odd valued and satisfy
\begin{equation}
	\{a_m, a_n\}=0 \,, \quad \{\bar{b}_m, \bar{b}_n\}=0 \,, \quad \{a_m, \bar{b}_n\}= 1 \,.
\end{equation}
Given this expansion, the path integral measure over $\Psi$ and $\bar{\Psi}$ can be defined as:
\begin{equation}
	\cD \Psi \cD \bar{\Psi} = \prod_m \dd a_m \dd \bar{b}_m \,.
\end{equation}
Taking the transformation \eqref{chiral-symmetry} to be infinitesimal and local
\begin{equation}
	\Psi^{\prime}(x) = (\mathds{1} + i\epsilon(x) \gamma^5)\Psi(x) \, ,
\end{equation}
one computes by orthonormality of the eigenspinors $\psi_m$
\begin{equation}
	\begin{aligned}
	 a_m^\prime &= \sum_n \int \dd^4 x \, \epsilon(x) \bar{\psi}_m (\mathds{1} + i\epsilon(x) \gamma^5) \psi_n(x) a_n \, \\
	 &\eqdef \sum_n \left (\delta_{mn} + C_{mn} \right ) a_n .
	\end{aligned}
\end{equation}
The path integral measure picks up the Jacobian factor\footnote{The inverse of $\mathcal{J}$ appears as the integration is over Grassmannian variables.} 
\begin{equation}
	\cD \Psi^\prime \cD \bar{\Psi}^\prime = \mathcal{J}^{-2} \, \cD \Psi \cD \bar{\Psi} \,,
\end{equation}
where $\mathcal{J}$ is the determinant of the transformation $(\mathds{1} + C)$. As we are considering an infinitesimal transformation, we can simplify the expression 
\begin{equation}
	\mathcal{J} = \det (\mathds{1} + C) = \det [\exp(C)] = \exp(\textrm{Tr}\, C) \, ,
\end{equation}
where we have made use of the standard identity $\det \circ \exp \equiv \exp \circ \mathrm{Tr}$. Hence,
\begin{equation}
	\log \mathcal{J} = i \int \dd^4 x \, \epsilon(x) \sum_m \bar{\psi}_m(x) \gamma^5 \psi_m(x) \, .
\end{equation}
Due to the ultraviolet divergences originating from the sum over the eigenspinors, $\psi_m$, the integral is not well-defined. Consequently, this must be regulated whilst preserving gauge invariance. This calculation can be found in standard references such as \cite{Fujikawa:1980eg, Alvarez-Gaume:1985zzv, Peskin:1995ev, Nakahara:2003nw}. After renormalisation, one finds that,
\begin{equation}
	\mathcal{J} = \exp \left[-i \int \dd^4 x \, \epsilon(x)\left(\frac{e^2}{32 \pi^2} \epsilon^{\mu \nu \lambda \sigma} F_{\mu \nu} F_{\lambda \sigma}(x)\right)\right] \, .
\end{equation}
Therefore, after the transformation \eqref{chiral-symmetry}, the partition function takes the form:
\begin{equation}
	\mathcal{Z} = \int \cD \Psi \cD \bar{\Psi} \exp \left[i \int \dd^4 x\, \left(\bar{\Psi}(i\slashed{D}) \Psi + \epsilon(x)\left\{\partial_\mu j_A^{\mu}+\frac{e^2}{16 \pi^2} \epsilon^{\mu \nu \lambda \sigma} F_{\mu \nu} F_{\lambda \sigma}\right\}\right)\right] \, .
\end{equation}
By varying the exponential with respect to the parameter $\epsilon(x)$ one arrives at the chiral anomaly
\begin{equation}
	\pd_\mu j_A^\mu = - \frac{e^2}{16 \pi^2} \epsilon^{\mu \nu \lambda \sigma} F_{\mu \nu} F_{\lambda \sigma} \, .
\end{equation}

\section{Twisted Covariant Form Hierarchies}

In the recent work on the classification of supersymmetric solutions of supergravity theories, two primary methods have been utilised: the ``bilinears method'' and the ``spinorial geometry method''. The bilinears method transforms the Killing spinor equations into a series of conditions on the Killing spinor bilinears, which are typically constructed from either a Dirac or Hermitian inner product\footnote{While a Dirac or Hermitian inner product is typical, any spin-invariant inner product can be used.} of the Killing spinors. These conditions are solved after the transformation of the Killing spinor equations in terms of the Killing spinor bilinears. The spinorial geometry method solves the Killing spinor equations directly. For further details, see \cite{review} for a review of this work.

In \cite{gptcfh} it has been found that the conditions imposed by the gravitino Killing spinor equation on the Killing spinor bilinears for all supergravity theories can be arranged into a generalisation of the conformal Killing-Yano equation with respect to a connection, $\mathcal{\nabla}^\mathcal{F}$, which is twisted by the fluxes $\mathcal{F}$ of the theory. The symmetry structure comprised of the collection of the Killing spinor bilinears, together with their associated generalisation of the conformal Killing-Yano equation, is known as a \emph{twisted covariant form hierarchy} (TCFH) of the supergravity theory. 

An important property of the TCFH structure is that it is not unique, as there is an ambiguity in the construction of the TCFH connection, although two classes of TCFHs are ubiquitously defined. Consequently, each supergravity theory has an associated family of TCFHs.

Given that Killing-Yano forms, which are a subclass of conformal Killing-Yano forms, generate symmetries of spinning particles on supersymmetric backgrounds, it is therefore pertinent to make use of the TCFH to investigate the conditions for which the Killing spinor bilinears reduce to Killing-Yano forms. This question forms the basis for the work set out in chapters \ref{TCFH-General-Theory-Backgrounds} and \ref{TCFH-AdS-Backgrounds}.

\subsection{Conformal Killing-Yano Equations and Killing Spinors}

Recall the conformal Killing-Yano equation,
\begin{equation}
	\nabla_X \omega = \frac{1}{k + 1} \, i_X \dd \omega - \frac{1}{n - k + 1} \, \alpha_X \wedge \delta \omega \,,
\end{equation}
as defined earlier. One should note that the conformal Killing-Yano equation can be generalised by substituting the Levi-Civita connection, $\nabla$, for a torsionful connection $\nabla^H = \nabla + \frac{1}{2} H$, where $H$ is a skew-symmetric torsion tensor. When performing such a generalisation, the exterior derivative and co-derivative must be substituted with $\dd^H$ and $\delta^H$, respectively.

Consider now the Killing spinor equation\footnote{Such Killing spinor equations appear on Anti-de Sitter AdS$_n$, de Sitter dS$_n$ and $n$-sphere S$^n$ geometries.}:
\begin{equation} \label{sphere-KSE}
	\mathcal{D}_M \epsilon \defeq \nabla_M \epsilon + \lambda \, \Gamma_M \epsilon = 0 \,,
\end{equation}
where $\lambda \in \bC$ is a constant. It can be seen that the Killing spinor bilinears associated with this Killing spinor equation satisfy the conformal Killing-Yano equation. First, construct the $k$-form bilinears:
\begin{equation}
	\phi^k = \frac{1}{k!} \Dirac{\epsilon}{\Gamma_{N_1 \dots N_k} \epsilon} \dd x^{N_1} \wedge \dots \wedge \dd x^{N_k} \,,
\end{equation}
where $\epsilon$ is a solution to \eqref{sphere-KSE} and $\Dirac{\cdot}{\cdot}$ is the Dirac inner product. A direct computation yields
\begin{equation} \label{sphere-CKY}
	\nabla_X \phi^k=\left(\pm \bar{\lambda}-(-1)^k \lambda\right) i_X \phi^{k+1}+\left(\pm \bar{\lambda}+(-1)^k \lambda\right) \alpha_X \wedge \phi^{k-1} \,,
\end{equation}
where one takes the positive or negative case for Lorentzian or Euclidean signature manifolds, respectively. By skew-symmetrising \eqref{sphere-CKY}, one arrives at the exterior derivative term of the conformal Killing-Yano equation, and by taking the contraction with the metric tensor, one arrives at the term involving the adjoint of the exterior derivative. Therefore, one sees that the conditions imposed by the Killing spinor equations on the Killing spinor bilinears, $\phi^k$, lead to a conformal Killing-Yano equation. While we have established this result for supergravities with a Killing spinor equation of the type \eqref{sphere-KSE}, in general, the Killing spinor equations of generic supergravity theories impose significantly more complicated conditions, and, as such, we require a necessary generalisation of the conformal Killing-Yano conditions to encapsulate the full geometric structure.

\subsection{Twisted Form Hierarchies and Conformal Killing-Yano Equations} \label{sec:TCFH-CKY}

Consider an $n$-dimensional manifold, $M$, equipped with a metric, $g$, of arbitrary signature $(r,s)$ and let $\Lambda^*_c(M)$ be the complexified bundle of all forms on $M$. Let $\mathcal{F} \in \Gamma(\oplus^m \Lambda^*_c(M))$ be a multi-form, i.e. $\mathcal{F}$ is a collection of $m$ independent complex forms with potentially different degrees. A \emph{twisted covariant form hierarchy} \cite{gptcfh}, twisted by $\mathcal{F}$, is a collection of forms $\{\chi^p\} \in \Gamma(\oplus^l \Lambda^*_c(M))$, where the forms $\{\chi^p\}$ can also have varying degree $p$, that satisfy:
\begin{equation} \label{TCFH}
	\nabla_{X}^{\mathcal{F}}\left(\left\{\chi^{p}\right\}\right)=i_{X} \mathcal{P}\left(\mathcal{F},\left\{\chi^{p}\right\}\right)+\alpha_{X} \wedge \mathcal{Q}\left(\mathcal{F},\left\{\chi^{p}\right\}\right) \, ,
\end{equation}
where $\mathcal{P}, \mathcal{Q}: \Gamma(\Lambda^*_c(M)) \to \Gamma(\Lambda^*_c(M))$. The multi-forms $\mathcal{P}$ and $\mathcal{Q}$ are constructed from algebraic operations between the forms $\{\chi^p\}$ and $\mathcal{F}$ such as the wedge product and inner derivation and their corresponding adjoints. $\nabla^\mathcal{F}$ is the \emph{twisted covariant hierarchy connection} which acts on $\Gamma(\oplus^l \Lambda^*_c(M))$; this connection is constructed from the Levi-Civita connection and $\mathcal{F}$ and is not necessarily degree preserving. In what follows, the collection of forms $\{\chi^p\}$ will be identified as the set of algebraically independent Killing spinor bilinears and $\mathcal{F}$ will be built from the supergravity $p$-form field strengths. 

Equation \eqref{TCFH} is analogous to \eqref{sphere-CKY} and leads to a generalisation of the conformal Killing-Yano equation. As previously, one can express the right-hand side of \eqref{TCFH} in terms of the left-hand side; by skew-symmetrising \eqref{TCFH}, one arrives at an exterior derivative type term, and by taking the contraction with the spacetime metric one arrives at an adjoint type term,
\begin{equation} \label{TCFH-CKY}
	\left( \nabla_{X}^{\mathcal{F}} \left\{ \chi^{q} \right\} \right) |_{p} = \frac{1}{p+1} \left( i_{X} \dd^{\mathcal{F}} \left( \left\{ \chi^{q} \right\} \right) \right) |_{p} - \frac{1}{n-p+1} \alpha_{X} \wedge \left( \delta^{\mathcal{F}} \left( \left\{ \chi^{q} \right\} \right) \right) |_{p-1} \, ,
\end{equation}
where $(\dots)|_p$ denotes the restriction to $p$-forms. The $\dd^{\mathcal{F}}$ operator is the twisted exterior derivative and $\delta^{\mathcal{F}}$ is the adjoint of $\dd^{\mathcal{F}}$, these are both built from the TCFH connection, $\nabla_{X}^{\mathcal{F}}$, and are constructed through the skew-symmetrisation and contraction with the metric as outlined above. 

Equation \eqref{TCFH-CKY} represents a generalisation of the conformal Killing-Yano equation, and while \eqref{TCFH-CKY} is implied by \eqref{TCFH}, the converse is not true unless one imposes that
\begin{equation}
	\left. \frac{1}{p+1} \left( i_X d^{\mathcal{F}} \left( \left\{ \chi^q \right\} \right) \right) \right|_p 
	= \left. \left( i_X \mathcal{P} \right) \right|_p, \quad 
	-\left. \frac{1}{n-p+1} \left( \delta^{\mathcal{F}} \left( \left\{ \chi^q \right\} \right) \right) \right|_{p-1} 
	= \left. \mathcal{Q} \right|_{p-1} \,.
\end{equation}
As such, \eqref{TCFH-CKY} is a coarser relation than \eqref{TCFH}, thus one might expect there to be solutions of \eqref{TCFH} that are not given by solutions of \eqref{TCFH-CKY}. 

Given its importance to the work that follows in this thesis, we present a sketch of the proof that the Killing spinor equations of all supergravity theories are associated with a family of TCFHs. This was first presented in \cite{gptcfh}, where the full details can be found.

First, consider the general structure of the supercovariant connection of an arbitrary supergravity theory:
\begin{equation}
	\mathcal{D}_X = \nabla_X + c \left(i_X \mathcal{H}\right) + c\left(\alpha_X \wedge \mathcal{G}\right) \, ,
\end{equation}
where $\nabla_X$ is the Levi-Civita connection. The term $c$ denotes the Clifford algebra element associated with the multi-forms $i_X \mathcal{H} = \sum_p i_X H^p$ and $\mathcal{G}=\sum_p G^p$. Each $H^p$ and $G^p$ represent the $p$-form supergravity field strengths of the theory. 

Given that the following series of steps is linear in the field strengths $H^p$ and $G^p$, without loss of generality, one can take $\mathcal{H}$ and $\mathcal{G}$ as single forms of degree $l$, i.e. $\mathcal{H} = H$ and $\mathcal{G} = G$. Now, consider the Killing spinor bilinears $\{\chi^p\}$ constructed from a spin-invariant inner product $\Spin{\cdot}{\cdot}$,
\begin{equation}
	\chi^p = \frac{1}{p!}\Spin{\epsilon}{\Gamma_{A_1 \dots A_p} \epsilon} \e^{A_1} \wedge \dots \wedge \e^{A_p} \,,
\end{equation}
where $\epsilon$ is a Killing spinor, $\mathcal{D}_M \epsilon = 0$ and $\{\e^A\}$ is a (pseudo)-orthonormal frame adapted to the spacetime metric. The Killing spinor equations imply that
\begin{equation}
	\begin{aligned}
		\nabla_X \chi^p =& -\frac{1}{p!} \left( \Spin{c(i_X H) \epsilon}{\Gamma_{A_1 \ldots A_p} \epsilon} + \Spin{\epsilon}{\Gamma_{A_1 \ldots A_p} c(i_X H) \epsilon} \right) \e^{A_1} \wedge \cdots \wedge \e^{A_p} \\
		&- \frac{1}{p!} \left( \Spin{c(\alpha_X \wedge G) \epsilon}{\Gamma_{A_1 \ldots A_p} \epsilon} + \Spin{\epsilon}{\Gamma_{A_1 \ldots A_p} c(\alpha_X \wedge G) \epsilon} \right) \e^{A_1} \wedge \cdots \wedge \e^{A_p} \,.
	\end{aligned}
\end{equation}
After significant Clifford algebra manipulation and making extensive use of the Hermiticity properties of the inner product and the definition of the Killing spinor bilinears, one arrives at the expression:
\begin{equation} \label{TCFH-proof}
	\begin{aligned}
	 \nabla_X \chi^p +& \left( \sum_q \left( c_q^1 (i_X H) \cdot \chi^q + c_q^5 (i_X G) \cdot \chi^q + \tilde{c}_q^1 (i_X \bar{H}) \cdot \chi^q + \tilde{c}_q^5 (i_X G) \cdot \chi^q \right) \right) \bigg|_p \\
		= -& \left( i_X \left( \sum_q \left( c_q^4 G \cdot \chi^q + \tilde{c}_q^4 \bar{G} \cdot \chi^q \right) \right) \right) \bigg|_p - \alpha_X \wedge \left( \sum_q \left( c_q^2 G \cdot \chi^q + \tilde{c}_q^2 \bar{G} \cdot \chi^q \right) \right) \bigg|_{p-1}
	\end{aligned}
\end{equation}
where $c_q^1$, $\tilde{c}_q^1$, $c_q^2$, $\tilde{c}_q^2$, $c_q^4$, $\tilde{c}_q^4$, $c_q^5$, and $\tilde{c}_q^5$ are combinatorial coefficients that depend on $p$, $l$, and the inner product $\Spin{\cdot}{\cdot}$. The terms $\bar{H}$ and $\bar{G}$ are the complex conjugates of the field strengths $H$ and $G$, respectively. The product $\varphi^k \cdot \xi^m$ is a multi-index contraction and denotes contractions of the type
\begin{equation}
	(\varphi^k \cdot \xi^m)_{N_1 \ldots N_p} \defeq \frac{1}{s!} \varphi^{M_1 \ldots M_s}{}_{[N_1 \ldots N_{k - s}} \xi_{|M_1 \ldots M_s|N_{k - s + 1} \ldots N_p]} \,,
\end{equation} 
where $p = k + m - 2s$. Equation \eqref{TCFH-proof} defines a TCFH where one makes the following identifications,
\begin{equation}
	\begin{gathered}
	 \nabla_X^\mathcal{F} \chi^p \defeq \nabla_X \chi^p + \left( \sum_q \left( c_q^1 (i_X H) \cdot \chi^q + c_q^5 (i_X G) \cdot \chi^q + \tilde{c}_q^1 (i_X \bar{H}) \cdot \chi^q + \tilde{c}_q^5 (i_X G) \cdot \chi^q \right) \right) \bigg|_p \,, \\
		\left. \left( i_X \mathcal{P} \right) \right|_p \defeq - \left( i_X \left( \sum_q \left( c_q^4 G \cdot \chi^q + \tilde{c}_q^4 \bar{G} \cdot \chi^q \right) \right) \right) \bigg|_p \,, \\
		\left. \mathcal{Q} \right|_{p-1} \defeq - \alpha_X \wedge \left( \sum_q \left( c_q^2 G \cdot \chi^q + \tilde{c}_q^2 \bar{G} \cdot \chi^q \right) \right) \bigg|_{p-1} \,,
	\end{gathered}
\end{equation}
where $\mathcal{F} = \{\mathcal{H}, \mathcal{G}\}_{\mathrm{ind}}$ are the linearly independent supergravity field strengths. It should be noted that the TCFH connection, $\nabla^\mathcal{F}$, is a connection on $\Gamma(\oplus^l \Lambda^*_c(M))$ and satisfies all four axioms required to be a connection. 

There is an ambiguity in the choice of the TCFH connection, which gives rise to a family of possible TCFHs for each supergravity theory. This ambiguity arises from the exterior derivative type term in \eqref{TCFH-CKY}; there may be terms $\mathcal{F} \cdot \chi^p$ for which
\begin{equation}
	i_X (\mathcal{F} \cdot \chi^p) = i_X \mathcal{F} \cdot \chi^p \,,
\end{equation}
i.e. that the Killing spinor bilinear is fully contracted with a field strength, leaving the free indices on the flux. In this case, such terms can contribute to either the connection or to the multi-form $\mathcal{P}$ as part of the right-hand side of the generalised conformal Killing-Yano equation. If all such terms are included in the connection, then the TCFH connection is known as \emph{maximal} and is denoted by $\nabla^\mathcal{F}$. Conversely, if all such terms are included in the multi-form $\mathcal{P}$, then the TCFH connection is known as \emph{minimal} and is denoted by $\mathcal{D}^\mathcal{F}$. These choices represent the extreme ends of the family of TCFHs, and there will be many possible intermediate selections of connection.

It is clear that a characterising feature of a TCFH is the set of bilinears; one can choose to include in this set both the Killing spinor bilinears $\{\chi^p\}$ and their Hodge duals. If one makes this choice to include the Hodge duals of the Killing spinor bilinears in the set, then the Hodge duality operation on $\{\chi^p\}$ acts as an automorphism of the hierarchy, and in this case, the TCFH will be twisted by $\mathcal{F}$. If one chooses not to include the Hodge duals of the Killing spinor bilinears\footnote{This is the approach chosen for the work that follows in this thesis.} in this set, then the hierarchy will be twisted with respect to both $\mathcal{F}$ and its dual $\star \mathcal{F}$.

\subsection{A Type IIA Supergravity Example}

While the full TCFH of type IIA supergravity will be presented in chapter \ref{TCFH-General-Theory-Backgrounds}, it is illustrative to consider a relevant example. The TCFH for the one-form Killing spinor bilinear $\tilde{k}$, with respect to the minimal connection, can be computed and is given by
\begin{equation}
	\begin{split}
		\mathcal{D}_M^\mathcal{F} \tilde k_N \defeq& \nabla_M \tilde{k}_N - \frac{1}{2} e^\Phi F_{MP}\omega^P{}_N - \frac{1}{12}e^\Phi G_{MPQR}\tilde{\zeta}^{PQR}{}_N \\
		=& -\frac{1}{2}H_{MNP}k^P + \frac{1}{4}e^\Phi g_{MN}S\tilde{\sigma} +\frac{1}{8}e^\Phi g_{MN} F_{PQ}\omega^{PQ} -\frac{1}{2}e^\Phi F_{[M|P|}\omega^P{}_{N]} \\
		&+ \frac{1}{4\cdot 4!}e^\Phi g_{MN} G_{P_1\dots P_4}\tilde{\zeta}^{P_1 \dots P_4} - \frac{1}{12}e^\Phi G_{[M|PQR|}\tilde{\zeta}^{PQR}{}_{N]}~ \, ,
	\end{split}
\end{equation}
where $\tilde{\sigma}$, $k$, $\omega$ and $\tilde{\zeta}$ are the 0-, 1-, 2- and 4-form Killing spinor bilinears and $S$, $F$, $H$ and $G$ are the (massive) IIA supergravity field strengths. For full details on field content and the Killing spinor bilinears, see chapter \ref{TCFH-General-Theory-Backgrounds}. Observe that the minimal TCFH connection, $\mathcal{D}_M^\mathcal{F}$, is twisted with respect to $\mathcal{F} = \{F, G\}$ and that this connection does not preserve degree. This gives rise to a generalised conformal Killing-Yano equation for this bilinear,
\begin{equation}
	\mathcal{D}_M^\mathcal{F} \tilde k_N = \mathcal{D}_{[M}^\mathcal{F} \tilde k_{N]} + \frac{1}{10} g_{MN}\mathcal{D}^\mathcal{F\,P} \tilde k_P \,.
\end{equation}
The TCFH with respect to the maximally twisted connection, $\nabla_M^\mathcal{F}$, for this bilinear is given by:
\begin{equation}
	\begin{split}
		\nabla_M^\mathcal{F} \tilde k_N \defeq& \nabla_M \tilde{k}_N - \frac{1}{2} e^\Phi F_{MP}\omega^P{}_N + \frac{1}{2}H_{MNP}k^P - \frac{1}{12}e^\Phi G_{MPQR}\tilde{\zeta}^{PQR}{}_N \\
		=& \frac{1}{4}e^\Phi g_{MN}S\tilde{\sigma} +\frac{1}{8}e^\Phi g_{MN} F_{PQ}\omega^{PQ} -\frac{1}{2}e^\Phi F_{[M|P|}\omega^P{}_{N]} \\
		&+ \frac{1}{4\cdot 4!}e^\Phi g_{MN} G_{P_1\dots P_4}\tilde{\zeta}^{P_1 \dots P_4} - \frac{1}{12}e^\Phi G_{[M|PQR|}\tilde{\zeta}^{PQR}{}_{N]}~ \, ,
	\end{split}
\end{equation}
where we see that the maximal connection is twisted with respect to $\mathcal{F} = \{F, H, G\}$. The associated generalised conformal Killing-Yano equation with respect to the maximal connection for this bilinear is
\begin{equation}
	\nabla_M^\mathcal{F} \tilde k_N = \nabla_{[M}^\mathcal{F} \tilde k_{N]} + \frac{1}{10} g_{MN}\nabla^\mathcal{F\,P} \tilde k_P \,.
\end{equation}

\chapter{TCFH, Hidden Symmetries and Type IIA Backgrounds} \label{TCFH-General-Theory-Backgrounds}

\section{Introduction}

In this chapter we present the twisted covariant form hierarchy (TCFH) of IIA supergravity and discuss some of the properties of the TCFH connections $\mathcal{D}^\mathcal{F}$, such as their holonomy, on generic as well as on some special supersymmetric backgrounds. As a consequence we demonstrate that the form bilinears of these theories satisfy a conformal Killing-Yano (CKY) equation with respect to $\mathcal{D}^\mathcal{F}$ in agreement with the general result of \cite{gptcfh}. Another purpose of this chapter is to give the Killing-Stäckel (KS) tensors of type IIA branes\footnote{Brane solutions have been instrumental in the understanding of string dualities \cite{hulltown, town}.} \cite{funstring, ns5, callan-string, d5, d5hs, d3, d7, d8a, d8} and to use them to prove the complete integrability of the geodesic flow of those solutions that are spherically symmetric, i.e. those that depend on a harmonic function with one centre. In addition, the Killing-Yano (KY) tensors that square to the KS tensors of these backgrounds will be given and the symmetries of spinning particles propagating on these backgrounds will be explored. Furthermore we shall investigate the conditions required for the TCFH to yield symmetries for particle and string probes propagating in common sector and D-brane backgrounds. Finally we shall compare the results we have obtained from the point of view of KS and KY tensors with those that arise from the TCFHs.

To investigate under which conditions the (Killing spinor) form bilinears generate symmetries for certain probe actions propagating in type IIA supersymmetric backgrounds, we shall match the conditions required for certain probe actions to be invariant under transformations generated by form bilinears with those imposed on them by the TCFHs. For the common sector of type II theories, it is shown that all form bilinears which are covariantly constant with respect to a connection with torsion given by the NS-NS 3-form field strength generate symmetries for string and spinning particle probes propagating on these backgrounds. Common sector backgrounds also admit form bilinears which are not covariantly constant and instead satisfy a general TCFH. These form bilinears may not generate symmetries for probes propagating in common sector backgrounds but nevertheless are part of their geometric structure. In particular the form bilinears of the fundamental string and NS5-brane solutions that are allowed to depend on multi-centre harmonic functions have been computed. It has been found that the type II fundamental string solution admits $2^7$ covariantly constant independent form bilinears while the type II NS5-brane solution admits $2^5$ covariantly constant independent form bilinears. All these forms generate (hidden) symmetries for probe string and spinning particle actions propagating on these backgrounds.

A similar analysis is presented for all type IIA D-branes. In particular, the form bilinears of all D-branes are computed. It is found that the requirement for these to generate symmetries for spinning particle probes propagating on these backgrounds is rather restrictive. This is due to the difficulties of constructing probe actions which exhibit appropriate form couplings. Nevertheless all type IIA D-branes, which may depend on multi-centre harmonic functions, admit form bilinears which generate symmetries for spinning particle probe actions. It turns out that all such form bilinears have components only along the worldvolume directions of the D-branes. A comparison of the symmetries we have found generated by the KS and KY tensors and those generated by the form bilinears in type IIA brane backgrounds will be presented in the conclusions.

This chapter is organised as follows. In section \ref{iiatcfhs} we give the TCFH of IIA supergravity and discuss some of the properties of the TCFH connections. In section \ref{brane-integrability} we present the KS and KY tensors of all IIA branes. In addition, we prove the complete integrability of the geodesic flow in all IIA branes that depend on a harmonic function with one centre by presenting all the independent conserved charges which are in involution. In section \ref{common-sector-tcfh}, we demonstrate that all covariantly constant form bilinears with respect to a connection with skew-symmetric torsion generate symmetries for certain probe string and particle actions propagating on common sector backgrounds. In addition, we explicitly give all the covariantly constant form bilinears for the type II fundamental string and NS5-brane solutions. In section \ref{IIA-Dbranes} we identify the form bilinears that generate symmetries for spinning particle actions propagating on IIA D-brane backgrounds. In appendix \ref{IIA-bilinears} we give all the form bilinears of type II common sector branes and IIA D-branes.

\section{The TCFH of (Massive) IIA Supergravity}\label{iiatcfhs}

The Killing spinor equations (KSEs) of massive IIA supergravity \cite{romans} are given by the vanishing conditions of the supersymmetry variations of the gravitino and dilatino fields evaluated at the locus that all fermions are set to zero. The KSE associated with the gravitino field is a parallel transport equation for the supercovariant connection $\mathcal{D}$. In the string frame, this is given by
\begin{equation} \label{iiasc}
	\begin{split}
		\mathcal{D}_M \defeq \nabla_M &+ \frac{1}{8}H_{MPQ}\Gamma^{PQ}\Gamma_{11} + \frac{1}{8}e^\Phi S \Gamma_M \\
		&+ \frac{1}{16}e^\Phi F_{PQ} \Gamma^{PQ}\Gamma_M\Gamma_{11} + \frac{1}{8\cdot 4!}e^\Phi G_{P_1 \dots P_4} \Gamma^{P_1\dots P_4}\Gamma_M~,
	\end{split}	
\end{equation}
see e.g. \cite{eric}, where $H$ is the NS-NS 3-form field strength, $\Phi$ is the dilaton, and $F$ and $G$ are the 2-form and 4-form R-R field strengths, respectively. In addition, $\nabla$ is the Levi-Civita connection induced on the spinor bundle and $S = e^\Phi m$, where $m$ is a constant which is non-zero in massive IIA and vanishes in the standard IIA supergravity. Furthermore $\Gamma$ denotes the gamma matrices which satisfy the Clifford algebra relation $\Gamma_A\Gamma_B + \Gamma_B \Gamma_A = 2 \eta_{AB}$ and in our conventions $\Gamma_{11} \defeq \Gamma_{012\dots 9}$. In what follows, we shall not make a sharp distinction between spacetime and frame indices but we shall always assume that the indices of gamma matrices are frame indices. It turns out that $\mathcal{D}$ is a connection on the spin bundle over the spacetime associated with the Majorana (real) representation of $\mathfrak{spin} (9,1)$. The (reduced) holonomy of $\mathcal{D}$ for generic backgrounds is $SL(32, \bR)$ \cite{gpdt}, see \cite{hull, duff, gpdtx} for the computation of the holonomy of the supercovariant connection of 11-dimensional supergravity.

The Killing spinors $\epsilon$ satisfy the gravitino KSE, $\mathcal{D}\epsilon = 0$, as well as the dilatino KSE which is an algebraic equation. Backgrounds that admit such Killing spinors are special and both the spacetime metric and fluxes are suitably restricted, see \cite{gpug1, gpug2, gpug3} where the IIA KSEs have been solved for one Killing spinor. The TCFHs are associated with the gravitino KSE which we shall focus on in what follows.

Given $N$ Killing spinors $\epsilon^r$, $r = 1,\dots, N$, one can construct the form bilinears
\begin{equation}\label{fbil}
	\phi^{rs} = \frac{1}{k!}\Dirac{\epsilon^r}{\Gamma_{A_1\dots A_k}\epsilon^s}\, \e^{A_1}\wedge\dots\wedge \e^{A_k}~,
\end{equation}
where $\Dirac{\cdot}{\cdot}$ denotes that Dirac inner product and $\e^A$ is a suitable spacetime frame, $g_{MN}= \eta_{AB} e^A_M e^B_N$. As
\begin{equation}
	\nabla_M \phi^{rs}_{A_1\dots A_k} = \Dirac{\nabla_M \epsilon^r}{\Gamma_{A_1\dots A_k}\epsilon^s} + \Dirac{\epsilon^r}{\Gamma_{A_1\dots A_k} \nabla_M\epsilon^s}~,
\end{equation}
one can use the gravitino KSE, $\mathcal{D}\epsilon = 0$, and \eqref{iiasc} to express the right-hand side of the above equation in terms of the fluxes and form bilinears of the theory. In \cite{gptcfh} has been shown that these equations can be organised as TCFH.

Using the reality condition on $\epsilon$, there are form bilinears which are either symmetric or skew-symmetric in the exchange of spinors $\epsilon^r$ and $\epsilon^s$ in \eqref{fbil}. As a consequence the TCFH of the IIA supergravity factorises in two parts. A basis in form bilinears, up to a Hodge duality\footnote{Our conventions are given in appendix \ref{Conventions}.} operation, which are symmetric in the exchange of the two Killing spinors $\epsilon^r$ and $\epsilon^s$ is
\begin{equation}\label{symiia}
	\begin{gathered}
		\tilde{\sigma}^{rs} = \Dirac{\epsilon^r}{\Gamma_{11}\epsilon^s} , \quad k^{rs} = \Dirac{\epsilon^r}{\Gamma_N\epsilon^s} \, \e^N , \quad \tilde{k}^{rs} = \Dirac{\epsilon^r}{\Gamma_N\Gamma_{11}\epsilon^s} \, \e^N , \\
		\omega^{rs} = \frac{1}{2} \Dirac{\epsilon^r}{\Gamma_{NR} \epsilon^s} \,\e^N \wedge \e^R, \quad
		\tilde{\zeta}^{rs} = \frac{1}{4!} \Dirac{\epsilon^r}{\Gamma_{N_1 \dots N_4}\Gamma_{11} \epsilon^s}\, \e^{N_1} \wedge \dots \wedge \e^{N_4}, \\
		\tau^{rs} = \frac{1}{5!} \Dirac{\epsilon^r}{\Gamma_{N_1 \dots N_5}\epsilon^s}\, \e^{N_1} \wedge \dots \wedge \e^{N_5}~.
	\end{gathered}
\end{equation}
A direct computation reveals that the TCFH is
\begin{equation}\label{iiatcfha}
	\begin{split}
		\mathcal{D}^\mathcal{F}_M\tilde\sigma \defeq&	\nabla_M \tilde{\sigma} \\
		=& -\frac{1}{4} H_{MPQ}\omega^{PQ} + \frac{1}{4} e^\Phi S \tilde{k}_M - \frac{1}{4} e^\Phi F_{MP} k^P + \frac{1}{4 \cdot 5!}\hodge{G}_{MP_1\dots P_5}\tau^{P_1\dots P_5}~,
	\end{split}
\end{equation}
\begin{equation}\label{iiatcfhb}
	\begin{split}
		\mathcal{D}_M^\mathcal{F} k_N \defeq& \nabla_M k_N \\
		=& -\frac{1}{2}H_{MNP}\tilde{k}^P + \frac{1}{4}e^\Phi S \omega_{MN} + \frac{1}{8} e^\Phi F_{PQ}\tilde{\zeta}^{PQ}{}_{MN} +\frac{1}{4}e^\Phi F_{MN}\tilde{\sigma} \\
		&-\frac{1}{4\cdot 4!}e^\Phi \hodge{G}_{MNP_1\dots P_4}\tilde{\zeta}^{P_1 \dots P_4} + \frac{1}{8}e^\Phi G_{MNPQ}\omega^{PQ}~,
	\end{split}
\end{equation}
\begin{equation}\label{iiatcfhc}
	\begin{split}
		\mathcal{D}_M^\mathcal{F} \tilde k_N \defeq& \nabla_M \tilde{k}_N - \frac{1}{2} e^\Phi F_{MP}\omega^P{}_N - \frac{1}{12}e^\Phi G_{MPQR}\tilde{\zeta}^{PQR}{}_N \\
		=& -\frac{1}{2}H_{MNP}k^P + \frac{1}{4}e^\Phi g_{MN}S\tilde{\sigma} +\frac{1}{8}e^\Phi g_{MN} F_{PQ}\omega^{PQ} -\frac{1}{2}e^\Phi F_{[M|P|}\omega^P{}_{N]} \\
		&+ \frac{1}{4\cdot 4!}e^\Phi g_{MN} G_{P_1\dots P_4}\tilde{\zeta}^{P_1 \dots P_4} - \frac{1}{12}e^\Phi G_{[M|PQR|}\tilde{\zeta}^{PQR}{}_{N]}~,
	\end{split}
\end{equation}
\begin{equation}\label{iiatcfhd}
	\begin{split}
		\mathcal{D}_M^\mathcal{F}\omega_{NR} \defeq& \nabla_M \omega_{NR} +\frac{1}{4}H_{MPQ}\tilde{\zeta}^{PQ}{}_{NR}+ e^\Phi F_{M[N}\tilde{k}_{R]} - \frac{1}{12} e^\Phi G_{MP_1P_2P_3}\tau^{P_1P_2P_3}{}_{NR} \\
		=& \frac{1}{2}H_{MNR}\tilde{\sigma}+ \frac{1}{2}e^\Phi S g_{M[N}k_{R]} + \frac{3}{4} e^\Phi F_{[MN}\tilde{k}_{R]} + \frac{1}{2} e^\Phi g_{M[N}F_{R]P}\tilde{k}^P \\
		&-\frac{1}{4 \cdot 5!} e^\Phi \hodge{F}_{MNRP_1\dots P_5}\tau^{P_1\dots P_5} + \frac{1}{2\cdot 4!}e^\Phi g_{M[N} G_{|P_1\dots P_4|}\tau^{P_1 \dots P_4}{}_{R]} \\
		&- \frac{1}{8}e^\Phi G_{[M|P_1 P_2 P_3|}\tau^{P_1 P_2 P_3}{}_{NR]} - \frac{1}{4}e^\Phi G_{MNRP} k^P ~,
	\end{split}
\end{equation}
\begin{equation}\label{iiatcfhe}
	\begin{split}
		\mathcal{D}_M^\mathcal{F} \tilde{\zeta}_{N_1 \dots N_4} \defeq& \nabla_M \tilde{\zeta}_{N_1 \dots N_4} -\frac{1}{3}\hodge{H}_{M[N_1N_2N_3|PQR|}\tilde{\zeta}^{PQR}{}_{N_4]} - 3 H_{M[N_1 N_2}\omega_{N_3 N_4]} \\
		&+ \frac{1}{2} e^\Phi F_{MP}\tau^P{}_{N_1 \dots N_4} + \frac{1}{2}e^\Phi \hodge{G}_{M[N_1N_2|PQR|}\tau^{PQR}{}_{N_3N_4]} \\
		&+ 2e^\Phi G_{M[N_1N_2N_3}\tilde{k}_{N_4]} \\
		=& \frac{1}{12}g_{M[N_1}\hodge{H}_{N_2 N_3 N_4] P_1 \dots P_4}\tilde{\zeta}^{P_1 \dots P_4}- \frac{5}{12} \hodge{H}_{[MN_1N_2N_3|PQR|}\tilde{\zeta}^{PQR}{}_{N_4]} \\
		&+ \frac{1}{4 \cdot 5!}e^\Phi \hodge{S}_{MN_1\dots N_4P_1\dots P_5}\tau^{P_1\dots P_5} -\frac{1}{2} e^\Phi g_{M[N_1}F_{|PQ|}\tau^{PQ}{}_{N_2N_3N_4]} \\
		&+ \frac{5}{8} e^\Phi F_{[M|P|}\tau^P{}_{N_1 \dots N_4]} +\frac{5}{12}e^\Phi \hodge{G}_{[MN_1N_2|PQR|}\tau^{PQR}{}_{N_3N_4]} \\
		&+ 3e^\Phi g_{M[N_1}F_{N_2N_3}k_{N_4]} - \frac{1}{8}e^\Phi g_{M[N_1}\hodge{G}_{N_2 N_3|P_1 \dots P_4|}\tau^{P_1 \dots P_4}{}_{N_4]} \\
		&+ \frac{1}{4}e^\Phi \hodge{G}_{MN_1\dots N_4P}k^P + \frac{5}{4}e^\Phi G_{[MN_1N_2N_3}\tilde{k}_{N_4]}+ e^\Phi g_{M[N_1}G_{N_2N_3N_4]P}\tilde{k}^P ~,
	\end{split}
\end{equation}
\begin{equation}\label{iiatcfhf}
	\begin{split}
		\mathcal{D}_M^\mathcal{F} \tau_{N_1 \dots N_5} \defeq& \nabla_M \tau_{N_1 \dots N_5} - \frac{5}{6} \hodge{H}_{M[N_1N_2N_3|PQR|}\tau^{PQR}{}_{N_4 N_5]} - \frac{5}{2}e^\Phi F_{M[N_1}\tilde{\zeta}_{N_2 \dots N_5]} \\
		&-\frac{5}{2}e^\Phi \hodge{G}_{M[N_1N_2N_3|PQ|}\tilde{\zeta}^{PQ}{}_{N_4N_5]} +5e^\Phi G_{M[N_1N_2N_3}\omega_{N_4N_5]} \\
		=& -\frac{5}{4} \hodge{H}_{[MN_1N_2N_3|PQR|}\tau^{PQR}{}_{N_4 N_5]} + \frac{5}{12} g_{M[N_1}\hodge{H}_{N_2N_3N_4|P_1\dots P_4|}\tau^{P_1\dots P_4}{}_{N_5]} \\
		&-\frac{1}{4\cdot 4!}e^\Phi \hodge{S}_{MN_1\dots N_5 P_1 \dots P_4}\tilde{\zeta}^{P_1\dots P_4} + \frac{1}{8} e^\Phi \hodge{F}_{MN_1\dots N_5}{}^{PQ}\omega_{PQ} \\
		&- 5e^\Phi g_{M[N_1}F_{N_2|P|}\tilde{\zeta}^P{}_{N_3N_4N_5]} - \frac{15}{4} e^\Phi F_{[MN_1}\tilde{\zeta}_{N_2\dots N_5]} \\
		&- \frac{15}{8}e^\Phi \hodge{G}_{[MN_1N_2N_3|PQ|}\tilde{\zeta}^{PQ}{}_{N_4N_5]} - \frac{1}{4}e^\Phi \hodge{G}_{MN_1\dots N_5}\tilde{\sigma} \\
		&- \frac{5}{6 }e^\Phi g_{M[N_1}\hodge{G}_{N_2N_3N_4|PQR|}\tilde{\zeta}^{PQR}{}_{N_5]} + \frac{15}{4} e^\Phi G_{[MN_1N_2N_3}\omega_{N_4N_5]} \\
		&+ 5 e^\Phi g_{M[N_1}G_{N_2N_3N_4|P|}\omega^P{}_{N_5]} ~,
	\end{split}
\end{equation}
where for simplicity we have suppressed the $r, s$ indices on the form bilinears which count the different Killing spinors. The connection $\mathcal{D}^\mathcal{F}$ is the minimal connection of the TCFH, see \ref{sec:TCFH-CKY} for the definition. As it has been explained in the introduction, the above TCFH implies that the form bilinears \eqref{symiia} satisfy a generalisation of the CKY with respect to the connection $\mathcal{D}^\mathcal{F}$. As expected $k$ is Killing, $\nabla_{(M} k_{N)} = 0$.

A basis in the form bilinears, up to a Hodge duality operation, which are skew-symmetric in the exchange of the two Killing spinors is
\begin{equation}\label{iiaskew}
	\begin{gathered}
		\sigma^{rs} = \Dirac{\epsilon^r}{\epsilon^s}~, \quad \tilde\omega^{rs} = \frac{1}{2} \Dirac{\epsilon^r}{\Gamma_{NR}\Gamma_{11}\epsilon^s} \, \e^N \wedge \e^R~, \\
		\pi^{rs} = \frac{1}{3!} \Dirac{\epsilon^r}{\Gamma_{NRS} \epsilon^s} \, \e^N \wedge \e^R \wedge \e^S~, \quad \tilde\pi^{rs} = \frac{1}{3!} \Dirac{\epsilon^r}{\Gamma_{NRS}\Gamma_{11}\epsilon^s} \, \e^N \wedge \e^R \wedge \e^S~, \\
		\zeta^{rs} = \frac{1}{4!} \Dirac{\epsilon^r}{\Gamma_{N_1 \dots N_4} \epsilon^s} \, \e^{N_1} \wedge \dots \wedge \e^{N_4}~.
	\end{gathered}
\end{equation}
The associated TCFH with respect to the minimal connection is
\begin{equation}
	\begin{split}
		\mathcal{D}^\mathcal{F}_M\sigma \defeq& \nabla_M \sigma \\
		=& - \frac{1}{4} H_{MPQ} \tilde \omega^{PQ} - \frac{1}{8} e^\Phi F_{PQ} \tilde\pi^{PQ}{}_M + \frac{1}{4!} e^\Phi G_{MPQR} \pi^{PQR} ~,
	\end{split}
\end{equation}
\begin{equation}\label{iiatcfha1}
	\begin{split}
		\mathcal{D}^\mathcal{F}_M \tilde\omega_{NR} \defeq&	\nabla_M \tilde\omega_{NR} +\frac{1}{4} H_{MPQ} \zeta^{PQ}{}_{NR} + \frac{1}{2}e^\Phi F_{MP} \pi^P{}_{NR} - \frac{1}{2} e^\Phi G_{M[N|PQ|} \tilde\pi^{PQ}{}_{R]} \\
		=& \frac{1}{2} H_{MNR} \sigma + \frac{1}{4} e^\Phi S \tilde\pi_{MNR} -\frac{1}{4}e^\Phi g_{M[N}F_{|PQ|}\pi^{PQ}{}_{R]} + \frac{3}{4} e^\Phi F_{[M|P|} \pi^P{}_{NR]} \\
		&- \frac{1}{4!} e^\Phi \hodge{G}_{MNRP_1P_2P_3}\pi^{P_1P_2P_3} - \frac{1}{12} e^\Phi g_{M[N}G_{R]P_1P_2P_3}\tilde\pi^{P_1P_2P_3} \\
		&- \frac{3}{8}e^\Phi G_{[MN|PQ|} \tilde\pi^{PQ}{}_{R]} ~,
	\end{split}
\end{equation}
\begin{equation}\label{iiatcfha2}
	\begin{split}
		\mathcal{D}^\mathcal{F}_M \pi_{NRS} \defeq& \nabla_M \pi_{NRS} +\frac{3}{2} H_{M[N|P|} \tilde\pi^P{}_{RS]} -\frac{3}{2} e^\Phi F_{M[N} \tilde\omega_{RS]} - \frac{3}{4}e^\Phi G_{M[N|PQ|} \zeta^{PQ}{}_{RS]} \\
		=& \frac{1}{4} e^\Phi S\zeta_{MNRS}- \frac{1}{4 \cdot 4!} e^\Phi \hodge{F}_{MNRSP_1 \dots P_4} \zeta^{P_1\dots P_4} - \frac{3}{2} e^\Phi g_{M[N} F_{R|P|} \tilde \omega^P{}_{S]} \\
		&- \frac{3}{2} e^\Phi F_{[MN} \tilde\omega_{RS]} - \frac{1}{4} e^\Phi G_{MNRS} \sigma + \frac{1}{8} e^\Phi \hodge{G}_{MNRSPQ} \tilde\omega^{PQ} \\
		&- \frac{1}{4} e^\Phi g_{M[N} G_{R|P_1P_2P_3|}\zeta^{P_1P_2P_3}{}_{S]} - \frac{3}{4}e^\Phi G_{[MN|PQ|}\zeta^{PQ}{}_{RS]} ~,
	\end{split}
\end{equation}
\begin{equation}\label{iiatcfha3}
	\begin{split}
		\mathcal{D}^\mathcal{F}_M \tilde\pi_{NRS} \defeq& \nabla_M \tilde\pi_{NRS} + \frac{3}{2} H_{M[N|P|} \pi^P{}_{RS]} - \frac{1}{2} e^\Phi F_{MP} \zeta^P{}_{NRS} \\
		&+\frac{3}{2} e^\Phi G_{M[NR|P|} \tilde\omega^P{}_{S]} - \frac{1}{4}e^\Phi \hodge{G}_{M[NR|P_1P_2P_3|}\zeta^{P_1P_2P_3}{}_{S]} \\
		=& + \frac{3}{4} e^\Phi S g_{M[N} \tilde\omega_{RS]} + \frac{1}{2} e^\Phi F_{MP} \zeta^P{}_{NRS} + \frac{3}{8} e^\Phi g_{M[N} F_{|PQ|} \zeta^{PQ}{}_{RS]} \\
		&- e^\Phi F_{[M|P|} \zeta^P{}_{NRS]} - \frac{3}{4}e^\Phi g_{M[N}F_{RS]} \sigma - \frac{3}{8} e^\Phi g_{M[N}G_{RS]PQ} \tilde \omega^{PQ} \\
		&+ e^\Phi G_{[MNR|P|} \tilde\omega^P{}_{S]} + \frac{1}{32} e^\Phi g_{M[N} \hodge{G}_{RS]P_1 \dots P_4}\zeta^{P_1\dots P_4} \\
		&- \frac{1}{6} e^\Phi \hodge{G}_{[MNR|P_1P_2P_3|}\zeta^{P_1P_2P_3}{}_{S]} ~,
	\end{split}
\end{equation}
\begin{equation}\label{iiatcfha4}
	\begin{split}
		\mathcal{D}^\mathcal{F}_M \zeta_{N_1 \dots N_4} \defeq& \nabla_M \zeta_{N_1 \dots N_4} - \frac{1}{3} \hodge{H}_{M[N_1N_2N_3|PQR|}\zeta^{PQR}{}_{N_4]} - 3 H_{M[N_1N_2} \tilde\omega_{N_3N_4]} \\
		&+ 2 e^\Phi F_{M[N_1} \tilde\pi_{N_2 N_3 N_4]} + 3 e^\Phi G_{M[N_1N_2|P|}\pi^P{}_{N_3N_4]} \\
		&+ e^\Phi \hodge{G}_{M[N_1N_2N_3|PQ|}\tilde\pi^{PQ}{}_{N_4]} \\
		=& \frac{1}{12} g_{M[N_1} \hodge{H}_{N_2 N_3 N_4] P_1 \dots P_4} \zeta^{P_1\dots P_4} - \frac{5}{12} \hodge{H}_{[MN_1N_2N_3|PQR|}\zeta^{PQR}{}_{N_4]} \\
		&+ e^\Phi S g_{M[N_1} \pi_{N_2N_3N_4]} + \frac{1}{4!} e^\Phi \hodge{F}_{MN_1 \dots N_4 PQR} \pi^{PQR} \\
		&+ 3 e^\Phi g_{M[N_1} F_{N_2|P|}\tilde\pi^P{}_{N_3N_4]} + \frac{5}{2} e^\Phi F_{[MN_1}\tilde\pi_{N_2N_3N_4]} \\
		&+ \frac{1}{6} e^\Phi g_{M[N_1} \hodge{G}_{N_2N_3N_4]PQR}\tilde\pi^{PQR} + \frac{5}{8} e^\Phi \hodge{G}_{[MN_1N_2N_3|PQ|}\tilde\pi^{PQ}{}_{N_4]} \\
		&- \frac{3}{2} e^\Phi g_{M[N_1} G_{N_2N_3|PQ|}\pi^{PQ}{}_{N_4]} + \frac{5}{2} e^\Phi G_{[MN_1N_2|P|} \pi^P{}_{N_3N_4]} ~.
	\end{split}
\end{equation}
As in the previous case, a consequence of the TCFH above is that the forms \eqref{iiaskew} satisfy a generalisation of the CKY equation with respect to the connection $\mathcal{D}^\mathcal{F}$. Later we shall demonstrate that in some cases the forms \eqref{symiia} and \eqref{iiaskew} generate symmetries in string and particle actions probing some IIA backgrounds.

The factorisation of the domain that the minimal TCFH connection $\mathcal{D}^\mathcal{F}$ acts as in \eqref{symiia} and \eqref{iiaskew} can be understood as follows. The product of two Majorana representations $\Delta_{32}$ in terms of forms is $\otimes^2 \Delta_{32} = \Lambda^*(\bR^{9,1})$. Therefore the form bilinears of all spinor span all spacetime forms. Therefore generically the TCFH connection acts on the space of all spacetime forms. However we have seen that the TCFH connection preserves the forms which are symmetric (skew-symmetric) in the exchange of the two Killing spinors, i.e. it preserves that symmetrised $S^2\,(\Delta_{32})$ and skew-symmetrised $\Lambda^2\,(\Delta_{32})$ subspaces of the product. As $\mathrm{dim}\, S^2 ( \Delta_{32}) = 528$ and $\mathrm{dim}\, \Lambda^2 ( \Delta_{32}) = 496$, the (reduced) holonomy of $\mathcal{D}^\mathcal{F}$ is included in $GL(528)\times GL(496)$. In fact the holonomy\footnote{Note though that the (reduced) holonomy of the maximal TCFH connection, see \cite{gptcfh} for the definition, is included in $GL(528)\times GL(496)$.} of the minimal connection $\mathcal{D}^\mathcal{F}$ reduces further to $SO(9,1)\times GL(517)\times GL(495)$ as it acts trivially on the scalars $\sigma$ and $\tilde\sigma$ and does not mix $k$ with the other from bilinears. Of course the holonomy of $\mathcal{D}^\mathcal{F}$ reduces even further for special backgrounds.

\section{Particles and Integrability of Type IIA Branes} \label{brane-integrability}

Before we proceed to investigate the symmetries of particle and string probes generated by the TCFHs of type IIA supergravity, we shall construct the KS and KY tensors of type IIA brane solutions which to our knowledge have not presented before. We shall use these to argue that the geodesic flow of some of these solutions is completely integrable and we shall give the associated independent conserved charges in involution.

\subsection{Integrability and Separability}

Returning to the particle system described by the action \eqref{Free-Particle-Action}, the conserved charges \eqref{Hidden-Symmetry-Charge} can be written as functions on phase space, $T^*M$, as
\begin{equation}\label{kscharge}
	Q(d) = d^{N_1\cdots N_k} p_{N_1} \dots p_{N_k}~,
\end{equation}
where $p_M$ is the conjugate momentum of $x^M$. It turns out that if $Q(d)$ and $Q(e)$ are conserved charges associated with KS tensors $d$ and $e$, then $\{Q(d), Q(e)\}_{\mathrm{PB}}$ is associated with the KS tensor given in terms of the Nijenhuis-Schouten bracket
\begin{equation}
	([d, e]_{\mathrm{NS}})^{N_1\cdots N_{k + \ell - 1}} = k d^{M(N_1\cdots N_{k - 1}} \partial_{M} e^{N_k\cdots N_{k + \ell - 1})} - \ell e^{M(N_1\cdots N_{\ell - 1}} \partial_{M} d^{N_k\cdots N_{k + \ell - 1})}~,
\end{equation}
of $d$ and $e$. Therefore, one has
\begin{equation}\label{nsb}
	\{Q(d), Q(e)\}_{\mathrm{PB}} = Q(-[d, e]_{\mathrm{NS}})~.
\end{equation}
Observe that if $d$ is a vector, then $[d, e]_{\mathrm{NS}} = \mathcal{L}_d e$, i.e. the Nijenhuis-Schouten bracket is the Lie derivative of $e$ with respect to the vector field $d$. So two charges are in involution provided that the Nijenhuis-Schouten bracket of the associated KS tensors vanishes.

Completely integrable systems are special. There are difficulties in both finding conserved charges in involution and in proving that they are independent. For example if $Q(d)$ and $Q(e)$ are conserved charges, $Q(d) Q(e)$ is not an independent conserved charge, as its inclusion in the map $Q: P\rightarrow \bR^n$ does not alter its rank. However for the geodesic flow described by the action \eqref{Free-Particle-Action} that we shall investigate below, there is a simplifying feature. The spacetimes we shall be considering admit a non-abelian group of isometries. For every isometry generated by a Killing vector field $K_r$, there is an associated conserved charge
\begin{equation}
	Q_r = K_r^M p_M~.
\end{equation}
Of course these charges may not be in involution. However note that the charges $Q_r$ written in phase space do not depend on the spacetime metric. They only depend on the way that the isometry group acts on the spacetime. Typically there are many metrics for which $Q_r$ are constants of motion for the action \eqref{Free-Particle-Action}. Of course any polynomial of $Q_r$ is also conserved and is independent from the metric of the particle system. We shall refer to these charges as {\it orbital} to emphasise their independence from the spacetime metric. In many occasions, it is possible to find polynomials of $Q_r$ which are independent and are in involution. Suppose that one can find $n - 1$ such independent (polynomial) orbital charges in involution and the Hamiltonian,
\begin{equation}
	H = \frac{1}{2} g^{MN} p_M p_N~,
\end{equation}
is independent from the orbital charges. Then the geodesic flow is completely integrable because the orbital charges will Poisson commute with the Hamiltonian. Of course the Hamiltonian depends on the spacetime metric. To distinguish the conserved charges which depend on the spacetime metric from the orbital ones we shall refer to former as {\it Hamiltonian}. We shall demonstrate that this strategy of proving complete integrability of a geodesic flow based on non-abelian isometries is particularly effective whenever the non-abelian group of isometries has a principal orbit in a spacetime of codimension of at most one. The complete integrability of geodesic flows on homogeneous manifolds has been extensively investigated in the mathematics literature, see e.g. \cite{thimm}.

\subsubsection{An example}

Before we proceed to investigate the KS and KY tensors and the integrability of the geodesic flow on some type IIA backgrounds, let us present an example. The standard example is that of the Kerr black hole. However more suitable for the examples that follow is to consider $\bR^{2n}$ with a conformally flat metric
\begin{equation}\label{cflmetr}
	g = h(|y|) \delta_{ij} \dd y^i \dd y^j~,
\end{equation}
where $|y|$ is the length of the coordinate $y$ with respect to the Euclidean norm and $h>0$.

A direct computation reveals that the following tensors
\begin{equation}
	d_{i_1\dots i_k} = h^{k}(|y|) ~ y^{j_1}\dots y^{j_q} a_{j_1\dots j_q, i_1\dots i_k}~,
\end{equation}
are KS tensors provided that the coefficients $a$ are constant and satisfy
\begin{equation}
	a_{(j_1\dots j_q, i_1)\dots i_k} = a_{j_1\dots (j_q, i_1\dots i_k)} = 0~.
\end{equation}
For each of these KS tensors, there is an associated conserved charge $Q(d)$ given in \eqref{kscharge} of the geodesic flow on $\bR^{2n}$ with metric \eqref{cflmetr}. These generate an infinite dimensional symmetry algebra for the action \eqref{Free-Particle-Action} with metric \eqref{cflmetr} which is isomorphic to the Poisson algebra of $Q(d)$'s up to terms proportional to the equations of motion, i.e. the algebra of symmetry transformations is isomorphic on-shell to the Poisson bracket algebra of the charges. The conserved charges $Q(d)$ may neither be independent nor in involution.

Next let us turn to find the KY and CCKY tensors on $\bR^{2n}$ with metric \eqref{cflmetr}. After some computation, one finds that
\begin{equation}
	\alpha = h^{\frac{k}{2}} i_Y \varphi~, \quad \beta= h^{\frac{k + 2}{2}} Y\wedge \varphi~,
\end{equation}
are KY and CCKY forms, respectively, for any constant $k$-form $\varphi$ on $\bR^{2n}$, where $Y$ is either the vector field $Y = y^i\partial_i$ or the one-form $Y = y_i \dd y^i$; it is clear from the context what $Y$ denotes in each case.

For each KY tensor above, one can construct the infinitesimal variation \eqref{svar} which is a symmetry of the action \eqref{SUSY-Free-Particle-Action}. However the commutator of two such infinitesimal transformations does not close to an infinitesimal transformation of the same type. Typically, the right-hand side of the commutator will involve a term polynomial in $Dx$ as well as a term which is linear in the velocity $\dot x$. A systematic exploration of such commutators in a related context can be found in \cite{phgp1, phgp2}.

Next let us turn to investigate the integrability of the geodesic flow of the metric \eqref{cflmetr}. The geodesic equations can be easily integrated in angular coordinates. However it is instructive to provide a symmetry argument for the complete integrability of the geodesic equations.
The isometry group of the above backgrounds is $SO(2n)$. The Killing vector fields are
\begin{equation}\label{rotvf}
	k_{ij} = y_i \partial_j - y_j \partial_i~, \quad i<j~,
\end{equation}
where $y_i = y^i$. The associated conserved charges are
\begin{equation}\label{rotcc}
	Q_{ij} = Q(k_{ij}) = y_i p_j - y_j p_i~.
\end{equation}
Notice that all these conserved charges are orbital as they do not depend on the metric \eqref{cflmetr}. As $\mathcal{L}_{k_{ij}} g = 0$, one can show that $Q_{ij}$ commutes with the Hamiltonian $H= \frac{1}{2} h^{-1} \delta^{ij} p_i p_j$, i.e. $\{H, Q_{ij}\}_{\mathrm{PB}} = 0$.

The conserved charges $Q_{ij}$ are not in involution as $\{Q(k_{ij}), Q({k_{pq}})\}_{\mathrm{PB}} = Q([k_{ij}, k_{pq}])$. However using these, one can verify that the $2n - 1$ orbital conserved charges
\begin{equation}\label{casichso2n}
	D_m = \frac{1}{4} \sum_{i,j \geq 2n + 1 - m} (Q_{ij})^2~, \quad m= 2,3, \dots, 2n~,
\end{equation}
are in involution. These together with the Hamiltonian $H = \frac{1}{2} h^{-1} \delta^{ij} p_i p_j$ give $2n$ charges in involution. Therefore the geodesic flow of the metric \eqref{cflmetr} is completely (Liouville) integrable.

An alternative way to think about the complete integrability of the geodesic flow on $\bR^{2n}$ with metric \eqref{cflmetr} is to consider it as a motion along the round sphere $S^{2n - 1}$ in $\bR^{2n}$ and as a motion along the radial direction $r = |y|$. For this write the metric \eqref{cflmetr} as
\begin{equation}
	g = h(r) (\dd r^2+ r^2 g(S^{2n - 1}))~,
\end{equation}
where $g(S^{2n - 1})$ is the metric on the round $S^{2n - 1}$ sphere. It is well known that the vector fields \eqref{rotvf} are tangential to $S^{2n - 1}$ and leave the round metric on $S^{2n - 1}$ invariant. The associated conserved charges are as in \eqref{rotcc} and they are functions of $T^*S^{2n - 1}$, i.e. they do not depend on the radial direction $p_r$ of the momentum $p$. One can proceed to define \eqref{casichso2n} and in turn show that the geodesic flow on $S^{2n - 1}$ is completely integrable. Notice that $D_{2n}$ is the Hamiltonian of the geodesic flow on $S^{2n - 1}$. All these charges including the Hamiltonian on $S^{2n - 1}$ are orbital as they do not depend on the metric \eqref{cflmetr}.
As there are $2n - 1$ independent charges in involution associated with the geodesic flow on $S^{2n - 1}$, the addition of the Hamiltonian $H= \frac{1}{2} h^{-1} \delta^{ij} p_i p_j$ of the geodesic flow on $\bR^{2n}$ gives $2n$ independent conserved charges in involution proving the complete integrability of the geodesic flow of the metric \eqref{cflmetr}.

This construction can be reversed engineered and generalised. In particular consider a metric on a $n$-dimensional manifold $M^n$
\begin{equation}
	g(M^n) = \dd z^2 + g(N^{n - 1})(z)~,
\end{equation}
where $z$ is a coordinate and $g(N^{n - 1})(z)$ is a metric on the submanifold $N^{n - 1}$ of $M^n$ which may depend on $z$. Suppose now there is a group of isometries on $M^n$ which has as a principal orbit $N^{n - 1}$. Clearly the associated conserved charges $Q = K^M p_M$, for each Killing vector field $K$, will be functions on $T^*N$. If one is able to find orbital conserved charges $D_m$, $m = 1,\dots, n - 1$ in involution, then the geodesic flow on $M^n$ will be completely integrable after the inclusion of the Hamiltonian $H$ of the geodesic flow on $M^n$ as an additional conserved charge. This is because $H$ is a function on $T^*M^n$ and so it is independent from $D_m$ which are functions on $T^*N^{n - 1}$. Moreover $\{D_m, H\}_{\mathrm{PB}} = 0$ as $D_m$ are constructed as polynomials of the conserved charges associated with the isometries on $M^n$. This argument will be repeatedly used to prove complete integrability of geodesic flows of brane backgrounds and clearly can be adapted to all manifolds which have a principal orbit of codimension at most one with respect to a group action.

\subsection{D-branes}

\subsubsection{The KS and CCKY Tensors of D-branes}

The metric of type II Dp-branes in the string frame \cite{d5, d5hs, d3, d7, d8a, d8} is
\begin{equation}\label{dbrane}
	g = h^{-\frac{1}{2}} \sum_{a,b = 0}^p \eta_{ab} \dd \sigma^a \dd \sigma^b+ h^{\frac{1}{2}} \sum_{i,j = 1}^{9 - p} \delta_{ij} \dd y^i \dd y^j~,
\end{equation}
where $p = 0,\dots, 8$ with $p$ even for IIA D-branes, $\sigma^a$ are the worldvolume coordinates, $y^i$ are the transverse coordinates and $h = h(y)$ is a harmonic function $\delta^{ij} \partial_i\partial_j h = 0$. Apart from the metric, the solutions depend on a non-vanishing dilaton field and an appropriate form field strength which we suppress. For planar branes located at different points $y_s$ in $\bR^{9 - p}$, one takes for $p\leq 6$
\begin{equation}\label{mhf}
	h = 1 + \sum_{s} \frac{q_s}{|y - y_s|^{7 - p}}~,
\end{equation}
where $|\cdot|$ is the Euclidean norm in $\bR^{9 - p}$ and $q_s$ is a constant proportional to the charge density of the branes. The solution is invariant under the action of the Poincar\'e group, $SO(p,1)\ltimes \bR^{p,1}$, acting on the worldvolume coordinates $\sigma^a$. If the harmonic function is chosen such that $h = h(|y|)$\footnote{The harmonic function is $h = 1 + \frac{q}{|y|^{7 - p}}$ for $p = 0, \dots, 6$, $h = 1+ q \log |y|$ for $ = 7$ and $h = 1+ q |y|$ for $p = 8$.}, then the solution will be invariant under the action of $SO(9 - p)$ group acting on the transverse coordinates $y$.

Considering the Dp-branes \eqref{dbrane} with $h = h(|y|)$, the KS tensors which are invariant under the worldvolume symmetry of the solution are
\begin{equation}
	d_{a_1 \dots a_{2m} i_1 \dots i_k} = h^{\frac{1}{4}(k - m)}(|y|)\,\, y^{j_1} \dots y^{j_q} a_{j_1 \dots j_q, i_1\dots i_k} \eta_{(a_1 a_2} \dots \eta_{a_{2m - 1} a_{2m})}~,
\end{equation}
provided that the constant coefficients $a$ satisfy
\begin{equation}
	a_{(j_1 \dots j_q, i_1) \dots i_k} = a_{j_1 \dots (j_q, i_1 \dots i_k)} = 0~.
\end{equation}
Each of these KS tensors will generate a symmetry of the relativistic particle action \eqref{Free-Particle-Action}. As a result each such action on a D-brane background admits an infinite number of symmetries. The algebra of the associated transformations is on-shell isomorphic to that of the Poisson bracket algebra of the associated charges.

To investigate the symmetries of the spinning particles \eqref{SUSY-Free-Particle-Action} propagating on D-branes, it suffices to find the KY tensors of these backgrounds. For this, one begins with an ansatz which respects the worldvolume isometries of the solutions. As the KY tensors are dual to CCKY ones, let us focus on the latter. It turns out that
\begin{equation}
	\beta(\varphi) = h^{\frac{k + 1 - p}{4}}(|y|)\,\, Y \wedge \varphi \wedge \dvol(\bR^{p,1})~,
\end{equation}
is a CCKY tensor for any constant $k$-form $\varphi$ on $\bR^{8 - p}$, where $\dvol(\bR^{p,1})$ is the volume form of $\bR^{p,1}$ with respect to the flat metric and $Y = \delta_{ij} y^i \dd y^j$. Therefore Dp-branes admit $2^{8 - p}$ linearly independent KY forms each generating a symmetry of the action \eqref{SUSY-Free-Particle-Action} of spinning particle probes in these backgrounds. The associated conserved charges are given in \eqref{ccalpha}.

\subsubsection{Complete Integrability of Geodesic Flow}

The geodesic flow on all Dp-brane backgrounds with $h = h(|y|)$ is completely integrable. Of course one can separate the geodesic equation in angular variables. Here we shall give all the charges which are in involution. As we have already mentioned, the isometry group of such a Dp-brane solution is $SO(p,1)\ltimes \bR^{p,1}\times SO(9 - p)$. Such a group has a codimension one principal orbit $\bR^{p,1}\times S^{8 - p}$ in the Dp-brane background. In particular, the Killing vectors generated by the translations along the worldvolume coordinates are $k_a = \partial_a$ and those generated by $SO(9 - p)$ rotations on the transverse coordinates are
\begin{equation}\label{rotvfx}
	k_{ij} = y_i \partial_j - y_j\partial_i~, \quad i<j~,
\end{equation}
where $y_i = y^i$. The associated conserved charges written in terms of the momenta are
\begin{equation}
	Q_a = p_a~, ~~~ Q_{ij} = Q(k_{ij}) = y_i p_j - y_j p_i~.
\end{equation}
These charges are not in involution. However, one can verify that the 9 conserved charges
\begin{equation}
	Q_a~, ~~~D_m = \frac{1}{4} \sum_{i,j\geq 10 - p - m} (Q_{ij})^2~, ~~~m = 2,3, \dots, 9 - p~,
\end{equation}
are all orbital, independent and in involution. These together with the Hamiltonian of \eqref{Free-Particle-Action} yield 10 charges in involution and the geodesic flow on all such Dp-brane solutions is completely integrable.

\subsection{Common Sector Branes}

\subsubsection{KS and KY Tensors of Common Sector Branes}

The metric of the fundamental string solution \cite{funstring} is
\begin{equation}\label{fstring}
	g = h^{-1} \eta_{ab} \dd \sigma^a \dd \sigma^b+ \delta_{ij} \dd y^i \dd y^i~,
\end{equation}
where $a,b = 0,1$ and $i,j = 1,\dots 8$ and $h$ is a harmonic function on $\bR^8$, $\delta^{ij} \partial_i\partial_j h = 0$. We have suppressed the other two fields of the solution the dilaton and 3-form field strength.

As for D-branes consider the fundamental string solution with $h = h(|y|) = 1 + \frac{q}{|y|^6}$. Such a solution admits the same isometry group as that of D1-brane. Then one can demonstrate that the KS tensors that preserve the worldvolume symmetry of the fundamental string are
\begin{equation}
	d_{a_1 \dots a_{2m}i_1 \dots i_k} = h^{-m}(|y|) y^{j_1} \dots y^{j_q} a_{j_1\dots j_q, i_1\dots i_k} \eta_{(a_1a_2} \dots \eta_{a_{2m - 1} a_{2m})}~,
\end{equation}
provided that the constant coefficients satisfy $a_{(j_1\dots j_q, i_1)\dots i_k} = a_{j_1\dots (j_q, i_1\dots i_k)} = 0$. As a result a relativistic particle whose dynamics is described by the action \eqref{Free-Particle-Action} on such a background admits an infinite number of symmetries generated by these KS tensors.

After some computation, one can verify that CCKY forms of the fundamental string solution are
\begin{equation}
	\beta(\varphi) = h^{-1}(|y|)\, Y \wedge \varphi \wedge \dd \sigma^0\wedge \dd \sigma^1~,
\end{equation}
for any constant $k$-form $\varphi$ on $\bR^8$, where $Y = \delta_{ij} y^i \dd y^j$. These give rise to $2^7$ linearly independent dual KY forms which generate symmetries for a spinning particle with action \eqref{SUSY-Free-Particle-Action} propagating on this background.

The metric of the NS5-brane solution \cite{ns5, callan-string} is
\begin{equation}\label{ns5}
	g = \eta_{ab} \dd \sigma^a \dd \sigma^b + h \delta_{ij} \dd y^i \dd y^j~,
\end{equation}
where $a,b = 0,\dots, 5$, $i,j = 1,2,3,4$ and $h$ is a harmonic function on $\bR^4$. We have again suppressed the dilaton and 3-form fields of the solution. For $h = h(|y|) = 1 + \frac{q}{|y|^2}$, the solution has the same isometry group as that of the D5-brane.

As for the fundamental string solution above, the KS tensors that preserve the worldvolume symmetry of the NS5-brane are
\begin{equation}
	d_{a_1\dots a_{2m}i_1\dots i_k} = h^{k}(|y|)\,\, y^{j_1}\dots y^{j_q} a_{j_1\dots j_q, i_1\dots i_k} \eta_{(a_1a_2}\dots \eta_{a_{2m - 1} a_{2m})}~,
\end{equation}
provided that the constant tensors $a$ satisfy $a_{(j_1\dots j_q, i_1)\dots i_k} = a_{j_1\dots (j_q, i_1\dots i_k)} = 0$. Therefore the action \eqref{Free-Particle-Action} of a relativistic particle action propagating in this background admits an infinite number of symmetries generated by these KS tensors.

The CCKY forms of the NS5-brane are
\begin{equation}
	\beta(\varphi) = h^{\frac{k + 2}{2}}(|y|)\,\, Y\wedge \varphi\wedge \dvol(\bR^{5,1})~,
\end{equation}
for any constant $k$-form $\varphi$ on $\bR^4$, where $Y = \delta_{ij} y^i \dd y^j$ and $\dvol(\bR^{5,1})$ is the volume form of the worldvolume of the NS5-branes with respect to the flat metric. These give rise to $2^3$ linearly independent dual KY forms that generated the symmetries of a spinning particle with action \eqref{SUSY-Free-Particle-Action} propagating on the background.

\subsubsection{Complete Integrability of Geodesic Flow}

Consider a relativistic particle propagating on the fundamental string solution with $h = h(|y|)$. The worldsheet translations and transverse coordinate $SO(8)$ rotations give rise to the conserved charges
\begin{equation}
	Q_a = p_a~, \quad a = 0,1~; \qquad Q_{ij}= y_i p_j - y_j p_i~, \quad i,j = 1, \dots,8~,
\end{equation}
respectively. From these one can construct the following nine independent, orbital, conserved charges
\begin{equation}
	Q_a~, ~~~D_m = \frac{1}{4} \sum_{i,j\geq 9 - m} (Q_{ij})^2~, \quad m = 2, \dots, 8~,
\end{equation}
which are independent and in involution. These together with the Hamiltonian of the relativistic particle \eqref{Free-Particle-Action} lead to the integrability of the geodesic flow on the fundamental string background.

Similarly, the conserved charges of a relativistic particle propagating on a NS5-brane background associated with the worldvolume translations and transverse $SO(4)$ rotations are
\begin{equation}
	Q_a= p_a~, ~~~a = 0, \dots, 5~; \quad Q_{ij}= y_i p_j - y_j p_i~, \quad i,j = 1,2,3,4~.
\end{equation}
These give rise to nine independent, orbital, conserved charges
\begin{equation}
	Q_a~, ~~~D_m = \frac{1}{4} \sum_{i,j\geq 5 - m} (Q_{ij})^2~, ~~~m = 2,\dots, 4~,
\end{equation}
which are independent and in involution. These together with the Hamiltonian of the relativistic particle imply the complete integrability of the geodesic flow of NS5-brane.

\section{Common Sector and TCFHs} \label{common-sector-tcfh}

The simplest sector to explore the TCFH of type IIA supergravity is the common sector. For this sector, all fields vanish apart from the metric, dilaton and the NS-NS 3-form field strength $H$, $\dd H = 0$. A direct inspection of the TCFH of type IIA supergravity reveals that some of the spinor bilinears are covariantly constant with respect to a connection with skew-symmetric torsion while some others satisfy a more general TCFH. The former are well known, especially in the context of string compactifications, and have been extensively investigated in the sigma model approach to string theory. They generate additional supersymmetries of the worldvolume actions as well as W-type of symmetries \cite{phgp1, phgp2}. Here we shall demonstrate that string probes on all common sector supersymmetric solutions admit W-type of symmetries generated by the form bilinears.

\subsection{IIA Common Sector}

\subsubsection{The TCFH}

The TCFH of the common sector can be written as
\begin{equation}\label{iiacovf}
	\nabla_M \phi_{N_1 \dots N_p} -\frac{p}{2} H^P{}_{M[N_1} \tilde\phi_{|P|\dots N_p]} = 0~,~~\nabla_M \tilde\phi_{N_1 \dots N_p} -\frac{p}{2} H^P{}_{M[N_1} \phi_{|P|\dots N_p]} = 0~,
\end{equation}
for $\phi= k, \pi, \tau$ and
\begin{equation}\label{iaom}
\nabla_M \tilde{\sigma} = -\frac{1}{4}H_{MPQ}\omega^{PQ} ~, \quad 		\nabla_M \omega_{NR} +\frac{1}{4}H_{MPQ}\tilde{\zeta}^{PQ}{}_{NR} = \frac{1}{2}H_{MNR}\tilde{\sigma} ~ ,
\end{equation}
\begin{equation}
	\begin{split}
		&\nabla_M \tilde{\zeta}_{N_1 \dots N_4} -\frac{1}{3} \hodge{H}_{M[N_1N_2N_3|PQR|}\tilde{\zeta}^{PQR}{}_{N_4]} - 3 H_{M[N_1 N_2}\,\omega_{N_3 N_4]} = \\
		&\quad\frac{1}{12}g_{M[N_1}\hodge{H}_{N_2 N_3 N_4] P_1 \dots P_4}\tilde{\zeta}^{P_1 \dots P_4} - \frac{5}{12} \hodge{H}_{[MN_1N_2N_3|PQR|}\tilde{\zeta}^{PQR}{}_{N_4]}~,
	\end{split}
\end{equation}
\begin{equation}\label{iiatom}
	\nabla_M \sigma = - \frac{1}{4} H_{MPQ} \tilde \omega^{PQ} ~ , ~~~\nabla_M \tilde\omega_{NR} + \frac{1}{4} H_{MPQ} \zeta^{PQ}{}_{NR} = \frac{1}{2} H_{MNR} \sigma ~ ,
\end{equation}
\begin{equation}\label{iiiatom}
	\begin{split}
		&\nabla_M \zeta_{N_1 \dots N_4} - \frac{1}{3} \hodge{H}_{M[N_1N_2N_3|PQR|}\zeta^{PQR}{}_{N_4]}- 3 H_{M[N_1N_2} \tilde\omega_{N_3N_4]}= \\
		&\qquad \frac{1}{12} g_{M[N_1} \hodge{H}_{N_2 N_3 N_4] P_1 \dots P_4} \zeta^{P_1\dots P_4} - \frac{5}{12} \hodge{H}_{[MN_1N_2N_3|PQR|}\zeta^{PQR}{}_{N_4]} ~.
	\end{split}
\end{equation}
These can be easily derived from the general IIA TCFH in section \ref{iiatcfhs} upon setting all other fields apart from the metric, dilaton and NS-NS 3-form to zero.

It is clear from the TCFH that $k^{\pm rs} = k^{rs}\pm \tilde k^{rs}$, $\pi^{\pm rs} = \pi^{rs}\pm \tilde \pi^{rs}$ and $\tau^{\pm rs}= \tau^{rs}\pm \tilde \tau^{rs}$ are covariantly constant
\begin{equation}\label{ccf}
	\nabla^{(\pm)}k^{\pm rs} = \nabla^{(\pm)} \pi^{\pm rs} = \nabla^{(\pm)} \tau^{\pm rs} = 0~,
\end{equation}
 with respect to the connections
\begin{equation}\label{nablapm}
	\nabla^{(\pm)} = \nabla\pm \frac{1}{2} H~.
\end{equation}
These are the forms that have mostly been explored in the literature. Although the rest do not satisfy such a straightforward condition they are nevertheless part of the geometric structure of the common sector backgrounds. A consequence of the TCFH above is that the (reduced) holonomy of the connection\footnote{Note that the TCFH connection as stated above is not the minimal on $k$ and $\tilde k$.} of a generic common sector background is included in $SO(9,1)\times SO(9,1)\times GL(255)\times GL(255)$. The subgroup $SO(9,1)\times SO(9,1)$ is the holonomy of the connections $\nabla^{(\pm)}$ as expected for the common sector. Here in addition we have demonstrated that the holonomy of the TCFH connection factorizes because of the way that it acts on the 2- and 4-form bilinears yielding the $GL(255)\times GL(255)$ subgroup.

\subsubsection{Probe Hidden Symmetries Generated by the TCFH}

After identifying the 3-form coupling $C = \dd b$ of the probe actions \eqref{pact} and \eqref{sact} with the 3-form field strength $H$ of common sector backgrounds, $C = H$, the conditions on the form bilinears $k^{\pm rs}$, $\pi^{\pm rs}$ and $\tau^{\pm rs}$ imposed by the TCFH \eqref{ccf} coincide with those in \eqref{conp} and \eqref{conm} as required for the invariance of these probe actions. Therefore the $\nabla^{(\pm)}$-covariantly constant form bilinears $k^{\pm rs}$, $\pi^{\pm rs}$ and $\tau^{\pm rs}$ generate symmetries for the particle \eqref{pact} and string \eqref{sact} probe actions. These are given by the infinitesimal transformations
\begin{equation}\label{iiainf}
	\begin{split}
		\delta x^M =& \epsilon^{(\pm)}_{ rs}(k^{\pm rs})^M~, \quad \delta x^M= \epsilon^{(\pm)}_{ rs} (\pi^{\pm rs})^M{}_{PQ} D_\pm x^P D_\pm x^Q~, \\
		\delta x^M =& \epsilon^{(\pm)}_{ rs} (\tau^{\pm rs})^M{}_{N_1\dots N_4} D_\pm x^{N_1}\dots D_\pm x^{N_4}~,
	\end{split}
\end{equation}
where $\epsilon^{(\pm)}_{ rs}$ are the infinitesimal parameters.

Similarly after identifying $C$ with $H$ the spinning particle probes described by the action \eqref{1part} are invariant under symmetries generated by the $\nabla^{(+)}$-covariantly constant forms $k^{+ rs}$, $\pi^{+ rs}$ and $\tau^{+ rs}$. The infinitesimal variations are given as in \eqref{iiainf} after replacing the worldsheet superfields with the worldline ones and the superspace derivative $D_+$ with $D$. The $\nabla^{(-)}$-covariantly constant forms $k^{- rs}$, $\pi^{- rs}$ and $\tau^{- rs}$ also generate symmetries for the spinning particle probe with action given in \eqref{1part} but now with the coupling $C$ identified with $-H$, $C = -H$.

The interpretation of the rest of the form bilinears satisfying the TCFH conditions \eqref{iaom}-\eqref{iiiatom} as generators of symmetries of worldvolume probe actions is not apparent. For generic common sector backgrounds, these bilinears do not generate symmetries for the probe actions we have considered here. Nevertheless, they may generate symmetries for probes on some special backgrounds, as some terms in the TCFH may vanish and so the remaining TCFH conditions can be interpreted as invariance conditions of some worldvolume probe action.

\subsubsection{Hidden Symmetries of Probes on Common Sector IIA Branes}

We have demonstrated that particle and string probes in common sector backgrounds exhibit a large number of symmetries generated by the $\nabla^{(\pm)}$-covariantly constant forms $k^{\pm rs}$, $\pi^{\pm rs}$ and $\tau^{\pm rs}$. To present some examples, we shall explore the symmetries generated by the form bilinears of the fundamental string and NS5-brane. For this, we have to compute the form bilinears of these two backgrounds.

To begin, let us assume that the worldsheet directions of the fundamental string are along $05$. Then the Killing spinors of the solution can be written as $\epsilon = h^{-\frac{1}{4}} \epsilon_0$, where $\epsilon_0$ is a constant spinor that satisfies the condition $\Gamma_0\Gamma_5\Gamma_{11}\epsilon_0 = \pm \epsilon_0$ with the gamma matrices in a frame basis\footnote{This will be the case for the conditions on the Killing spinors of all brane solutions that we shall investigate from now on.}. The metric of the solution is given in \eqref{fstring} after changing the worldvolume directions from $01$ to $05$ and taking $h$ to be any harmonic function on $\bR^8$, e.g. $h$ can be a multi-centred harmonic function as in \eqref{mhf} for $p = 1$. The choice of worldsheet directions we have made for the string above may be thought as unconventional. However, it turns out that such a choice is split with the basis used in spinorial geometry \cite{uggp} to construct realisations of Clifford algebras in terms of forms; for a review on spinorial geometry techniques see \cite{review}. We shall use spinorial geometry to solve the condition on $\epsilon_0$ and so this labelling of the coordinates is convenient.

Indeed choosing the plus sign in the condition on $\epsilon_0$ and using the realisation of spinors in terms of forms\footnote{In spinorial geometry the Dirac spinors of $\mathfrak{spin}(9,1)$ are identified with $\Lambda^*(\bC^5)$. The Gamma matrices are realised on $\Lambda^*(\bC^5)$ using the exterior multiplication and inner derivation operations with respect to a Hermitian basis $(e_1, \dots, e_5)$ in $\bC^5$. The Majorana spinors satisfy the reality condition $\Gamma_{6789} *\epsilon = \epsilon$. For more details see e.g. appendix B of \cite{review}.} write $\epsilon_0 = \eta+ e_5\wedge \lambda$, where $\eta$ and $\lambda$ are constant Majorana $\mathfrak{spin}(8)$ spinors. Then the condition $\Gamma_0\Gamma_5\Gamma_{11}\epsilon_0= \epsilon_0$ restricts $\eta$ and $\lambda$ to be positive chirality Majorana-Weyl spinors of $\mathfrak{spin}(8)$, i.e. $\eta, \lambda \in \Delta_8^{+}\equiv \Lambda^{\mathrm{ev}}(\bR\langle e_1, e_2, e_3, e_4\rangle)$. Thus the most general solution of $\Gamma_0\Gamma_5\Gamma_{11}\epsilon_0= \epsilon_0$ is
\begin{equation}\label{fsol}
	\epsilon_0 = \eta+ e_5 \wedge \lambda~,
\end{equation}
where $\eta$ and $\lambda$ are positive chirality Majorana-Weyl spinors of $\mathfrak{spin}(8)$.

Using \eqref{fsol} one can easily express all the form bilinears of the fundamental string background in terms of the form bilinears of $\eta$ and $\lambda$. The explicit expressions have been collected in appendix \ref{common-sector-bilinears}. Using these one finds that
\begin{equation}
	\begin{split}
		k^{+rs} =& 2 h^{-\frac{1}{2}}\Herm{\eta^r}{\eta^s} (\e^0 - \e^5)~, \quad k^{-rs}= h^{-\frac{1}{2}} \Herm{\lambda^r}{\lambda^s} (\e^0 + \e^5)~, \\
		\pi^{+rs} =& h^{-\frac{1}{2}} \Herm{\eta^r}{\Gamma_{ij}\eta^s} (\e^0 - \e^5)\wedge \e^i\wedge \e^j~, \quad \pi^{-rs}= h^{-\frac{1}{2}}\Herm{\lambda^r}{\Gamma_{ij} \lambda^s} (\e^0 + \e^5)\wedge \e^i\wedge \e^j~, \\
		\tau^{+rs} =& \frac{2}{4!}h^{-\frac{1}{2}}\Herm{\eta^r}{\Gamma_{ijk\ell} \eta^s} (\e^0 - \e^5)\wedge \e^i\wedge \e^j\wedge \e^k\wedge \e^\ell~, \\
		\tau^{-rs} =& \frac{2}{4!} h^{-\frac{1}{2}} \Herm{\lambda^r}{\Gamma_{ijk\ell} \lambda^s} (\e^0 + \e^5)\wedge \e^i\wedge \e^j\wedge \e^k\wedge \e^\ell~,
	\end{split}
\end{equation}
where $(\e^0, \e^5, \e^i)$ is a pseudo-orthonormal frame for the metric \eqref{fstring}, i.e. $g = -(\e^0)^2+ (\e^5)^2 + \sum_i (\e^i)^2$, and $\Herm{\cdot}{\cdot}$ is the $\mathfrak{spin}(8)$-invariant (Hermitian) inner product on $\Delta_8^{+}$. Both $k^{\pm rs}$ are along the worldvolume directions and Killing. This implies that both $k$ and $\tilde k$ are Killing as well. This is expected for $k$ but not for $\tilde k$. Nevertheless $\tilde k$ is Killing because the fundamental string is a special background. Observe that the $\nabla^{(+)}$- ($\nabla^{(-)}$-) parallel form bilinears are left- (right-) handed from the string worldvolume perspective as indicated by their dependence on the worldsheet lightcone directions.

It remains to compute the bilinears of $\mathfrak{spin}(8)$ Majorana-Weyl spinors $\eta$ and $\lambda$. These can be obtained using the decomposition of the product of two positive chirality Majorana-Weyl representations $\Delta_8^{+}$ in terms of forms on $\bR^8$ as
\begin{equation}\label{repdec}
	\Delta_8^{+}\otimes \Delta_8^+ = \Lambda^{0}(\bR^8)\oplus \Lambda^{2}(\bR^8)\oplus \Lambda^{4+}(\bR^8)~,
\end{equation}
where $\Lambda^{4+}(\bR^8)$ are the self-dual 4-forms on $\bR^8$. As $\eta$ and $\lambda$ are in $\Delta_8^{+}$ and otherwise unrestricted, their bilinears span all 0-, 2- and self-dual 4-forms in $\bR^8$. As a consequence, the string probe \eqref{sact} and particle probe \eqref{pact} actions are invariant under $2^7$ independent symmetries.

Next let us turn to the symmetries of probes on the NS5-brane background. Choosing the worldvolume of the NS5-brane along the $012567$ directions, the Killing spinors $\epsilon = \epsilon_0$ of the background satisfy the condition $\Gamma_{3489}\Gamma_{11}\epsilon_0 = \pm\epsilon_0$, where $\epsilon_0$ is a constant Majorana spinor. The metric of the solution is given in \eqref{ns5} after changing the worldvolume directions from $012345$ to $012567$ for similar reasons as those explained for the fundamental string above and after taking $h$ to be a harmonic function on $\bR^4$ as in \eqref{mhf} for $p = 5$. Choosing the plus sign, the condition $\Gamma_{3489}\Gamma_{11}\epsilon_0 = \epsilon_0$ can be solved using spinorial geometry. It is convenient to first solve this condition for Dirac spinors and then impose the reality condition on $\epsilon$. The solution can be expressed as
\begin{equation}\label{ns5s}
	\epsilon = \eta^1 + e_{34}\wedge \lambda^1+ e_3\wedge \eta^2 + e_4\wedge \lambda^2~,
\end{equation}
where $\eta$ and $\lambda$ are positive chirality Weyl spinors of $\mathfrak{spin}(5,1)$, i.e. $\eta, \lambda\in \Delta^+_{(6)}\equiv \Lambda^{\mathrm{ev}}(\bC\langle e_1,e_2, e_5\rangle)$. Imposing the reality condition on $\epsilon$, $\Gamma_{6789}*\epsilon = \epsilon$, one finds that
\begin{equation}
	\lambda^1 = -\Gamma_{67} (\eta^1)^*~, \quad \lambda^2 = -\Gamma_{67} (\eta^2)^*~.
\end{equation}
So the Killing spinor $\epsilon$ is completely determined by the (complex) positive chirality $\mathfrak{spin}(5,1)$ spinors $\eta^1$ and $\eta^2$.

Using \eqref{ns5s}, one can easily compute all the form bilinears of the NS5-brane background and express them in terms of the form bilinears of $\eta^1$ and $\eta^2$. All these can be found in appendix \ref{common-sector-bilinears}.

In particular the $\nabla^{(\pm)}$-covariantly constant spinor bilinears are
\begin{equation}
	k^{+rs} = 4 \mathrm{Re}\Dirac{\eta^{1r}}{\Gamma_a\eta^{1s}} \e^a~, \quad k^{-rs} = 4\mathrm{Re}\Dirac{\eta^{2r}}{\Gamma_a\eta^{2s}} ~\e^a~,
\end{equation}
\begin{equation}
	\begin{split}
		\pi^{+rs} &= \frac{2}{3}\mathrm{Re}\Dirac{\eta^{1r}}{\Gamma_{abc} \eta^{1s}}\, \e^a\wedge \e^b\wedge \e^c \\
		&- 4\mathrm{Re}\Dirac{\eta^{1r}}{\Gamma_{a} \lambda^{1s}}\, (\e^3\wedge \e^4 -\e^8\wedge \e^9)\wedge \e^a \\
		&- 4\mathrm{Im}\Dirac{\eta^{1r}}{\Gamma_{a} \eta^{1s}}\, (\e^3\wedge \e^8 +\e^4\wedge \e^9)\wedge \e^a \\
		&- 4\mathrm{Im}\Dirac{\eta^{1r}}{\Gamma_{a} \lambda^{1s}}\, (\e^3\wedge \e^9 -\e^4\wedge \e^8)\wedge \e^a~,
	\end{split}
\end{equation}
\begin{equation}
	\begin{split}
		\pi^{-rs} &= \frac{2}{3}\mathrm{Re}\Dirac{\eta^{2r}}{\Gamma_{abc} \eta^{2r}}\, \e^a\wedge \e^b\wedge \e^c \\
		&+ 4\mathrm{Re}\Dirac{\eta^{2r}}{\Gamma_{a} \lambda^{2s}}\, (\e^3\wedge \e^4 +\e^8\wedge \e^9)\wedge \e^a \\
		&+ 4\mathrm{Im}\Dirac{\eta^{2r}}{\Gamma_{a} \eta^{2s}}\, (\e^3\wedge \e^8 -\e^4\wedge \e^9)\wedge \e^a \\
		&+ 4\mathrm{Im}\Dirac{\eta^{2r}}{\Gamma_{a} \lambda^{2s}}\, (\e^3\wedge \e^9 +\e^4\wedge \e^8)\wedge \e^a~,
	\end{split}
\end{equation}
\begin{equation}
	\begin{split}
		\tau^{+rs} &= k^{+rs}\wedge \e^3\wedge \e^4\wedge \e^8\wedge \e^9 \\
		&-\frac{2}{3}\mathrm{Re}\Dirac{\eta^{1r}}{\Gamma_{abc} \lambda^{1s}}\, (\e^3\wedge \e^4 -\e^8\wedge \e^9)\wedge \e^a\wedge \e^b\wedge \e^c \\
		&-\frac{2}{3}\mathrm{Im}\Dirac{\eta^{1r}}{\Gamma_{abc} \eta^{1s}}\, (\e^3\wedge \e^8 +\e^4\wedge \e^9)\wedge \e^a\wedge \e^b\wedge \e^c \\
		&-\frac{2}{3}\mathrm{Im}\Dirac{\eta^{1r}}{\Gamma_{abc} \lambda^{1s}}\, (\e^3\wedge \e^9 -\e^4\wedge \e^8)\wedge \e^a\wedge \e^b\wedge \e^c \\
		&+\frac{4}{5!}\mathrm{Re}\Dirac{\eta^{1r}}{\Gamma_{a_1\dots a_5}\eta^{1s}}~\e^{a_1}\wedge\dots\wedge \e^{a_5}~,
	\end{split}
\end{equation}
\begin{equation}
\begin{split}
	\tau^{-rs} &= - k^{-rs}\wedge \e^3\wedge \e^4\wedge \e^8\wedge \e^9 \\
	&+\frac{2}{3}\mathrm{Re}\Dirac{\eta^{2r}}{\Gamma_{abc} \lambda^{2s}} (\e^3\wedge \e^4 +\e^8\wedge \e^9)\wedge \e^a\wedge \e^b\wedge \e^c \\
	&+\frac{2}{3}\mathrm{Im}\Dirac{\eta^{2r}}{\Gamma_{abc} \eta^{2s}} (\e^3\wedge \e^8 -\e^4\wedge \e^9)\wedge \e^a\wedge \e^b\wedge \e^c \\
	&+\frac{2}{3}\mathrm{Im}\Dirac{\eta^{2r}}{\Gamma_{abc} \lambda^{2s}} (\e^3\wedge \e^9 +\e^4\wedge \e^8)\wedge \e^a\wedge \e^b\wedge \e^c \\
	&+\frac{4}{5!}\mathrm{Re}\Dirac{\eta^{2r}}{\Gamma_{a_1\dots a_5}\eta^{2s}}~\e^{a_1}\wedge\dots\wedge \e^{a_5}~,
\end{split}
\end{equation}
where $a,b,c = 0,1,2,5,6,7$ are the worldvolume directions, $(\e^a, \e^3, \e^4, \e^8, \e^9)$ is a pseudo~orthonormal frame for the metric \eqref{ns5}, $\Dirac{\cdot}{\cdot}$ is the $\mathfrak{spin}(5,1)$ invariant Dirac inner product and $\epsilon_{3489} = 1$. Both $k^{\pm rs}$ are along the worldvolume directions of the brane and are Killing. This in turn implies that both $k$ and $\tilde k$ are Killing as well. Again $\tilde k$ is Killing because the NS5-brane is a special background. The 3- and 5-forms have mixed components along both worldvolume and transverse directions. Note that the anti-self-dual and self-dual 2-forms along the transverse directions contribute to $\nabla^{(+)}$ and $\nabla^{(-)}$ covariantly constant forms, respectively.

Therefore the NS5-brane form bilinears have been expressed in terms of those of two positive chirality Weyl $\mathfrak{spin}(5,1)$ spinors. The decomposition of two positive chirality Weyl $\mathfrak{spin}(5,1)$ representations, $\Delta^+_4$, into forms on $\bC^6$ is given by
\begin{equation}
	\otimes^2 \Delta^+_{4}= \Lambda^1(\bC^6)\oplus \Lambda^{3+} (\bC^6)
\end{equation}
Therefore the string probe with action \eqref{sact} and particle probe with action \eqref{pact} are invariant under $2^5$ symmetries counted over the reals. To see this, observe that from the decomposition above all 1- and self-dual 3-forms along the NS5-brane worldvolume are spanned by these spinors. So there are $6 + 10 = 2^4$ independent symmetries generated by the $\nabla^{(+)}$-covariantly constant forms and similarly for the $\nabla^{(-)}$-covariantly constant forms yielding $2^5$ in total. These generate a symmetry algebra of W-type \cite{phgp1, phgp2}. For the remaining form bilinears in appendix \ref{common-sector-bilinears}, there is not a straightforward way to relate them to symmetries of particle or string probe actions.

\section{IIA D-branes} \label{IIA-Dbranes}

There is no classification of IIA supersymmetric backgrounds. So to give more examples for which the TCFH can be interpreted as invariance condition for probe particle and string actions under symmetries generated by the form bilinears, we shall turn to some special solutions and in particular to the D-branes\footnote{A consequence of this investigation is that we shall find all form bilinears of the type IIA D-brane solutions.}. It is convenient to organise the investigation in electric-magnetic brane pairs as the non-vanishing fields that appear in TCFH are the same. The TCFH for each D-brane pair can be easily found from that of the IIA TCFH given in \eqref{iiatcfha}-\eqref{iiatcfhf} and \eqref{iiatcfha1}-\eqref{iiatcfha4} upon setting all the form field strengths to zero apart from those associated to the D-brane under investigation.

\subsection{D0- and D6-branes}

\subsubsection{D0-branes}

The Killing spinors of the D0-brane are given by $\epsilon= h^{-\frac{1}{8}} \epsilon_0$, where $\epsilon_0$ is a constant spinor restricted as $\Gamma_0\Gamma_{11}\epsilon_0 = \pm \epsilon_0$, the worldline is along the 0-th direction and $h$ is a multi-centred harmonic function as in \eqref{mhf} for $p = 0$. Choosing the plus sign and using spinorial geometry \cite{uggp}, one can solve this condition by setting
\begin{equation}\label{d0sol}
	\epsilon_0 = \eta-e_5\wedge \Gamma_{11} \eta~,
\end{equation}
where $\eta\in \Lambda^*(\bR\langle e_1, \dots, e_4\rangle)$ and the reality condition is imposed by $\Gamma_{6789}*\eta = \eta$. Using this, one can compute the form bilinears. These are given in appendix \ref{Dbrane-bilinears}.

As expected $k$ is a Killing vector. As a result $k$ generates a symmetry in all probe actions \eqref{sact}, \eqref{pact} and \eqref{1part} after setting the form couplings to zero. It also generates a symmetry in the probe action of \cite{gpeb} with the 2-form coupling; the D0-brane 2-form field strength $F = F_{0i}\, \e^0\wedge \e^i$ is invariant under the action of $k$. An investigation of the TCFH for the rest of the form bilinears using that $F_{0i}\not = 0$ reveals that these do not generate symmetries for the probe actions we have been considering. Because of this we postpone a more detailed analysis of the TCFH for later and in particular for the D6- and D2-branes.

\subsubsection{D6-brane}

Choosing the transverse directions of the D6-brane along $549$, the Killing spinor $\epsilon = h^{-\frac{1}{8}} \epsilon_0$ satisfies the condition
\begin{equation}\label{d6pro}
	\Gamma_{549} \Gamma_{11}\epsilon_0 = \pm\epsilon_0~,
\end{equation}
where $\epsilon_0$ is a constant spinor and $h$ is a multi-centred harmonic function as in \eqref{mhf} with $p = 6$. To solve this condition with the plus sign using spinorial geometry, set
\begin{equation}\label{d6sol}
	\epsilon_0 = \eta + e_4\wedge \lambda~,
\end{equation}
where $\eta, \lambda\in \Lambda^*(\bC\langle e_1, e_2, e_3, e_5\rangle)$. Then the condition \eqref{d6pro} gives
\begin{equation}\label{d6elcon}
	\Gamma_5 \Gamma_{11}\eta = -i\eta~, \quad \Gamma_5 \Gamma_{11}\lambda = i\lambda~.
\end{equation}
One can proceed to expand $\eta$ and $\lambda$ as $\eta = \eta^1 + e_5\wedge \eta^2$ and $\lambda = \lambda^1 + e_5\wedge \lambda^2$ in which case the conditions \eqref{d6elcon} give $\eta^2 = i \Gamma_{11} \eta^1$ and $\lambda^2 = -i \Gamma_{11}\lambda^1$, where $\eta^1, \lambda^1 \in \Lambda^*(\bC\langle e_1, e_2, e_3\rangle)$ are the independent spinors. However if one proceeds in this way the form bilinears will not be manifestly worldvolume Lorentz covariant, as the $0$-th direction will be separated from the rest. Because of this, we shall not solve \eqref{d6elcon} and do the computation with $\eta$ and $\lambda$. After the computation of the form bilinears, one can substitute in the formulae the solution of \eqref{d6elcon} in terms of $\eta^1$ and $\lambda^1$. However this is not necessary for the purpose of this chapter. It remains to impose the reality condition on $\epsilon_0$. This gives $\eta = -i \Gamma_{678}*\lambda$ or equivalently $\lambda = -i \Gamma_{678}*\eta$. The form bilinears are given in appendix \ref{Dbrane-bilinears}.

The TCFH for $k$ on a background with a 2-form field strength is
\begin{equation}\label{ftcfh1}
	\begin{split}
		\nabla_M k_N = \frac{1}{8} e^\Phi F_{PQ}\tilde{\zeta}^{PQ}{}_{MN} + \frac{1}{4} e^\Phi F_{MN}\tilde{\sigma}~.
	\end{split}
\end{equation}
As expected $k$ generates isometries and so symmetries in all the probe actions \eqref{pact}, \eqref{sact} and \eqref{1part} with vanishing form couplings. It also generates a symmetry for the probe action of \cite{gpeb} with the 2-form coupling, as the D6-brane 2-form field strength $F = \frac{1}{2} F_{ij}\, \e^i\wedge \e^j$ is invariant under the action of $k$. In what follows we shall be mostly concerned with the symmetries generated by the form bilinears for the probe action \eqref{1part}. The invariance of this action imposes the weakest conditions on the form bilinears amongst all probe actions that we have been investigating.

Next consider the $\tilde k$ and $\omega$ bilinears on a background with a 2-form field strength. The TCFH for these is
\begin{equation}\label{ftcfh2}
	\begin{split}
		\nabla_M \tilde{k}_N- \frac{1}{2} e^\Phi F_{MP}\omega^P{}_N = \frac{1}{8}e^\Phi g_{MN} F_{PQ}\omega^{PQ} -\frac{1}{2}e^\Phi F_{[M|P|}\omega^P{}_{N]}~,
	\end{split}
\end{equation}
\begin{equation}\label{ftcfh3}
	\begin{split}
		\nabla_M \omega_{NR} + e^\Phi F_{M[N}\tilde{k}_{R]} &= \frac{3}{4} e^\Phi F_{[MN}\tilde{k}_{R]} + \frac{1}{2} e^\Phi g_{M[N}F_{R]P}\tilde{k}^P \\
		&- \frac{1}{4 \cdot 5!} e^\Phi \hodge{F}_{MNRP_1\dots P_5}\tau^{P_1\dots P_5}~.
	\end{split}
\end{equation}
For $\tilde k$ to generate symmetries in probe action \eqref{1part} with $C = 0$, it must be a KY tensor. As for D6-branes $F_{ij}\not = 0$, the term proportional to the spacetime metric in the first of the equations above must vanish. This requires that $\omega_{ij} = 0$. Then from the expressions of the form bilinears of D6-brane in appendix \ref{Dbrane-bilinears} and \eqref{d6elcon}, one concludes that $\tilde k = 0$. Therefore $\tilde k$ does not generate symmetries for the probe action \eqref{1part}.

Similarly for $\omega$ to generate a symmetry for probe action \eqref{1part} with $C = 0$, one finds from the last TCFH above that $\tilde k = 0$. Then from the expressions for the D6-brane form bilinears in appendix \ref{Dbrane-bilinears}, this implies that $\omega_{ij} = 0$ or equivalently
\begin{equation}
	{\Dirac{\eta^r}{\Gamma_{11} \lambda^s}} = \mathrm{Im} \Dirac{\eta^r}{\eta^s} = 0~.
\end{equation}
Then
\begin{equation}
	\omega = \frac{1}{2} \omega_{ab}\, \e^a\wedge \e^b= h^{-\frac{1}{4}} \mathrm{Re}{\Dirac{\eta^r}{\Gamma_{ab}\eta^s}}\, \e^a \wedge \e^b~,
\end{equation}
is a KY form and generates a (hidden) symmetry for the probe action \eqref{1part} with $C = 0$. Note that there are Killing spinors for which $\omega\not = 0$ even though $\omega_{ij} = 0$. Indeed, take $\eta^r = \eta^s = 1 + e_1+ e_5\wedge (i 1-ie_1)$.

The TCFH for the bilinears $\tilde \omega$ and $\pi$ is
\begin{equation}\label{ftcfh4}
	\begin{split}
		\nabla_M \tilde\omega_{NR} + \frac{1}{2}e^\Phi F_{MP} \pi^P{}_{NR}= -\frac{1}{4}e^\Phi g_{M[N}F_{|PQ|}\pi^{PQ}{}_{R]} + \frac{3}{4} e^\Phi F_{[M|P|} \pi^P{}_{NR]}~,
	\end{split}
\end{equation}
\begin{equation}\label{ftcfh5}
	\begin{split}
		\nabla_M \pi_{NRS} - &\frac{3}{2} e^\Phi F_{M[N} \tilde\omega_{RS]}= - \frac{1}{4 \cdot 4!} e^\Phi \hodge{F}_{MNRSP_1 \dots P_4} \zeta^{P_1\dots P_4} \\
		&- \frac{3}{2} e^\Phi g_{M[N} F_{R|P|} \tilde \omega^P{}_{S]} - \frac{3}{2} e^\Phi F_{[MN} \tilde\omega_{RS]}~.
	\end{split}
\end{equation}
For $\tilde \omega$ to be a KY form and so generate a symmetry in the probe action \eqref{1part} with $C = 0$, $\pi_{aij} = 0$. As it can be seen from the D6-brane bilinears in appendix \ref{Dbrane-bilinears} after using \eqref{d6elcon}, this implies that $\tilde \omega = 0$ and so $\tilde \omega$ does not generate any symmetries. Turning to $\pi$, one finds that this is a KY tensor provided that $\tilde\omega = 0$ which implies that $\pi_{aij} = 0$ or equivalently
\begin{equation}\label{bibix}
	{\Dirac{\eta^r}{\Gamma_a \Gamma_{5} \lambda^s}} = \mathrm{Im}{\Dirac{\eta^r}{\Gamma_a \eta^s}} = 0~.
\end{equation}
The remaining components of $\pi$,
\begin{equation}\label{bibixx}
	\pi = \frac{1}{3!} \pi_{abc} \e^a\wedge \e^b\wedge \e^c = \frac{1}{3} h^{-\frac{1}{4}} \mathrm{Re}{\Dirac{\eta^r}{\Gamma_{abc} \eta^s}} \e^a \wedge \e^b \wedge \e^c~,
\end{equation}
generate a (hidden) symmetry for the probe action \eqref{1part} with $C = 0$. There are Killing spinors such that they satisfy \eqref{bibix} and $\pi\not = 0$, e.g. $\eta^r = 1 + e_1+ i e_5\wedge (1-e_1)$ and $\eta^s= i(1-e_1)-e_5\wedge (1 + e_1)$.

From now on to simplify the analysis that follows on the symmetries generated by TCFHs for all IIA D-branes, we shall only mention the components of the form bilinears that are required to vanish in order for some others become KY forms. In particular, we shall not give the explicit expressions for the vanishing components of the form bilinears and those of the KY forms in terms of the Killing spinors as we have done in e.g. \eqref{bibix} and \eqref{bibixx}, respectively. These can be easily read from the expressions of the form bilinears of D-branes given in appendix \ref{Dbrane-bilinears}.

The TCFH for the bilinears $\zeta$ and $\tilde \pi$ is
\begin{equation}\label{ftcfh6}
	\begin{split}
		\nabla_M \tilde\pi_{NRS}-&\frac{1}{2} e^\Phi F_{MP} \zeta^P{}_{NRS} = \frac{3}{8} e^\Phi g_{M[N} F_{|PQ|} \zeta^{PQ}{}_{RS]} \\
		&- e^\Phi F_{[M|P|} \zeta^P{}_{NRS]} - \frac{3}{4}e^\Phi g_{M[N}F_{RS]} \sigma~,
	\end{split}
\end{equation}
\begin{equation}\label{ftcfh7}
	\begin{split}
		\nabla_M \zeta_{N_1 \dots N_4}& +2 e^\Phi F_{M[N_1} \tilde\pi_{N_2 N_3 N_4]} = \frac{1}{4!} e^\Phi \hodge{F}_{MN_1 \dots N_4 PQR} \pi^{PQR} \\
		&+ 3 e^\Phi g_{M[N_1} F_{N_2|P|}\tilde\pi^P{}_{N_3N_4]} + \frac{5}{2} e^\Phi F_{[MN_1}\tilde\pi_{N_2N_3N_4]} ~ .
	\end{split}
\end{equation}
A similar analysis to the one presented above reveals that $\tilde \pi$ does not generate symmetries in the probe actions we have been considering. While for $\zeta$ to be a KY form, and so generate a (hidden) symmetry for the probe action \eqref{1part} with $C = 0$, one requires that $\tilde \pi = 0$. This in turn implies that $\zeta_{abij} = 0$. So there is the possibility that $\zeta = \frac{1}{24} \zeta_{a_1\dots a_4} \e^{a_1}\wedge\dots\wedge \e^{a_4}$ is a KY form. But one can verify after some computation\footnote{To prove this one uses spinorial geometry techniques and the freedom to choose a pseudo-orthonormal frame to find a representative for $\eta^{1r}$. Then one solves for all conditions arising from $\zeta_{abij} = 0$. This restricts $\eta^{1s}$ and leads to $\zeta = 0$. It is a lengthy computation that will not be presented here.} that there are not Killing spinors such that $\zeta_{abij} = 0$ with $\zeta\not = 0$.

The TCFH for $\tilde \zeta$ and $\tau$ is
\begin{equation}\label{ftcfh8}
	\begin{split}
		\nabla_M \tilde{\zeta}_{N_1 \dots N_4} +& \frac{1}{2} e^\Phi F_{MP}\tau^P{}_{N_1 \dots N_4} = -\frac{1}{2} e^\Phi g_{M[N_1}F_{|PQ|}\tau^{PQ}{}_{N_2N_3N_4]} + \frac{5}{8} e^\Phi F_{[M|P|}\tau^P{}_{N_1 \dots N_4]} \\
			&+ 3e^\Phi g_{M[N_1}F_{N_2N_3}k_{N_4]}~,
	\end{split}
\end{equation}
\begin{equation}\label{ftcfh9}
	\begin{split}
		\nabla_M \tau_{N_1 \dots N_5} &-\frac{5}{2}e^\Phi F_{M[N_1}\tilde{\zeta}_{N_2 \dots N_5]}= \frac{1}{8} e^\Phi \hodge{F}_{MN_1\dots N_5}{}^{PQ}\omega_{PQ} \\
		&-5e^\Phi g_{M[N_1}F_{N_2|P|}\tilde{\zeta}^P{}_{N_3N_4N_5]} - \frac{15}{4} e^\Phi F_{[MN_1}\tilde{\zeta}_{N_2\dots N_5]}~.
	\end{split}
\end{equation}
For $\tilde \zeta$ to generate a symmetry, the above TCFH requires $k_a = 0$ and $\tau_{abcij} = 0$. These imply that $\tilde \zeta = 0$ and so this bilinear does not generate a symmetry. It turns out that $\tau$ is a KY form provided that $\tilde\zeta_{abci} = 0$. As a result $\tau_{abcij} = 0$. The remaining non-vanishing components of $\tau$, $\tau = \frac{1}{5!} \tau_{a_1\dots a_5} \e^{a_1}\wedge\dots\wedge \e^{a_5}$ potentially generates a (hidden) symmetry of the probe action \eqref{1part} with $C = 0$. But after some computation one can verify that there are no Killing spinors such that $\tau_{abcij} = 0$ such that $\tau\not = 0$.

It is clear from the TCFH in \eqref{ftcfh1}-\eqref{ftcfh3}, \eqref{ftcfh4}, \eqref{ftcfh5}, \eqref{ftcfh6}, \eqref{ftcfh7}, \eqref{ftcfh8} and \eqref{ftcfh9} that the holonomy of the minimal connection reduces for backgrounds with only a 2-form field strength. In particular, the (reduced) holonomy of the minimal connection reduces to a subgroup of $SO(9,1)\times GL(55)\times GL(165)\times GL(330)\times GL(462)$. For completeness we state the TCFH on the scalar bilinears
\begin{equation}
	\nabla_M \tilde{\sigma} = - \frac{1}{4} e^\Phi F_{MP}k^P~,~~ \nabla_M \sigma = - \frac{1}{8} e^\Phi F_{PQ} \tilde\pi^{PQ}{}_M~.
\end{equation}
These give a trivial contribution to the holonomy of the minimal connection.

To summarise the results of this section, we have concluded as a consequence of the TCFH that there are Killing spinors such that $k$, $\omega$ and $\pi$, which have non-vanishing components only along the worldvolume directions of the D6-brane, are KY forms. Therefore they generate symmetries for the probe described by the action \eqref{1part} with $C = 0$ in a D6-brane background. This is the case for any multi-centred harmonic function $h$ that the D6-brane solution depends on.

\subsection{D2 and D4-branes}

\subsubsection{D2 brane}

Choosing the worldvolume directions of the D2-brane along $051$, the Killing spinors $\epsilon = h^{-\frac{1}{8}} \epsilon_0$ of the solution satisfy the condition
\begin{equation}\label{d2scon}
	\Gamma_{051}\epsilon_0 = \pm\epsilon_0~,
\end{equation}
where $\epsilon_0$ is a constant spinor and $h$ is given in \eqref{mhf} for $p = 2$. To solve this condition with the plus sign using spinorial geometry, set
\begin{equation}\label{d2solscon}
	\epsilon_0 = \eta+ e_5\wedge \lambda~,
\end{equation}
to find that the remaining restrictions on $\eta$ and $\lambda$ are
\begin{equation}
	\Gamma_1\eta = \eta~, \quad \Gamma_1\lambda = \lambda~,
\end{equation}
where $\eta, \lambda\in \Lambda^*(\bR \langle e_1, e_2, e_3, e_4\rangle)$; the reality condition is imposed with $\Gamma_{6789}*\eta = \eta$ and $\Gamma_{6789}*\lambda = \lambda$. As in the D6-brane case, the remaining condition on $\eta$ and $\lambda$ can be solved by setting $\eta = \eta^1+ e_1\wedge \eta^1$ and $\lambda = \lambda^1 + e_1\wedge \lambda^1$, where $\eta^1, \lambda^1\in \Lambda^*(\bR \langle e_2, e_3, e_4\rangle)$ label the independent solutions of \eqref{d2scon}. However, we shall perform the computation of the form bilinears using \eqref{d2solscon} as otherwise their expression will not be manifestly covariant along the transverse directions of the D2-brane, e.g. the 6-th direction will have to be treated separately from the rest. The form bilinears of the D2-brane can be found in appendix \ref{Dbrane-bilinears}.

D2-branes exhibit a non-vanishing 4-form field strength $G_{015i}\not = 0$. As the probe actions we have been considering do not exhibit such a coupling, the only remaining coupling is that of the spacetime metric. Therefore for the form bilinears to generate a symmetry, they must be KY forms. To investigate which of the form bilinears are KY, we shall organise the TCFH according to the domain that the minimal connection acts on. As expected the TCFH
\begin{equation}\label{iiatcfhbx1}
	\begin{split}
		\nabla_M k_N = -\frac{1}{4\cdot 4!}e^\Phi \hodge{G}_{MNP_1\dots P_4}\tilde{\zeta}^{P_1 \dots P_4} + \frac{1}{8}e^\Phi G_{MNPQ}\omega^{PQ}~,
	\end{split}
\end{equation}
implies that $k$ is a Killing 1-form. As a result it generates symmetries in all probe action \eqref{pact}, \eqref{sact} and \eqref{1part} after setting $b = C=0$.

Next observe that
\begin{equation}\label{iiatcfhcx2}
	\nabla_M \tilde{k}_N - \frac{1}{12}e^\Phi G_{MPQR}\tilde{\zeta}^{PQR}{}_N= \frac{1}{4\cdot 4!}e^\Phi g_{MN} G_{P_1\dots P_4}\tilde{\zeta}^{P_1 \dots P_4} - \frac{1}{12}e^\Phi G_{[M|PQR|}\tilde{\zeta}^{PQR}{}_{N]}~,
\end{equation}
\begin{equation}\label{iiatcfhex3}
	\begin{split}
		&\nabla_M \tilde{\zeta}_{N_1 \dots N_4} +\frac{1}{2}e^\Phi \hodge{G}_{M[N_1N_2|PQR|}\tau^{PQR}{}_{N_3N_4]}+ 2e^\Phi G_{M[N_1N_2N_3}\tilde{k}_{N_4]} \\
		&= - \frac{1}{8}e^\Phi g_{M[N_1}\hodge{G}_{N_2 N_3|P_1 \dots P_4|}\tau^{P_1 \dots P_4}{}_{N_4]} + \frac{5}{12}e^\Phi \hodge{G}_{[MN_1N_2|PQR|}\tau^{PQR}{}_{N_3N_4]} \\
		&+\frac{1}{4}e^\Phi \hodge{G}_{MN_1\dots N_4P}k^P + \frac{5}{4}e^\Phi G_{[MN_1N_2N_3}\tilde{k}_{N_4]} + e^\Phi g_{M[N_1}G_{N_2N_3N_4]P}\tilde{k}^P ~.
	\end{split}
\end{equation}
\begin{equation}\label{iiatcfhfx4}
	\begin{split}
		&\nabla_M \tau_{N_1 \dots N_5}-\frac{5}{2}e^\Phi \hodge{G}_{M[N_1N_2N_3|PQ|}\tilde{\zeta}^{PQ}{}_{N_4N_5]} +5e^\Phi G_{M[N_1N_2N_3}\omega_{N_4N_5]} \\
		&= - \frac{15}{8}e^\Phi \hodge{G}_{[MN_1N_2N_3|PQ|}\tilde{\zeta}^{PQ}{}_{N_4N_5]} - \frac{1}{4}e^\Phi \hodge{G}_{MN_1\dots N_5}\tilde{\sigma} \\ 
		&- \frac{5}{6 }e^\Phi g_{M[N_1}\hodge{G}_{N_2N_3N_4|PQR|}\tilde{\zeta}^{PQR}{}_{N_5]} + \frac{15}{4} e^\Phi G_{[MN_1N_2N_3}\omega_{N_4N_5]} \\
		&+ 5 e^\Phi g_{M[N_1}G_{N_2N_3N_4|P|}\omega^P{}_{N_5]} ~,
	\end{split}
\end{equation}
\begin{equation}\label{iiatcfhdx5}
	\begin{split}
		&\nabla_M \omega_{NR} - \frac{1}{12} e^\Phi G_{MP_1P_2P_3}\tau^{P_1P_2P_3}{}_{NR} = \frac{1}{2\cdot 4!}e^\Phi g_{M[N} G_{|P_1\dots P_4|}\tau^{P_1 \dots P_4}{}_{R]} \\
		&- \frac{1}{8}e^\Phi G_{[M|P_1 P_2 P_3|}\tau^{P_1 P_2 P_3}{}_{NR]} - \frac{1}{4}e^\Phi G_{MNRP} k^P ~,
	\end{split}
\end{equation}
and so the minimal connection acts on the domain of $\tilde k$, $\tilde\zeta$, $\tau$ and $\omega$ form bilinears. Using that for D2-branes $G_{015i}\not = 0$ and the explicit expression for the form bilinears in appendix \ref{Dbrane-bilinears}, one finds that the TCFH implies that the form bilinears $\tilde k$, $\tilde\zeta$ and $\tau$ cannot be KY tensors. So these do not generate a symmetry in probe actions. On the other hand for $\omega$ to be a KY tensor, the TCFH implies that $\tau_{abcij} = 0$. This in turn implies that $\omega_{ij} = 0$. As a result $\omega = \frac{1}{2} \omega_{ab} \e^a\wedge \e^b$ is a KY form and generates a (hidden) symmetry in the probe action \eqref{1part}. The condition $\tau_{abcij} = 0$ on the Killing spinors and the expression for $\omega_{ab}$ in terms of Killing spinors can be easily read from the expressions of these form bilinears in appendix \ref{Dbrane-bilinears}. There are Killing spinors such that $\tau_{abcij} = 0$ and $\omega\not = 0$. For example set $\eta^r = \lambda^r$ and $\eta^s = \lambda^s$ with $\Herm{\eta^r}{\eta^s}\not = 0$.

The TCFH on the remaining form bilinears is
\begin{equation}\label{iiatcfha2x1}
	\begin{split}
		\nabla_M \pi_{NRS} -& \frac{3}{4}e^\Phi G_{M[N|PQ|} \zeta^{PQ}{}_{RS]}= - \frac{1}{4} e^\Phi G_{MNRS} \sigma +\frac{1}{8} e^\Phi \hodge{G}_{MNRSPQ} \tilde\omega^{PQ} \\
		&- \frac{1}{4} e^\Phi g_{M[N} G_{R|P_1P_2P_3|}\zeta^{P_1P_2P_3}{}_{S]} - \frac{3}{4}e^\Phi G_{[MN|PQ|}\zeta^{PQ}{}_{RS]} ~,
	\end{split}
\end{equation}
\begin{equation}\label{iiatcfha4x2}
	\begin{split}
		\nabla_M \zeta_{N_1 \dots N_4} &+ 3 e^\Phi G_{M[N_1N_2|P|}\pi^P{}_{N_3N_4]}+ e^\Phi \hodge{G}_{M[N_1N_2N_3|PQ|}\tilde\pi^{PQ}{}_{N_4]} \\
		&= \frac{1}{6} e^\Phi g_{M[N_1} \hodge{G}_{N_2N_3N_4]PQR}\tilde\pi^{PQR}
	+ \frac{5}{8} e^\Phi \hodge{G}_{[MN_1N_2N_3|PQ|}\tilde\pi^{PQ}{}_{N_4]} \\
		&- \frac{3}{2} e^\Phi g_{M[N_1} G_{N_2N_3|PQ|}\pi^{PQ}{}_{N_4]} + \frac{5}{2} e^\Phi G_{[MN_1N_2|P|} \pi^P{}_{N_3N_4]} ~,
	\end{split}
\end{equation}
\begin{equation}\label{iiatcfha3x3}
	\begin{split}
		\nabla_M \tilde\pi_{NRS} &+\frac{3}{2} e^\Phi G_{M[NR|P|} \tilde\omega^P{}_{S]}
		- \frac{1}{4}e^\Phi \hodge{G}_{M[NR|P_1P_2P_3|}\zeta^{P_1P_2P_3}{}_{S]} \\
		&=- \frac{3}{8} e^\Phi g_{M[N}G_{RS]PQ} \tilde \omega^{PQ}+ e^\Phi G_{[MNR|P|} \tilde\omega^P{}_{S]} \\
		&+ \frac{1}{32} e^\Phi g_{M[N} \hodge{G}_{RS]P_1 \dots P_4}\zeta^{P_1\dots P_4} - \frac{1}{6} e^\Phi \hodge{G}_{[MNR|P_1P_2P_3|}\zeta^{P_1P_2P_3}{}_{S]} ~,
	\end{split}
\end{equation}
\begin{equation}\label{iiatcfha1x4}
	\begin{split}
		\nabla_M \tilde\omega_{NR} &- \frac{1}{2} e^\Phi G_{M[N|PQ|} \tilde\pi^{PQ}{}_{R]}= - \frac{1}{4!} e^\Phi \hodge{G}_{MNRP_1P_2P_3}\pi^{P_1P_2P_3} \\
		&- \frac{1}{12} e^\Phi g_{M[N}G_{R]P_1P_2P_3}\tilde\pi^{P_1P_2P_3} - \frac{3}{8}e^\Phi G_{[MN|PQ|} \tilde\pi^{PQ}{}_{R]} ~.
	\end{split}
\end{equation}
Requiring that these form bilinears must be KY tensors, the above TCFH together with the explicit expressions for the D2-brane form bilinears in \ref{Dbrane-bilinears} reveal that $\zeta= \tilde \pi = \tilde \omega = 0$. For $\pi$ to be a KY form, the TCFH implies that $ \zeta_{ijab} = 0$ which in turn gives $\pi_{ija} = 0$. The remaining non-vanishing component of $\pi$, $\pi = \frac{1}{3!} \pi_{abc} \e^a\wedge \e^b\wedge \e^c$, is a KY tensor and generates a (hidden) symmetry in the probe action \eqref{1part} with $C = 0$. Again the expression of the conditions $ \zeta_{ijab} = 0$ and that of $\pi$ in terms of the Killing spinors can be found in appendix \ref{Dbrane-bilinears}. There are Killing spinors such that $ \zeta_{ijab} = 0$ and $ \pi\not = 0$. Indeed set $\lambda^r = -\eta^r$, $\lambda^s = \eta^s$ and $\eta^r = \eta^s = 1 + e_{234}+ e_1\wedge (1 + e_{234})$.

It is clear that the holonomy of the minimal connection of the TCFH with only the 4-form field strength reduces. In particular, the reduced holonomy is included in $SO(9,1)\times GL(517)\times GL(495)$. For completeness we give the TCFH on the scalars as
\begin{equation}
		\nabla_M \tilde{\sigma} = \frac{1}{4 \cdot 5!}\hodge{G}_{MP_1\dots P_5}\tau^{P_1\dots P_5}~,~~\nabla_M \sigma = \frac{1}{4!} e^\Phi G_{MPQR} \pi^{PQR} ~,
\end{equation}
which give a trivial contribution in the holonomy of the minimal connection.

To summarise the results of this section, we have shown that there are choices of Killing spinors such that $\omega$ and $\pi$, with non-vanishing components only along the worldvolume directions of the D2-brane, are KY tensors. Therefore these bilinears generate (hidden) symmetries for a probe described by the action \eqref{1part} with $C = 0$ on all D2-brane backgrounds, including those that depend on a multi-centred harmonic function $h$.

\subsubsection{D4 brane}

Choosing the transverse directions of the D4-brane as $23849$, the Killing spinors $\epsilon = h^{-\frac{1}{8}} \epsilon_0$ of the solution satisfy the condition
\begin{equation}\label{d4ks}
	\Gamma_{23849}\epsilon_0 = \pm\epsilon_0~,
\end{equation}
where $\epsilon_0$ is a constant spinor and $h$ is a harmonic function as in \eqref{mhf} for $p = 4$. To solve this condition with the plus sign using spinorial geometry write
\begin{equation}\label{d4sol}
	\epsilon_0 = \eta^1 + e_{34}\wedge \eta^2+ e_3\wedge \lambda^1+ e_4\wedge \lambda^2~,
\end{equation}
where $\eta, \lambda \in \Lambda^*(\bC\langle e_5, e_1, e_2\rangle)$. Substituting this into \eqref{d4ks}, one finds that
\begin{equation}\label{rd4ks}
	\Gamma_2\eta^{1} = -\eta^{1}~, \quad \Gamma_2\lambda^{1} = -\lambda^{1}~,
\end{equation}
and similarly for $\eta^2$ and $\lambda^2$. The reality condition on $\epsilon$ implies that $\eta^1 = \Gamma_{67}* \eta^2$ and $\lambda^1 = \Gamma_{67}* \lambda^2$. The remaining conditions \eqref{rd4ks} can be solved by setting $\eta^1 = \rho-e_2\wedge \rho$, where $\rho\in \Lambda^*\langle e_5, e_1\rangle$, and similarly for the rest of the spinors. However as for the D2-brane, we shall not do this as otherwise the expression for the form bilinears will not be manifestly covariant in the worldvolume directions because the 6-th direction will have to be treated separately from the rest. The form bilinears of the D4-brane can be expressed in terms of those of $\eta$ and $\lambda$ spinors. Their expressions can be found in appendix \ref{Dbrane-bilinears}.

As in the D2-brane case, the form bilinears generate symmetries in the probe actions we have been considering provided that they are KY forms. This condition requires that certain terms in the TCFH must vanish. Using that for the D4-brane solution $G_{ijkl}\not = 0$ and the explicit expression of the form bilinears in appendix \ref{Dbrane-bilinears}, one finds after a detailed analysis of the TCFH that only $k$, $\tilde \zeta$, $\tau$, $\tilde \omega$ and $\pi$ can be KY tensors while the rest of the bilinears vanish. In particular, as expected, $k$ is Killing and so generates a symmetry for the probe actions we have been considering.

For $\tilde \zeta$ to be a KY tensor, the TCFH requires that $\tilde k = 0$, $\tau_{ija_1a_2a_3} = 0$ and $\tau_{a_1\dots a_5} = 0$. These imply that $\tilde \zeta_{ija_1a_2} = 0$. The non-vanishing component of $\tilde \zeta$, $\tilde\zeta = \frac{1}{4!} \tilde\zeta_{a_1\dots a_4} \e^{a_1}\wedge \dots \wedge \e^{a_4}$, generates a (hidden) symmetry for the probe action \eqref{1part} with $C = 0$.
Similarly for $\tau$ to be a KY form, the TCFH requires that $\omega = 0$ and $\tilde\zeta_{ijab} = 0$. These imply that $\tau = \frac{1}{5!} \tau_{a_1\dots\tau_5} \e^{a_1}\wedge \dots \wedge \e^{a_5}$ is a KY form and generates a (hidden) symmetry for the probe action \eqref{1part} with $C = 0$.

For $\tilde \omega$ to be a KY form the TCFH requires that $\tilde \pi_{ijk} = 0$, which in turn implies that $\tilde \omega_{ij} = 0$. The remaining component of $\tilde \omega = \frac{1}{2} \tilde\omega_{ab} \e^a\wedge \e^b$ is a KY tensor and generates a symmetry for probe action \eqref{1part} with $C = 0$. There are Killing spinors such that $\tilde \pi_{ijk} = 0$ while $\tilde \omega\not = 0$. Indeed take $\eta^{1r} = x 1 + y e_1-e_2\wedge (x1 + y e_1)$ and $\eta^{1s} = -i x 1 + i y e_1-e_2\wedge (-ix 1 + iy e_1)$, $x,y\in \bC-\{0\}$ and $\lambda^{1r} = \lambda^{1s} = 0$.

Similarly for $\pi$ to be a KY form, the TCFH requires that $\zeta_{aijk} = 0$, which in turn gives $\pi_{aij} = 0$. Then $\pi = \frac{1}{3!} \pi_{abc} \e^a\wedge \e^b\wedge \e^c$ is a KY form and generates a (hidden) symmetry for the probe action \eqref{1part} with $C = 0$. In all the above cases, the explicit expressions for the vanishing conditions on some of the components of the form bilinears, as well as the expressions of KY forms in terms of the Killing spinors, can be easily read from the results of appendix \ref{Dbrane-bilinears} and so they will not be repeated here. For $\tilde \zeta$, $\tau$ and $\pi$ we have not verified whether there exist Killing spinors such that these are non-vanishing KY forms. A preliminary investigation has revealed that they do not exist.

To summarise the results of this section, there are Killing spinors such that $k$, and $\tilde \omega$ with non-vanishing components only along the worldvolume directions of D4-brane, are KY tensors. Therefore, they generate (hidden) symmetries for the probe described by the action \eqref{1part} with $C = 0$ on any D4-brane background depending of a harmonic function $h$ as in \eqref{mhf} for $p = 4$.

\subsection {D8-brane}

To derive the TCFH on D8-brane type of backgrounds set all the IIA form fields strengths to zero apart from $S$. Then the IIA TCFH in section \ref{iiatcfhs} reduces to
\begin{equation}
	\nabla_M \tilde{\sigma} = \frac{1}{4}e^\Phi S \tilde{k}_M~, \quad 
		\nabla_M k_N = \frac{1}{4}e^\Phi S \omega_{MN} ~, \quad 
		\nabla_M \tilde{k}_N = \frac{1}{4}e^\Phi g_{MN}S\tilde{\sigma}~,
\end{equation}
\begin{equation}
	 \nabla_M \omega_{NR} = \frac{1}{2}e^\Phi S g_{M[N}k_{R]}~, \quad 
	\nabla_M \tilde{\zeta}_{N_1 \dots N_4} = \frac{1}{4 \cdot 5!}e^\Phi \hodge{S}_{MN_1\dots N_4P_1\dots P_5}\tau^{P_1\dots P_5} ~,
\end{equation}
\begin{equation}
	\begin{gathered}
		\nabla_M \tau_{N_1 \dots N_5} =- \frac{1}{4\cdot 4!}e^\Phi \hodge{S}_{MN_1\dots N_5 P_1 \dots P_4}\tilde{\zeta}^{P_1\dots P_4}~,~~ \nabla_M \sigma = 0 ~,~~ \\
		\nabla_M \tilde\omega_{NR} = \frac{1}{4} e^\Phi S \tilde\pi_{MNR}~,
	\end{gathered}
\end{equation}
\begin{equation}
	\begin{gathered}
		\nabla_M \pi_{NRS}= \frac{1}{4} e^\Phi S\zeta_{MNRS}~,~~
			\nabla_M \tilde\pi_{NRS}= \frac{3}{4} e^\Phi S g_{M[N} \tilde\omega_{RS]}~,~~ \\
			\nabla_M \zeta_{N_1 \dots N_4} = e^\Phi S g_{M[N_1} \pi_{N_2N_3N_4]}~.
	\end{gathered}
\end{equation}
It is clear from this that $k$, $\tilde\zeta$, $\tau$, $\tilde\omega$ and $\pi$ are KY tensors and generate a (hidden) symmetry of the probe action \eqref{1part} with $C = 0$. Note that all these form bilinears $k$, $\tilde\zeta$, $\tau$, $\tilde\omega$ and $\pi$ have components only along the worldvolume directions of the D8-brane. Notice also that the (reduced) holonomy of the minimal TCFH connection is included in $SO(9,1)$.

To find an explicit expression of the form bilinears of D8-brane solution choose the worldvolume directions along $012346789$. The Killing spinors $\epsilon= h^{-\frac{1}{8}} \epsilon_0$ of the solution satisfy the condition $\Gamma_5\epsilon_0 = \pm \epsilon_0$, where $\epsilon_0$ is a constant spinor and $h = 1 + \sum_\ell q_\ell |y-y_\ell|$. Taking the plus sign, this condition can be solved using spinorial geometry by setting
\begin{equation}\label{d8sol}
	\epsilon_0 = \eta+ e_5\wedge \eta~,
\end{equation}
where $\eta\in \Lambda^*(\bR\langle e_1, e_2, e_3, e_4\rangle)$ after imposing the reality condition $\Gamma_{6789} * \eta = \eta$. Using the solution for $\epsilon_0$ above, one can easily compute the form bilinears of D8-brane in terms of those of $\eta$. Their expressions can be found in appendix \ref{Dbrane-bilinears}. Imposing the condition that the remaining form bilinears $\tilde k$, $\zeta$, $\omega$ and $\tilde \pi$ must be KY forms, the TCFH together with their explicit expressions in \ref{Dbrane-bilinears} imply that they should vanish. Therefore they do not generate symmetries for probe actions. However as a consequence of the TCFH above $\tilde k$, $\zeta$, $\omega$ and $\tilde \pi$ are CCKY forms and so their spacetimes duals are KY forms.

\chapter{TCFHs and Hidden Symmetries of Type IIA AdS Backgrounds} \label{TCFH-AdS-Backgrounds}

\section{Introduction}

In this chapter we present the TCFH on the internal spaces of all warped AdS backgrounds of (massive) IIA supergravity \cite{romans}. In addition some of their properties are explored which include the reduced holonomy of the minimal connection for generic supersymmetric backgrounds. Next we investigate the question on whether some of the form bilinears generate symmetries for spinning particles propagating on such backgrounds. It is demonstrated that this is the case for a class of AdS backgrounds constructed using ansatze that include the near horizon geometries of some IIA intersecting brane configurations. This work completes the construction of TCFHs for all AdS backgrounds of type II supergravities in 10- and 11-dimensions.

This chapter has been organised as follows. In sections \ref{AdS2-TCFH}, \ref{AdS3-TCFH} and \ref{AdS4-TCFH}, the TCFH of warped IIA AdS$_k$, $k = 2,3,4$ backgrounds are presented. This also includes the investigation of some of the properties of the TCFH connections, such as their holonomy. In section \ref{AdSn-TCFH}, the TCFH of warped IIA AdS$_k$, $k = 5,6,7$ backgrounds are given. In section \ref{AdS-Probes}, we present some explicit examples where the TCFH generates symmetries for spinning particles propagating on the internal space of AdS$_2$ and AdS$_3$ backgrounds.

\section{The TCFH of Warped AdS\texorpdfstring{\textsubscript{2}}{2} Backgrounds} \label{AdS2-TCFH}

The approach that we shall follow below to construct the TCFHs on the internal spaces of all warped AdS backgrounds of massive IIA supergravity is based on the solution of the KSEs of the theory presented in \cite{ggkpiia, bgpiia}. In these works the KSEs of the theory are integrated over the AdS subspace of warped AdS backgrounds without any additional assumptions on the form of the Killing spinors. Then the remaining independent KSEs on the internal space of the AdS backgrounds are identified. A similar procedure is used for the field equations of the theory. The main advantage of this method is that it does not involve additional assumptions, such as a certain factorisation of Killing spinors, and so it is general. For a comparison of the different methods to solve the KSEs of warped AdS backgrounds see \cite{adsdes}.

\subsection{Fields and Killing Spinors}

Let $\Phi$ be the dilaton, and $G$, $H$, $F$ be the 4-, 3- and 2-form field strengths of (massive) IIA supergravity, respectively. The bosonic fields of a warped AdS$_2$ background, AdS$_2 \times_w M^8$, with internal space $M^8$ can be expressed as follows
\begin{equation}
	\begin{gathered}
		g = 2\, \e^+ \e^- + g(M^8), \quad
		G = \e^+ \wedge \e^- \wedge X + Y, \quad H = \e^+ \wedge \e^- \wedge W + Z, \\
		F = N \, \e^+ \wedge \e^- + P, \quad S = m e^\Phi, \quad \Phi = \Phi,
	\end{gathered}
\end{equation}
where $\Phi$ is a function on $M^8$, $\Phi \in C^\infty(M^8)$, $g(M^8)$ is a metric on the internal space $M^8$, and $N \in C^\infty(M^8)$, $W \in \Omega^1(M^8)$, $X, P \in \Omega^2(M^8)$, $Z \in \Omega^3(M^8)$ and $Y \in \Omega^4(M^8)$. For simplicity, we have denoted the spacetime dilaton and its restriction on $M^8$ with the same symbol. Moreover, $m$ is a constant\footnote{Viewing $m$ as a field and in the presence of D8-brane sources, $m$ can be taken as piecewise constant. The same applies in the description of other AdS backgrounds but we shall not elaborate on this below.} that is non-zero in massive IIA and vanishes in standard IIA supergravity. We have also introduced the pseudo-orthonormal (co-)frame
\begin{equation}\label{ads2frame}
		\e^+ = \dd u~, \quad \e^- = \dd r -2 r A^{-1} \dd A - \frac{1}{2} r^2 \ell^{-2} A^{-2} \dd u~, \quad \e^i = e^i{}_J \dd y^J~,
\end{equation}
on AdS$_2 \times_w M^8$, where $A\in C^\infty(M^8)$ is the warp factor, $\e^i$ is an orthonormal frame on $M^8$ that depends only on the coordinates $y$ of $M^8$, $g(M^8) = \delta_{ij} \e^i \e^j$, and $\ell$ is the radius of AdS\textsubscript{2}. Moreover $(u, r)$ are the remaining coordinates of the spacetime. It can be seen after a coordinate transformation that the spacetime metric $g$ can be put into the standard warped form $g = A^2 g_{\ell}(AdS_2) + g(M^8)$, where $g_{\ell}(AdS_2)$ is the standard metric on AdS$_2$ with radius $\ell$.

The KSEs of massive IIA supergravity for warped AdS$_2$ backgrounds have been integrated over the $(u,r)$ coordinates in \cite{ggkpiia, bgpiia}. In such a case, the Killing spinors can be expressed as $\epsilon = \epsilon(u,r, \eta_\pm)$, where $\eta_\pm$ are spinors that depend only on the coordinates of $M^8$ and satisfy $\Gamma_\pm\eta_\pm = 0$, where the gamma matrices $(\Gamma_+, \Gamma_-, \Gamma_i)$ are taken with respect to the frame \eqref{ads2frame}. The precise expression for $\epsilon$ in terms of $u,r$ and $\eta_\pm$, which can be found in appendix \ref{sec:Warped-AdS-Backgrounds}, is not essential in what follows and so it will not be presented here.
Furthermore, the conditions that gravitino KSE imposes on $\eta_\pm$ along $M^8$ are
\begin{equation} \label{AdS2KSE}
	\mathcal{D}_m^{(\pm)} \eta_\pm = 0~,
\end{equation}
where
\begin{equation} \label{AdS2_Connection}
	\begin{split}
			\mathcal{D}_m^{(\pm)}\eta_\pm = \nabla_m \eta_\pm\, \pm & \frac{1}{2} A^{-1} \partial_{m} A \,\eta_\pm \mp \frac{1}{16} \slashed{X} \Gamma_{m} \eta_\pm + \frac{1}{8 \cdot 4 !} \slashed{Y} \Gamma_{m} \eta_\pm + \frac{1}{8} S \Gamma_{m} \eta_\pm \\
		&+ \Gamma_{11}\left(\mp \frac{1}{4} W_{m} \eta_\pm + \frac{1}{8} \slashed{Z}_{m} \eta_\pm \pm \frac{1}{8} N \Gamma_{m} \eta_\pm - \frac{1}{16} \slashed{P} \Gamma_{m} \eta_\pm \right)~,
	\end{split}
\end{equation}
is the supercovariant connection\footnote{See appendix \ref{Conventions} for our conventions.} on $M^8$, $m = 1,\dots,8$ and $\nabla$ is the spin connection associated with the metric $g(M^8)$. These are clearly parallel transport equations for $\eta_\pm$. The Killing spinors $\eta_\pm$ satisfy additional conditions \cite{bgpiia} arising from the dilatino KSE of massive IIA supergravity. But these additional conditions are not essential for the TCFH below; however they will be used later when we discuss examples. For completeness we include these conditions in appendix \ref{sec:Warped-AdS-Backgrounds} as well as provide a brief summary in the relevant example.

\subsection{The TCFH on \texorpdfstring{M\textsuperscript{8}}{M8}} \label{sec:TCFH}

It has been demonstrated in \cite{gptcfh} that the conditions imposed on the Killing spinor bilinears by the gravitino KSE of any supergravity theory
can be organised as a TCFH. Here we shall focus on the TCFH associated with the form bilinears on $M^8$ constructed from the Killing spinors $\eta_\pm$ satisfying the KSEs \eqref{AdS2KSE}. Given two such Killing spinors $\eta_\pm^r$ and $\eta^s_\pm$, one can define the $k$-form bilinears
\begin{equation}
	\phi_\pm^{rs} = \frac{1}{k!} \Herm{\eta_\pm^r}{\Gamma_{i_1 \dots i_k} \eta_\pm^s}\, \e^{i_1} \wedge \dots \wedge \e^{i_k}~,\quad \tilde\phi_\pm^{rs} = \frac{1}{k!} \Herm{\eta_\pm^r}{\Gamma_{i_1 \dots i_k} \Gamma_{11} \eta_\pm^s}\, \e^{i_1} \wedge \dots \wedge \e^{i_k}~,
\label{fbi}
\end{equation}
where $\Herm{\cdot}{\cdot}$ denotes the spin-invariant inner product on $M^8$ for which the spacelike gamma matrices are Hermitian while the time-like ones are anti-Hermitian.

Because of the reality condition on $\eta_\pm$, which follows from that of IIA Killing spinors, the form bilinears are either symmetric or skew-symmetric on the exchange of $\eta^r$ and $\eta^s$. A basis in the space of form bilinears\footnote{Note that the form bilinears constructed from $\eta_+$ and $\eta_-$ spinors vanish.} on $M^8$, up to Hodge duality\footnote{Our conventions are given in appendix \ref{Conventions}.}, which are symmetric in the exchange of Killing spinors is
\begin{equation} \label{AdS2_Sym_Bilinears}
	\begin{gathered}
		f^{rs}_\pm = \Herm{\eta^r_\pm}{\eta^s_\pm} , \quad \tilde{f}^{rs}_\pm = \Herm{\eta^r_\pm}{\Gamma_{11} \eta^s_\pm} , \quad k^{rs}_\pm = \Herm{\eta^r_\pm}{\Gamma_ i\eta^s_\pm} \, \e^i~, \\
		\tilde{\pi}^{rs}_\pm = \frac{1}{3!} \Herm{\eta^r_\pm}{\Gamma_ {ijk}\Gamma_{11} \eta^s_\pm} \, \e^i \wedge \e^j \wedge \e^k, \quad \zeta^{rs}_\pm = \frac{1}{4!} \Herm{\eta^r_\pm}{\Gamma_{i_1 \dots i_4}\eta^s_\pm}\, \e^{i_1} \wedge \dots \wedge \e^{i_4}~.
	\end{gathered}
\end{equation}

To find the TCFH associated to the above form bilinears note that
\begin{equation}
	\nabla_m \phi_\pm{}^{rs}_{i_1 \dots i_k} = \Herm{\nabla_m \eta_\pm^r}{\Gamma_{i_1 \dots i_k} \eta_\pm^s} + \Herm{\eta_\pm^r}{\Gamma_{i_1 \dots i_k} \nabla_m \eta_\pm^s}~,
\end{equation}
and similarly for $\tilde \phi_\pm{}^{rs}$. Then using the KSEs \eqref{AdS2KSE}, one can replace in the right-hand-side of the above equation the derivatives on the spinors in terms of a Clifford algebra element constructed from the fluxes of the theory. After some extensive Clifford algebra computation, one can demonstrate that the right-hand-side can always be organised as a TCFH.

In particular, the TCFH of the form bilinears \eqref{AdS2_Sym_Bilinears}, with respect to the minimal connection\footnote{See \ref{sec:TCFH-CKY} for the definition.} $\mathcal{D}^\mathcal{F}$ is
\begin{equation}
	\begin{split}
		\mathcal{D}^\mathcal{F}_m f_\pm \defeq& \nabla_m f_\pm \\
		=& \mp A^{-1} \partial_{m} A \, f_\pm \mp \frac{1}{4} X_{mp}k_\pm{}^p \pm \frac{1}{4!} \hodge{Y}_{mpqr} \tilde{\pi}_\pm{}^{pqr} \\
		&- \frac{1}{4}Sk_\pm{}_m \pm\frac{1}{2}W_m\tilde{f}_\pm-\frac{1}{8}P_{pq}\tilde{\pi}_\pm{}^{pq}{}_m~,
	\end{split}
\end{equation}

\begin{equation}
	\begin{split}
		\mathcal{D}^\mathcal{F}_m \tilde{f}_\pm \defeq& \nabla_m \tilde{f}_\pm \\
		=& \mp A^{-1} \partial_{m} A \, \tilde{f}_\pm \mp \frac{1}{8} X_{pq}\tilde{\pi}_\pm{}^{pq}{}_m - \frac{1}{4!} Y_{mpqr} \tilde{\pi}_\pm{}^{pqr} \\
		&\pm \frac{1}{2} W_m f_\pm \mp \frac{1}{4}Nk_\pm {}_m - \frac{1}{4} P_{mp}k_\pm{}^p~,
	\end{split}
\end{equation}

\begin{equation}
	\begin{split}
		\mathcal{D}^\mathcal{F}_m k_\pm{}_i \defeq& \nabla_m k_\pm{}_i + \frac{1}{12}Y_{mpqr}\zeta_\pm{}^{pqr}{}_i + \frac{1}{4}Z_{mpq}\tilde{\pi}_\pm{}^{pq}{}_i \\
		=& \mp A^{-1} \partial_{m} A \, k_\pm{}_i \, \mp \frac{1}{8}X_{pq}\zeta_\pm{}^{pq}{}_{mi} \mp \frac{1}{4} X_{mi} f_\pm - \frac{1}{4 \cdot 4!} \delta_{mi}Y_{p_1 \dots p_4} \zeta_\pm{}^{p_1 \dots p_4} \\
		&+ \frac{1}{12}Y_{[m|pqr|}\zeta_\pm {}^{pqr}{}_{i]} - \frac{1}{4}\delta_{mi} Sf_\pm \pm \frac{1}{4}\delta_{mi} N \tilde{f}_\pm \mp \frac{1}{4 \cdot 4!} \hodge{P}_{mip_1\dots p_4}\zeta_\pm{}^{p_1 \dots p_4} \\
		&+ \frac{1}{4} P_{mi}\tilde{f}_\pm~,
	\end{split}
\end{equation}

\begin{equation}
	\begin{split}
		\mathcal{D}^\mathcal{F}_m \tilde{\pi}_\pm{}_{ijk} \defeq& \nabla_m \tilde{\pi}_\pm{}_{ijk} + \frac{1}{4} \hodge{X}_{m[ij|pqr|}\zeta_\pm{}^{pqr}{}_{k]} \pm \frac{3}{4} \hodge{Y}_{m[i|pq|}\zeta_\pm{}^{pq}{}_{jk]} \pm \frac{3}{4}\hodge{Z}_{m[ij|pq|}\tilde\pi_\pm{}^{pq}{}_{k]} \\
		&- \frac{3}{2}Z_{m[ij}k_\pm{}_{k]} - \frac{1}{2}P_{mp}\zeta_\pm{}^p{}_{ijk} \\
		=& \mp A^{-1} \partial_{m} A \, \tilde{\pi}_\pm{}_{ijk} \, \pm \frac{3}{4}\delta_{m[i}X_{jk]}\tilde{f}_\pm -\frac{1}{32} \delta_{m[i}\hodge{X}_{jk]p_1 \dots p_4}\zeta_\pm{}^{p_1 \dots p_4} \\
		&+ \frac{1}{6} \hodge{X}_{[mij|pqr|}\zeta_\pm{}^{pqr}{}_{k]} \pm \frac{1}{4}\hodge{Y}_{mijk} f_\pm + \frac{1}{4} Y_{mijk}\tilde{f}_\pm \\ 
		&\pm \frac{1}{4}\delta_{m[i}\hodge{Y}_{j|pqr|}\zeta_\pm {}^{pqr}{}_{k]} \pm \frac{3}{4}\hodge{Y}_{[mi|pq|}\zeta_\pm{}^{pq}{}_{jk]} \\
		&\pm\frac{1}{4 \cdot 4!} \hodge{S}_{mijkp_1\dots p_4}\zeta_\pm {}^{p_1\dots p_4} \pm \frac{1}{4} \delta_{m[i}\hodge{Z}_{jk]pqr} \tilde\pi_\pm {}^{pqr} \\
		& \pm \hodge{Z}_{[mij|pqr|}\tilde\pi_\pm{}^{pq}{}_{k]} \pm \frac{1}{4} N\zeta_\pm{}_{mijk} + \frac{3}{8} \delta_{m[i|}P_{pq|}\zeta_\pm{}^{pq}{}_{jk]} \\
		&- P_{[m|p|}\zeta_\pm{}^p{}_{ijk]} - \frac{3}{4}\delta_{m[i}P_{jk]}f_\pm~,
	\end{split}
\end{equation}

\begin{equation}
	\begin{split}
		\mathcal{D}^\mathcal{F}_m \zeta_\pm{}_{i_1 \dots i_4} \defeq& \nabla_m \zeta_\pm{}_{i_1 \dots i_4} -\hodge{X}_{m[i_1i_2i_3|pq|}\tilde\pi_\pm{}^{pq}{}_{i_4]} - 2 Y_{m[i_1i_2i_3}k_\pm{}_{i_4]} \pm \,3 \hodge{Y}_{m[i_1i_2|p|}\tilde\pi_\pm{}^p{}_{i_3i_4]} \\
		&+ \frac{1}{2 \cdot 4!}W_m \epsilon_{i_1 \dots i_4}{}^{j_1 \dots j_4}\zeta_\pm{}_{j_1 \dots j_4} \pm \frac{3}{2} \hodge{Z}_{m[i_1i_2|pq|}\zeta_\pm{}^{pq}{}_{i_3i_4]} + 2 P_{m[i_1}\tilde\pi_\pm{}_{i_2i_3i_4]} \\
		=& \mp A^{-1} \partial_{m} A \, \zeta_\pm{}_{i_1 \dots i_4} \, \pm 3\delta_{m[i_1}X_{i_2i_3}k_\pm{}_{i_4]} -\frac{1}{6}\delta_{m[i_1}\hodge{X}_{i_2i_3i_4]pqr}\tilde\pi_\pm{}^{pqr} \\
		&- \frac{5}{8}\hodge{X}_{[mi_1i_2i_3|pq|}\tilde\pi_\pm{}^{pq}{}_{i_4]} - \delta_{m[i_1}Y_{i_2i_3i_4]p}k_\pm{}^p - \frac{5}{4}Y_{[mi_1i_2i_3} k_\pm{}_{i_4]} \\
		&\pm \frac{5}{2} \hodge{Y}_{[mi_1i_2|p|}\tilde\pi_\pm{}^p{}_{i_3i_4]} \mp \frac{3}{2}\delta_{m[i_1}\hodge{Y}_{i_2i_3|pq|}\tilde\pi_\pm{}^{pq}{}_{i_4]} \\
		&\pm \frac{1}{24} \hodge{S}_{mi_1\dots i_4 pqr}\tilde\pi_\pm{}^{pqr} \pm \delta_{m[i_1}\hodge{Z}_{i_2 i_3|pqr|}\zeta_\pm{}^{pqr}{}_{i_4]} \pm \frac{5}{2}\hodge{Z}_{[mi_1i_2|pq|}\zeta_\pm{}^{pq}{}_{i_3i_4]} \\
		&\mp N\delta_{m[i_1}\tilde\pi_\pm{}_{i_2i_3i_4]} \mp \frac{1}{4} \hodge{P}_{mi_1\dots i_4 p}k_\pm{}^p + 3\delta_{m[i_1}P_{i_2|p|}\tilde\pi_\pm{}^p{}_{i_3i_4]} + \frac{5}{2}P_{[mi_1}\tilde\pi_\pm{}_{i_2i_3i_4]}~,
	\end{split}
\end{equation}
where for simplicity we have suppressed the $r,s$ indices on the form bilinears that label the different Killing spinors. It is clear that the above conditions on the form bilinears are of the form of a TCFH.

A basis in the space of form bilinears on $M^8$, up to Hodge duality, which are skew-symmetric in the exchange of $\eta^r$ and $\eta^s$ is the following
\begin{equation} \label{AdS2_Skew_Bilinears}
	\begin{gathered}
		\tilde{k}^{rs}_{\pm} = \Herm{\eta^r_\pm}{\Gamma_i \Gamma_{11} \eta^s_\pm} \, \e^i~, \quad \omega^{rs}_\pm = \frac{1}{2} \Herm{\eta^r_\pm}{\Gamma_{ij}\eta^s_\pm} \, \e^i \wedge \e^j~, \\
		\tilde{\omega}^{rs}_\pm = \frac{1}{2} \Herm{\eta^r_\pm}{\Gamma_{ij}\Gamma_{11}\eta^s_\pm} \, \e^i \wedge \e^j~, \quad	\pi^{rs}_\pm = \frac{1}{3!} \Herm{\eta^r_\pm}{\Gamma_ {ijk}\eta^s_\pm} \, \e^i \wedge \e^j \wedge \e^k~.
	\end{gathered}
\end{equation}
The associated TCFH with respect to the minimal connection, $\mathcal{D}^\mathcal{F}$, is given by
\begin{equation}
	\begin{split}
		\mathcal{D}^\mathcal{F}_m \tilde k_\pm{}_i \defeq& \nabla_m \tilde k_\pm{}_i \pm \frac{1}{2} X_{mp} \tilde\omega_\pm{}^p{}_i + \frac{1}{4} Z_{mpq} \pi_\pm{}^{pq}{}_i - \frac{1}{2} P_{mp} \omega_\pm{}^p{}_i \\
		=& \mp A^{-1} \partial_{m} A \, \tilde{k}_\pm{}_i \, \mp \frac{1}{8} \delta_{mi} X_{pq} \tilde\omega_\pm{}^{pq} \pm \frac{1}{2} X_{[m|p|}\tilde\omega_\pm{}^p{}_{i]} \\
		& \mp \frac{1}{8} \hodge{Y}_{mipq} \omega_\pm {}^{pq} -\frac{1}{8} Y_{mipq} \tilde\omega_\pm{}^{pq} - \frac{1}{4} S \tilde\omega_\pm{}_{mi} \\
		& \pm \frac{1}{4} N \omega_\pm{}_{mi} + \frac{1}{8} \delta_{mi}P_{pq} \omega_\pm{}^{pq} - \frac{1}{2} P_{[m|p|}\omega_\pm{}^p{}_{i]}~,
	\end{split}
\end{equation}

\begin{equation}
	\begin{split}
		\mathcal{D}^\mathcal{F}_m \omega_\pm{}_{ij} \defeq& \nabla_m \omega_\pm{}_{ij} \pm \frac{1}{2} X_{mp} \pi_\pm{}^p{}_{ij} + \frac{1}{2} Y_{m[i|pq|}\pi_\pm{}^{pq}{}_{j]} \mp \frac{1}{2} W_m \tilde\omega_{\pm ij} \\
		&+ Z_{m[i|p|}\tilde\omega_\pm{}^p{}_{j]} + P_{m[i} \tilde k_{\pm j]} \\
		=& \mp A^{-1} \partial_{m} A \, \omega_\pm{}_{ij} \, \mp \frac{1}{4} \delta_{m[i}X_{|pq|}\pi_\pm{}^{pq}{}_{j]} \pm \frac{3}{4} X_{[m|p|}\pi_\pm{}^p{}_{ij]} \\
		& \mp \frac{1}{4} \hodge{Y}_{mijp} \tilde k_\pm{}^p + \frac{1}{12} \delta_{m[i}Y_{j]pqr}\pi_\pm{}^{pqr} + \frac{3}{8} Y_{[mi|pq|}\pi_\pm{}^{pq}{}_{j]} \\
		&- \frac{1}{4} S \pi_{\pm mij} \mp \frac{1}{2} N \delta_{m[i}\tilde k_{\pm j]} \pm \frac{1}{4!} \hodge{P}_{mijpqr} \pi_\pm{}^{pqr} \\
		&+ \frac{1}{2} \delta_{m[i}P_{j]p}\tilde k_\pm{}^p + \frac{3}{4} P_{[mi} \tilde k_{\pm j]}~,
	\end{split}
\end{equation}

\begin{equation}
	\begin{split}
		\mathcal{D}^\mathcal{F}_m \tilde\omega_\pm{}_{ij} \defeq& \nabla_m \tilde\omega_\pm{}_{ij} \pm X_{m[i} \tilde k_{\pm j]} \mp \frac{1}{2} \hodge{Y}_{m[i|pq|}\pi_\pm{}^{pq}{}_{j]} \mp \frac{1}{2} W_m \omega_{\pm ij} \\
		&+ Z_{m[i|p|}\omega_\pm{}^p{}_{j]} + \frac{1}{2} P_{mp} \pi_\pm{}^p{}_{ij} \\
		=& \mp A^{-1} \partial_{m} A \, \tilde\omega_\pm{}_{ij} + \frac{1}{4!} \hodge{X}_{mijpqr} \pi_\pm{}^{pqr} \pm \frac{1}{2} \delta_{m[i} X_{j]p} \tilde k_\pm{}^p\pm \frac{3}{4} X_{[mi} \tilde k_{\pm j]} \\
		& \mp \frac{1}{12} \delta_{m[i} \hodge{Y}_{j]pqr} \pi_\pm{}^{pqr} \mp \frac{3}{8} \hodge{Y}_{[mi|pq|}\pi_\pm{}^{pq}{}_{j]} + \frac{1}{4} Y_{mijp} \tilde k_\pm{}^p \\
		&- \frac{1}{2} S \delta_{m[i} \tilde k_{\pm j]} \mp \frac{1}{4} N \pi_{\pm mij} - \frac{1}{4} \delta_{m[i}P_{|pq|} \pi_\pm{}^{pq}{}_{j]} + \frac{3}{4} P_{[m|p|}\pi_\pm{}^p{}_{ij]}~,
	\end{split}
\end{equation}

\begin{equation}
	\begin{split}
		\mathcal{D}^\mathcal{F}_m \pi_{\pm}{}_{ijk} \defeq& \nabla_m \pi_{\pm}{}_{ijk} \pm \frac{3}{2} X_{m[i}\omega_{\pm jk]} - \frac{3}{2} Y_{m[ij|p|}\omega_\pm{}^p{}_{k]}\mp \frac{3}{2} \hodge{Y}_{m[ij|p|} \tilde\omega_\pm{}^p{}_{k]} \\
		&\pm \frac{3}{4} \hodge{Z}_{m[ij|pq|} \pi_\pm{}^{pq}{}_{k]} - \frac{3}{2} Z_{m[ij} \tilde k_{\pm k]} - \frac{3}{2} P_{m[i} \tilde\omega_{\pm jk]}\\
		=& \mp A^{-1} \partial_{m} A \, \pi_{\pm}{}_{ijk} - \frac{1}{8} \hodge{X}_{mijkpq} \tilde \omega_\pm{}^{pq} \pm \frac{3}{2} \delta_{m[i}X_{j|p|}\omega_\pm{}^p{}_{k]} \pm \frac{3}{2} X_{[mi}\omega_{\pm jk]} \\
		&+ \frac{3}{8} \delta_{m[i}Y_{jk]pq}\omega_\pm{}^{pq} - Y_{[mij|p|}\omega_\pm{}^p{}_{k]} \pm \frac{3}{8} \delta_{m[i}\hodge{Y}_{jk]pq}\tilde\omega_\pm{}^{pq} \mp \hodge{Y}_{[mij|p|} \tilde\omega_\pm{}^p{}_{k]} \\
		&- \frac{3}{4} S \delta_{m[i}\omega_{\pm jk]} \pm\frac{1}{4} \delta_{m[i} \hodge{Z}_{jk]pqr} \pi_\pm{}^{pqr} \pm \hodge{Z}_{[mij|pq|}\pi_\pm{}^{pq}{}_{k]} \pm \frac{3}{4} N \delta_{m[i} \tilde \omega_{\pm jk]} \\
		&\pm \frac{1}{8} \hodge{P}_{mijkpq} \omega_\pm{}^{pq} - \frac{3}{2} \delta_{m[i}P_{j|p|} \tilde \omega_\pm{}^p{}_{k]} - \frac{3}{2} P_{[mi} \tilde\omega_{\pm jk]}~,
	\end{split}
\end{equation}
where for simplicity we have suppressed the $r,s$ indices on the form bilinears that label the different Killing spinors. Again the above conditions on the form bilinears have been organised as those of a TCFH. 

As it is apparent from the analysis above, the domain of the minimal TCFH connection $\mathcal{D}^\mathcal{F}$ can be identified with $\Omega^*(M^8)$. This is the span of $\phi$ and the Hodge dual of $\tilde\phi$ form bilinears\footnote{Note that $\zeta$ and $\tilde\zeta$ are Hodge duals and so only $\zeta$ is chosen to belong in the basis.}. This domain factorises into the space of symmetric form bilinears, \eqref{AdS2_Sym_Bilinears} and the space of skew-symmetric form bilinears, \eqref{AdS2_Skew_Bilinears}. This can be understood as follows. The spinors $\eta_\pm$ can be viewed as Majorana $\mathfrak{spin}(8)$ spinors. The product of two Majorana $\mathfrak{spin}(8)$ representations, $\Delta_{16}$, decomposes as
\begin{equation}
	\otimes^2 \Delta_{16} = \Lambda^*(\mathbb{R}^8)~,
\end{equation}
and so the space of form bilinears spans all forms over $M^8$, where $\oplus_{k = 0}^4 \Lambda^k(\mathbb{R}^8)$ is associated with the span of $\phi$ form bilinears while $\oplus_{k = 5}^8 \Lambda^k(\mathbb{R}^8)$ is associated with the span of the Hodge duals of the $\tilde\phi$ form bilinears. Indeed, we note that $\textrm{dim}(\otimes^2 \Delta_{16}) = 2^4 \cdot 2^4 = \textrm{dim}(\Lambda^*(\mathbb{R}^8))$. Thus $\mathcal{D}^\mathcal{F}$ acts on the space of all forms on $M^8$. However, we see that the minimal TCFH connection preserves the subspaces of form bilinears that are symmetric and skew-symmetric in the exchange of the two Killing spinors respectively, i.e. it preserves the symmetrised $S^2(\Delta_{16})$ and skew-symmetrised $\Lambda^2(\Delta_{16})$ subspaces of $\otimes^2 \Delta_{16}$. Therefore, the reduced holonomy of $\mathcal{D}^\mathcal{F}$ will be contained within the connected component\footnote{The reduced holonomy of a connection is by definition connected. So from now on when we refer to a group in the context of reduced holonomy we shall consider only its connected component even if this is not explicitly mentioned. } of $GL(136) \times GL(120)$. However, the reduced holonomy of the minimal TCFH connection reduces further to $GL(134) \times GL(120)$ as it acts with partial derivatives on the scalars $f$ and $\tilde{f}$ and so their contribution to the holonomy is trivial.

\section{The TCFH of Warped AdS\texorpdfstring{\textsubscript{3}}{3} Backgrounds} \label{AdS3-TCFH}

\subsection{Fields and Killing Spinors}

The bosonic fields of warped AdS$_3$ backgrounds, AdS$_3 \times_w M^7$, with internal space $M^7$ of massive IIA supergravity can be expressed as
\begin{equation}\label{ads3back}
	\begin{gathered}
		g = 2\, \e^+ \e^- + (\e^z)^2 + g(M^7), \\
		G = \e^+ \wedge \e^- \wedge \e^z \wedge X + Y, \quad H = W\, \e^+ \wedge \e^- \wedge \e^z + Z, \\
		F = F, \quad S = m e^\Phi, \quad \Phi = \Phi,
	\end{gathered}
\end{equation}
where $m$ is a constant, $g(M^7)$ is a metric on $M^7$, $\Phi, W \in C^\infty(M^7)$, $X \in \Omega^1(M^7)$, $F \in \Omega^2(M^7)$, $Z \in \Omega^3(M^7)$ and $Y \in \Omega^4(M^7)$. Note that the Bianchi identities imply that either $S = 0$ or $W = 0$. From now on, to simplify the notation, whenever a form field strength has non-vanishing components only along the internal space, the components along the internal space will be denoted with the same symbol as the spacetime field, e.g. as in $F = F$ in \eqref{ads3back}. Further,
\begin{equation}
	\begin{gathered}
		\e^+ = \dd u, \quad \e^- = \dd r - \frac{2}{\ell} r \dd z - 2 r A^{-1} \dd A, \quad \e^z = A \dd z~, \quad \e^i = e^i{}_J \dd y^J~,
	\end{gathered}
\end{equation}
is a pseudo-orthonormal frame on AdS$_3 \times_w M^7$ with $g(M^7) = \delta_{ij} \e^i \e^j$, where $y$ are the coordinates of the internal space and $(u, r, z)$ are the remaining coordinates of spacetime. After a coordinate transformation, the spacetime metric takes the standard warped form $g = A^2 g_\ell(AdS_3) + g(M^7)$ with warp factor $A$, $A\in C^\infty(M^7)$, where $g_\ell(AdS_3)$ is the standard metric on AdS$_3$ of radius $\ell$.

As in the previous case, the KSEs of warped AdS\textsubscript{3} backgrounds can be integrated over the coordinates $(u,r,z)$, see \cite{bgpiia}. The Killing spinors can be written schematically as $\epsilon = \epsilon(u,r,z, \sigma_\pm, \tau_\pm)$, where the spinors $\sigma_\pm$ and $\tau_\pm$ depend only on the coordinates of $M^7$ and satisfy $\Gamma_\pm\sigma_\pm = \Gamma_\pm \tau_\pm = 0$. Moreover, the gravitino KSE implies that $\mathcal{D}_m^{(\pm)} \chi_\pm = 0$, where
\begin{equation}\label{supconeads3}
	\begin{split}
			\mathcal{D}_m^{(\pm)} = \nabla_m \, \pm & \frac{1}{2} A^{-1} \partial_{m} A \, + \frac{1}{8} \slashed{Z}_{m} \Gamma_{11} + \frac{1}{8} S \Gamma_{m} \\&+ \frac{1}{16} \slashed{F}\Gamma_{m}\Gamma_{11} + \frac{1}{192} \slashed{Y} \Gamma_{m} \pm \frac{1}{8} \slashed{X} \Gamma_{z m}~,
	\end{split}
\end{equation}
is the supercovariant connection along the internal space $M^7$, $m = 1,\dots, 7$, $\nabla$ is the spin connection associated with the metric $g(M^7)$ and $\chi_\pm$ stands for either $\sigma_\pm$ or $\tau_\pm$.

The Killing spinors $\chi_\pm$ satisfy two algebraic KSEs \cite{bgpiia} in addition to the gravitino KSE along $M^7$. One of these is induced by the dilatino KSE of massive IIA supergravity. The other arises during the integration of the gravitino KSE of massive IIA supergravity over the $z$ spacetime coordinate. We shall not describe these here as they are not essential for the description of the TCFH on $M^7$. They are necessary for the correct counting of Killing spinors in the examples that follow however they have been presented in \ref{sec:Warped-AdS-Backgrounds}.

For warped AdS$_3$ backgrounds, the $\sigma_\pm$ and $\tau_\pm$ spinors are independent, i.e. there is no a priori Clifford algebra operation that relates the $\sigma_\pm$ solutions of the KSEs to the $\tau_\pm$ ones. A well known consequence of this is that the symmetry superalgebra of warped AdS$_3$ backgrounds factorises into a left and right sector that commute with each other. As we shall mention later, this is no longer the case for warped AdS$_k$, $k>3$, backgrounds where the $\sigma_\pm$ and $\tau_\pm$ Killing spinors are related with Clifford algebra operations.

\subsection{The TCFH on \texorpdfstring{M\textsuperscript{7}}{M7}}

Given Killing spinors $\chi_\pm^r$ and $\chi_\pm^s$, the form bilinears on $M^7$ can be constructed as for AdS$_2$ backgrounds in \eqref{fbi} with $\eta_\pm$ replaced with $\chi_\pm$. However there are differences. One is that now $\e^i$ is an orthonormal frame on $M^7$ instead on $M^8$ as was the case for AdS$_2$ backgrounds. The other is that one can also insert in addition to $\Gamma_{11}$ the gamma matrix $\Gamma_z$ in the form bilinears. Again, the reality condition on $\chi_\pm$ implies that the form bilinears are either symmetric or skew-symmetric in the exchange of $\chi_\pm^r$ and $\chi_\pm^s$.

A basis in the space of form bilinears\footnote{The TCFHs associated with the form bilinears constructed from the pairs $(\sigma_+, \tau_+)$ and $(\sigma_+, \sigma_+)$ (and $(\sigma_-, \tau_-)$ and $(\sigma_-, \sigma_-)$) are identical as the supercovariant connection \eqref{supconeads3} on $\sigma_\pm$ is identical to that on $\tau_\pm$. So it is sufficient to consider only the TCFHs of the form bilinears constructed from the pairs $(\sigma_+,\sigma_+)$ and $(\sigma_-,\sigma_-)$.} on $M^7$, up to Hodge duality, which are symmetric in the exchange of Killing spinors $\chi_\pm^r$ and $\chi_\pm^s$ is
\begin{equation} \label{AdS3_Sym_Bilinears}
	\begin{gathered}
		f^{rs}_\pm = \Herm{\chi^r_\pm}{\chi^s_\pm} , \quad \tilde{f}^{rs}_\pm = \Herm{\chi^r_\pm}{\Gamma_{11} \chi^s_\pm} , \quad	\hat f^{rs}_{\pm} = \Herm{\chi^r_\pm}{\Gamma_ z\chi^s_\pm} , \\
		k^{rs}_\pm = \Herm{\chi^r_\pm}{\Gamma_ i\chi^s_\pm} \, \e^i, \quad	 \mathring \omega^{rs}_{\pm} = \frac{1}{2} \Herm{\chi^r_\pm}{\Gamma_ {ijz}\Gamma_{11} \chi^s_\pm} \, \e^i \wedge \e^j , \\
		\tilde{\pi}^{rs}_\pm = \frac{1}{3!} \Herm{\chi^r_\pm}{\Gamma_ {ijk}\Gamma_{11} \chi^s_\pm} \, \e^i \wedge \e^j \wedge \e^k, \quad		\hat \pi^{rs}_{\pm} = \frac{1}{3!} \Herm{\chi^r_\pm}{\Gamma_{ijkz}\chi^s_\pm} \, \e^{i} \wedge \e^j \wedge \e^k , \\
		\mathring \pi^{rs}_{\pm} = \frac{1}{3!} \Herm{\chi^r_\pm}{\Gamma_{ijkz}\Gamma_{11}\chi^s_\pm} \, \e^{i} \wedge \e^j \wedge \e^k~.
	\end{gathered}
\end{equation}
The computation of the TCFH follows the steps described in section \ref{sec:TCFH}. In particular the TCFH expressed in terms of the minimal connection, $\mathcal{D}^\mathcal{F}$, is

\begin{equation}
	\begin{split}
		\mathcal{D}^\mathcal{F}_m f_\pm \defeq& \nabla_m f_\pm \\
		=& \mp A^{-1} \partial_{m} A \, f_\pm - \frac{1}{4}Sk_\pm{}_m -\frac{1}{8}F_{pq}\tilde{\pi}_\pm{}^{pq}{}_m \pm \frac{1}{8}\hodge{Y}_{mpq}\mathring \omega_{\pm}{}^{pq} \pm \frac{1}{4}X_m \hat f_{\pm}~,
	\end{split}
\end{equation}

\begin{equation}
	\begin{split}
		\mathcal{D}^\mathcal{F}_m \tilde{f}_\pm \defeq& \nabla_m \tilde{f}_\pm \\
		=& \mp A^{-1} \partial_{m} A \, \tilde{f}_\pm - \frac{1}{4} F_{mp}k_\pm{}^p - \frac{1}{4!} Y_{mpqr} \tilde{\pi}_\pm{}^{pqr} \mp \frac{1}{4} X_p \mathring \omega_{\pm}{}^p{}_m~,
	\end{split}
\end{equation}

\begin{equation}
	\begin{split}
		\mathcal{D}^\mathcal{F}_m \hat f_{\pm} \defeq& \nabla_m \hat f_{\pm} \\
		=& \mp A^{-1} \partial_{m} A \, \hat f_{\pm} -\frac{1}{4} Z_{mpq}\mathring \omega_{\pm}{}^{pq} + \frac{1}{8} F_{pq}\mathring \pi_{\pm}{}^{pq}{}_m - \frac{1}{4!}Y_{mpqr}\hat \pi_{\pm}{}^{pqr} \pm \frac{1}{4}X_m f_\pm~,
	\end{split}
\end{equation}

\begin{equation}
	\begin{split}
		\mathcal{D}^\mathcal{F}_m k_\pm{}_i \defeq& \nabla_m k_\pm{}_i + \frac{1}{4}Z_{mpq}\tilde{\pi}_\pm{}^{pq}{}_i \mp \frac{1}{4}\hodge{Y}_{mpq}\mathring \pi_{\pm}{}^{pq}{}_i\\
		=& \mp A^{-1} \partial_{m} A \, k_\pm{}_i \, -\frac{1}{4}\delta_{mi} Sf_\pm \mp \frac{1}{4!} \hodge{F}_{mipqr}\hat \pi_{\pm}{}^{pqr} + \frac{1}{4} F_{mi}\tilde{f}_\pm \\
		& \mp \frac{1}{4!} \delta_{mi} \hodge{Y}_{pqr} \mathring \pi_{\pm}{}^{pqr} \mp \frac{1}{4} \hodge{Y}_{[m|pq|}\mathring \pi_{\pm}{}^{pq}{}_{i]} \pm \frac{1}{4}X_{p}\hat \pi_{\pm}{}^{p}{}_{mi}~,
	\end{split}
\end{equation}

\begin{equation}
	\begin{split}
		\mathcal{D}^\mathcal{F}_m \mathring \omega_{\pm}{}_{ij} \defeq& \nabla_m \mathring \omega_{\pm}{}_{ij} \mp \frac{1}{2}\hodge{Z}_{m[i|pq|}\tilde\pi_\pm{}^{pq}{}_{j]} - \frac{1}{2} F_{mp}\hat \pi_{\pm}{}^p{}_{ij} + \frac{1}{2} Y_{m[i|pq|} \mathring \pi_{\pm}{}^{pq}{}_{j]} \\
		=& \mp A^{-1} \partial_{m} A \, \mathring \omega_{\pm}{}_{ij} \mp \frac{1}{6} \delta_{m[i}\hodge{Z}_{j]pqr}\tilde\pi_\pm{}^{pqr} \mp \frac{3}{4}\hodge{Z}_{[mi|pq|}\tilde\pi_\pm{}^{pq}{}_{j]} \\
		 &+ \frac{1}{2} Z_{mij}\hat f_{\pm} -\frac{1}{4} S\mathring \pi_{\pm}{}_{mij} + \frac{1}{4}\delta_{m[i|}F_{pq|}\hat \pi_{\pm}{}^{pq}{}_{j]} \\
		 &- \frac{3}{4}F_{[m|p|}\hat \pi_{\pm}{}^p{}_{ij]} + \frac{1}{12} \delta_{m[i}Y_{j]pqr} \mathring \pi_{\pm}{}^{pqr} + \frac{3}{8} Y_{[mi|pq|}\mathring \pi_{\pm}{}^{pq}{}_{j]} \\
		&\pm \frac{1}{4} \hodge{Y}_{mij}f_\pm + \frac{1}{4!} \hodge{X}_{mijpqr}\hat \pi_{\pm}{}^{pqr} \mp \frac{1}{2} \delta_{m[i}X_{j]}\tilde{f}_\pm~,
	\end{split}
\end{equation}

\begin{equation}
	\begin{split}
		\mathcal{D}^\mathcal{F}_m \tilde{\pi}_{\pm}{}_{ijk} \defeq& \nabla_m \tilde{\pi}_{\pm}{}_{ijk} \mp \frac{3}{2} \hodge{Z}_{m[ij|p|}\mathring \omega_{\pm}{}^p{}_{k]} - \frac{3}{2}Z_{m[ij}k_\pm{}_{k]} \pm \frac{3}{4} \hodge{F}_{m[ij|pq|}\mathring \pi_{\pm}{}^{pq}{}_{k]} \\
		&\pm \frac{3}{2}\hodge{Y}_{m[i|p|}\hat \pi_{\pm}{}^p{}_{jk]} \pm \frac{1}{2}X_m \mathring \pi_{\pm}{}_{ijk} \\
		=& \mp A^{-1} \partial_{m} A \, \tilde{\pi}_{\pm}{}_{ijk} \mp 2 \hodge{Z}_{[mij|p|}\mathring \omega_{\pm}{}^p{}_{k]} \pm \frac{3}{4} \delta_{m[i}\hodge{Z}_{jk]pq}\mathring \omega_{\pm}{}^{pq} \\
		&\pm \frac{1}{4!} \hodge{S}_{mijkpqr}\hat \pi_{\pm}{}^{pqr} \pm \frac{1}{8} \delta_{m[i}\hodge{F}_{jk]pqr}\mathring \pi_{\pm}{}^{pqr} \pm \frac{1}{2} \hodge{F}_{[mij|pq|}\mathring \pi_{\pm}{}^{pq}{}_{k]} \\
		&- \frac{3}{4}\delta_{m[i}F_{jk]}f_\pm \mp \frac{3}{4}\delta_{m[i}\hodge{Y}_{j|pq|}\hat \pi_{\pm}{}^{pq}{}_{k]} \pm \frac{3}{2} \hodge{Y}_{[mi|p|}\hat \pi_{\pm}{}^p{}_{jk]} \\
		&+ \frac{1}{4}Y_{mijk}\tilde{f}_\pm \pm X_{[m}\mathring \pi_{\pm}{}_{ijk]} \pm \frac{3}{4} \delta_{m[i|}X_{p|}\mathring \pi_{\pm}{}^p{}_{jk]}~,
	\end{split}
\end{equation}

\begin{equation}
	\begin{split}
		\mathcal{D}^\mathcal{F}_m \hat \pi_{\pm}{}_{ijk} \defeq& \nabla_m \hat \pi_{\pm}{}_{ijk} + \frac{3}{2} Z_{m[i|p|}\mathring \pi_{\pm}{}^p{}_{jk]} + \frac{3}{2}F_{m[i}\mathring \omega_{\pm}{}_{jk]} \mp \frac{3}{2}\hodge{Y}_{m[i|p|}\tilde\pi_\pm{}^p{}_{jk]} \\
		=& \mp A^{-1} \partial_{m} A \, \hat \pi_{\pm}{}_{ijk} \mp \frac{1}{4!} \hodge{S}_{mijkpqr}\tilde\pi_\pm{}^{pqr} \pm \frac{1}{4} \hodge{F}_{mijkp}k_\pm{}^p \\
		&+ \frac{3}{2}F_{[mi}\mathring \omega_{\pm}{}_{jk]} + \frac{3}{2}\delta_{m[i}F_{j|p|}\mathring \omega_{\pm}{}^p{}_{k]} \pm \frac{3}{4} \delta_{m[i} \hodge{Y}_{j|pq|} \tilde\pi_\pm{}^{pq}{}_{k]} \\
		&\mp \frac{3}{2} \hodge{Y}_{[mi|p|} \tilde\pi_\pm{}^p{}_{jk]} + \frac{1}{4}Y_{mijk}\hat f_{\pm} - \frac{1}{8}\hodge{X}_{mijkpq}\mathring \omega_{\pm}{}^{pq} \pm \frac{3}{2}\delta_{m[i}X_jk_{\pm}{}_{k]}~,
	\end{split}
\end{equation}

\begin{equation}
	\begin{split}
		\mathcal{D}^\mathcal{F}_m \mathring \pi_{\pm}{}_{ijk} \defeq& \nabla_m \mathring \pi_{\pm}{}_{ijk} + \frac{3}{2} Z_{m[i|p|}\hat \pi_{\pm}{}^p{}_{jk]} \pm \frac{3}{4} \hodge{F}_{m[ij|pq|}\tilde\pi_\pm{}^{pq}{}_{k]} - \frac{3}{2} Y_{m[ij|p|} \mathring \omega_{\pm}{}^p{}_{k]} \\
		&\mp \frac{3}{2} \hodge{Y}_{m[ij}k_{\pm}{}_{k]} \pm \frac{1}{2} X_m \tilde\pi_\pm{}_{ijk} \\
		=& \mp A^{-1} \partial_{m} A \, \mathring \pi_{\pm}{}_{ijk} - \frac{3}{4} S \delta_{m[i}\mathring \omega_{\pm}{}_{jk]} + \frac{3}{4} \delta_{m[i} F_{jk]}\hat f_{\pm} \pm \frac{1}{8} \delta_{m[i}\hodge{F}_{jk]pqr}\tilde\pi_\pm{}^{pqr} \\
		& \pm \frac{1}{2} \hodge{F}_{[mij|pq|} \tilde\pi_\pm{}^{pq}{}_{k]} + \frac{3}{8} \delta_{m[i}Y_{jk]pq}\mathring \omega_{\pm}{}^{pq} - Y_{[mij|p|} \mathring \omega_{\pm}{}^p{}_{k]} \\
		&\mp \frac{3}{4} \delta_{m[i}\hodge{Y}_{jk]p}k_\pm{}^p \mp \hodge{Y}_{[mij}k_\pm{}_{k]} \pm X_{[m}\tilde\pi_\pm{}_{ijk]} \pm \frac{3}{4} \delta_{m[i|} X_{p|}\tilde\pi_\pm{}^p{}_{jk]}~,
	\end{split}
\end{equation}
where for simplicity we have suppressed the $r,s$ indices on the form bilinears that label the different Killing spinors.

Similarly a basis in the space of Killing spinor bilinears of AdS$_3 \times_w M^7$, up to Hodge duality, which are skew-symmetric in the exchange of Killing spinors is
\begin{equation} \label{AdS3_Skew_Bilinears}
	\begin{gathered}
		\mathring f^{rs}_{\pm} = \Herm{\chi^r_\pm}{\Gamma_{z} \Gamma_{11} \chi^s_\pm}~, \quad \tilde k^{rs}_{\pm} = \Herm{\chi^r_\pm}{\Gamma_{i}\Gamma_{11}\chi^s_\pm} \, \e^i , \quad \hat k^{rs}_{\pm} = \Herm{\chi^r_\pm}{\Gamma_{iz}\chi^s_\pm} \, \e^i~, \\
		\mathring k^{rs}_{\pm} = \Herm{\chi^r_\pm}{\Gamma_{iz}\Gamma_{11}\chi^s_\pm} \, \e^i~, \quad \omega^{rs}_\pm = \frac{1}{2} \Herm{\chi^r_\pm}{\Gamma_{ij}\chi^s_\pm} \, \e^i \wedge \e^j~, \\
		\tilde{\omega}^{rs}_\pm = \frac{1}{2} \Herm{\chi^r_\pm}{\Gamma_{ij}\Gamma_{11}\chi^s_\pm} \, \e^i \wedge \e^j~, \quad \hat \omega^{rs}_{\pm} = \frac{1}{2} \Herm{\chi^r_\pm}{\Gamma_ {ijz} \chi^s_\pm} \, \e^i \wedge \e^j~, \\
		\pi^{rs}_\pm = \frac{1}{3!} \Herm{\chi^r_\pm}{\Gamma_ {ijk} \chi^s_\pm} \, \e^i \wedge \e^j \wedge \e^k~.
	\end{gathered}
\end{equation}
The associated TCFH on $M^7$ with respect to the minimal connection, $\mathcal{D}^\mathcal{F}$, reads

\begin{equation}
	\begin{split}
		\mathcal{D}^\mathcal{F}_m \mathring f_{\pm} \defeq& \nabla_m \mathring f_{\pm} \\
		=& \mp A^{-1} \partial_{m} A \, \mathring f_{\pm} - \frac{1}{4} Z_{mpq} \hat \omega_{\pm}{}^{pq} - \frac{1}{4} S \mathring k_{\pm m} \\
		&+ \frac{1}{4} F_{mp} \hat k_{\pm}{}^p \mp \frac{1}{8} \hodge{Y}_{mpq} \omega_\pm{}^{pq} \mp \frac{1}{4} X_p \tilde\omega_{\pm}{}^p{}_m~,
	\end{split}
\end{equation}

\begin{equation}
	\begin{split}
		\mathcal{D}^\mathcal{F}_m \tilde k_\pm{}_i \defeq& \nabla_m \tilde k_\pm{}_i - \frac{1}{2} F_{mp} \omega_\pm{}^p{}_i \pm \frac{1}{2} X_m \mathring k_{\pm i} \\
		=& \mp A^{-1} \partial_{m} A \, \tilde k_\pm{}_i \, -\frac{1}{4} Z_{mpq} \pi_\pm{}^{pq}{}_i - \frac{1}{4} S \tilde\omega_{\pm mi} + \frac{1}{8} \delta_{mi} F_{pq} \omega_{\pm}{}^{pq} \\
		&- \frac{1}{2} F_{[m|p|} \omega_\pm{}^p{}_{i]} \mp \frac{1}{4} \hodge{Y}_{mip} \hat k_{\pm}{}^p - \frac{1}{8} Y_{mipq} \tilde\omega_\pm{}^{pq} \pm \frac{1}{4} \delta_{mi} X_p \mathring k_{\pm}{}^p \\
		 &\pm \frac{1}{2} X_{[m} \mathring k_{\pm i]}~,
	\end{split}
\end{equation}

\begin{equation}
	\begin{split}
		\mathcal{D}^\mathcal{F}_m \hat k_{\pm}{}_i \defeq& \nabla_m \hat k_{\pm}{}_i \\
		=& \mp A^{-1} \partial_{m} A \, \hat k_{\pm}{}_i \, - \frac{1}{2} Z_{mip} \mathring k_{\pm}{}^p - \frac{1}{4} S \hat \omega_{\pm mi} \mp \frac{1}{4!} \hodge{F}_{mipqr} \pi_{\pm}{}^{pqr} - \frac{1}{4} F_{mi} \mathring f_{\pm} \\
		&\pm \frac{1}{4} \hodge{Y}_{mip} \tilde k_\pm{}^p - \frac{1}{8} Y_{mipq} \hat \omega_{\pm}{}^{pq} \pm \frac{1}{4} X_p \pi_\pm{}^p{}_{mi}~,
	\end{split}
\end{equation}

\begin{equation}
	\begin{split}
		\mathcal{D}^\mathcal{F}_m \mathring k_{\pm}{}_i \defeq& \nabla_m \mathring k_{\pm}{}_i + \frac{1}{2} F_{mp} \hat \omega_{\pm}{}^p{}_i \pm \frac{1}{4} \hodge{Y}_{mpq} \pi_\pm{}^{pq}{}_i \pm \frac{1}{2} X_m \tilde k_{\pm i} \\
		=& \mp A^{-1} \partial_{m} A \, \mathring k_{\pm}{}_i \, - \frac{1}{2} Z_{mip} \hat k_{\pm}{}^p - \frac{1}{4} S \delta_{mi} \mathring f_{\pm} - \frac{1}{8} \delta_{mi} F_{pq} \hat \omega_{\pm}{}^{pq} \\
		&+ \frac{1}{2} F_{[m|p|} \hat \omega_{\pm}{}^p{}_{i]} \pm \frac{1}{4!} \delta_{mi} \hodge{Y}_{pqr} \pi_\pm{}^{pqr} \pm \frac{1}{4} \hodge{Y}_{[m|pq|} \pi_\pm{}^{pq}{}_{i]} \\
		&\pm \frac{1}{4} \delta_{mi} X_p \tilde k_{\pm}{}^p \pm \frac{1}{2} X_{[m} \tilde k_{\pm i]}~,
	\end{split}
\end{equation}

\begin{equation}
	\begin{split}
		\mathcal{D}^\mathcal{F}_m \omega_{\pm ij} \defeq& \nabla_m \omega_{\pm ij} + Z_{m[i|p|} \tilde \omega_\pm{}^p{}_{j]} + F_{m[i} \tilde k_{\pm j]} + \frac{1}{2} Y_{m[i|pq|} \pi_\pm{}^{pq}{}_{j]} \mp \frac{1}{2} X_m \hat \omega_{\pm ij} \\
		=& \mp A^{-1} \partial_{m} A \, \omega_{\pm ij} \, - \frac{1}{4} S \pi_{\pm mij} \pm \frac{1}{8} \hodge{F}_{mijpq} \hat \omega_{\pm}{}^{pq} \\
		&+ \frac{1}{2} \delta_{m[i} F_{j]p} \tilde k_\pm{}^p + \frac{3}{4} F_{[mi} \tilde k_{\pm j]} + \frac{1}{12} \delta_{m[i} Y_{j]pqr} \pi_\pm{}^{pqr} \\
		&+ \frac{3}{8} Y_{[mi|pq|} \pi_\pm{}^{pq}{}_{j]} \mp \frac{1}{4} \hodge{Y}_{mij} \mathring f_{\pm} \mp \frac{1}{2} \delta_{m[i} X_{|p|} \hat \omega_{\pm}{}^p{}_{j]} \\
		&\mp \frac{3}{4} X_{[m} \hat \omega_{\pm ij]}~,
	\end{split}
\end{equation}

\begin{equation}
	\begin{split}
		\mathcal{D}^\mathcal{F}_m \tilde\omega_{\pm ij} \defeq& \nabla_m \tilde\omega_{\pm ij} + Z_{m[i|p|} \omega_\pm{}^p{}_{j]} + \frac{1}{2} F_{mp} \pi_\pm{}^p{}_{ij} \pm \hodge{Y}_{m[i|p|} \hat \omega_{\pm}{}^p{}_{j]} \\
		=& \mp A^{-1} \partial_{m} A \, \tilde\omega_{\pm ij} \, - \frac{1}{2} S \delta_{m[i} \tilde k_{\pm j]} - \frac{1}{4} \delta_{m[i} F_{|pq|} \pi_\pm{}^{pq}{}_{j]} \\
		&+ \frac{3}{4} F_{[m|p|} \pi_\pm{}^p{}_{ij]} \mp \frac{1}{4} \delta_{m[i} \hodge{Y}_{j]pq} \hat \omega_{\pm}{}^{pq} \pm \frac{3}{4} \hodge{Y}_{[mi|p|} \hat \omega_{\pm}{}^p{}_{j]} + \frac{1}{4} Y_{mijp} \tilde k_{\pm}{}^p \\
		&- \frac{1}{4!} \hodge{X}_{mijpqr} \pi_\pm{}^{pqr} \mp \frac{1}{2} \delta_{m[i} X_{j]} \mathring f_{\pm}~,
	\end{split}
\end{equation}

\begin{equation}
	\begin{split}
		\mathcal{D}^\mathcal{F}_m \hat \omega_{\pm ij} \defeq& \nabla_m \hat \omega_{\pm ij} \mp \frac{1}{2} \hodge{Z}_{m[i|pq|} \pi_\pm{}^{pq}{}_{j]} - F_{m[i} \mathring k_{\pm j]} \mp \hodge{Y}_{m[i|p|} \tilde \omega_\pm{}^p{}_{j]} \mp \frac{1}{2} X_m \omega_{\pm ij} \\
		=& \mp A^{-1} \partial_{m} A \, \hat \omega_{\pm ij} \, \mp	 \frac{1}{6} \delta_{m[i} \hodge{Z}_{j]pqr} \pi_\pm{}^{pqr} \mp \frac{3}{4} \hodge{Z}_{[mi|pq|} \pi_\pm{}^{pq}{}_{j]} \\
		&+ \frac{1}{2} Z_{mij} \mathring f_{\pm} - \frac{1}{2} S \delta_{m[i} \hat k_{\pm j]} \pm \frac{1}{8} \hodge{F}_{mijpq} \omega_\pm{}^{pq} - \frac{1}{2} \delta_{m[i} F_{j]p} \mathring k_{\pm}{}^p \\
		&- \frac{3}{4} F_{[mi} \mathring k_{\pm j]} + \frac{1}{4} Y_{mijp} \hat k_{\pm}{}^p \pm \frac{1}{4} \delta_{m[i} \hodge{Y}_{j]pq} \tilde \omega_\pm{}^{pq} \mp \frac{3}{4} \hodge{Y}_{[mi|p|} \tilde\omega_\pm{}^p{}_{j]} \\
		&\mp \frac{1}{2} \delta_{m[i} X_{|p|} \omega_\pm{}^p{}_{j]} \mp \frac{3}{4} X_{[m} \omega_{\pm ij]}~,
	\end{split}
\end{equation}

\begin{equation}
	\begin{split}
		\mathcal{D}^\mathcal{F}_m \pi_{\pm ijk} \defeq& \nabla_m \pi_{\pm ijk} \mp \frac{3}{2} \hodge{Z}_{m[ij|p|} \hat \omega_{\pm}{}^p{}_{k]} - \frac{3}{2} Z_{m[ij} \tilde k_{\pm k]} - \frac{3}{2} F_{m[i} \tilde \omega_{\pm jk]} \\
		&- \frac{3}{2} Y_{m[ij|p|} \omega_\pm{}^p{}_{k]} \pm \frac{3}{2} \hodge{Y}_{m[ij} \mathring k_{\pm k]} \\
		=& \mp A^{-1} \partial_{m} A \, \pi_{\pm ijk} \, \pm \frac{3}{4} \delta_{m[i} \hodge{Z}_{jk]pq} \hat \omega_{\pm}{}^{pq} \mp 2 \hodge{Z}_{[mij|p|} \hat \omega_{\pm}{}^p{}_{k]} \\
		&- \frac{3}{4} S \delta_{m[i} \omega_{\pm jk]} \pm \frac{1}{4} \hodge{F}_{mijkp} \hat k_{\pm}{}^p - \frac{3}{2} \delta_{m[i} F_{j|p|} \tilde \omega_{\pm}{}^p{}_{k]} - \frac{3}{2} F_{[mi} \tilde \omega_{\pm jk]} \\
		&+ \frac{3}{8} \delta_{m[i} Y_{jk]pq} \omega_\pm{}^{pq} - Y_{[mij|p|} \omega_\pm{}^p{}_{k]} \pm \frac{3}{4} \delta_{m[i} \hodge{Y}_{jk]p} \mathring k_{\pm}{}^p \pm \hodge{Y}_{[mij} \mathring k_{\pm k]} \\
		&+ \frac{1}{8} \hodge{X}_{mijkpq} \tilde \omega_\pm{}^{pq} \pm \frac{3}{2} \delta_{m[i} X_{j} \hat k_{\pm k]}~,
	\end{split}
\end{equation}
where, again, for simplicity we have suppressed the $r,s$ indices on the form bilinears\footnote{From now on, we shall always suppress the $r,s$ indices on the form bilinears that label the different Killing spinors in all the TCFHs below.}.

Upon using Hodge duality on $M^7$, the domain of $\mathcal{D}^\mathcal{F}$ can be identified with $\Omega^*(M^7)\oplus \Omega^*(M^7)$. Moreover it is clear from the TCFH above that the domain of $\mathcal{D}^\mathcal{F}$ factorises into the space of symmetric form bilinears, \eqref{AdS3_Sym_Bilinears}, and the space of skew-symmetric form bilinears, \eqref{AdS3_Skew_Bilinears}. To understand this observe that the 16-dimensional Majorana representation, $\Delta_{16}$, of $\mathfrak{spin}(8)$ decomposes under $\mathfrak{spin}(7)$ into a sum of two 8-dimensional Majorana representations, $\Delta_{8}$. In turn the product of two $\Delta_{16}$ viewed as representations of $\mathfrak{spin}(7)$ decompose as
\begin{equation}
	\otimes^2 \Delta_{16} = \Lambda^*(\mathbb{R}^7) \oplus \Lambda^*(\mathbb{R}^7)~.
\end{equation}
Indeed, we note that $\textrm{dim}(\otimes^2 \Delta_{16}) = 2^4 \cdot 2^4 = 2\, \textrm{dim}(\Lambda^*(\mathbb{R}^7))$. However, we see that the minimal TCFH connection preserves the symmetrised $S^2(\Delta_{16})$ and skew-symmetrised $\Lambda^2(\Delta_{16})$ subspaces of $\otimes^2 \Delta_{16}$. Therefore, the reduced holonomy of $\mathcal{D}^\mathcal{F}$ will be contained within $GL(136) \times GL(120)$. However, the reduced holonomy of the minimal TCFH connection reduces further to a subgroup of $GL(133) \times SO(7) \times GL(112)$ as it acts with partial derivatives on the scalars $f$, $\tilde{f}$, $\hat f$ and $\mathring{f}$, and with the Levi-Civita connection on $\hat{k}$.

\section{The TCFH of Warped AdS\texorpdfstring{\textsubscript{4}}{4} Backgrounds} \label{AdS4-TCFH}

\subsection{Fields and Killing Spinors}

As in the previous cases, the bosonic fields of warped AdS$_4$ backgrounds, AdS$_4 \times_w M^6$, with internal space $M^6$ of massive IIA supergravity can be expressed as
\begin{equation}
	\begin{gathered}
		g = 2\, \e^+ \e^- + (\e^z)^2 + (\e^x)^2 + g(M^6)~, \quad G = X \, \e^+ \wedge \e^- \wedge \e^z \wedge \e^x + Y~, \\
		H = H~, \quad F = F~, \quad S = m e^\Phi~, \quad \Phi = \Phi~,
	\end{gathered}
\end{equation}
where $g(M^6)$ is a metric on $M^6$, $m$ is a constant, $\Phi, X \in C^\infty(M^6)$, $F \in \Omega^2(M^6)$, $H \in \Omega^3(M^6)$ and $Y \in \Omega^4(M^6)$. Further,
\begin{equation}
	\begin{gathered}
		\e^+ = \dd u, \quad \e^- = \dd r - r \frac{2}{\ell} \dd z - 2 r A^{-1} \dd A~, \\
		\e^z= A \dd z~, \quad \e^x= A e^{z/\ell} \dd x~, \quad \e^i = e^i{}_J \dd y^J~,
	\end{gathered}
\end{equation}
is a pseudo-orthonormal frame on AdS$_4 \times_w M^6$ with $g(M^6) = \delta_{ij}\e^i \e^j$, where $y$ are the coordinates of $M^6$ and $(u, r, z, x)$ are the remaining coordinates of spacetime. As in previous cases after a coordinate transformation the spacetime metric $g$ can be put into standard warped form $g = A^2 g_\ell(AdS_4) + g(M^6)$, where $A$ is the warp factor, $A\in C^\infty(M^6)$, and $g_\ell(AdS_4)$ is the standard metric on AdS$_4$ with radius $\ell$.

Integrating the KSEs of massive IIA supergravity along the coordinates $(u,r,z,x)$, one finds that the Killing spinors can be expressed as $\epsilon = \epsilon(u,r,z,x, \sigma_\pm, \tau_\pm)$, where $\sigma_\pm$ and $\tau_\pm$ are spinors that depend only on the coordinates of $M^6$ and $\Gamma_\pm\sigma_\pm = \Gamma_\pm\tau_\pm = 0$ \cite{bgpiia}. Furthermore, the gravitino KSE restricts $\sigma_\pm$ and $\tau_\pm$ along $M^6$ as $\mathcal{D}_m^{(\pm)} \chi_\pm = 0$, where $\chi_\pm$ stands for either $\sigma_\pm$ or $\tau_\pm$ and
\begin{equation}
	\begin{split}
			\mathcal{D}_m^{(\pm)} = \nabla_m \, \pm & \frac{1}{2} A^{-1} \partial_{m} A + \frac{1}{8} \slashed{H}_{m}\Gamma_{11} + \frac{1}{8} S \Gamma_{m} \\
			+& \frac{1}{16} \slashed{F} \Gamma_{m} \Gamma_{11} + \frac{1}{192} \slashed{Y} \Gamma_{m} \mp \frac{1}{8} X \Gamma_{z x m}~,
	\end{split}
\end{equation}
with $\nabla_m$, $m = 1,\dots,6$, the spin connection of $g(M^6)$. The Killing spinors satisfy two additional algebraic KSEs. One is associated to the dilatino KSE of massive IIA supergravity and the other arises as a consequence of the integration of the gravitino KSE over $z$. Both are essential for identifying the Killing spinors of an AdS$_4$ background but they do not contribute in the computation of TCFH on $M^6$. As a result, they will not be summarised here.

Unlike for warped AdS$_3$ backgrounds, the $\sigma_\pm$ and $\tau_\pm$ Killing spinors are related by a Clifford algebra operation. In particular, if $\sigma_\pm$ is a Killing spinor, then $\Gamma_{zx}\sigma_\pm$ is a $\tau_\pm$ Killing spinor, i.e. it solves all three Killing spinor equations that the $\tau_\pm$ Killing spinors satisfy \cite{bgpiia}. Using this, one can demonstrate that the Killing spinors of AdS$_4$ backgrounds come in multiples of four.

\subsection{The TCFH on \texorpdfstring{M\textsuperscript{6}}{M6}}

The computation of the TCFH of warped AdS$_4$ backgrounds is similar to that of warped AdS$_2$ and AdS$_3$ cases that have already been described in some detail. Because of this we shall be brief. A basis in the space of Killing spinor form bilinears\footnote{We could have considered a more general class of bilinears like for example those that contain either a single insertion of $\Gamma_z$ or a single insertion of $\Gamma_x$, i.e. $\Herm{\chi^r_\pm}{\Gamma_z\chi^s_\pm}$ and $\Herm{\chi^r_\pm}{\Gamma_x\chi^s_\pm}$ for scalars and similarly for higher degree forms. However, the choices of form bilinears below will suffice.} on $ M^6$, up to Hodge duality, which are symmetric in the exchange of Killing spinors $\chi^r_\pm$ and $\chi_\pm^s$ is
\begin{equation} \label{AdS4_Sym_Bilinears}
	\begin{gathered}
		f^{rs}_\pm = \Herm{\chi^r_\pm}{\chi^s_\pm}~, \quad \tilde{f}^{rs}_\pm = \Herm{\chi^r_\pm}{\Gamma_{11} \chi^s_\pm}~, \quad	k^{rs}_\pm = \Herm{\chi^r_\pm}{\Gamma_ i\chi^s_\pm} \, \e^i~, \\
		\mathring k^{rs}_{\pm} = \Herm{\chi^r_\pm}{\Gamma_ {izx}\Gamma_{11} \chi^s_\pm} \, \e^i~, \quad \hat \omega^{rs}_{\pm} = \frac{1}{2} \Herm{\chi^r_\pm}{\Gamma_{ijzx}\chi^s_\pm} \, \e^{i} \wedge \e^j~, \\
		\mathring \omega^{rs}_{\pm} = \frac{1}{2} \Herm{\chi^r_\pm}{\Gamma_{ijzx}\Gamma_{11}\chi^s_\pm} \, \e^{i} \wedge \e^j~, \quad \tilde{\pi}^{rs}_\pm = \frac{1}{3!} \Herm{\chi^r_\pm}{\Gamma_ {ijk}\Gamma_{11} \chi^s_\pm} \, \e^i \wedge \e^j \wedge \e^k~,
	\end{gathered}
\end{equation}
where again $\chi_\pm$ stands for either $\sigma_\pm$ or $\tau_\pm$. After some computation, the TCFH is
\begin{equation}
	\begin{split}
		\mathcal{D}^\mathcal{F}_m f_\pm \defeq& \nabla_m f_\pm \\
		=& \mp A^{-1} \partial_{m} A \, f_\pm - \frac{1}{4}Sk_\pm{}_m -\frac{1}{8}F_{pq}\tilde{\pi}_\pm{}^{pq}{}_m \mp \frac{1}{4}\hodge{Y}_{mp}\mathring k_{\pm}{}^{p}~,
	\end{split}
\end{equation}

\begin{equation}
	\begin{split}
		\mathcal{D}^\mathcal{F}_m \tilde{f}_\pm \defeq& \nabla_m \tilde{f}_\pm \\
		=& \mp A^{-1} \partial_{m} A \, \tilde{f}_\pm - \frac{1}{4} F_{mp}k_\pm{}^p - \frac{1}{4!} Y_{mpqr} \tilde{\pi}_\pm{}^{pqr} \mp \frac{1}{4} X\mathring k_{\pm}{}_m~,
	\end{split}
\end{equation}

\begin{equation}
	\begin{split}
		\mathcal{D}^\mathcal{F}_m k_\pm{}_i \defeq& \nabla_m k_\pm{}_i + \frac{1}{4}H_{mpq}\tilde{\pi}_\pm{}^{pq}{}_i \mp \frac{1}{2} \hodge{Y}_{mp}\mathring \omega_{\pm}{}^p{}_i	\\
		=& \mp A^{-1} \partial_{m} A \, k_\pm{}_i \, -\frac{1}{4}\delta_{mi} Sf_\pm \pm \frac{1}{8} \hodge{F}_{mipq}\hat \omega_{\pm}{}^{pq} + \frac{1}{4} F_{mi}\tilde{f}_\pm \\
		&\pm \frac{1}{8} \delta_{mi}\hodge{Y}_{pq}\mathring \omega_{\pm}{}^{pq} \mp \frac{1}{2}\hodge{Y}_{[m|p|}\mathring \omega_{\pm}{}^p{}_{i]} \mp \frac{1}{4}X\hat \omega_{\pm}{}_{mi}~,
	\end{split}
\end{equation}

\begin{equation}
	\begin{split}
		\mathcal{D}^\mathcal{F}_m \mathring k_{\pm}{}_i \defeq& \nabla_m \mathring k_{\pm}{}_i \mp \frac{1}{4} \hodge{H}_{mpq} \tilde\pi_\pm{}^{pq}{}_i - \frac{1}{2} F_{mp}\hat \omega_{\pm}{}^p{}_i \\
		=& \mp A^{-1} \partial_{m} A \, \mathring k_{\pm}{}_i \, \mp \frac{1}{12} \delta_{mi} \hodge{H}_{pqr} \tilde\pi_\pm{}^{pqr} \mp \frac{1}{2} \hodge{H}_{[m|pq|}\tilde\pi_\pm{}^{pq}{}_{i]} \\
		&-\frac{1}{4}S\mathring \omega_{\pm}{}_{mi} + \frac{1}{8}\delta_{mi} F_{pq} \hat \omega_{\pm}{}^{pq} - \frac{1}{2}F_{[m|p|}\hat \omega_{\pm}{}^p{}_{i]} \\
		&\mp \frac{1}{4}\hodge{Y}_{mi}f_\pm - \frac{1}{8}Y_{mipq} \mathring \omega_{\pm}{}^{pq} \pm \frac{1}{4}X\delta_{mi}\tilde{f}_\pm~,
	\end{split}
\end{equation}

\begin{equation}
	\begin{split}
		\mathcal{D}^\mathcal{F}_m \hat \omega_{\pm}{}_{ij} \defeq& \nabla_m \hat \omega_{\pm}{}_{ij} + H_{m[i|p|}\mathring \omega_{\pm}{}^p{}_{j]} + F_{m[i}\mathring k_{\pm}{}_{j]} \mp \frac{1}{2} \hodge{Y}_{mp}\tilde\pi_\pm{}^p{}_{ij} \\
		=& \mp A^{-1} \partial_{m} A \, \hat \omega_{\pm}{}_{ij} \mp \frac{1}{24} \hodge{S}_{mijpqr}\tilde\pi_\pm{}^{pqr} \pm\frac{1}{4}\hodge{F}_{mijp}k_\pm{}^p \\
		&+ \frac{1}{2}\delta_{m[i}F_{j]p}\mathring k_{\pm}{}^p	+ \frac{3}{4}F_{[mi}\mathring k_{\pm}{}_{j]} \mp \frac{3}{4}\hodge{Y}_{[m|p|} \tilde\pi_\pm{}^p{}_{ij]} \\
		&\pm \frac{1}{4} \delta_{m[i}\hodge{Y}_{|pq|}\tilde\pi_\pm{}^{pq}{}_{j]} \pm \frac{1}{2}X\delta_{m[i}k_\pm{}_{j]}~,
	\end{split}
\end{equation}

\begin{equation}
	\begin{split}
		\mathcal{D}^\mathcal{F}_m \mathring \omega_{\pm}{}_{ij} \defeq& \nabla_m \mathring \omega_{\pm}{}_{ij} + H_{m[i|p|}\hat \omega_{\pm}{}^p{}_{j]} \pm \frac{1}{2} \hodge{F}_{m[i|pq|}\tilde\pi_\pm{}^{pq}{}_{j]} \mp \hodge{Y}_{m[i}k_\pm{}_{j]} \\
		=& \mp A^{-1} \partial_{m} A \, \mathring \omega_{\pm}{}_{ij} - \frac{1}{2} S \delta_{m[i}\mathring k_{\pm}{}_{j]} \pm \frac{3}{8} \hodge{F}_{[mi|pq|}\tilde\pi_\pm{}^{pq}{}_{j]} \\
		&\pm \frac{1}{12} \delta_{m[i}\hodge{F}_{j]pqr}\tilde\pi_\pm{}^{pqr} \mp \frac{1}{2} \delta_{m[i}\hodge{Y}_{j]p}k_\pm{}^p \mp \frac{3}{4} \hodge{Y}_{[mi}k_\pm{}_{j]} \\
		&+ \frac{1}{4}Y_{mijp}\mathring k_{\pm}{}^p \pm \frac{1}{4}X\tilde\pi_\pm{}_{mij}~,
	\end{split}
\end{equation}

\begin{equation}
	\begin{split}
		\mathcal{D}^\mathcal{F}_m \tilde{\pi}_{\pm}{}_{ijk} \defeq& \nabla_m \tilde{\pi}_{\pm}{}_{ijk} \mp \frac{3}{2} \hodge{H}_{m[ij}\mathring k_{\pm}{}_{k]} - \frac{3}{2}H_{m[ij}k_\pm{}_{k]} \pm \frac{3}{2} \hodge{F}_{m[ij|p|}\mathring \omega_{\pm}{}^p{}_{k]} \\
		&\mp \frac{3}{2}\hodge{Y}_{m[i}\hat \omega_{\pm}{}_{jk]} \\
		=& \mp A^{-1} \partial_{m} A \, \tilde{\pi}_{\pm}{}_{ijk} \mp \frac{3}{2}\delta_{m[i}\hodge{H}_{jk]p}\mathring k_{\pm}{}^p \mp 2 \hodge{H}_{[mij}\mathring k_{\pm}{}_{k]} \mp \frac{1}{8} \hodge{S}_{mijkpq} \hat \omega_{\pm}{}^{pq} \\
		&- \frac{3}{4}\delta_{m[i}F_{jk]}f_\pm \mp \frac{3}{8} \delta_{m[i}\hodge{F}_{jk]pq}\mathring \omega_{\pm}{}^{pq} \pm \hodge{F}_{[mij|p|}\mathring \omega_{\pm}{}^p{}_{k]} \\
		&\mp \frac{3}{2}\delta_{m[i}\hodge{Y}_{j|p|} \hat \omega_{\pm}{}^p{}_{k]} \mp \frac{3}{2}\hodge{Y}_{[mi}\hat \omega_{\pm}{}_{jk]} + \frac{1}{4}Y_{mijk}\tilde{f}_\pm \mp \frac{3}{4}X\delta_{m[i}\mathring \omega_{\pm}{}_{jk]}~,
	\end{split}
\end{equation}
where $\mathcal{D}^\mathcal{F}$ is the minimal connection.

Similarly, a basis in the space of form bilinears on $M^6$, up to Hodge duality, which are skew-symmetric in the exchange of Killing spinors $\chi_\pm^r$ and $\chi_\pm^s$ is
\begin{equation} \label{AdS4_Skew_Bilinears}
	\begin{gathered}
		\hat f^{rs}_{\pm} = \Herm{\chi^r_\pm}{\Gamma_{zx}\chi^s_\pm}~, \quad \mathring f^{rs}_{\pm} = \Herm{\chi^r_\pm}{\Gamma_{zx}\Gamma_{11}\chi^s_\pm}~, \\
		\hat k^{rs}_{\pm} = \Herm{\chi^r_\pm}{\Gamma_{izx}\chi^s_\pm} \, \e^i~, \quad \tilde k^{rs}_\pm = \Herm{\chi^r_\pm}{\Gamma_i \Gamma_{11}\chi^s_\pm} \, \e^i~, \\
		\omega^{rs}_\pm = \frac{1}{2} \Herm{\chi^r_\pm}{\Gamma_{ij}\chi^s_\pm} \, \e^i \wedge \e^j~, \quad \tilde{\omega}^{rs}_\pm = \frac{1}{2} \Herm{\chi^r_\pm}{\Gamma_{ij}\Gamma_{11}\chi^s_\pm} \, \e^i \wedge \e^j~, \\
		\pi^{rs}_\pm = \frac{1}{3!} \Herm{\chi^r_\pm}{\Gamma_ {ijk} \chi^s_\pm} \, \e^i \wedge \e^j \wedge \e^k~.
	\end{gathered}
\end{equation}
The associated TCFH is
\begin{equation}
	\begin{split}
		\mathcal{D}^\mathcal{F}_m \hat f_{\pm} \defeq& \nabla_m \hat f_{\pm} \\
		=& \mp A^{-1} \partial_{m} A \, \hat f_{\pm} \, - \frac{1}{4} S \hat k_{\pm m} \mp \frac{1}{4!} \hodge{F}_{mpqr} \pi_\pm{}^{pqr} \pm \frac{1}{4} \hodge{Y}_{mp} \tilde k_\pm{}^p~,
	\end{split}
\end{equation}

\begin{equation}
	\begin{split}
		\mathcal{D}^\mathcal{F}_m \mathring f_{\pm} \defeq& \nabla_m \mathring f_{\pm} \\
		=& \mp A^{-1} \partial_{m} A \, \mathring f_{\pm} \, - \frac{1}{4} F_{mp} \hat k_{\pm}{}^p \pm \frac{1}{8} \hodge{Y}_{pq} \pi_\pm{}^{pq}{}_m \pm \frac{1}{4} X \tilde k_{\pm m}~,
	\end{split}
\end{equation}

\begin{equation}
	\begin{split}
		\mathcal{D}^\mathcal{F}_m \tilde k_\pm{}_i \defeq& \nabla_m \tilde k_\pm{}_i + \frac{1}{4} H_{mpq} \pi_\pm{}^{pq}{}_i - \frac{1}{2} F_{mp} \omega_\pm{}^p{}_i \\
		=& \mp A^{-1} \partial_{m} A \, \tilde{k}_\pm{}_i \, - \frac{1}{4} S \omega_{\pm mi} + \frac{1}{8} \delta_{mi} F_{pq} \omega_\pm{}^{pq} - \frac{1}{2} F_{[m|p|} \omega_\pm{}^p{}_{i]} \\
		&\pm \frac{1}{4} \hodge{Y}_{mi} \hat f_{\pm} - \frac{1}{8} Y_{mipq} \tilde\omega_\pm{}^{pq} \mp \frac{1}{4} X \delta_{mi} \mathring f_{\pm}~,
	\end{split}
\end{equation}

\begin{equation}
	\begin{split}
		\mathcal{D}^\mathcal{F}_m \hat k_{\pm}{}_i \defeq& \nabla_m \hat k_{\pm}{}_i \mp \frac{1}{4} \hodge{H}_{mpq} \pi_\pm{}^{pq}{}_i \pm \frac{1}{2} \hodge{Y}_{mp} \tilde\omega_\pm{}^p{}_i \\
		=& \mp A^{-1} \partial_{m} A \, \hat k_{\pm}{}_i \, \mp \frac{1}{12} \delta_{mi} \hodge{H}_{pqr} \pi_\pm{}^{pqr} \mp \frac{1}{2} \hodge{H}_{[m|pq|}\pi_\pm{}^{pq}{}_{i]} \\
		&- \frac{1}{4} S \delta_{mi} \hat f_{\pm} \mp \frac{1}{8} \hodge{F}_{mipq} \omega_\pm{}^{pq} + \frac{1}{4} F_{mi} \mathring f_{\pm} \mp \frac{1}{8} \delta_{mi} \hodge{Y}_{pq} \tilde\omega_\pm{}^{pq} \\
		&\pm \frac{1}{2} \hodge{Y}_{[m|p|}\tilde\omega_\pm{}^p{}_{i]} \pm \frac{1}{4} X \omega_{\pm mi}~,
	\end{split}
\end{equation}

\begin{equation}
	\begin{split}
		\mathcal{D}^\mathcal{F}_m \omega_{\pm ij} \defeq& \nabla_m \omega_{\pm ij} + H_{m[i|p|} \tilde\omega_\pm{}^p{}_{j]} + F_{m[i} \tilde k_{\pm j]} + \frac{1}{2} Y_{m[i|pq|} \pi_\pm{}^{pq}{}_{j]} \\
		=& \mp A^{-1} \partial_{m} A \, \omega_{\pm ij} \, - \frac{1}{4} S \pi_{\pm mij} \mp \frac{1}{4} \hodge{F}_{mijp} \hat k_{\pm}{}^p + \frac{1}{2} \delta_{m[i} F_{j]p} \tilde k_\pm{}^p \\
		&+ \frac{3}{4} F_{[mi} \tilde k_{\pm j]} + \frac{1}{12} \delta_{m[i} Y_{j]pqr} \pi_\pm{}^{pqr} + \frac{3}{8} Y_{[mi|pq|} \pi_\pm{}^{pq}{}_{j]} \mp \frac{1}{2} X \delta_{m[i} \hat k_{\pm j]}~,
	\end{split}
\end{equation}

\begin{equation}
	\begin{split}
		\mathcal{D}^\mathcal{F}_m \tilde\omega_{\pm ij} \defeq& \nabla_m \tilde\omega_{\pm ij} + H_{m[i|p|} \omega_\pm{}^p{}_{j]} + \frac{1}{2} F_{mp} \pi_\pm{}^p{}_{ij} \pm \hodge{Y}_{m[i} \hat k_{\pm j]} \\
		=& \mp A^{-1} \partial_{m} A \, \tilde\omega_{\pm ij} \, - \frac{1}{2} S \delta_{m[i} \tilde k_{\pm j]} - \frac{1}{4} \delta_{m[i}F_{|pq|} \pi_\pm{}^{pq}{}_{j]} \\
		&+ \frac{3}{4} F_{[m|p|} \pi_\pm{}^p{}_{ij]} \pm \frac{1}{2} \delta_{m[i} \hodge{Y}_{j]p} \hat k_{\pm}{}^p \pm \frac{3}{4} \hodge{Y}_{[mi} \hat k_{\pm j]} \\
		&+ \frac{1}{4} Y_{mijp} \tilde k_\pm{}^p - \frac{1}{4!} \hodge{X}_{mijpqr} \pi_\pm{}^{pqr}~,
	\end{split}
\end{equation}

\begin{equation}
	\begin{split}
		\mathcal{D}^\mathcal{F}_m \pi_{\pm ijk} \defeq& \nabla_m \pi_{\pm ijk} \mp \frac{3}{2} \hodge{H}_{m[ij} \hat k_{\pm k]} - \frac{3}{2} H_{m[ij} \tilde k_{\pm k]} - \frac{3}{2} F_{m[i} \tilde \omega_{\pm jk]} - \frac{3}{2} Y_{m[ij|p|} \omega_\pm{}^p{}_{k]} \\
		=& \mp A^{-1} \partial_{m} A \, \pi_{\pm ijk} \, \mp \frac{3}{2} \delta_{m[i} \hodge{H}_{jk]p} \hat k_{\pm}{}^p \mp 2 \hodge{H}_{[mij} \hat k_{\pm k]} - \frac{3}{4} S \delta_{m[i} \omega_{\pm jk]} \\
		&\mp \frac{1}{4} \hodge{F}_{mijk} \hat f_{\pm} - \frac{3}{2} \delta_{m[i} F_{j|p|} \tilde \omega_\pm{}^p{}_{k]} - \frac{3}{2} F_{[mi} \tilde\omega_{\pm jk]} + \frac{3}{8} \delta_{m[i} Y_{jk]pq} \omega_\pm{}^{pq} \\
		&- Y_{[mij|p|} \omega_\pm{}^p{}_{k]} \mp \frac{3}{4} \delta_{m[i} \hodge{Y}_{jk]} \mathring f_{\pm}
+ \frac{1}{8} \hodge{X}_{mijkpq} \tilde\omega_\pm{}^{pq}~,
	\end{split}
\end{equation}
where, again, $\mathcal{D}^\mathcal{F}$ is the minimal connection.

The domain that the minimal TCFH connection $\mathcal{D}^\mathcal{F}$ acts factorises into the space of symmetric form bilinears, \eqref{AdS4_Sym_Bilinears}, and the space of skew-symmetric form bilinears, \eqref{AdS4_Skew_Bilinears} in the exchange of the two Killing spinors $\chi_\pm^r$ and $\chi_\pm^s$. A direct counting of dimensions reveals that the reduced holonomy of $\mathcal{D}^\mathcal{F}$ must be contained in $GL(64) \times GL(64)$. But as $\mathcal{D}^\mathcal{F}$ acts trivially on the scalars $f$, $\tilde{f}$, $\hat{f}$ and $\mathring{f}$, its reduced holonomy is contained in $GL(62) \times GL(62)$.

\section{The TCFH of Warped AdS\texorpdfstring{\textsubscript{n}, $n \geq 5$}{n, n >= 5} Backgrounds} \label{AdSn-TCFH}

\subsection{Fields and Killing Spinors}

The bosonic fields of warped AdS$_n$, AdS$_n \times_w M^{10 - n}$, $n \geq 5$, backgrounds with internal space $M^{10-n}$ of (massive) IIA backgrounds can be written as follows
\begin{equation}
	\begin{gathered}
		g = 2\, \e^+ \e^- + (\e^z)^2 + \sum\limits_{a = 1}^{n - 3} (\e^a)^2 + g(M^{10 - n})~, \\
		G = G, \quad H = H, \quad F = F, \quad S = m e^\Phi , \quad \Phi = \Phi~,
	\end{gathered}
\end{equation}
where $g(M^{10-n})$ is a metric on $M^{10-n}$, $m$ is a constant, $\Phi \in C^\infty(M^{10 - n})$, $F \in \Omega^2(M^{10 - n})$, $H \in \Omega^3(M^{10 - n})$ and $G \in \Omega^4(M^{10 - n})$. For sufficiently large $n$, some of the fluxes may vanish; for example $G$ vanishes for $n \geq 7$. Further,
\begin{equation}
	\begin{gathered}
		\e^+ = \dd u, \quad \e^- = \dd r - \frac{2}{\ell} r \dd z - 2 r A^{-1} \dd A, \\
		\e^z = A \dd z~, \quad \e^a = A e^{z/\ell} \dd x^a~, \quad\e^i = e^i{}_J \dd y^J~,
	\end{gathered}
\end{equation}
is a pseudo-orthonormal frame on AdS$_n \times_w M^{10 - n}$ with $g(M^{10-n}) = \delta_{ij} \e^i \e^j$, where $y$ are coordinates on $M^{10-n}$ and $(u, r, z, x^a)$ are the remaining coordinates of the spacetime. As in previous cases, $A\in C^\infty(M^{10-n})$ is the warp factor and after a coordinate transformation the spacetime metric $g$ can be written in the usual warped form involving the standard metric on AdS$_n$ of radius $\ell$.

Again the Killing spinors of these backgrounds can be expressed as $\epsilon = \epsilon(u,r, z, x^a, \sigma_\pm, \tau_\pm)$, where $\sigma_\pm$ and $\tau_\pm$ depend only on the coordinates of $M^{10-n}$ and $\Gamma_\pm\sigma_\pm = \Gamma_\pm \tau_\pm = 0$ \cite{bgpiia}. Furthermore, the gravitino KSE along $M^{10-n}$ requires that $\mathcal{D}_m^{(\pm)} \chi_\pm = 0$ with
\begin{equation}
	\begin{split}
			\mathcal{D}_m^{(\pm)} = \nabla_m \pm & \frac{1}{2} A^{-1} \partial_{m} A + \frac{1}{8} \slashed{H}_{m}\Gamma_{11} + \frac{1}{8} S \Gamma_{m} \\
			&+ \frac{1}{16} \slashed{F} \Gamma_{m} \Gamma_{11} + \frac{1}{192} \slashed{G} \Gamma_{m}~,
	\end{split}
\end{equation}
where $\nabla_m$, $m = 1,\dots, 10-n$, is the spin connection of $g(M^{10 - n})$ and $\chi_\pm$ stands for either $\sigma_\pm$ or $\tau_\pm$.

TCFH of warped AdS$_n$ backgrounds will be stated below for each $n$, $5\leq n\leq 7$. As the computation is similar to those that have already been described in previous cases, we shall simply state the results.

\subsection{The TCFH of Warped AdS\texorpdfstring{\textsubscript{5}}{5} Backgrounds}

A basis in the space of form bilinears\footnote{As for warped AdS$_4$ backgrounds a more general class of form bilinears can be considered but the choices below for all AdS$_n$, $n\geq 5$, backgrounds will suffice.} on $M^5$, up to Hodge duality, which are symmetric in the exchange of Killing spinors $\chi_\pm^r$ and $\chi_\pm^s$ is
\begin{equation} \label{AdS5_Sym_Bilinears}
	\begin{gathered}
		f^{rs}_\pm = \Herm{\chi^r_\pm}{\chi^s_\pm}~, \quad 	 \tilde{f}^{rs}_\pm = \Herm{\chi^r_\pm}{\Gamma_{11} \chi^s_\pm}~, \quad	\mathring f^{rs}_{\pm} = \Herm{\chi^r_\pm}{\Gamma_ {zx_1x_2}\Gamma_{11} \chi^s_\pm}~, \\
		 k^{rs}_\pm = \Herm{\chi^r_\pm}{\Gamma_ i\chi^s_\pm} \, \e^i~, \quad \hat k^{rs}_{\pm} = \Herm{\chi^r_\pm}{\Gamma_{izx_1x_2}\chi^s_\pm} \, \e^{i}~, \\
		\mathring k^{rs}_{\pm} = \Herm{\chi^r_\pm}{\Gamma_{izx_1x_2}\Gamma_{11}\chi^s_\pm} \, \e^{i}~, \quad \hat \omega^{rs}_{\pm} = \frac{1}{2} \Herm{\chi^r_\pm}{\Gamma_ {ijzx_1x_2} \chi^s_\pm} \, \e^i \wedge \e^j~.
	\end{gathered}
\end{equation}
The TCFH is

\begin{equation}
	\begin{split}
		\mathcal{D}^\mathcal{F}_m f_\pm \defeq& \nabla_m f_\pm \\
		=& \mp A^{-1} \partial_{m} A \, f_\pm - \frac{1}{4}Sk_\pm{}_m \pm \frac{1}{8}\hodge{F}_{mpq}\hat \omega_{\pm}{}^{pq} \mp \frac{1}{4}\hodge{G}_{m}\mathring f_{\pm}~,
	\end{split}
\end{equation}

\begin{equation}
	\begin{split}
		\mathcal{D}^\mathcal{F}_m \tilde{f}_\pm \defeq& \nabla_m \tilde{f}_\pm \\
		=& \mp A^{-1} \partial_{m} A \, \tilde{f}_\pm - \frac{1}{4} F_{mp}k_\pm{}^p \pm \frac{1}{4}\hodge{G}_p \hat \omega_{\pm}{}^p{}_m~,
	\end{split}
\end{equation}

\begin{equation}
	\begin{split}
		\mathcal{D}^\mathcal{F}_m \mathring f_{\pm} \defeq& \nabla_m \mathring f_{\pm} \\
		=& \mp A^{-1} \partial_{m} A \, \mathring f_{\pm} -\frac{1}{4}H_{mpq}\hat \omega_{\pm}{}^{pq} -\frac{1}{4}S\mathring k_{\pm}{}_m + \frac{1}{4} F_{mp}\hat k_{\pm}{}^{p} \mp \frac{1}{4} \hodge{G}_{m}f_\pm~,
	\end{split}
\end{equation}

\begin{equation}
	\begin{split}
		\mathcal{D}^\mathcal{F}_m k_\pm{}_i \defeq& \nabla_m k_\pm{}_i \mp \frac{1}{2}\hodge{H}_{mp}\hat \omega_{\pm}{}^p{}_i \pm \frac{1}{2} \hodge{G}_{m}\mathring k_{\pm}{}_i \\
		=& \mp A^{-1} \partial_{m} A \, k_\pm{}_i \, \mp \hodge{H}_{[m|p|}\hat \omega_{\pm}{}^p{}_{i]} \pm \frac{1}{4}\delta_{mi}\hodge{H}_{pq}\hat \omega_{\pm}{}^{pq} -\frac{1}{4} \delta_{mi} Sf_\pm \\
		&\pm \frac{1}{4} \hodge{F}_{mip}\hat k_{\pm}{}^{p} + \frac{1}{4} F_{mi}\tilde{f}_\pm \pm \frac{1}{4} \delta_{mi}\hodge{G}_{p}\mathring k_{\pm}{}^{p} \pm \frac{1}{2}\hodge{G}_{[m}\mathring k_{\pm}{}_{i]}~,
	\end{split}
\end{equation}

\begin{equation}
	\begin{split}
		\mathcal{D}^\mathcal{F}_m \hat k_{\pm}{}_i \defeq& \nabla_m \hat k_{\pm}{}_i \\
		=& \mp A^{-1} \partial_{m} A \, \hat k_{\pm}{}_i \, - \frac{1}{2} H_{mip} \mathring k_{\pm}{}^p - \frac{1}{4}S\hat \omega_{\pm}{}_{mi} \mp \frac{1}{4} \hodge{F}_{mip}k_\pm{}^p \\
		&- \frac{1}{4} F_{mi} \mathring f_{\pm} - \frac{1}{8} G_{mipq} \hat \omega_{\pm}{}^{pq}~,
	\end{split}
\end{equation}

\begin{equation}
	\begin{split}
		\mathcal{D}^\mathcal{F}_m \mathring k_{\pm}{}_i \defeq& \nabla_m \mathring k_{\pm}{}_i + \frac{1}{2} F_{mp} \hat \omega_{\pm}{}^p{}_i \pm \frac{1}{2} \hodge{G}_m k_\pm{}_i \\
		=& \mp A^{-1} \partial_{m} A \, \mathring k_{\pm}{}_i \, - \frac{1}{2} H_{mip} \hat k_{\pm}{}^p - \frac{1}{4}\delta_{mi} S\mathring f_{\pm} - \frac{1}{8} \delta_{mi}F_{pq}\hat \omega_{\pm}{}^{pq} \\
		&+ \frac{1}{2} F_{[m|p|} \hat \omega_{\pm}{}^p{}_{i]} \pm \frac{1}{4} \delta_{mi} \hodge{G}_p k_\pm{}^p \pm \frac{1}{2} \hodge{G}_{[m} k_\pm{}_{i]}~,
	\end{split}
\end{equation}

\begin{equation}
	\begin{split}
		\mathcal{D}^\mathcal{F}_m \hat \omega_{\pm}{}_{ij} \defeq& \nabla_m \hat \omega_{\pm}{}_{ij} \mp \hodge{H}_{m[i} k_\pm{}_{j]} - F_{m[i} \mathring k_{\pm}{}_{j]} \\
		=& \mp A^{-1} \partial_{m} A \, \hat \omega_{\pm}{}_{ij} \, \mp \delta_{m[i} \hodge{H}_{j]p} k_\pm{}^p \mp \frac{3}{2} \hodge{H}_{[mi} k_\pm{}_{j]} \\
		&+ \frac{1}{2} H_{mij} \mathring f_{\pm} - \frac{1}{2} S \delta_{m[i} \hat k_{\pm}{}_{j]} \pm \frac{1}{4} \hodge{F}_{mij} f_\pm - \frac{1}{2} \delta_{m[i} F_{j]p} \mathring k_{\pm}{}^p \\
		&- \frac{3}{4} F_{[mi} \mathring k_{\pm}{}_{j]} \pm \frac{1}{2} \delta_{m[i} \hodge{G}_{j]} \tilde{f}_\pm + \frac{1}{4} G_{mijp} \hat k_{\pm}{}^p~,
	\end{split}
\end{equation}
where $\nabla$ is the frame connection of $g(M^5)$.

A basis in the space of form bilinears on $M^5$, up to Hodge duality, which are skew-symmetric in the exchange of $\chi^r$ and $\chi^s$ is
\begin{equation} \label{AdS5_Skew_Bilinears}
	\begin{gathered}
		\hat f^{rs}_{\pm} = \Herm{\chi^r_\pm}{\Gamma_{zx_1x_2}\chi^s_\pm}~, \quad \tilde k^{rs}_\pm = \Herm{\chi^r_\pm}{\Gamma_i \Gamma_{11}\chi^s_\pm} \, \e^i~, \\
		\omega^{rs}_\pm = \frac{1}{2} \Herm{\chi^r_\pm}{\Gamma_{ij}\chi^s_\pm} \, \e^i \wedge \e^j~, \quad \tilde{\omega}^{rs}_\pm = \frac{1}{2} \Herm{\chi^r_\pm}{\Gamma_{ij}\Gamma_{11}\chi^s_\pm} \, \e^i \wedge \e^j~, \\
		\mathring \omega^{rs}_{\pm} = \frac{1}{2} \Herm{\chi^r_\pm}{\Gamma_{ijzx_1x_2}\Gamma_{11}\chi^s_\pm} \, \e^i \wedge \e^j~.
	\end{gathered}
\end{equation}
The TCFH is
\begin{equation}
	\begin{split}
		\mathcal{D}^\mathcal{F}_m \hat f_{\pm} \defeq& \nabla_m \hat f_{\pm} \\
		=& \mp A^{-1} \partial_{m} A \, \hat f_{\pm} \, - \frac{1}{4} H_{mpq} \mathring \omega_{\pm}{}^{pq} \mp \frac{1}{8} \hodge{F}_{mpq} \omega_\pm{}^{pq} \pm \frac{1}{4} \hodge{G}_p \tilde\omega_\pm{}^p{}_m~,
	\end{split}
\end{equation}

\begin{equation}
	\begin{split}
		\mathcal{D}^\mathcal{F}_m \tilde k_\pm{}_i \defeq& \nabla_m \tilde k_\pm{}_i \mp \frac{1}{2} \hodge{H}_{mp} \mathring \omega_{\pm}{}^p{}_i - \frac{1}{2} F_{mp} \omega_\pm{}^p{}_i \\
		=& \mp A^{-1} \partial_{m} A \, \tilde{k}_\pm{}_i \, \pm \frac{1}{4} \delta_{mi} \hodge{H}_{pq} \mathring \omega_{\pm}{}^{pq} \mp \hodge{H}_{[m|p|} \mathring \omega_{\pm}{}^p{}_{i]} \\
		&- \frac{1}{4} S \tilde\omega_{\pm mi} + \frac{1}{8} \delta_{mi} F_{pq} \omega_\pm{}^{pq} - \frac{1}{2} F_{[m|p|} \omega_\pm{}^p{}_{i]} - \frac{1}{8} G_{mipq} \tilde\omega_\pm{}^{pq}~,
	\end{split}
\end{equation}

\begin{equation}
	\begin{split}
		\mathcal{D}^\mathcal{F}_m \omega_\pm{}_{ij} \defeq& \nabla_m \omega_\pm{}_{ij} + H_{m[i|p|} \tilde\omega_\pm{}^p{}_{j]} + F_{m[i} \tilde k_{\pm j]} \pm \frac{1}{2} \hodge{G}_m \mathring \omega_{\pm ij} \\
		=& \mp A^{-1} \partial_{m} A \, \omega_\pm{}_{ij} \, \pm \frac{1}{8} \hodge{S}_{mijpq} \mathring \omega_{\pm}{}^{pq} \mp \frac{1}{4} \hodge{F}_{mij} \hat f_{\pm} + \frac{1}{2} \delta_{m[i} F_{j]p} \tilde k_\pm{}^p \\
		&+ \frac{3}{4} F_{[mi} \tilde k_{\pm j]} \pm \frac{1}{2} \delta_{m[i}\hodge{G}_{|p|} \mathring \omega_{\pm}{}^p{}_{j]} \pm \frac{3}{4} \hodge{G}_{[m}\mathring \omega_{\pm ij]}~,
	\end{split}
\end{equation}

\begin{equation}
	\begin{split}
		\mathcal{D}^\mathcal{F}_m \tilde\omega_\pm{}_{ij} \defeq& \nabla_m \tilde\omega_\pm{}_{ij} + H_{m[i|p|}\omega_\pm{}^p{}_{j]} \pm \hodge{F}_{m[i|p|} \mathring \omega_{\pm}{}^p{}_{j]} \\
		=& \mp A^{-1} \partial_{m} A \, \tilde\omega_\pm{}_{ij} - \frac{1}{2} S\delta_{m[i} \tilde k_{\pm j]} \mp \frac{1}{4} \delta_{m[i} \hodge{F}_{j]pq} \mathring \omega_{\pm}{}^{pq} \\
		&\pm \frac{3}{4} \hodge{F}_{[mi|p|} \mathring \omega_{\pm}{}^p{}_{j]} \pm \frac{1}{2} \delta_{m[i} \hodge{G}_{j]} \mathring f_{\pm} + \frac{1}{4} G_{mijp} \tilde k_\pm{}^p~,
	\end{split}
\end{equation}

\begin{equation}
	\begin{split}
		\mathcal{D}^\mathcal{F}_m \mathring \omega_{\pm}{}_{ij} \defeq& \nabla_m \mathring \omega_{\pm}{}_{ij} \mp \hodge{H}_{m[i} \tilde k_{\pm j]} \mp \hodge{F}_{m[i|p|} \tilde \omega_\pm{}^p{}_{j]} \pm \frac{1}{2} \hodge{G}_m \omega_{\pm ij} \\
		=& \mp A^{-1} \partial_{m} A \, \mathring \omega_{\pm}{}_{ij} \mp \delta_{m[i} \hodge{H}_{j]p} \tilde k_\pm{}^p \mp \frac{3}{2} \hodge{H}_{[mi} \tilde k_{\pm j]} + \frac{1}{2} H_{mij} \hat f_{\pm} \\
		& \pm \frac{1}{8} \hodge{S}_{mijpq} \omega_\pm{}^{pq} \pm \frac{1}{4} \delta_{m[i} \hodge{F}_{j]pq} \tilde\omega_\pm{}^{pq} \mp \frac{3}{4} \hodge{F}_{[mi|p|}\tilde\omega_\pm{}^p{}_{j]} \\
		& \pm \frac{1}{2} \delta_{m[i} \hodge{G}_{|p|} \omega_\pm{}^p{}_{j]} \pm \frac{3}{4} \hodge{G}_{[m}\omega_{\pm ij]}~.
	\end{split}
\end{equation}
As the domain of the TCFH minimal connection, $\mathcal{D}^\mathcal{F}$, factorises on the symmetric and skew-symmetric form bilinears under the exchange of $\chi_\pm^r$ and $\chi_\pm^s$ and after taking into account the details of the action of $\mathcal{D}^\mathcal{F}$ on the forms, one concludes that the reduced holonomy of $\mathcal{D}^\mathcal{F}$ is included in $GL(20) \times SO(5) \times GL(35)$.

\subsection{The TCFH of Warped AdS\texorpdfstring{\textsubscript{6}}{6} Backgrounds}

A basis in the space of form bilinears on $ M^4$, up to Hodge duality, which are symmetric in the exchange of $\chi_\pm^r$ and $\chi_\pm^s$ is
\begin{equation} \label{AdS6_Sym_Bilinears}
	\begin{gathered}
		f^{rs}_\pm = \Herm{\chi^r_\pm}{\chi^s_\pm}~, \quad \tilde{f}^{rs}_\pm = \Herm{\chi^r_\pm}{\Gamma_{11} \chi^s_\pm} , \quad \hat f^{rs}_{\pm} = \Herm{\chi^r_\pm}{\Gamma_ {zx_1x_2x_3}\chi^s_\pm}~, \\
		\mathring f^{rs}_{\pm} = \Herm{\chi^r_\pm}{\Gamma_ {zx_1x_2x_3}\Gamma_{11} \chi^s_\pm} , \quad k^{rs}_\pm = \Herm{\chi^r_\pm}{\Gamma_ i\chi^s_\pm} \, \e^i~, \quad \hat k^{rs}_{\pm} = \Herm{\chi^r_\pm}{\Gamma_ {izx_1x_2x_3} \chi^s_\pm} \, \e^i~.
	\end{gathered}
\end{equation}
The TCFH is
\begin{equation}
	\begin{split}
		\mathcal{D}^\mathcal{F}_m f_{\pm} \defeq& \nabla_m f_{\pm} \\
		=& \mp A^{-1} \partial_{m} A \, f_\pm - \frac{1}{4}Sk_\pm{}_m \mp \frac{1}{4}\hodge{F}_{mp}\hat k_{\pm}{}^{p}~,
	\end{split}
\end{equation}

\begin{equation}
	\begin{split}
		\mathcal{D}^\mathcal{F}_m \tilde{f}_\pm \defeq& \nabla_m \tilde{f}_\pm \\
		=& \mp A^{-1} \partial_{m} A \, \tilde{f}_\pm - \frac{1}{4} F_{mp}k_\pm{}^p \pm \frac{1}{4}\hodge{G} \hat k_{\pm}{}_m~,
	\end{split}
\end{equation}

\begin{equation}
	\begin{split}
		\mathcal{D}^\mathcal{F}_m \hat f_{\pm} \defeq& \nabla_m \hat f_{\pm} \\
		=& \mp A^{-1} \partial_{m} A \, \hat f_{\pm} -\frac{1}{4}S\hat k_{\pm}{}_m \mp \frac{1}{4} \hodge{F}_{mp}k_\pm{}^p~,
	\end{split}
\end{equation}

\begin{equation}
	\begin{split}
		\mathcal{D}^\mathcal{F}_m \mathring f_{\pm} \defeq& \nabla_m \mathring f_{\pm} \\
		=& \mp A^{-1} \partial_{m} A \, \mathring f_{\pm} -\frac{1}{4}F_{mp}\hat k_{\pm}{}^p \pm \frac{1}{4} \hodge{G} k_\pm{}_m~,
	\end{split}
\end{equation}

\begin{equation}
	\begin{split}
		\mathcal{D}^\mathcal{F}_m k_\pm{}_i \defeq& \nabla_m k_\pm{}_i \mp \frac{1}{2} \hodge{H}_m \hat k_{\pm}{}_i \\
		=& \mp A^{-1} \partial_{m} A \, k_\pm{}_i \, \mp \frac{1}{2} \delta_{mi} \hodge{H}_p \hat k_{\pm}{}^p \mp \hodge{H}_{[m}\hat k_{\pm}{}_{i]} -\frac{1}{4}\delta_{mi} Sf_\pm \\
		&\mp \frac{1}{4} \hodge{F}_{mi}\hat f_{\pm} + \frac{1}{4} F_{mi}\tilde{f}_\pm \mp \frac{1}{4} \delta_{mi}\hodge{G}\mathring f_{\pm}~,
	\end{split}
\end{equation}

\begin{equation}
	\begin{split}
		\mathcal{D}^\mathcal{F}_m \hat k_{\pm}{}_i \defeq& \nabla_m \hat k_{\pm}{}_i \mp \frac{1}{2} \hodge{H}_m k_\pm{}_i \\
		=& \mp A^{-1} \partial_{m} A \, \hat k_{\pm}{}_i \, \mp \frac{1}{2} \delta_{mi}\hodge{H}_p k_\pm{}^p \mp \hodge{H}_{[m}k_\pm{}_{i]} - \frac{1}{4}\delta_{mi} S \hat f_{\pm} \\
		&\mp \frac{1}{4}\hodge{F}_{mi} f_\pm + \frac{1}{4} F_{mi} \mathring f_{\pm} \mp \frac{1}{4} \delta_{mi} \hodge{G} \tilde{f}~,
	\end{split}
\end{equation}
where $\nabla$ is the spin connection of $g(M^4)$.

A basis in the space of form bilinears on $ M^4$, up to Hodge duality, which are skew-symmetric in the exchange of $\chi_\pm^r$ and $\chi_\pm^s$ is
\begin{equation} \label{AdS6_Skew_Bilinears}
	\begin{gathered}
		\tilde{k}^{rs}_{\pm} = \Herm{\chi^r_\pm}{\Gamma_i \Gamma_{11} \chi^s_\pm} \, \e^i~, \quad \mathring k^{rs}_{\pm} = \Herm{\chi^r_\pm}{\Gamma_{izx_1x_2x_3}\Gamma_{11}\chi^s_\pm} \, \e^i~, \\
		\omega^{rs}_\pm = \frac{1}{2} \Herm{\chi^r_\pm}{\Gamma_{ij}\chi^s_\pm} \, \e^i \wedge \e^j~, \quad \tilde{\omega}^{rs}_\pm = \frac{1}{2} \Herm{\chi^r_\pm}{\Gamma_{ij}\Gamma_{11}\chi^s_\pm} \, \e^i \wedge \e^j~.
	\end{gathered}
\end{equation}
The TCFH is
\begin{equation}
	\begin{split}
		\mathcal{D}^\mathcal{F}_m \tilde k_\pm{}_i \defeq& \nabla_m \tilde k_\pm{}_i \pm \frac{1}{2} \hodge{H}_m \mathring k_{\pm i} - \frac{1}{2} F_{mp} \omega_\pm{}^p{}_i \\
		=& \mp A^{-1} \partial_{m} A \, \tilde{k}_\pm{}_i \, \pm \frac{1}{2} \delta_{mi} \hodge{H}_p \mathring k_{\pm}{}^p \pm \hodge{H}_{[m} \mathring k_{\pm i]} - \frac{1}{4} S \omega_{\pm mi} \\
		&+ \frac{1}{8} \delta_{mi} F_{pq} \omega_\pm{}^{pq} - \frac{1}{2} F_{[m|p|} \omega_\pm{}^p{}_{i]} - \frac{1}{8} G_{mipq} \tilde\omega_\pm{}^{pq}~,
	\end{split}
\end{equation}

\begin{equation}
	\begin{split}
		\mathcal{D}^\mathcal{F}_m \mathring k_{\pm i} \defeq& \nabla_m \mathring k_{\pm i} \pm \frac{1}{2} \hodge{H}_m \tilde k_{\pm i} \pm \frac{1}{2} \hodge{F}_{mp} \tilde \omega_\pm{}^p{}_i \\
		=& \mp A^{-1} \partial_{m} A \, \mathring k_{\pm i} \, \pm \frac{1}{2} \delta_{mi} \hodge{H}_p \tilde k_{\pm}{}^p \pm \hodge{H}_{[m} \tilde k_{\pm i]} \mp \frac{1}{8} \hodge{S}_{mipq}\omega_\pm{}^{pq} \\
		&\mp \frac{1}{8} \delta_{mi} \hodge{F}_{pq} \tilde \omega_\pm{}^{pq} \pm \frac{1}{2} \hodge{F}_{[m|p|} \tilde\omega_\pm{}^p{}_{i]} \mp \frac{1}{4} \hodge{G} \omega_{\pm mi}~,
	\end{split}
\end{equation}

\begin{equation}
	\begin{split}
		\mathcal{D}^\mathcal{F}_m \omega_\pm{}_{ij} \defeq& \nabla_m \omega_\pm{}_{ij} + H_{m[i|p|}\tilde \omega_\pm{}^p{}_{j]} + F_{m[i} \tilde k_{\pm j]} \\
		=& \mp A^{-1} \partial_{m} A \, \omega_\pm{}_{ij} \, \mp \frac{1}{4} \hodge{S}_{mijp} \mathring k_{\pm}{}^p + \frac{1}{2} \delta_{m[i} F_{j]p} \tilde k_\pm{}^p \\
		&+ \frac{3}{4} F_{[mi} \tilde k_{\pm j]} \pm \frac{1}{2} \hodge{G} \delta_{m[i} \mathring k_{\pm j]}~,
	\end{split}
\end{equation}

\begin{equation}
	\begin{split}
		\mathcal{D}^\mathcal{F}_m \tilde\omega_\pm{}_{ij} \defeq& \nabla_m \tilde\omega_\pm{}_{ij} + H_{m[i|p|} \omega_\pm{}^p{}_{j]} \pm \hodge{F}_{m[i}\mathring k_{\pm j]} \\
		=& \mp A^{-1} \partial_{m} A \, \tilde\omega_\pm{}_{ij} - \frac{1}{2} S \delta_{m[i} \tilde k_{\pm j]} \pm \frac{1}{2} \delta_{m[i} \hodge{F}_{j]p} \mathring k_{\pm}{}^p \\
		&\pm \frac{3}{4} \hodge{F}_{[mi}\mathring k_{\pm j]} + \frac{1}{4} G_{mijp} \tilde k_\pm{}^p~.
	\end{split}
\end{equation}
Notice that the minimal TCFH connection, $\mathcal{D}^\mathcal{F}$, acts on the form bilinears $k_\pm + \hat k_\pm$ and $k_\pm- \hat k_\pm$ as a connection gauging a scale symmetry of the type $k\pm\hat k\rightarrow s^{\pm 1} (k\pm\hat k)$, $s\in \bR-\{0\}$. Therefore the reduced holonomy of the minimal TCFH connection, $\mathcal{D}^\mathcal{F}$, is included in $SO(5)\times GL(1) \times GL(20)$.

\subsection{The TCFH of Warped AdS\texorpdfstring{\textsubscript{7}}{7} Backgrounds}

A basis in the space of form bilinears on $ M^3$, up to Hodge duality, which are symmetric in the exchange of $\chi_\pm^r$ and $\chi_\pm^s$ is
\begin{equation} \label{AdS7_Sym_Bilinears}
	\begin{gathered}
		f^{rs}_\pm = \Herm{\chi^r_\pm}{\chi^s_\pm}~, \quad \tilde{f}^{rs}_\pm = \Herm{\chi^r_\pm}{\Gamma_{11} \chi^s_\pm}~, \quad \hat f^{rs}_{\pm} = \Herm{\chi^r_\pm}{\Gamma_ {zx_1\dots x_4} \chi^s_\pm}~, \\
		k^{rs}_\pm = \Herm{\chi^r_\pm}{\Gamma_ i\chi^s_\pm} \, \e^i~.
	\end{gathered}
\end{equation}
The TCFH is
\begin{equation}
	\begin{split}
		\mathcal{D}^\mathcal{F}_m f_{\pm} \defeq& \nabla_m f_{\pm} \\
		=& \mp A^{-1} \partial_{m} A \, f_\pm - \frac{1}{4}Sk_\pm{}_m \mp \frac{1}{4}\hodge{F}_{m}\hat f_{\pm}~,
	\end{split}
\end{equation}

\begin{equation}
	\begin{split}
		\mathcal{D}^\mathcal{F}_m \tilde{f}_\pm \defeq& \nabla_m \tilde{f}_\pm \\
		=& \mp A^{-1} \partial_{m} A \, \tilde{f}_\pm - \frac{1}{4} F_{mp}k_\pm{}^p~,
	\end{split}
\end{equation}

\begin{equation}
	\begin{split}
		\mathcal{D}^\mathcal{F}_m \hat f_{\pm} \defeq& \nabla_m \hat f_{\pm} \\
		=& \mp A^{-1} \partial_{m} A \, \hat f_{\pm} \, \pm \frac{1}{2} \hodge{H} k_\pm{}_m \mp \frac{1}{4} \hodge{F}_m f_\pm~,
	\end{split}
\end{equation}

\begin{equation}
	\begin{split}
		\mathcal{D}^\mathcal{F}_m k_\pm{}_i \defeq& \nabla_m k_\pm{}_i \\
		=& \mp A^{-1} \partial_{m} A \, k_\pm{}_i \, \mp \frac{1}{2}\delta_{mi} \hodge{H} \hat f_{\pm} -\frac{1}{4}\delta_{mi} Sf_\pm + \frac{1}{4} F_{mi}\tilde{f}_\pm~,
	\end{split}
\end{equation}
where $\nabla$ is the spin connection of $g(M^3)$.

A basis in the space of form bilinears of $ M^3$, up to Hodge duality, which are skew-symmetric in the exchange of $\chi_\pm^r$ and $\chi_\pm^s$ is
\begin{equation} \label{AdS7_Skew_Bilinears}
	\begin{gathered}
		\mathring f^{rs}_{\pm} = \Herm{\chi^r_\pm}{\Gamma_{zx_1\dots x_4}\Gamma_{11}\chi^s_\pm} , \quad \tilde{k}^{rs}_{\pm} = \Herm{\chi^r_\pm}{\Gamma_i \Gamma_{11} \chi^s_\pm} \, \e^i , \\
		\hat k^{rs}_{\pm} = \Herm{\chi^r_\pm}{\Gamma_{izx_1\dots x_4}\chi^s_\pm} \, \e^i , \quad \mathring k^{rs}_{\pm} = \Herm{\chi^r_\pm}{\Gamma_{izx_1\dots x_4}\Gamma_{11}\chi^s_\pm} \, \e^i~,
	\end{gathered}
\end{equation}
The TCFH is

\begin{equation}
	\begin{split}
		\mathcal{D}^\mathcal{F}_m \mathring f_{\pm} \defeq& \nabla_m \mathring f_{\pm} \\
		=& \mp A^{-1} \partial_{m} A \, \mathring f_{\pm} \pm \frac{1}{2} \hodge{H} \tilde k_{\pm m} - \frac{1}{4} \mathring k_{\pm m} + \frac{1}{4} F_{mp} \hat k_{\pm}{}^p~,
	\end{split}
\end{equation}

\begin{equation}
	\begin{split}
		\mathcal{D}^\mathcal{F}_m \tilde k_{\pm i} \defeq& \nabla_m \tilde k_{\pm i} \mp \frac{1}{2} \hodge{F}_m \mathring k_{\pm i} \\
		=& \mp A^{-1} \partial_{m} A \, \tilde k_{\pm i} \mp \frac{1}{2} \hodge{H} \delta_{mi} \mathring f_{\pm} \mp \frac{1}{4} \hodge{S}_{mip} \hat k_{\pm}{}^p \\
		& \mp \frac{1}{4} \delta_{mi} \hodge{F}_p \mathring k_{\pm}{}^p \mp \frac{1}{2} \hodge{F}_{[m} \mathring k_{\pm i]}~,
	\end{split}
\end{equation}

\begin{equation}
	\begin{split}
		\mathcal{D}^\mathcal{F}_m \hat k_{\pm i} \defeq& \nabla_m \hat k_{\pm i} \\
		=& \mp A^{-1} \partial_{m} A \, \hat k_{\pm i} - \frac{1}{2} H_{mip} \mathring k_{\pm}{}^p \pm \frac{1}{4} \hodge{S}_{mip} \tilde k_\pm{}^p - \frac{1}{4} F_{mi} \mathring f_{\pm}~,
	\end{split}
\end{equation}

\begin{equation}
	\begin{split}
		\mathcal{D}^\mathcal{F}_m \mathring k_{\pm i} \defeq& \nabla_m \mathring k_{\pm i} \mp \frac{1}{2} \hodge{F}_{m} \tilde k_{\pm i} \\
		=& \mp A^{-1} \partial_{m} A \, \mathring k_{\pm i} - \frac{1}{2} H_{mip} \hat k_{\pm}{}^p - \frac{1}{4} S \delta_{mi} \mathring f_{\pm} \\
		&\mp \frac{1}{4} \delta_{mi} \hodge{F}_p \tilde k_\pm{}^p \mp \frac{1}{2} \hodge{F}_{[m} \tilde k_{\pm i]}~.
	\end{split}
\end{equation}
As in the previous AdS$_6$ case, observe that the the minimal TCFH connection, $\mathcal{D}^\mathcal{F}$, acts on $\tilde k\pm \mathring k$ like gauging an additional gauge symmetry. Therefore the reduced holonomy of the minimal TCFH connection, $\mathcal{D}^\mathcal{F}$, is included in $ SO(3) \times SO(3) \times GL(1)$.

\section{Symmetries of Probes, AdS Backgrounds and TCFHs} \label{AdS-Probes}

\subsection{Probes and Symmetries}

The dynamics of relativistic and spinning particles propagating on warped AdS backgrounds, AdS$_n\times_w M^{10-n}$, have been investigated in detail in \cite{epbgp}. Here we shall summarise some key properties of the dynamics of spinning particles which are relevant for the examples that we shall present below. As we shall consider examples for which the warp factor is constant, the action of spinning particles propagating on the spacetime factorises to an action on AdS$_n$ and an action on the internal space $M^{10-n}$. The latter can be written as
\begin{equation}
	S_M = -\frac{i}{2} \int\, \dd t\, \dd \theta\, \gamma_{IJ} Dy^I \partial_t y^J~,
\end{equation}
where $y = y(t,\theta)$ is a worldline superfield, $(t,\theta)$ are the worldline coordinates, $\gamma$ is the internal space metric and $D^2 = i\partial_t$. Of course if $M^{10-n}$ is the product of two or more other manifolds, then the action $S_M$ factorises further into actions associated to each manifold in the product.

It turns out that the infinitesimal variation
\begin{equation}
	\delta y^I = \epsilon \alpha^I{}_{J_1\dots J_{m-1}} Dy^{J_1}\cdots Dy^{J_{m-1}}~,
\end{equation}
associated with a $m$-form $\alpha$ on $M^{10-n}$ is a (hidden) symmetry of $S_M$, if and only if $\alpha$ is a (standard) KY form, where $\epsilon$ is an infinitesimal parameter. Below we shall present several examples of IIA AdS backgrounds where KY forms arise as a consequence of the TCFH on their internal spaces. In this way, we shall provide a link between TCFHs and conservation laws of probes propagating on such backgrounds.

\subsection{Examples of TCFH and KY Forms}

There are many IIA AdS backgrounds that we can consider, see e.g. \cite{FR, romans, jose, cvetic, lust1, lust2, passias1, rosa, jggpx}. As the aim is to provide some examples of backgrounds for which the TCFHs give rise to symmetries for spinning particle probes, we shall not be comprehensive and instead focus on AdS backgrounds that arise as near horizon geometries of intersecting branes \cite{gppt, aat, jgkt}, see also \cite{bps}. In the analysis that follows, we shall present an ansatz which includes the near horizon geometry of intersecting branes under consideration and proceed to demonstrate that the associated TCFH gives rise to KY forms on the internal space. In turn these generate symmetries for spinning particle probes and so demonstrate a relation between TCFHs and probe symmetries.

The formulae for the reduced field equations and KSEs on the internal space of a warped AdS background that we shall use to construct the AdS solutions suitable for our purposes can be found in appendix \ref{sec:Warped-AdS-Backgrounds} and \cite{bgpiia}. As it has already been mentioned, these have been obtained after suitably solving the field equations and KSEs of the theory over the AdS subspace and identifying the remaining equations on the internal space of these backgrounds. Here we shall typically quote the relevant parts of these equations -- for the derivation and the full expressions of these equations the reader should consult the original reference.

\subsubsection{An AdS\texorpdfstring{$_3$}{3} Solution from a Fundamental String on a NS5-brane}

An example of an AdS$_3$ solution arises as the near horizon geometry of a fundamental string on a NS5-brane background. This configuration has played a prominent role in a microscopic string theory counting of entropy for extreme black holes \cite{strominger, callan}. An ansatz which includes such a solution is
\begin{equation}\label{F1NS5}
	g = g_\ell (AdS_3) + g (\bR^4) + g (S^3) \quad H = p\, \dvol_\ell(AdS_3) + q\, \dvol(S^3)~,
\end{equation}
the dilaton is constant, $\Phi = \textrm{const}$, and the rest of the fields are set to zero, where $g_\ell (AdS_3)$ ($g(S^3)$) and $\dvol_\ell(AdS_3)$ ($\dvol(S^3)$) are the standard metric and associated volume form of AdS$_3$ ($S^3$) of radius $\ell$ (unit radius), respectively, $g(\bR^4)$ is the Euclidean metric of $\bR^4$ and $p, q \in \mathbb{R}$. From here on we shall adopt the same conventions for the AdS$_n$ $(S^k$) metric and volume form in all the examples below -- $g(\bR^m)$ will always denote the Euclidean metric on $\bR^m$. Note that $\bR^4$ can be replaced with any Ricci flat manifold, like for example $K_3$, but the choice of $\bR^4$ suffices for the purpose of this example. Moreover as the warp factor $A$ is constant and the radius $\ell$ of AdS$_3$ has been kept arbitrary, so without loss of generality, we have set $A = 1$. Furthermore, the radius of $S^3$ has been set to 1 after possibly an overall rescaling of the spacetime metric and $H$.

To find a solution based on the ansatz \eqref{F1NS5}, one has to determine $p,q$ and $\ell$ after solving the field and KSEs on the $\bR^4\times S^3$ internal space. As the IIA 4-form flux vanishes, one has that $X = Y = 0$. Moreover a direct comparison of \eqref{ads3back} with \eqref{F1NS5} reveals that $p = W$ and $Z = q\, \dvol(S^3)$ .

To determine the remaining constants $q$ and $\ell$, one first considers the field equation of the dilaton $\Phi$,
\begin{equation}
	\nabla^2 \Phi = - \frac{1}{12} Z^2 + \frac{1}{2} W^2 \equiv 0~,
\end{equation}
which implies that $q^2 = W^2 = p^2$. Next, the Einstein field equations along the $S^3$ directions and the field equation of the warp factor
\begin{equation}
	\begin{gathered}
		R_{\alpha \beta}^{S^3} = \frac{1}{4} Z_{\alpha \gamma \delta}Z^{\gamma \delta}{}_{\beta} \equiv 2\, \delta_{\alpha \beta} \,, \\
	\nabla^2 \log A = - \frac{2}{\ell^2} + \frac{1}{2} W^2 \equiv 0~,
	\end{gathered}
\end{equation}
respectively yield $p^2 = W^2 = 4$ and $\ell = 1$, i.e. the AdS$_3$ and $S^3$ subspaces have the same radius and $p,q = \pm2$.

Turning attention to the KSEs, and focusing for simplicity on those on $\sigma_+$, the dilatino KSE, $\mathcal{A}^{(+)} \sigma_+ = 0$, with
\begin{equation}
	\mathcal{A}^{(+)} = \frac{1}{12} \slashed{Z} \Gamma_{11} - \frac{1}{2} W \Gamma_z \Gamma_{11}~,
\end{equation}
gives the condition $\Gamma_{(3)} \Gamma_z \sigma_+ = -\sigma_+$ provided we choose\footnote{The treatment of $p = -q$ case follows from that of $p = q$ in a straightforward manner.} $p = q$, where $\Gamma_{(3)}$ is the product of the three gamma matrices along the orthonormal directions tangent to the three sphere. The additional algebraic KSE, , $\Xi_+ \sigma_+ = 0$, which can be found in appendix \ref{sec:Warped-AdS-Backgrounds} with
\begin{equation}
	\Xi_+ = - \frac{1}{2 \ell} + \frac{1}{4} W \Gamma_{11}~,
\end{equation}
that arises from the integration of the gravitino KSE along the $z$ directions, results in the condition $\Gamma_{11} \sigma_+ = \sigma_+$, where we have chosen $p = 2$. Therefore, we find that $\sigma_+ $ is a spacetime chiral spinor. The solution with $p = -2$ can be investigated in a similar way to that for $p = 2$.

The gravitino KSE \eqref{supconeads3} along $\bR^4$ shows that the Killing spinors $\sigma_+$ satisfy the condition $\nabla_i^{\bR^4} \sigma_+ = 0$ and so do not depend on the coordinates of $\bR^4$. Furthermore, the gravitino KSE along $S^3$ can be written as:
\begin{equation}\label{f1ns5s3}
	\nabla_\alpha^{S^3} \sigma_+ + \frac{1}{2} \Gamma_\alpha \Gamma_z \sigma_+ = 0~,
\end{equation}
where we have made use of the conditions $\Gamma_{(3)} \Gamma_z \sigma_+ = -\sigma_+$ and $\Gamma_{11} \sigma_+ = \sigma_+$. This does not impose further constraints on $\sigma_+$. Therefore the only conditions on $\sigma_+$ are $\Gamma_{(3)} \Gamma_z \sigma_+ = -\sigma_+$ and $\Gamma_{11} \sigma_+ = \sigma_+$ and so $\sigma_+$ has 4 independent components. A similar analysis of the KSEs on $\sigma_-$ and $\tau_\pm$ spinors yields another 12 independent Killing spinors and so the solution preserves 1/2 of the supersymmetry as expected. Note that if $\bR^4$ is replaced by $K_3$ or any other 4-dimensional hyper-Kähler manifold $Q^4$ and the orientation of $Q^4$ is chosen to be compatible with the conditions $\Gamma_{(3)} \Gamma_z \sigma_+ = -\sigma_+$ and $\Gamma_{11} \sigma_+ = \sigma_+$, the solution will again preserve 1/2 of supersymmetry. The spinors $\sigma_\pm$ and $\tau_\pm$ will be covariantly constant with respect to the spin connection of the hyper-Kähler metric on $X^4$.

A consequence of \eqref{f1ns5s3} is that the bilinears
\begin{equation}\label{s3bi0}
	(k^{rs}_\pm)_\alpha =\Herm{\sigma^r_\pm}{\Gamma_\alpha \sigma^s_\pm}~, \quad (\omega^{rs}_\pm)_{\alpha\beta} =\Herm{\sigma^r_\pm}{\Gamma_{\alpha\beta} \sigma^s_\pm}~, \quad (\pi^{rs}_\pm)_{\alpha\beta\gamma} =\Herm{\sigma^r_\pm}{\Gamma_{\alpha\beta\gamma} \sigma^s_\pm}~,
\end{equation}
are CCKY forms on $S^3$, while the bilinears
\begin{equation}\label{s3bi}
	\begin{gathered}
	(\hat k^{rs}_\pm)_\alpha =\Herm{\sigma^r_\pm}{\Gamma_\alpha \Gamma_z \sigma^s_\pm}~, \quad (\hat \omega^{rs}_\pm)_{\alpha\beta} =\Herm{\sigma^r_\pm}{\Gamma_{\alpha\beta}\Gamma_z \sigma^s_\pm}~, \\
	(\hat\pi^{rs}_\pm)_{\alpha\beta\gamma} =\Herm{\sigma^r_\pm}{\Gamma_{\alpha\beta\gamma} \Gamma_z\sigma^s_\pm}~,
	\end{gathered}
\end{equation}
are KY forms on $S^3$. The latter generate symmetries for spinning particle actions on $S^3$.

\subsubsection{An AdS\texorpdfstring{$_2$}{2} Solution from Intersecting D2- and D4-branes}

An ansatz which includes the near horizon geometry of two D2- and two D4-branes intersecting on a 0-brane is
\begin{equation}\label{d2d2d4d4}
	g = g_\ell(AdS_2) + g(S^2) + g(\bR^2) + g(\bR^4)~, \quad G= \dvol_\ell(AdS_2)\wedge \alpha + \dvol(S^2)\wedge \beta~,
\end{equation}
with constant dilaton $\Phi$ and all other remaining fields set to zero, where $\ell$ is the radius of AdS$_2$ and $\alpha$ and $\beta$ are constant 2-forms on $\bR^4$.

Assuming that $\bR^4 = \bR\langle(\e_3,\e_4, \e_5, \e_6)\rangle$, there is an $SO(4)$ transformation such that the form $\alpha$ can be written as $\alpha = p\, \e^3\wedge \e^4 + q\, \e^5\wedge \e^6$. The isotropy group $SO(2)\times SO(2)$ of $\alpha$ can then be used to choose $\beta$ without loss of generality as
\begin{equation}\label{betaeqn}
	\beta = r\, \e^3\wedge \e^4 + s\, \e^5\wedge \e^6 + a\, \e^3\wedge \e^5 + b\, \e^4\wedge \e^6 + c\, \e^4\wedge \e^5~,
\end{equation}
where all components of $\alpha$ and $\beta$ are constants in $\bR$.

The Einstein equations along $\bR^4$ (with the two indices distinct) imply that $c r = c s = c b = c a = 0$. Thus if $c\not=0$, $r = s = b = a = 0$. Then the remaining Einstein equations along $\bR^4$ give that $p = q = 0$. Finally, the dilatino KSE for the ansatz \eqref{d2d2d4d4} is
\begin{equation}
\left( - \frac{1}{8}\slashed{X} + \frac{1}{4\cdot 4!}\slashed{Y} \right)\eta_+ = 0~,
\end{equation}
and gives $c = 0$. Therefore all fluxes vanish for this case, so to proceed we take $c = 0$.

Setting $c = 0$, the dilatino KSE as well as the gravitino KSE along $\bR^4$ can be written for the fluxes \eqref{d2d2d4d4} as
\begin{equation}
	\begin{aligned}
		\big(&-p + q I_1 + \Gamma_{(2)} (-r + s I_1)-a I_2-b I_1 I_2\big)\eta_+ = 0~, \\
		\big(&-p + q I_1 + \Gamma_{(2)} (-r + s I_1)-a I_2-b I_1 I_2\big)\Gamma_\mu \eta_+ = 0~, \quad \mu = 3,4,5,6
	\end{aligned}
\end{equation}
where $I_1 = \Gamma_{3456}$, $I_2 = \Gamma_{(2)} \Gamma_{45}$, $\Gamma_{(2)}$ is the product of two gamma matrices along orthonormal directions tangent to $S^2$ and we have taken $\eta_+$ to be constant along $\bR^4$. Separating the Hermitian and anti-Hermitian components of the above equations and using that $I_1 \Gamma_\mu = -\Gamma_\mu I_1$ as well as the commutation relations of $\Gamma_\mu$ with $I_2$, one finds that $r,s = 0$ and
\begin{equation}
	(q I_1 + p)\eta_+ = 0~, \quad (b I_1-a)\eta_+ = 0~, \quad (a I_2 + p)\eta_+ = 0~.
\end{equation}
These can be solved by restricting $\eta_+$ to the eigenspaces of $I_1$ and $I_2$. In turn, one finds that $p,q,a,b$ are proportional to each other with proportionality factor of a sign. Therefore in all cases, $a^2 = b^2 = p^2 = q^2$. A similar analysis holds for the $\eta_-$ Killing spinors. As each eigenspace of $I_1$ and $I_2$ on either $\eta_+$ or $\eta_-$ has dimension 4, there are 8 Killing spinors that solve the above KSEs.

After using that $S^2$ has radius 1, the Einstein equation along $S^2$ reveals that $a^2 = 1$. In turn the field equation for the warp factor $A$ gives $\ell = 1$. Therefore AdS$_2$ and $S^2$ have the same radius. All the remaining field equations are satisfied.

As the gravitino KSE along $\bR^2$ is satisfied, it remains to explore the gravitino KSE along $S^2$. This can be written as
\begin{equation}
	\nabla_\alpha^{S^2}\eta_+ + \frac{p}{2} \Gamma_{34}\Gamma_\alpha \eta_+ = 0~.
\end{equation}
This does not impose any additional conditions on $\eta_+$ and the same applies for the corresponding equation on $\eta_-$. Therefore the solution preserves $1/4$ of supersymmetry. It follows from this that the 1- and 2-form bilinears along $S^2$ and their duals are either KY or CCKY forms. There are several KY forms. For example, one can easily show that $(k^{rs}_\pm)_\alpha= \Herm{\eta^r_\pm}{\Gamma_\alpha \eta^s_\pm}$ and $(\check k^{rs}_\pm)_\alpha= \Herm{\eta^r_\pm}{\Gamma_\alpha \Gamma_{12} \eta^s_\pm}$ are KY forms. The KY forms generate symmetries for spinning particles propagating on the internal space of these backgrounds.

The background can be generalised somewhat by replacing $\bR^4$ with any other 4-dimensional hyper-Kähler manifold $Q^4$. In such a case, $X$ and $Y$ are chosen as
\begin{equation}
	X = p^r \lambda_r~, \quad Y = \dvol(S^2)\wedge a^r \lambda_r~,
\end{equation}
where $\lambda$ are the 3 Kähler forms of $Q^4$ associated with the hyper-complex structure and $p^r$ and $a^r$ are constant 3-vectors. Under a frame $SO(4)$ rotation both $p^r$ and $a^r$ transform as $SO(3)$ vectors. Moreover, the field equation for the magnetic component of the 3-form field strength implies that $\delta_{rs} p^r a^s = 0$, i.e. they are orthogonal. In such a case, there is an $SO(4)$ rotation such that $p^r \lambda_r = \alpha$ with $p^2 = q^2$ and $a^r\lambda_r = \beta$ as in \eqref{betaeqn} with $r = s = c = 0$ and $a^2 = b^2$. Moreover the relative signs in the equalities $p = \pm q$ and $a = \pm b$ should be chosen such that $\alpha$ and $\beta$ have the same self-duality properties on $Q^4$. After that the previous analysis on $\bR^4$ can be repeated to solve both KSEs and field equations yielding a new solution preserving again $1/4$ of supersymmetry. The identification of the KY forms on $S^2$ can be done as for $Q^4 = \bR^4$.

\subsubsection{AdS\texorpdfstring{$_3$}{3} Solutions from Intersecting D2- and D4-branes}

An ansatz that includes the near horizon geometry AdS$_2$ of a D2- and a D4-brane intersecting on a 1-brane is
\begin{equation}\label{d2d4}
	g = g_\ell(AdS_3) + g(S^3) + g(\bR^4)~, \quad G= \dvol_\ell(AdS_3)\wedge \alpha + \dvol(S^3)\wedge \beta~,
\end{equation}
with constant dilaton $\Phi$ and all other remaining fields set to zero, where $\ell$ is the radius of AdS$_3$ and $\alpha$ and $\beta$ are constant 1-forms on $\bR^4$.

First notice that the field equation for the magnetic component of the NS 3-form implies that $\alpha\wedge \beta = 0$ and so $\alpha$ and $\beta$ are co-linear, i.e. they are proportional and so write $\beta= p\alpha$. Next the dilatino KSE on $\sigma_+$ and the algebraic KSE $\Xi_+ \sigma_+ = 0$ imply that
\begin{equation}\label{proads3}
	\left(\Gamma_{(3)} \Gamma_z + \frac{1}{p} \right)\sigma_+ = 0~, \quad \left(\frac{1}{\ell} + {\slashed \alpha} \right)\sigma_+ = 0~,
\end{equation}
where $\Gamma_{(3)}$ is the product of three gamma matrices along orthonormal tangent directions of $S^3$, i.e. the Clifford algebra element associated to $\dvol_\ell(AdS_3)$. The dilaton field equation gives $p = \pm 1$ and so $\alpha^2 = \beta^2$. Moreover the warp factor field equation yields $\alpha^2 = 4 \ell^{-2}$.

Turning to the Einstein equation along $S^3$, one finds that
\begin{equation}
	R^{S^3}_{\alpha\beta} = \frac{2}{\ell^2} \delta_{\alpha\beta}~.
\end{equation}
As $S^3$ has unit radius, one concludes that $\ell = 1$ and so $\alpha^2 = \beta^2 = 4$. Therefore AdS$_3$ and $S^3$ have the same radius. Furthermore, one can verify that all the remaining field equations and KSEs are satisfied apart from the gravitino KSE along $S^3$. This can be written using \eqref{proads3} as
\begin{equation}
	\big(\nabla^{S^3}_\gamma + \frac{1}{4} \Gamma_z \slashed{\alpha} \Gamma_\gamma\big)\sigma_+ = 0~,
\end{equation}
and gives no additional conditions on $\sigma_+$. A similar analysis holds for the remaining Killing spinors $\sigma_-$ and $\tau_\pm$. As a result, the solution preserves 1/2 of the supersymmetry.

To proceed one can consider the bilinears as in \eqref{s3bi} and \eqref{s3bi0} and proceed to demonstrate that these and their Hodge duals on $S^3$ are either KY or CCKY forms. The former generate symmetries for spinning probes on $S^3$. In particular $k_\pm$, $\star\omega_\pm$ and $\pi_\pm$ are KY forms on $S^3$.

\subsubsection{AdS\texorpdfstring{$_2$}{2} solutions from intersecting D2-branes and fundamental strings}

An ansatz that includes the near horizon geometry of two D2-branes and a fundamental string intersecting on a 0-brane is
\begin{equation}
	\begin{gathered}
		g = g_\ell(AdS_2) + g(S^3) + g(\bR^5)~, \\
		G = \dvol_\ell(AdS_2)\wedge X~, \quad H = \dvol_\ell(AdS_2)\wedge W~,
	\end{gathered}
\end{equation}
with constant dilaton $\Phi$ and all other remaining fields set to zero, where $X$ and $W$ are a 2-form and 1-form on $\bR^5$, respectively.

The field equation for the magnetic part of the 2-form field strength implies that $i_W X = 0$. The dilaton field equation gives $W^2 = 1/4\,\, X^2$ and the warp factor field equations can be expressed as $W^2 = \ell^{-2}$.

Taking $\bR^5 = \bR\langle(\e_1,\e_2, \dots, \e_5)\rangle$, there is a $SO(5)$ transformation, up to a possible relabelling of the basis, such that $X = \lambda_1\, \e^1\wedge \e^2 + \lambda_2\, \e^3\wedge \e^4$ and $\lambda_1, \lambda_2\in \bR$. Next if either $\lambda_1$ or $\lambda_2$ vanish together with $i_W X = 0$, one can show that the gravitino KSE on $\eta_+$ along $\bR^5$ becomes inconsistent. Therefore from now on, we take $\lambda_1, \lambda_2\not = 0$ and as $i_W X = 0$, we have $W = p\, \e^5$.
Using this, the dilatino KSE yields
\begin{equation}
	\left(\frac{1}{2} \lambda_1-\frac{1}{2} \lambda_2 \Gamma_{1234} + p \Gamma_{12} \Gamma_{11} \Gamma_5 \right)\eta_+ = 0~.
\end{equation}
This together with the gravitino KSE along $\bR^5$ imply that
\begin{equation}\label{d2d2f1pro}
	(\lambda_2 \Gamma_{1234} + \lambda_1)\eta_+ = 0~, \quad ( p \Gamma_{12} \Gamma_{11} \Gamma_5 + \lambda_1)\eta_+ = 0~.
\end{equation}
As a result $\lambda_1^2 = \lambda_2^2 = p^2 = W^2$.

Restricting the Einstein equation along $S^3$, which has unit radius, yields $\lambda_1^2 = 4$. The warp factor field equation in turn gives $\ell = 1/2$. Therefore the AdS$_2$ subspace has half the radius of the internal space $S^3$. It remains to explore the gravitino KSE along $S^3$. This can be rewritten as
\begin{equation}\label{d2d2f1grav}
	\big(\nabla_\alpha^{S^3} + \frac{1}{4} \lambda_1 \Gamma_{12} \Gamma_\alpha\big)\eta_+ = 0~.
\end{equation}
This does not impose any additional conditions on $\eta_+$. A similar analysis can be carried out for the $\eta_-$ Killing spinors. As a result the solution preserves 1/4 of supersymmetry as a consequence of the conditions \eqref{d2d2f1pro} on $\eta_+$ and the analogous conditions on $\eta_-$.

There are several form bilinears that one can consider on $S^3$ like for example those in \eqref{s3bi} and \eqref{s3bi0} and their duals on $S^3$. All of them are either KY or CCKY as a consequence of \eqref{d2d2f1grav}. In particular, $k_\pm$, $\star\omega_\pm$ and $\pi_\pm$ are KY forms and so generate symmetries for spinning particles propagating on $S^3$.

\chapter{W-symmetries, Anomalies and Heterotic Backgrounds} \label{W-symmetries-heterotic-backgrounds}

\section{Introduction}

It has been known for sometime \cite{phgpw1, Howe:1991vs, phgp1}, following earlier related work in \cite{odake, dvn}, that $\hat\nabla$-covariantly constant forms on a spacetime $M$ generate symmetries\footnote{As the existence of $\hat\nabla$-covariantly constant forms on a spacetime implies the reduction of the holonomy of $\hat\nabla$ to a subgroup of $SO(9,1)$, the associated symmetries of probes are refereed to as holonomy symmetries.} in heterotic probes propagating on $M$, where $\hat\nabla$ is a metric connection with torsion given by the 3-form field strength, $H$, of heterotic theory, see \eqref{hnabla}. The algebra of these symmetries is a W-algebra \cite{phgp1}; for other applications, see e.g. \cite{vafa, jfof, mg}. Typically, the heterotic probes are (1,0)-supersymmetric sigma models \cite{ewch} that exhibit a metric, a 2-form and gauge couplings, and have as a target space the spacetime $M$. A class of heterotic backgrounds with such $\hat\nabla$-covariantly constant forms are those that preserve a fraction of the spacetime supersymmetry --- the $\hat\nabla$-covariantly constant forms are the (Killing spinor) form bilinears. The geometry of all supersymmetric heterotic backgrounds has been investigated in \cite{hetgpug1, hetgpug2} and the form bilinears have been identified; for a review see \cite{review}. There are two classes of supersymmetric heterotic backgrounds which are distinguished by whether the holonomy group of $\hat\nabla$ is either compact or non-compact. The symmetries of heterotic probes propagating on supersymmetric backgrounds with a compact holonomy group have been investigated in \cite{lggpepb}. In particular, it has been demonstrated that the algebra of symmetries closes as a W-algebra\footnote{The structure constants of the algebra depend on the conserved currents of the symmetries.}, provided that suitable additional symmetries are included. In addition, it has been shown that the chiral anomalies of these symmetries are consistent and can be cancelled up to two loops in sigma model perturbation theory.

The purpose of this chapter is to explore the holonomy symmetries of sigma models on supersymmetric heterotic backgrounds with a non-compact holonomy group. The non-compact holonomy groups that arise are semi-direct products of a compact group, $G$, with $\bR^8$, $G\ltimes\bR^8$, where $G$ is one of the groups listed in \eqref{hol}. The distinguished feature of supersymmetric heterotic backgrounds with non-compact holonomy group is that they admit a null $\hat\nabla$-covariantly constant 1-form bilinear $K$ and the remaining form bilinears, $L$, are null along $K$, i.e. they satisfy \eqref{nk}. First, we demonstrate that the commutator of two holonomy symmetries generated by null $\hat\nabla$-covariantly constant forms closes as a W-algebra, see \eqref{walg}. To describe the commutator in more detail, we find that the space of null forms along $K$ on $M$, $\Omega^*_K(M)$, admits a generalisation of the exterior product, $\curlywedge$, and the inner product, $\bar\curlywedge$, operations on forms, see \eqref{twoop}. In particular, we find that the commutator of two holonomy symmetries generated by the null forms $L$ and $M$ closes on transformations generated by $K$ and $L\bar\curlywedge M$. Moreover, if $L$ and $M$ are $\hat\nabla$-covariantly constant, then $L\curlywedge M$ and $L\bar\curlywedge M$ are $\hat\nabla$-covariantly constant. Thus the space of all null $\hat\nabla$-covariantly constant forms, $\Omega^*_{\hat\nabla}(M)$, is closed under these two operations. In fact, $\Omega^*_{\hat\nabla}(M)$ is a Lie algebra with Lie bracket operation $\bar\curlywedge$, which underpins the W-algebra of holonomy symmetries. As a result, the W-algebra of holonomy symmetries is completely determined from the Lie algebra $\Omega^*_{\hat\nabla}(M)$.

Therefore, to explore the W-algebra of holonomy symmetries it suffices to determine the Lie algebra structure of $\Omega^*_{\hat\nabla}(M)$. For this is useful to first identify the Lie algebra of {\it fundamental forms}, $\mathfrak{f}$. For this, first consider the $\bar\curlywedge$-closure of a (minimal) collection of null forms on $M$ whose covariant constancy condition determines the holonomy of the connection $\hat\nabla$. As $K$ $\bar\curlywedge$-commutes with all other elements of $\Omega^*_{\hat\nabla}(M)$, it is convenient to exclude $K$ as an element of $\mathfrak{f}$. Then, $\Omega^*_{\hat\nabla}(M)$ is generated from $\mathfrak{f}$ upon taking $\curlywedge$-products of the elements of $\mathfrak{f}$ and including $K$. The Lie algebra structure of $\Omega^*_{\hat\nabla}(M)$ is unravelled after decomposing $\Omega^*_{\hat\nabla}(M)$ in representations of $\mathfrak{f}$.

After establishing some general results on the Lie algebra structure of $\Omega^*_{\hat\nabla}(M)$, a more detailed investigation is ensued on the Lie algebra structure of $\mathfrak{f}$ and $\Omega^*_{\hat\nabla}(M)$ for each background with holonomy listed in \eqref{hol}. The Lie algebras $\mathfrak{f}$ are tabulated in table \ref{tab:holonomy_groups}. Moreover, the Lie algebra structure of $\Omega^*_{\hat\nabla}(M)$ is determined in each case.

The investigation of chiral anomalies in sigma models has a long history \cite{gmpn, laggin, bny, sen, chpt2, phgpa}. More recently, the chiral anomalies of holonomy symmetries for sigma models with Euclidean signature manifolds as target spaces have been investigated in \cite{dlossa} and those of sigma models with supersymmetric heterotic backgrounds with compact holonomy group as target spaces have been explored in \cite{lggpepb}. After determining the W-algebra of holonomy symmetries of sigma models on supersymmetric heterotic backgrounds with non-compact holonomy groups, we find the associated chiral anomalies using Wess-Zumino consistency conditions \cite{zumino-book}. We demonstrate that the anomalies are consistent up to at least two loops in sigma model perturbation theory. At the same loop level, we find that the anomalies can be cancelled after an appropriate quantum correction to the covariantly constant forms, which is consistent with the Green-Schwarz anomaly cancellation mechanism \cite{mgjs}.

This chapter is organised as follows. In section \ref{sigma-model-symmetries}, after defining the operations $\curlywedge$ and $\bar\curlywedge$ on $\Omega^*_K(M)$ and $\Omega^*_{\hat\nabla}(M)$ and summarising some of the geometric properties of heterotic backgrounds with non-compact holonomy group, we determine the commutator \eqref{walg} of two holonomy symmetries. Moreover, we argue that the W-algebra of holonomy symmetries is underpinned by the Lie algebras $\mathfrak{f}$ and $\Omega^*_{\hat\nabla}(M)$. In section \ref{holonomy-symmetries-w-algebras}, first we establish some general properties of the Lie algebra structure on $\Omega^*_{\hat\nabla}(M)$ and then determine the Lie algebras $\mathfrak{f}$ and $\Omega^*_{\hat\nabla}(M)$ for each of the backgrounds with holonomy group $G\ltimes \bR^8$, where $G$ listed in equation \eqref{hol}. In section \ref{anomalies}, we give the anomalies of these holonomy symmetries, prove that they are consistent up to at least two-loops and discuss their cancellation.

\section{Geometry and Sigma Model Symmetries} \label{sigma-model-symmetries}

\subsection{Geometry of Backgrounds with Non-Compact Holonomy}
As it has already been mentioned, the geometry of a spacetime $M$, with metric $g$, of supersymmetric heterotic backgrounds can be characterised by the holonomy group\footnote{We assume that the structure group of the spacetime $M$ reduces to a subgroup of the connected component $SO(9,1)$ of the Lorentz group. $M$ is a spin manifold and we always refer to the reduced holonomy group. } of the metric connection, $\hat\nabla$, with skew-symmetric torsion given by the 3-form field strength $H$ of the theory. In particular, one has that
\begin{equation}\label{hnabla}
	\hat{\nabla}_\mu X^\nu = \nabla_\mu X^\nu + \frac{1}{2} H^\nu{}_{\mu\rho} X^\rho,
\end{equation}
where $\nabla$ is the Levi-Civita connection of $g$, $X$ is a vector field on $M$ and $\mu, \nu, \rho=0, \dots, 9$. For supersymmetric backgrounds, the holonomy group of $\hat\nabla$ is a subgroup of the isotropy group of Killing spinors in $\mathrm{Spin}(9,1)$. This is either compact or non-compact. The non-compact holonomy groups that can occur are $G\ltimes \bR^8$ with $G$ given by one of the following groups \cite{hetgpug1, hetgpug2}
\begin{equation}\label{hol}
	\begin{gathered}
		\mathrm{Spin}(7)\ (1),\quad SU(4)\ (2),\quad Sp(2)\ (3),\quad \times^2Sp(1)\ (4), \\
	Sp(1)\ (5),\quad U(1)\ (6),\quad \{1\}\ (8),
	\end{gathered}
\end{equation}
where in parenthesis is the number of Killing spinors for each case. The form bilinears, which are $\hat\nabla$-covariantly constant by construction, include a null 1-form denoted by $K$. The 1-form $K$ is no-where vanishing on the spacetime and is Killing. The remaining form bilinears $L$ are also null along $K$, i.e. they satisfy
\begin{equation}\label{nk}
	K \wedge L = i_K L = 0 \, ,
\end{equation}
where $i_K$ denotes the inner derivation\footnote{Our conventions for the inner derivation can be found in appendix \ref{Conventions}.} of $L$ with respect to $K$ now viewed as a vector field --- the index is raised with the spacetime metric $g$.

Let us denote the space of null forms of $M$ along $K$ with $\Omega^*_K(M)$. One way to describe the elements of $\Omega^*_K(M)$, and so the form bilinears, is to introduce a local pseudo-orthonormal null co-frame $(\e^-, \e^+, \e^i)$ on the spacetime with $\e^-=K$, i.e. $g=2 \e^+\e^-+\delta_{ij}\e^i \e^j$. Then the conditions\footnote{In the dual frame $(\e_-, \e_+, \e_i)$ to the co-frame $(\e^-, \e^+, \e^i)$, the vector field $K=\e^\mu_+\partial_\mu$.} $K\wedge L=i_KL=0$ can be solved to yield
\begin{equation}\label{exL}
	L = \frac{1}{\ell!}\, L_{-i_1\dots i_\ell}\, \e^-\wedge \e^{i_1}\wedge \dots \wedge \e^{i_\ell} \equiv \frac{1}{\ell!}\, L_{i_1\dots i_\ell}\, \e^{-i_1 \dots i_\ell} \, ,
\end{equation}
i.e. the only non-vanishing components of $L$ are $L_{-i_1\dots i_\ell}\equiv L_{i_1\dots i_\ell}$, where in the last expression we have simplified the notation for the wedge product of co-frame 1-forms. Note that to establish \eqref{exL}, we have used that $K$ is no-where vanishing on the spacetime.

Typically for form bilinears, the components of $L$ can be identified with those of the usual fundamental forms of $G$-structures in 8 dimensions, up to an equivalence that will be discussed below, where $G$ is given in \eqref{hol}. For example if $G=\mathrm{Spin}(7)$, the components of $L$ on an open set are identified with those of the fundamental self-dual 4-form of the group. Of course $L$ depends on all coordinates of $M$.

Prompted by this, we associate to every $L\in \Omega^{\ell+1}_K(M)$ a locally defined $\ell$-form $\tilde L_\alpha$ on each open set, $U_\alpha$, of $M$ such that
\begin{equation}
	\tilde{L}_\alpha = \frac{1}{\ell!}\, L_{i_1\dots i_\ell}\, \e^{i_1\dots i_\ell} \, ,
\end{equation}
i.e. $\tilde L_{i_1\dots i_\ell}= L_{i_1\dots i_\ell}$. Note that, unlike $L$, $\tilde L_\alpha$ is not a null form along $K$ as it has non-vanishing components along directions transverse to the lightcone. The form $\tilde L=\{\tilde L_\alpha\}$ is not globally defined on $M$. Instead at the intersection, $U_\alpha\cap U_\beta$, of two open sets $U_\alpha$ and $U_\beta$,
\begin{equation}\label{tpatch}
	\tilde{L}_\alpha = \tilde{L}_\beta + \e^-\wedge \tilde{N}_{\alpha\beta} \, ,
\end{equation}
where $\tilde N_{\alpha\beta}=\frac{1}{(\ell-1)!}N_{i_1\dots i_{\ell-1}} \e^{i_1\dots i_{\ell-1}}$. To establish this, we have used that $L_\alpha=\e^- \wedge \tilde L_\alpha=\e^- \wedge \tilde L_\beta=L_\beta$ on $U_\alpha\cap U_\beta$ and that $\e^-$ is no-where vanishing on the spacetime. Equivalently, the patching conditions of the pseudo-orthonormal co-frame under the structure group $G\ltimes\bR^8$ are
\begin{equation}
	\begin{aligned}
		\e_\alpha^- &= \e^-_\beta, \\
		\e^+_\alpha &= \e_\beta^+ - \frac{1}{2} q^2_{\alpha\beta} \e^-_\beta - ((q^t O)_{\alpha\beta})_j \e^j_\beta, \\
		\e^i_\alpha &= (O_{\alpha\beta})^i{}_j \e_\beta^j + q^i_{\alpha\beta} \e^-_\beta,
	\end{aligned}
\end{equation}
where $(O, q)\in G\ltimes\bR^8$, $q^t$ is the transposed on $q$ and $G\subset SO(8)$. It is clear from the last patching condition that $\tilde L$ transforms as \eqref{tpatch}, i.e. it does not transform as a form while $L$ does. In the following, the above relation between $L$ and $\tilde L$ will be referred to as $L$ is \emph{represented} by $\tilde L$ or equivalently $\tilde L$ \emph{represents} $L$.

We shall use the observations we have made above to define two algebraic operations on $\Omega^*_K(M)$. Indeed given $L,M\in \Omega^*_K(M)$, we define
\begin{equation}\label{twoop}
	\begin{gathered}
		L\bar{\curlywedge} M \equiv \e^-\wedge i_{\tilde L} \tilde M = \frac{1}{(\ell-1)! (m-1)!} \, L^j{}_{i_1\dots i_{\ell-1}} M_{j i_\ell \dots i_{\ell+m-2}}\, \e^{-i_1\dots i_{\ell+m-2}}, \\
		\\
		L\curlywedge M \equiv L\wedge \tilde M = \e^-\wedge \tilde L\wedge \tilde M,
	\end{gathered}
\end{equation}
where we have used $\tilde L$ and $\tilde M$ that represent $L$ and $M$, respectively. Although both operations\footnote{Note that if $L,M\in \Omega^*_K$, then $L\bar\wedge M=L\wedge M=0$.} $\bar{\curlywedge}$ and $\curlywedge$ are defined using local data, $L\bar{\curlywedge} M$ and $L\curlywedge M$ are globally defined forms on the spacetime. Moreover, if $L$ and $M$ are $\hat\nabla$-covariantly constant, then $L\curlywedge M$ and $L\bar{\curlywedge} M$ are $\hat\nabla$-covariantly constant as well. So clearly, the two operations described in \eqref{twoop} can be used to construct new $\hat\nabla$-covariantly constant forms from old ones.

Let us denote with, $\Omega^*_{\hat\nabla}(M)$, the vector space spanned by all null along $K$, $\hat\nabla$-covariantly constant forms of a spacetime $M$. As all such forms have odd degree for the backgrounds we shall be investigating, $\Omega^*_{\hat\nabla}(M)$ with bracket operation $\bar{\curlywedge}$ is a Lie algebra. As we shall demonstrate, the algebra of holonomy symmetries of sigma models with target spaces given by supersymmetric heterotic backgrounds with non-compact holonomy groups is determined by the Lie algebra $\Omega^*_{\hat\nabla}(M)$. Therefore, the Lie algebra structure of $\Omega^*_{\hat\nabla}(M)$ is of interest. Some of it is easily unravelled. First, $\Omega^1_{\hat\nabla}(M)$ is spanned by the null 1-form $K$ which commutes with all the remaining elements of $\Omega^*_{\hat\nabla}(M)$. In addition, $\Omega^9_{\hat\nabla}(M)$ is spanned by the Hodge dual form $E$ of $K$, $E = \star K$. Again $E$ commutes with the rest of the elements of $\Omega^*_{\hat\nabla}(M)$. Therefore both $K$ and $E$ are in the centre of $\Omega^*_{\hat\nabla}(M)$.

To further explore the Lie algebra structure of $\Omega^*_{\hat\nabla}(M)$, one defines the Lie algebra of fundamental forms, $\mathfrak{f}$, as described in the introduction. $\mathfrak{f}$ is a Lie subalgebra of $\Omega^*_{\hat\nabla}(M)$. In the supersymmetric backgrounds considered here, $\mathfrak{f}$ is spanned by form bilinears\footnote{An exception are the backgrounds with holonomy $U(1)\ltimes\bR^8$; this will be explained later.}. In all examples that we shall investigate, $\Omega^*_{\hat\nabla}(M)$ is generated by $\mathfrak{f}$ upon taking the $\curlywedge$ product of elements of $\mathfrak{f}$ and including $K$. Moreover, it turns out that, apart from backgrounds with holonomy $\mathrm{Spin}(7)\ltimes \bR^8$ and $SU(4)\ltimes \bR^8$, $\mathfrak{f}$ is generated by 3-forms. In all such cases, the algebraic structure of $\Omega^*_{\hat\nabla}(M)$, and so that of the symmetries of the sigma model, can be unravelled by decomposing $\Omega^*_{\hat\nabla}(M)$ into irreducible representations of $\mathfrak{f}$. The complete structure is presented later on a case by case basis.

Another ingredient needed in the analysis is a description of the geometry of supersymmetric heterotic backgrounds with non-compact holonomy \cite{hetgpug1, hetgpug2}. For our purposes, a brief outline suffices. In particular, after solving the Killing spinor equations of heterotic theory, the fields can be expressed as
\begin{equation}\label{mh}
	\begin{aligned}
		g &= 2\e^+ \e^- + \delta_{ij}\e^i\e^j \, , \\
		H &= H_{+-i} \e^{+-i} + \frac{1}{2} H_{+ij} \e^{+ij} + \frac{1}{2} H_{-ij} \e^{-ij} + \frac{1}{3!} H_{ijk} \e^{ijk} \\
	&= \dd\e^-\wedge \e^+ + \frac{1}{2} H_{-ij} \e^{-ij} + \frac{1}{3!} H_{ijk} \e^{ijk} \, , \\
		F &= F_{-i} \e^{-i} + \frac{1}{2} F_{ij} \e^{ij} \, ,
	\end{aligned}
\end{equation}
where $F$ is the curvature of the gauge sector and we have suppressed the gauge indices. Most of the components of $H$ given above are determined in terms of the metric and the form bilinears of the theory \cite{hetgpug1, hetgpug2}. Though, there is no need to give a detailed description. However, it is significant that the Killing spinor equations imply that $i_KH$ satisfies
\begin{equation}\label{invcon1}
	i_L(i_K H) = 0 \Longleftrightarrow H_{+\nu [\mu_1} L^\nu{}_{\mu_2\dots \mu_{\ell+1}]} = 0 \, ,
\end{equation}
for all fundamental forms $L$ of the holonomy group $G\ltimes\bR^8$. Using this and the $\hat\nabla$-covariantly constancy of $L$, one can show that the Lie derivative of $L$ with respect to $K$ vanishes
\begin{equation}\label{liedev}
	\mathcal{L}_K L = 0 \, .
\end{equation}
Therefore, all fundamental forms are invariant under the action of the vector field $K$. Furthermore, the gaugino Killing spinor equation implies that $i_K F=i_L F=0$. Also \eqref{liedev} holds for all $L\in \Omega^*_{\hat\nabla}(M)$, i.e. not only for the fundamental forms, provided that $i_L(i_K H)=0$.

\subsection{Holonomy Symmetries of Chiral Sigma Models}

A description of chiral 2-dimensional sigma models suitable for the analysis that follows has already been presented in \cite{lggpepb}. So we shall be brief. The fields of the sigma model are maps, $X$, from the worldsheet superspace $\Xi^{2|1}$, with coordinates $(\sigma^=, \sigma^{\pp}, \theta^+)$, into a spacetime $M$ and Grassmannian odd sections, $\psi$, of a vector bundle $S_-\otimes X^*E$ over $\Xi^{2|1}$. Here $S_-$ is the anti-chiral spinor bundle over $\Xi^{2|1}$ and $E$ is a vector bundle over $M$. An action for these fields \cite{ewch} is
\begin{equation}\label{act1}
	S=-i \int \dd^2\sigma \dd\theta^+ \Big ((g_{\mu\nu}+b_{\mu\nu}) D_+X^\mu \partial_=X^\nu+i h_{\mathrm{a}\mathrm{b}} \psi_-^\mathrm{a}\mathcal{D}_+\psi_-^\mathrm{b}\Big) \, ,
\end{equation}
where $g$ is a spacetime metric, $b$ is a locally defined 2-form on $M$, such that $H = \dd b$ is a globally defined 3-form. In addition, $D_+^2=i\partial_{\pp}$ and $h$ is a fibre metric on $E$,
\begin{equation}
	\mathcal{D}_+\psi_-^{\mathrm{a}} = D_+\psi_-^{\mathrm{a}} + D_+X^\mu \Omega_\mu{}^{\mathrm{a}}{}_{\mathrm{b}} \psi_-^{\mathrm{b}} \, ,
\end{equation}
where $\Omega$ is a connection on $E$ and $\mathcal{D}_\mu h_{\mathrm{a}\mathrm{b}}=0$. We shall refer to the part of the action with couplings $h$ and $\mathcal{D}$ as the gauge sector of the theory. Note that
\begin{equation}
	\delta S= -i \int \dd^2\sigma \dd\theta^+ \big(\delta X^\mu \mathcal{S}_\mu+\Delta \psi_-^{\mathrm{a}} \mathcal{S}_{\mathrm{a}}\big) \, ,
\end{equation}
where
\begin{equation}\label{feqns}
	\mathcal{S}_\mu=-2 g_{\mu\nu} \hat\nabla_= D_+ X^\nu-i \psi_-^{\mathrm{a}} \psi_-^{\mathrm{b}} D_+X^\nu F_{\mu\nu \mathrm{a}\mathrm{b}} \, ,~~
	\mathcal{S}^{\mathrm{a}}=2i\mathcal{D}_+\psi_-^{\mathrm{a}} \, ,
\end{equation}
are the field equations, $\Delta \psi_-^{\mathrm{a}} \equiv \delta \psi_-^{\mathrm{a}} + \delta X^\mu \Omega_\mu{}^{\mathrm{a}}{}_{\mathrm{b}} \psi_-^{\mathrm{b}}$ and $F$ is the curvature of the connection $\Omega$ of the gauge sector.

Before we proceed to describe the holonomy symmetries, we shall mention two sigma model symmetries that are relevant in the analysis of anomalies. First, in the background field method of quantising the theory \cite{honer, mukhi, friedan, phgpks, blasi}, it is convenient to express the quantum field in a frame basis. In such a case, if we write the metric as $g_{\mu\nu}= \eta_{AB} \e^A_\mu \e^B_\nu$, then the action of infinitesimal spacetime frame rotations will be
\begin{equation}\label{ftran2}
	\delta_\ell \e^A_\mu= \ell^A{}_B \e^B_\mu \, , \quad \delta_\ell \omega_\mu{}^A{}_B=-\partial_\mu \ell^A{}_B+ \ell^A{}_C\, \omega_\mu{}^C{}_B -\omega_\mu{}^A{}_C\, \ell^C{}_B \, ,
\end{equation}
where $\ell$ is the infinitesimal parameter and $\omega$ is a frame connection of the tangent bundle which we shall always assume preserves the spacetime metric.

Moreover, the fields and coupling constants of the gauge sector transform under infinitesimal gauge transformations as
\begin{equation}\label{ftran1}
	\begin{gathered}
		\delta_u \psi^{\mathrm{a}}_-= u^{\mathrm{a}}{}_{\mathrm{b}}\psi^{\mathrm{b}}_- \, , \\
		\delta_u\Omega_\mu{}^{\mathrm{a}}{}_{\mathrm{b}}=-\partial_\mu u^{\mathrm{a}}{}_{\mathrm{b}}+ u^{\mathrm{a}}{}_{\mathrm{c}}\, \Omega_\mu{}^{\mathrm{c}}{}_{\mathrm{b}} -\Omega_\mu{}^{\mathrm{a}}{}_{\mathrm{c}}\, u^{\mathrm{c}}{}_{\mathrm{b}} \, , \\
		\delta_u h_{\mathrm{a}\mathrm{b}}=- u^{\mathrm{c}}{}_{\mathrm{a}} h_{\mathrm{c}\mathrm{b}}- h_{\mathrm{a}\mathrm{c}} u^{\mathrm{c}}{}_{\mathrm{b}} \, ,
	\end{gathered}
\end{equation}
where $u$ is the infinitesimal parameter and the remaining fields and couplings of the theory remain inert.

The commutator of two spacetime frame rotations \eqref{ftran2} is $[\delta_\ell, \delta_{\ell'}]=\delta_{[\ell, \ell']}$, where $[\cdot, \cdot]$ is the usual commutator of two matrices. A similar result holds for the commutator of two gauge transformations \eqref{ftran1}. In addition to these, there is a gauge symmetry $\delta b_{\mu\nu}=(\dd m)_{\mu\nu}$ associated with the 2-form gauge potential $b$, where $m$ is a 1-form on the spacetime.

To describe the holonomy symmetries of the sigma model action \eqref{act1}, let $L$ be a $(\ell+1)$-form on the sigma model target space $M$ and consider the infinitesimal transformation
\begin{equation}\label{Gsym}
	\begin{aligned}
		\delta_L X^\mu &= a_L L^\mu{}_{\lambda_1\dots \lambda_\ell} D_+X^{\lambda_1}\dots D_+X^{\lambda_\ell} \equiv a_L L^\mu{}_L D_+X^L \, , \\
		\Delta_L\psi_-^\mathrm{a} &= 0 \, ,
	\end{aligned}
\end{equation}
where $a_L$ is the parameter of the transformation, chosen such that $\delta_L X^\mu$ is even under Grassmannian parity. The index $L$ is the multi-index $L=\lambda_1 \dots \lambda_\ell$ and $D_+X^L=D_+X^{\lambda_1}\cdots D_+X^{\lambda_\ell}$. Such a transformation \cite{phgpw1, Howe:1991vs, phgp1} leaves the action \eqref{act1} invariant provided that
\begin{equation}\label{invcon}
	\hat\nabla_{\nu} L_{\lambda_1\dots\lambda_{\ell+1}} = 0 \, , \quad F_{\nu[\lambda_1} L^\nu{}_{\lambda_2\dots \lambda_{\ell+1}]} = 0 \, ,
\end{equation}
i.e. $L$ is $\hat\nabla$-covariantly constant and $i_LF=0$. For form bilinears the former condition is satisfied as a consequence of the gravitino Killing spinor equations while the latter condition is a consequence of the gaugino Killing spinor equation. Moreover, the parameter $a_L$ satisfies $\partial_=a_L=0$, i.e. that $a_L=a_L(\sigma^{\pp}, \theta^+)$.

The commutator of two transformations \eqref{Gsym} on the field $X$ has been explored in detail in \cite{sven, phgp2}. Here, we shall summarise some of the key formulae. The commutator of two transformations \eqref{Gsym} on the field\footnote{In all cases considered here, the commutator of two holonomy symmetries on the field $\psi$ gives rise to a transformation with $\Delta_{LM}\psi=0$. So it will not be further investigated. } $X$ generated by the $(\ell+1)$-form $L$ and the $(m+1)$-form $M$ can be written as
\begin{equation}\label{comm}
	[\delta_L, \delta_M]X^\mu= \delta_{LM}^{(1)} X^\mu+\delta_{LM}^{(2)} X^\mu+\delta_{LM}^{(3)} X^\mu \, ,
\end{equation}
with
\begin{equation}
	\delta_{LM}^{(1)} X^\mu=a_M a_L N(L,M)^\mu{}_{LM} D_+X^{LM} \, ,
\end{equation}
\begin{equation}
	\begin{aligned}
		\delta_{LM}^{(2)} X_\mu = \big( -m a_M D_+a_L &(L\cdot M)_{\nu L_2, \mu M_2} \\
		&+ \ell (-1)^{(\ell+1) (m+1)} a_L D_+ a_M (L\cdot M)_{\mu L_2,\nu M_2}\big) D_+X^{\nu L_2M_2} \, ,
	\end{aligned}
\end{equation}
and
\begin{equation}
	\delta_{LM}^{(3)} X_\mu=-2i \ell m (-1)^\ell a_M a_L (L\cdot M)_{(\mu|L_2|, \nu )M_2} \partial_{\pp}X^\nu D_+X^{L_2M_2} \, ,
\end{equation}
where
\begin{equation}
	(L\cdot M)_{\lambda L_2,\mu M_2}=L_{\rho \lambda[L_2} M^\rho{}_{|\mu| M_2]} \, .
\end{equation}
The multi-index $M$ stands for $M=\mu_1\dots\mu_m$ while the multi-indices $L_2$ and $M_2$ stand for $L_2=\lambda_2\dots\lambda_\ell$ and $M_2=\mu_2\dots \mu_m$, respectively, and $N(L,M)$ is the Nijenhuis tensor of $L$ and $M$ which we shall not state here --- it is a generalisation of the standard Nijenhuis tensor of an almost complex structure. Using that $L$ and $M$ are $\hat\nabla$-covariantly constant, the Nijenhuis tensor can be re-expressed as
\begin{equation}
	N_{\mu LM} \dd x^{LM}= \Big(- (\ell+m+1)H_{[\mu|\nu\rho|} L^\nu{}_L M^\rho{}_{M]}+\ell m H^{\rho}{}_{\lambda_1\mu_1} ( L\cdot M)_{(\mu|L_2|, \rho)M_2}\Big)\, \dd x^{LM} \, ,
\end{equation}
in terms of $H$. The conserved current of a symmetry generated by the $(\ell+1)$-form $L$ is
\begin{equation}\label{curL}
	J_L=L_{\mu_1\dots\mu_{\ell+1}} D_+X^{\mu_1\dots\mu_{\ell+1}}~.
\end{equation}
It can easily be seen that $\partial_=J_L=0$ subject to the field equations \eqref{feqns}.

\subsection{The Commutator of Null Holonomy Symmetries}

The commutator of two symmetries generated by two $\hat\nabla$-covariantly constant forms $L$ and $M$ is significantly simplified whenever $L,M\in \Omega^*_K(M)$. But before we state this, one can prove, using $\mathcal{L}_K L=0$ and $i_KL=0$, that
\begin{equation}
	[\delta_K, \delta_L] X^\mu = 0 \, ,
\end{equation}
for any $\hat\nabla$-covariantly constant null form $L$ along $K$. Therefore, the symmetries generated by the Killing vector $K$ commutes with all other holonomy symmetries.

After some computation, using that $L,M \in \Omega^*_{\hat\nabla}(M)$, $\mathcal{L}_K L=\mathcal{L}_K M=0$ and the condition \eqref{invcon1}, the commutator of two symmetries generated by $L$ and $M$ can be expressed as
\begin{equation}\label{walg}
	[\delta_L, \delta_M] = \delta_K + \delta_{{L\bar{\curlywedge} M}} \, ,
\end{equation}
with $a_K=-\frac{\ell! m!}{ (\ell+m-1)!} D_+(a_L a_M J_{L\bar{\curlywedge} M})$ and $a_{{L\bar{\curlywedge} M}}=- \frac{\ell! m!}{ (\ell+m-2)!} a_L a_M\, D_+J_K$, where ${L\bar{\curlywedge} M}$ is given in \eqref{twoop} and is a globally defined form on the spacetime. Note that in all cases that we shall be considering, the forms that generate the holonomy symmetries have odd degree, and so the parameters $a_L, a_M$ of the transformations have even Grassmannian parity. As the structure constants of the algebra depend on the conserved currents of the associated symmetries, the algebra of variations closes as a W-algebra.

Furthermore, the W-algebra \eqref{walg} closes to transformations generated by $K$ and $L\bar{\curlywedge} M$. As $L\bar{\curlywedge} M\in \Omega^*_{\hat\nabla}(M)$ for all $L,M \in \Omega^*_{\hat\nabla}(M)$ and $i_{L\bar{\curlywedge} M} F=0$, $L\bar{\curlywedge} M$ generates a new symmetry for the sigma model action \eqref{act1}. Therefore, the commutator \eqref{walg} of the W-algebra of holonomy symmetries is determined by the Lie algebra $\Omega^*_{\hat\nabla}(M)$ with bracket operation $\bar{\curlywedge}$. So to determine the W-algebra of holonomy symmetries, it remains to identify the Lie algebra $\Omega^*_{\hat\nabla}(M)$ for each of the holonomy groups \eqref{hol}.

\section{The W-algebra of Null Holonomy Symmetries} \label{holonomy-symmetries-w-algebras}

\subsection{The W-algebra of Null Spin(7) and SU(4) Symmetries}

In both these cases, the space of fundamental forms $\mathfrak{f}$ contains forms of degree greater than three. This distinguishes them from the other holonomy groups stated in \eqref{hol} and so they are separately investigated.

\subsubsection{Spin(7)}

The holonomy symmetries in this case are generated by $K$ and a null 5-form spinor bilinear $L$ which is represented by the usual self-dual, $\mathrm{Spin}(7)$ invariant, fundamental 4-form $\tilde L$. Using the algebraic properties of $\tilde L$, one can demonstrate that $L\bar{\curlywedge} L=0$. Therefore $\mathfrak{f}=\bR\langle L\rangle$.

Moreover $L{\curlywedge} L$ is proportional to $E = \star K$. As $K, E$ are in the centre of $\Omega^*_{\hat\nabla}(M)$, one concludes that $\Omega^*_{\hat\nabla}(M)=\bR^3\langle K, L, E\rangle$ is abelian. As a result, the W-algebra of symmetries \eqref{walg} is abelian as well.

\subsubsection{SU(4)}

To describe the geometry, introduce a pseudo-hermitian co-frame $(\e^-, \e^+, \e^\alpha, \e^{\bar\alpha})$, $\alpha=1,\dots,4$, on the spacetime. In this co-frame, the metric is expressed as $g=2 \e^-\e^++2\delta_{\alpha\bar\beta} \e^\alpha \e^{\bar\beta}$ and the generators of the fundamental forms, other than $K=\e^-$, are given by
\begin{equation}
	\begin{aligned}
		I &= \frac{1}{2} I_{ij} \e^-\wedge \e^{ij} \equiv -i \delta_{\alpha\bar\beta} \e^{-\alpha\bar\beta} \, , \\
		L_1 &= \frac{1}{4!} (L_1)_{i_1\dots i_4} \e^{-i_1\dots i_4} \equiv \frac{1}{4!} \epsilon_{\alpha_1\dots \alpha_4} \e^{-\alpha_1\dots \alpha_4}+\frac{1}{4!} \epsilon_{\bar\alpha_1\dots \bar\alpha_4} \e^{-\bar\alpha_1\dots \bar\alpha_4} \, , \\
		L_2 &= \frac{1}{4!} (L_2)_{i_1\dots i_4} \e^{-i_1\dots i_4} \equiv -\frac{i}{4!} \epsilon_{\alpha_1\dots \alpha_4} \e^{-\alpha_1\dots \alpha_4}+\frac{i}{4!} \epsilon_{\bar\alpha_1\dots \bar\alpha_4} \e^{-\bar\alpha_1\dots \bar\alpha_4} \, .
	\end{aligned}
\end{equation}
Note that the conditions on $i_K H$ arising from $i_I i_KH= i_{L_1} i_K H=i_{L_2} i_K H=0$ in the pseudo-hermitian co-frame can be written as
$H_{+\alpha}{}^\alpha=H_{+\alpha\beta}=0$.

A straightforward calculation reveals that
\begin{equation}
	\begin{aligned}
		I\bar{\curlywedge} L_1 &= -4L_2 \, , \\
		I\bar{\curlywedge} L_2 &= 4L_1 \, , \\
		L_1\bar{\curlywedge} L_2 &= -\frac{3}{2} \frac{1}{6^2} I_{i_1i_2} I_{i_3i_4} I_{i_5 i_6} \e^{-i_1\dots i_6} = -\frac{1}{3} \curlywedge^3I \, .
	\end{aligned}
\end{equation}
where all the remaining $\bar{\curlywedge}$ operations amongst these form vanish. Therefore, closure of the Lie algebra of the fundamental forms $\mathfrak{f}$ requires the introduction of a new generator $\curlywedge^3I$ that commutes with $I$, $L_1$ and $L_2$. Therefore, $\curlywedge^3I$ is a central generator. In fact, $\mathfrak{f}=\hat {\mathfrak{e}}(2)$, where $\hat {\mathfrak{e}}(2)$ is the central extension of the Euclidean algebra, $\mathfrak{e}(2)=\mathfrak{so}(2)\oplus_s\bR^2$, with $I$ the generator of $\mathfrak{so}(2)$ and $L_1, L_2$ the generators of $\bR^2$.

To give an example where the W-algebra of holonomy symmetries is determined from the Lie algebra $\mathfrak{f}$, consider this case with $\mathfrak{f}=\hat {\mathfrak{e}}(2)$. Then, the commutators \eqref{walg} read
\begin{equation}
	\begin{aligned}\relax 
		[\delta_I, \delta_{L_1}] &= \frac{8}{5} D_+(a_L a_M J_{L_2})\delta_K+ 8 a_L a_M\, D_+J_K \,\delta_{{L_2}} \, , \\
		[\delta_I, \delta_{L_2}] &= -\frac{8}{5} D_+(a_L a_M J_{L_1})\delta_K- 8 a_L a_M\, D_+J_K \,\delta_{{L_1}} \, , \\
		[\delta_{L_1}, \delta_{L_2}] &= \frac{4}{15}\left(\frac{1}{7} D_+(a_L a_M J_{\curlywedge^3I})\delta_K+ a_L a_M\, D_+J_K \,\delta_{{\curlywedge^3I}}\right) \, .
	\end{aligned}
\end{equation}
with the remaining commutators to vanish. Notice that the structure constants depend on the currents $J_K$, $J_{L_1}$, $J_{L_2}$ and $J_{\curlywedge^3I}$.

In addition to the generators of $\mathfrak{f}$ and $K$, $\Omega^*_{\hat\nabla}(M)$ contains two more generators given by $\curlywedge^2 I$ and $\curlywedge^4 I$. The latter is proportional to $E$. In fact $L_1$, $L_2$ and $\curlywedge^2 I$ span the space of 5-form bilinears. Moreover, $L_1\curlywedge L_1$ and $L_2\curlywedge L_2$ are proportional to $E$ and $L_1\curlywedge L_2=0$. It turns out that $\curlywedge^2 I$ and $\curlywedge^4 I$ commute amongst themselves as well as with the generators of $\hat {\mathfrak{e}}(2)$. As a result, one concludes that $\Omega^*_{\hat\nabla}(M)=\hat {\mathfrak{e}}(2)\oplus \bR^3\langle K, \curlywedge^2 I, E\rangle$.

\subsection{The W-algebra of Remaining Null Holonomy Symmetries}\label{tpt}

For all the remaining backgrounds that are investigated below, $\mathfrak{f}$ is spanned by 3-forms. In addition, $\Omega^3_{\hat\nabla}(M)=\mathfrak{f}$ and $\Omega^7_{\hat\nabla}(M)$ is the Hodge dual space of $\mathfrak{f}$, $\Omega^7_{\hat\nabla}(M) = \star \mathfrak{f}$.

To determine the Lie algebra structure of $\Omega^*_{\hat\nabla}(M)$, note that $\Omega^*_{\hat\nabla}(M)$ decomposes into (irreducible) representations of $\mathfrak{f}$. Clearly $\mathfrak{f}$ acts on $\Omega^3_{\hat\nabla}(M)$ with the adjoint representation. It also acts with the same representation on $\Omega^7_{\hat\nabla}(M)$ as
\begin{equation}
	L\bar{\curlywedge} \star M = \star (L \bar{\curlywedge} M) \, ,
\end{equation}
for any $L,M\in \Omega^3_{\hat\nabla}(M)$. A calculation also reveals that
\begin{equation}
	L\bar{\curlywedge} M = 0 \, ,
\end{equation}
for any $L\in \Omega^5_{\hat\nabla}(M)$ and $M\in \Omega^7_{\hat\nabla}(M)$. So to determine the Lie algebra structure of $\Omega^*_{\hat\nabla}(M)$, it remains to compute $\Omega^3_{\hat\nabla}(M)\bar\curlywedge\Omega^5_{\hat\nabla}(M)$ and $\Omega^5_{\hat\nabla}(M)\bar\curlywedge\Omega^5_{\hat\nabla}(M)$. The former will be identified on a case by case basis. In computing the latter, we shall use the formula
\begin{equation}\label{LMNP}
	\begin{aligned}
		(L\curlywedge M)\bar\curlywedge (N\curlywedge P) &= (L\bar\curlywedge N)\curlywedge M \curlywedge P + (L\bar\curlywedge P)\curlywedge M \curlywedge N \\
		&+ (M\bar\curlywedge N)\curlywedge L \curlywedge P + (M\bar\curlywedge P)\curlywedge L \curlywedge N \, ,
	\end{aligned}
\end{equation}
where $L,M,N,P\in \Omega^3_{\hat\nabla}(M)$. For the backgrounds examined below, all the elements of $\Omega^5_{\hat\nabla}(M)$ can be written as the $\curlywedge$-product of two elements in $\Omega^3_{\hat\nabla}(M)$.

\subsubsection{Sp(2)}

The Lie algebra of fundamental forms $\mathfrak{f}$ is spanned by
\begin{equation}\label{sp2bi}
	I_1 = \frac{1}{2} (I_1)_{ij} \e^{-ij} \, , \quad I_2 = \frac{1}{2} (I_2)_{ij} \e^{-ij} \, , \quad I_3 = \frac{1}{2} (I_3)_{ij} \e^{-ij} \, .
\end{equation}
where their components are represented by the usual hypercomplex structure that characterises $Sp(2)$ in eight dimensions. In particular, $\tilde I_1^2=\tilde I_2^2=-{\bf 1}_{8\times 8}$, $\tilde I_1\tilde I_2+\tilde I_2 \tilde I_1=0$ and $\tilde I_3=\tilde I_1 \tilde I_2$. It is clear from this that
\begin{equation}
	I_r\bar\curlywedge I_s=-2 \epsilon_{rs}{}^t I_t~,
\end{equation}
and so $\mathfrak{f}=\mathfrak{sp}(1)$. Notice that $\mathfrak{sp}(2)\oplus \mathfrak{sp}(1)$ is a maximal subalgebra of $\mathfrak{so}(8)$.

To describe the Lie algebra structure of $\Omega^*_{\hat\nabla}(M)$, define
\begin{equation}
	I_{r_1\dots r_p}\equiv I_{r_1} \curlywedge \dots \curlywedge I_{r_p} \, , \quad p=1,2,3,4 \, .
\end{equation}
Next $\Omega^5_{\hat\nabla}(M)=\bR\langle I_{rs}\rangle$. As $I_{rs}=I_{sr}$ and
\begin{equation}
	I_r\bar{\curlywedge} I_{s_1\dots s_q} = -2q \epsilon_{r(s_1}{}^{t} I_{|t|\dots s_q)} = 2q \epsilon_{(s_1|r}{}^{t} I_{t|\dots s_q)}\, ,
\end{equation}
$\mathfrak{f}=\mathfrak{sp}(1)=\mathfrak{so}(3)$ acts on $\Omega^5_{\hat\nabla}(M)$ with the symmetric product of two vector representations of $\mathfrak{so}(3)$. This decomposes as $\Omega^5_{\hat\nabla}(M)= \Omega^5_{\bf 5}\oplus \Omega^5_{\bf 1}$, where $\Omega^5_{\bf 5}$ is the irreducible symmetric traceless representation and $\Omega^5_{\bf 1}$ is the trivial representation, which is spanned by the Casimir element $C=I_{11}+I_{22}+I_{33}$. In fact $C$ commutes with all elements of $\Omega^*_{\hat\nabla}(M)$ and is an element of the centre, together with $K$ and $E$.

It remains to compute $\Omega^5_{\hat\nabla}(M)\bar\curlywedge\Omega^5_{\hat\nabla}(M)$. For this, one can use the formula \eqref{LMNP} to deduce
\begin{equation}
	\begin{aligned}
		I_{r_1r_2}\bar\curlywedge I_{s_1s_2} &= -2 \epsilon_{r_1s_1}{}^t I_{tr_2s_2} - 2 \epsilon_{r_1s_2}{}^t I_{tr_2s_1} - 2 \epsilon_{r_2s_1}{}^t I_{tr_1s_2} - 2 \epsilon_{r_2s_2}{}^t I_{tr_1s_1} \\
		&= -16 (\epsilon_{r_1s_1}{}^t \delta_{r_2s_2} + \epsilon_{r_1s_2}{}^t \delta_{r_2s_1} + \epsilon_{r_2s_1}{}^t \delta_{r_1s_2} + \epsilon_{r_2s_2}{}^t \delta_{r_1s_1}) \star I_t \, ,
	\end{aligned}
\end{equation}
where we have used
\begin{equation}
	\begin{gathered}
		\star I_r = \frac{1}{4!} I_{rrr}~ \, , \quad \star I_r = \frac{1}{8} I_{ssr} \, , \quad r \neq s \, , \\
		I_{123} = 0 \, ,
	\end{gathered}
\end{equation}
which lead to
\begin{equation}
	I_{rst} = \frac{1}{3}(\delta_{rs} I_{ttt} + \delta_{rt} I_{sss} + \delta_{st} I_{rrr})~.
\end{equation}
This together with the results in the beginning of section \ref{tpt} completely determine the Lie algebra structure of $\Omega^*_{\hat\nabla}(M)$ and so the commutators \eqref{walg} of the associated holonomy symmetries.

\subsubsection{\texorpdfstring{$\times^2$Sp(1)}{Sp(1)xSp(1)}}
In an adapted pseudo-Hermitian co-frame $(\e^-, \e^+, \e^\alpha, \e^{\bar\alpha}; \alpha=1,\dots,4)$, to the $\times^2Sp(1)\ltimes \bR^8$ holonomy, the Lie algebra of fundamental forms $\mathfrak{f}$ is spanned by
\begin{equation}\label{IrJs}
	I_r = \frac{1}{2} (\tilde{I}_r)_{ij}\, \e^{-ij}, \quad J_r = \frac{1}{2} (\tilde{J}_r)_{ij}\, \e^{-ij} \, ,
\end{equation}
which are represented by
\begin{equation}
	\begin{aligned}
		\tilde{I}_1 &= -i(\e^{1\bar{1}}+\e^{2\bar{2}})~, & \tilde{I}_2 &= \e^{12}+\e^{\bar{1}\bar{2}}~, & \tilde{I}_3 &= -i(\e^{12}-\e^{\bar{1}\bar{2}}) \, , \\
		\tilde{J}_1 &= i(\e^{3\bar{3}}+\e^{4\bar{4}})~, & \tilde{J}_2 &= \e^{34}+\e^{\bar{3}\bar{4}}~, & \tilde{J}_3 &= i(\e^{34}-\e^{\bar{3}\bar{4}}) \, ,
	\end{aligned}
\end{equation}
and so $\tilde I_r \tilde I_s=-\delta_{rs} {\bf 1}_{4\times 4}+\epsilon_{rs}{}^t \tilde I_t$ and $\tilde J_r \tilde J_s=-\delta_{rs} {\bf 1}_{4\times 4}+\epsilon_{rs}{}^t \tilde J_t$. Thus, one has that
\begin{equation}\label{sp1sp1}
	I_r\bar\curlywedge I_s = -2 \epsilon_{rs}{}^t I_t\, , \quad J_r\bar\curlywedge J_s = -2 \epsilon_{rs}{}^t J_t\, , \quad I_r\bar\curlywedge J_s = 0\, ,
\end{equation}
and so $\mathfrak{f}=\Omega^3_{\hat\nabla}(M)=\oplus^2 \mathfrak{sp}(1)$. To see this, observe that the action of $\times^2Sp(1)$ on $\bR^8$ can be seen as two copies of the action of $Sp(1)$ on $\bR^4$ with each copy associated with a hyper-complex structure\footnote{It is conventional that if the (holonomy) group $Sp(1)$ is associated with the hyper-complex structure $(I_1,I_2, I_3)$ on $\bR^4$, then $\mathfrak{sp}(1)$ spans the subspace of (1,1)-forms with respect to $I_1$ on $\bR^4$ which are in addition $I_1$-traceless, i.e. $\mathfrak{sp}(1)$ spans the anti-self dual 2-form while the hyper-complex structure the self-dual 2-forms on $\bR^4$.} on $\bR^4$.

Next define the forms
\begin{equation}\label{LMN}
	\begin{gathered}
		L \equiv \curlywedge^2 I_1 = \curlywedge^2 I_2 = \curlywedge^2 I_3\, , \quad M \equiv \curlywedge^2 J_1 = \curlywedge^2 J_2 = \curlywedge^2 J_3\, , \\
		N_{rs} \equiv I_r\curlywedge J_s\, , \quad R_{r} \equiv I_r\curlywedge M = 4 \star I_r\, , \quad S_{r} \equiv L\curlywedge J_r = 4 \star J_r\, , \\
		E = \frac{1}{4}L \curlywedge M\, .
	\end{gathered}
\end{equation}
It can be shown that $\Omega^5_{\hat\nabla}(M)=\bR\langle L, M, N_{rs}\rangle$. It turns out that $L,M$ together with $K$ and $E$ are in the centre of $\Omega^*_{\hat\nabla}(M)$ while
\begin{equation}\label{comIN}
	I_r\bar{\curlywedge} N_{r's'}=-2\epsilon_{rr'}{}^{t'} N_{t's'}\, , \quad J_r\bar{\curlywedge} N_{r's'}=-2\epsilon_{rs'}{}^{t'} N_{r't'}\, .
\end{equation}
Thus $\Omega^5_{\hat\nabla}(M)$ decomposes under the action of $\mathfrak{f}$ as $\Omega^5_{\hat\nabla}(M)=\Omega^5_{\bf 9}\oplus \bR^2\langle L,M\rangle$, where $\Omega^5_{\bf 9}$ spanned by $N_{rs}$ is identified with the traceless symmetric product of two vector representations of $\mathfrak{so}(4)=\oplus^2 \mathfrak{sp}(1)$.

To specify the Lie algebra structure of $\Omega^*_{\hat\nabla}(M)$ it remains to determine $\Omega^5_{\hat\nabla}(M)\bar{\curlywedge}\Omega^5_{\hat\nabla}(M)$. As $L,M$ are in the centre of $\Omega^*_{\hat\nabla}(M)$, a straightforward computation using \eqref{LMNP} reveals that
\begin{equation}
	N_{rs}\bar{\curlywedge} N_{r's'}=-2\epsilon_{rr'}{}^{t'} \delta_{ss'} R_{t'}-2\epsilon_{ss'}{}^{t'} \delta_{rr'} S_{t'}\, .
\end{equation}
This together with the results in the beginning of section \ref{tpt} completely determine the Lie algebra structure $\Omega^*_{\hat\nabla}(M)$ and so the commutators \eqref{walg} of the W-algebra of holonomy symmetries.

\subsubsection{Sp(1)}

In addition to $I_r$ and $J_r$ \eqref{IrJs} fundamental forms of holonomy $\times^2Sp(1)\ltimes\bR^8$ backgrounds, the Lie algebra $\mathfrak{f}$ of holonomy $Sp(1)\ltimes\bR^8$ backgrounds contains the element $A$ which in the pseudo-hermitian basis of the previous section reads
\begin{equation}
	A = \e^{-13} + \e^{-24} + \e^{-\bar{1}\bar{3}} + \e^{-\bar{2}\bar{4}}\, .
\end{equation}
This is one of the additional 3-form bilinears associated with the holonomy $SU(2)\ltimes\bR^8$ backgrounds. Next define
\begin{equation}
	W_r \equiv I_r \bar{\curlywedge} A = -J_r \bar{\curlywedge} A \, .
\end{equation}
Then it turns out that $\mathfrak{f}=\bR\langle I_r, J_s, A, W_t\rangle$. The non-vanishing commutators are those already described for backgrounds with holonomy $\times^2 Sp(1)\ltimes\bR^8$ in \eqref{sp1sp1} and
\begin{equation}\label{comso5}
	\begin{aligned}
		I_r\bar{\curlywedge}A &= -J_r\bar{\curlywedge} A = W_r\, , & I_r\bar{\curlywedge} W_s &= -\delta_{rs} A - \epsilon_{rs}{}^t W_t\, , \\
		J_r\bar{\curlywedge} W_s &= \delta_{rs} A - \epsilon_{rs}{}^t W_t\, , & A\bar{\curlywedge} W_r &= 2 I_r - 2J_r\, , \\
		W_r\bar{\curlywedge} W_s &= -2\epsilon_{rs}{}^t (I_t + J_t)\, .
	\end{aligned}
\end{equation}
It turns out that $\mathfrak{f}=\Omega^2_{\hat\nabla}(M)=\mathfrak{so}(5)$. This is not unexpected. To see this, recall that for holonomy $Sp(2)\ltimes \bR^8$ backgrounds $\mathfrak{f}=\mathfrak{sp}(1)$. Note also that $\mathfrak{sp}(2)=\mathfrak{so}(5)$ and $\mathfrak{sp}(2)\oplus \mathfrak{sp}(1)$ is a maximal subgroup of $\mathfrak{so}(8)=\wedge^2\bR^8$. So if the invariant 2-forms under $\mathfrak{sp}(2)$ in $\wedge^2\bR^8$ span the Lie algebra $\mathfrak{sp}(1)$, then the invariant forms in $\wedge^2\bR^8$ under (the holonomy Lie algebra) $\mathfrak{sp}(1)$ span $\mathfrak{so}(5)$.

To describe the Lie algebra structure of $\Omega^*_{\hat\nabla}(M)$, one has to determine the representation of $\mathfrak{f}$ on $\Omega^5_{\hat\nabla}(M)$. For this observe that $\Omega^5_{\hat\nabla}(M)=\bR\langle L, M, A^2, P, Q, N_{rs}, Y_r, Z_r\rangle$, where $L,M,N$ are defined in \eqref{LMN}, $A^2\equiv \curlywedge^2 A$, $P\equiv \frac{1}{3} \delta^{rs} I_r \curlywedge W_s$, $Q\equiv \frac{1}{3} \delta^{rs} J_r \curlywedge W_s$ and $Y_r\equiv I_r \curlywedge A$ and $Z_r\equiv J_r \curlywedge A$. To establish this note the identities
\begin{equation}\label{relso5}
	\begin{aligned}
		W_r\curlywedge W_s &= (2\delta^{pq} N_{pq}+A^2)\delta_{rs}-N_{rs}-N_{sr}~, & I_r\curlywedge W_s &= P\delta_{rs}-\epsilon_{rs}{}^t Y_t~, \\
		J_r\curlywedge W_s &= Q\delta_{rs}+\epsilon_{rs}{}^t Z_t~, & A\curlywedge W_r &= -\epsilon_r{}^{st} N_{st}~.
	\end{aligned}
\end{equation}
It turns out that under the action of $\mathfrak{so}(5)$, $\Omega^5_{\hat\nabla}(M)$ decomposes as
\begin{equation}\label{decomsp1}
	\Omega^5_{\hat\nabla}(M) = \Omega^5_{\mathbf{14}} \oplus \Omega^5_{\mathbf{5}} \oplus \Omega^5_{\mathbf{1}}\, ,
\end{equation}
where $\Omega^5_{\bf 14}$ is the irreducible representation of $\mathfrak{so}(5)$ constructed as the symmetric and traceless product of two vector representations, $\Omega^5_{\bf 5}$ is the vector representation of $\mathfrak{so}(5)$ and the trivial representation $\Omega^5_{\bf 1}$ is spanned by the quadratic Casimir element of $\mathfrak{so}(5)$. In particular, $\Omega^5_{\bf 5}=\bR\langle Y_r+Z_r, P+Q, L-M\rangle$. Apart from a direct calculation, this result can be established after decomposing $\wedge^4\bR^8$ in $(\mathfrak{sp}(1)\oplus \mathfrak{sp}(2))\subset \mathfrak{so}(8)$ representations. Then $\Omega^5_{\hat\nabla}(M)$ can be identified with the span of the elements in $\wedge^4\bR^8$, which are invariant under the action of $\mathfrak{sp}(1)$. Decomposing this subspace into $\mathfrak{sp}(2)$ representations yields \eqref{decomsp1}, see e.g. \cite{salamon} proposition 9.2 and references within.

It is worth pointing out that the form bilinears do not span $\Omega^5_{\hat\nabla}(M)$. Indeed the 5-form bilinears are symmetric in the exchange of Killing spinors. As these backgrounds preserve five supersymmetries, the 5-form bilinears span an at most 15-dimensional vector space, while the dimension of $\Omega^5_{\hat\nabla}(M)$ is 20. Nevertheless, the additional elements of $\Omega^5_{\hat\nabla}(M)$ are $\hat\nabla$-covariantly constant and they should be included in the investigation as they generate holonomy symmetries in sigma model action \eqref{act1}.

To determine the Lie algebra structure of $\Omega^*_{\hat\nabla}(M)$, it remains to compute $\Omega^5_{\hat\nabla}(M)\bar\curlywedge\Omega^5_{\hat\nabla}(M)$ using \eqref{LMNP}. This is a straightforward computation and the result follows upon application of \eqref{comso5} and \eqref{relso5}. As the final formulae are not illuminating, they will not be presented here.

\subsubsection{U(1)}

The Lie algebra of fundamental forms is $\mathfrak{f}=\mathfrak{u}(4)$. This can be computed following the steps of a calculation similar to that performed for $Sp(1)\ltimes\bR^8$ backgrounds in the previous section. As this is elaborate, we shall present instead an alternative group theoretic justification for the assertion. First, the Lie subalgebra $\mathfrak{u}(1)$ of the holonomy group, viewed as a 1-dimensional subspace of $\mathfrak{so}(8)=\wedge^2\bR^8$, is spanned by a complex structure $U$ on $\bR^8$. The elements of $\wedge^2\bR^8$ that are invariant under the action of $U$, and so of $\mathfrak{u}(1)$, are the (1,1)-forms with respect to $U$. It is known that the latter span the Lie algebra $\mathfrak{u}(4)$, which is the subalgebra of $\mathfrak{so}(8)$ that leaves invariant $U$.

As the 3-form bilinears are skew-symmetric in the exchange of the two Killing spinors and these backgrounds preserve six supersymmetries, they span an at most 15-dimensional subspace in $\Omega^3_{\hat\nabla}(M)$. But as we have seen $\mathfrak{f}=\Omega^3_{\hat\nabla}(M)=\mathfrak{u}(4)$ has dimension 16. The additional $\hat\nabla$-covariantly constant form is represented by $U$ and it should be included in $\mathfrak{f}$ as it generates a holonomy symmetry for sigma model actions \eqref{act1}. The remaining symmetries are expected to be generated by form bilinears as $\wedge^2\bR^6=\mathfrak{so}(6)=\mathfrak{su}(4)$.

Similarly, the elements of $\Omega^5_{\hat\nabla}(M)$ are represented by the (2,2)-forms with respect to $U$ in $\wedge^4\bR^8$. Although, $\Omega^5_{\hat\nabla}(M)$ can be decomposed further into irreducible representations of $\mathfrak{u}(4)$ with this description of the action of $\mathfrak{f}$ on $\Omega^5_{\hat\nabla}(M)$ it suffices to specify the Lie algebra structure of $\Omega^*_{\hat\nabla}(M)$, as the action of $\mathfrak{u}(4)$ on (2,2)-forms is well known.

Again, $\Omega^5_{\hat\nabla}(M)$ contains more elements than those expected from counting 5-form bilinears. Indeed, the latter span an at most 21-dimensional space while $\Omega^5_{\hat\nabla}(M)$ has dimension 36. As it was mentioned in the previous case, all the elements of $\Omega^5_{\hat\nabla}(M)$ generate symmetries for the sigma model action \eqref{act1} and so they should be included in the description.

All elements in $\Omega^5_{\hat\nabla}(M)$ can be written as linear combinations of the $\curlywedge$-product of two elements in $\Omega^3_{\hat\nabla}(M)$. Indeed as $(\e^{-\alpha\bar\beta}; \alpha,\beta=1,2,3,4)$ is a basis in $\Omega^3_{\hat\nabla}(M)$, one has that $\e^{-\alpha_1\bar\beta_1}\curlywedge \e^{-\alpha_2\bar\beta_2}=\e^{-\alpha_1\bar\beta_1\alpha_2\bar\beta_2}$ span $\Omega^5_{\hat\nabla}(M)$. As a result, one can apply the formula \eqref{LMNP} to compute $\Omega^5_{\hat\nabla}(M)\bar\curlywedge\Omega^5_{\hat\nabla}(M)$. One finds that
\begin{equation}
	\begin{aligned}
		\!\! \e^{-\alpha_1\bar\beta_1\alpha_2\bar\beta_2}\bar\curlywedge \e^{-\alpha'_1\bar\beta'_1\alpha'_2\bar\beta'_2} &= -\Big(\big(\delta^{\alpha_1\bar\beta_1'} \e^{-\bar\beta_1\alpha_2\bar\beta_2 \alpha_1'\alpha_2'\bar\beta'_2} - \delta^{\alpha_1\bar\beta_2'} \e^{-\bar\beta_1\alpha_2\bar\beta_2 \alpha_1'\alpha_2'\bar\beta'_1}\big) - (\alpha_2\leftrightarrow \alpha_1)\Big) \\
		&- \Big(\big(\delta^{\bar\beta_1\alpha_1'} \e^{-\alpha_1\alpha_2\bar\beta_2\bar\beta_1'\alpha'_2\bar\beta'_2} - \delta^{\bar\beta_1\alpha_2'} \e^{-\alpha_1\alpha_2\bar\beta_2\bar\beta_1'\alpha'_1\bar\beta'_2}\big) - (\bar\beta_2\leftrightarrow \bar\beta_1)\Big)\, ,
	\end{aligned}
\end{equation}
This completes the description of the Lie algebra structure of $\Omega^*_{\hat\nabla}(M)$ and so that of the W-algebra commutators \eqref{walg} of holonomy symmetries.

\subsubsection{\{1\}}

In this case it is clear that in a co-frame adapted to the $\bR^8$ holonomy, $\Omega^*_{\hat\nabla}(M)$ is represented by the elements of the vector space $\wedge^*\bR^8$ including the even and odd degree forms. Keeping to the spirit of the discussion so far that $\Omega^*_{\hat\nabla}(M)$ has a Lie algebra structure, we shall restrict our analysis to the odd-degree $\hat\nabla$-covariantly constant forms. In particular, the Lie algebra of fundamental forms $\mathfrak{f}=\Omega^3_{\hat\nabla}(M)$ is represented by elements of $\wedge^2\bR^8=\mathfrak{so}(8)$ and so $\mathfrak{f}=\mathfrak{so}(8)$. The elements of $\mathfrak{f}$ are spanned by 2-form bilinears.

Furthermore, $\mathfrak{f}=\mathfrak{so}(8)$ acts on $\Omega^5_{\hat\nabla}(M)$ with the standard representation on $\wedge^4\bR^8$. This is a reducible representation and decomposes into the sum $\wedge^{4+}\bR^8\oplus \wedge^{4-}\bR^8$ spanned by the self-dual and anti-self-dual 4-forms, respectively. Only the former represent 5-form bilinears.

Finally, as all elements of $\Omega^5_{\hat\nabla}(M)$ can be written as a linear combination of $\curlywedge$-products of two elements in $\Omega^3_{\hat\nabla}(M)$, one can apply the formula \eqref{LMNP} to compute $\Omega^5_{\hat\nabla}(M)\bar\curlywedge\Omega^5_{\hat\nabla}(M)$. Indeed as $(\e^{-ij}; i<j, i,j=1,\dots, 8 )$ is a basis in $\Omega^3_{\hat\nabla}(M)$ and $\e^{-ij}\curlywedge\e^{-kl}=\e^{-ijkl}$ span $\Omega^5_{\hat\nabla}(M)$, one finds that
\begin{equation}
	\e^{-ijkl}\bar\curlywedge\e^{-i'j'k'l'} = 4 (\delta^{[i|i'|} \e^{jkl]j'k'l'-} - \delta^{[i|l'|} \e^{jkl]i'j'k'-} + \delta^{[i|k'|} \e^{jkl]l'i'j'-} - \delta^{[i|j'|} \e^{jkl]k'l'i'-})\, .
\end{equation}
This, together with the results in the beginning of section \ref{tpt}, complete the description of the Lie algebra structure of $\Omega^{\mathrm {odd}}_{\hat\nabla}(M)$ and so that of the commutators \eqref{walg} of holonomy symmetries.
\vskip 0.5cm

Some of the results of section \ref{holonomy-symmetries-w-algebras} are summarised in the table below.
\begin{table}[h]
	\centering
	\begin{tabular}{|c||c|c|c|c|c|c|c|}\hline
		$G$ & $\mathrm{Spin}(7)$ & $SU(4)$ & $Sp(2)$ & $\times^2Sp(1)$ & $Sp(1)$ & $U(1)$ & $\{1\}$ \\ \hline
		$\mathfrak{f}$ & $\bR$ & $\hat{\mathfrak{e}}(2)$ & $\mathfrak{sp}(1)$ & $\oplus^2\mathfrak{sp}(1)$ & $\mathfrak{so}(5)$ & $\mathfrak{u}(4)$ & $\mathfrak{so}(8)$ \\ \hline
	\end{tabular}
	\caption[Short caption]{In the first row, the $G$ subalgebras of holonomy groups $G\ltimes\bR^8$ of the supersymmetric heterotic backgrounds are stated. In the second row, the associated Lie algebras of the fundamental forms are given.}
	\label{tab:holonomy_groups}
\end{table}

\section{Anomalies} \label{anomalies}

\subsection{Anomaly Consistency Conditions}

Consider two classical symmetries of a theory generated by the infinitesimal variations $\delta_{a_1}$ and $\delta_{a_2}$ acting on the fields, where $a_1$ and $a_2$ are their parameters. Suppose that these symmetries are anomalous in the quantum theory with anomalies $\Delta(a_1)$ and $\Delta(a_2)$, respectively. If the commutator of these two symmetries is $[\delta_{a_1}, \delta_{a_2}]=\delta_{[a_1, a_2]}$, then the Wess-Zumino consistency condition of the anomalies of these symmetries is $\delta_{a_1} \Delta(a_2)-\delta_{a_2} \Delta(a_1)=\Delta([a_1, a_2])$, where $[a_1, a_2]$ denotes the parameter of the symmetry generated by the commutator and $\Delta([a_1, a_2])$ denotes the associated anomaly. If this condition is satisfied, the anomalies are consistent --- otherwise they are inconsistent. Anomaly consistency conditions can be used to determine the anomaly of a symmetry from other known anomalies of a theory.

The consistency conditions for holonomy anomalies have been extensively investigated in \cite{lggpepb}. Here, we shall summarise without explanation the main formulae and demonstrate that all holonomy anomalies of chiral sigma models on backgrounds with a non-compact holonomy group are consistent. To begin, the spacetime frame rotation \eqref{ftran2} and gauge sector transformation \eqref{ftran1} anomalies are given\footnote{For applications to string theory, replace $\hbar$ with $\alpha'$.} by
\begin{equation}\label{lanuan}
	\begin{aligned}
		\Delta(\ell) &= \frac{i\hbar}{4\pi} \int \dd^2\sigma \dd\theta^+ Q^1_2(\omega, \ell)_{\mu\nu} D_+X^\mu \partial_=X^\nu~, \\
		\Delta(u) &= -\frac{i\hbar}{4\pi} \int \dd^2\sigma \dd\theta^+ Q^1_2(\Omega, u)_{\mu\nu} D_+X^\mu \partial_=X^\nu~,
	\end{aligned}
\end{equation}
respectively, where the numerical coefficient in front of the expressions is determined after an explicit computation of the relevant part of the effective action \cite{chpt2}. The 2-form $Q_2^1(\omega, \ell)$ is determined by the descent equations \cite{zumino-book} starting from the 4-form, $P_4(R)= \mathrm{tr} (R(\omega)\wedge R(\omega))$, which is proportional to the first Pontryagin form of the spacetime, as
\begin{equation}
	\begin{gathered}
		\dd P_4(R) = 0\, ,\, \delta_\ell P_4(R) = 0 \Longrightarrow P_4(R) = \dd Q^0_3(\omega)\, , \\
		\dd \delta_\ell Q_3^0 = 0 \Longrightarrow \delta_\ell Q^0_3(\omega) = \dd Q^1_2(\ell, \omega)\, ,
	\end{gathered}
\end{equation}
and similarly for $Q^1_2(\Omega, u)$, where $R$ is the curvature of a frame connection $\omega$ of the spacetime. The connection $\omega$ in the expression for the anomaly is not uniquely specified as it can be altered with the addition of a suitable finite local counterterm in the effective action. This will be used later to prove the consistency of the anomalies.

The anomaly consistency conditions \cite{lggpepb} of frame rotation and gauge sector anomalies \eqref{lanuan} with the anomaly $\Delta(a_{{}_L})$ of a holonomy symmetry generated by the $\hat\nabla$-covariantly constant form $L$ yield
\begin{equation}\label{Lanom}
	\Delta(a_{{}_L}) = \frac{i\hbar}{4\pi} \int \dd^2\sigma \dd\theta^+\, Q_3^0(\omega, \Omega)_{\mu\nu\rho} \delta_L X^\mu D_+X^\nu \partial_=X^\rho + \Delta_{\mathrm{inv}}(a_{{}_L})\, ,
\end{equation}
where $ Q_3^0(\omega, \Omega)= Q_3^0(\omega)- Q_3^0(\Omega)$ and $\delta_\ell \Delta_{\mathrm {inv}}(a_{{}_L})=\delta_u \Delta_{\mathrm {inv}}(a_{{}_L})=0$. At one loop, we set $\Delta_{\mathrm {inv}}(a_{{}_L})=0$. Furthermore, checking the consistency condition between two holonomy anomalies\footnote{It turns out that the above consistency condition is more general. If the anomaly of two transformations $\delta_1$ and $\delta_2$ is given as in \eqref{Lanom}, then their mutual consistency condition will be given as in \eqref{LMcon} with $\delta_L=\delta_1$ and $\delta_M=\delta_2$.} generated by the $\hat\nabla$-covariantly constant forms $L$ and $M$, one finds that
\begin{equation}\label{LMcon}
	\begin{aligned}
		\delta_L \Delta(a_{{}_M}) - \delta_M \Delta(a_{{}_L}) &= \frac{i\hbar}{4\pi} \int \dd^2\sigma \dd\theta^+ Q_3^0(\omega, \Omega)_{\mu\nu\rho} [\delta_L, \delta_M] X^\mu D_+ X^\nu \partial_=X^\rho \\
		+& \frac{i\hbar}{4\pi} \int \dd^2\sigma \dd\theta^+ P_4(R, F)_{\mu\nu\rho\sigma} \delta_L X^\mu \delta_M X^\nu D_+ X^\rho \partial_=X^\sigma \, ,
	\end{aligned}
\end{equation}
where $P_4(R, F)=P_4(R)-P_4(F)$. The anomaly consistency conditions require that $\delta_L \Delta(a_{{}_M})-\delta_M \Delta(a_{{}_L})=\Delta([a_{{}_L}, a_{{}_M}])$, where here
\begin{equation}
	\Delta([a_{{}_L}, a_{{}_M}]) = \frac{i\hbar}{4\pi} \int \dd^2\sigma \dd\theta^+ Q_3^0(\omega, \Omega)_{\mu\nu\rho} [\delta_L, \delta_M] X^\mu D_+ X^\nu \partial_=X^\rho \,,
\end{equation}
with $[\delta_L, \delta_M] X^\mu$ given in \eqref{walg}. As a result it is clear that the consistency of two holonomy anomalies requires for the last term of \eqref{LMcon} to vanish.

As supersymmetric backgrounds with non-compact holonomy group admit a null $\hat\nabla$-parallel vector field $K$, the $\Delta(a_{{}_K})$ anomaly is consistent with any other holonomy anomaly $\Delta(a_{{}_M})$ provided that $i_K P_4(R, F)=0$. Indeed $i_KP_4(F)=0$ as $i_K F=0$. Therefore, it remains to choose the frame connection $\omega$ of the spacetime such that $i_KP_4(R)=0$. Such a connection can always be found. For example, one can choose the frame connection $\check \nabla$ whose torsion is $-H$. Then, the result follows from the restriction of the holonomy of $\hat\nabla$ to be a subgroup of $G\ltimes \bR^8$ and the Bianchi identity $\hat R_{\mu\nu, \rho\sigma}=\check R_{\rho\sigma, \mu\nu}$ for $\dd H=0$.

Assuming that the connection $\omega$ is chosen such that $i_K P_4(R, F)=0$, it is straightforward to observe that the last term in \eqref{LMcon} always vanishes for the holonomy transformations generated by $\hat\nabla$-covariantly constant forms that are null along $K$. This is a direct consequence of $i_K P_4(R, F)=0$ and the requirement that forms $L$ and $M$ are null along $K$. This can be verified by a short straightforward computation. As such, it can always be arranged for the anomalies of the holonomy symmetries to be consistent.

It should be stressed that the condition $i_K P_4(R, F)=0$ on $P_4(R, F)$ for the consistency of holonomy anomalies for backgrounds with a non-compact holonomy group is much weaker than the analogous conditions for backgrounds with compact holonomy groups \cite{lggpepb} or those for manifolds with a G-structure \cite{dlossa}. In particular, the application of the second condition in \eqref{invcon} for the gauge sector and the analogous condition for the curvature of spacetime are not required for the consistency of holonomy anomalies.

\subsection{Anomaly Cancellation}

There are two ways to cancel the anomaly of holonomy symmetries. One is to add to the effective action suitable finite local counterterms to cancel the anomalies. This method has been successful in cancelling \cite{sen, phgpa} the second supersymmetry anomaly generated by a complex structure on the spacetime and requires a refinement of the Poincar\'e lemma and possibly the existence of special coordinates on the spacetime, such as, for example, the local triviality of Hodge cohomology and complex coordinates. This is not the case here. The forms $L$ that generate the holonomy symmetries, in general, are not associated with the existence of such structures in a straightforward manner. The second method is to assume that the form $L$ that generates the holonomy symmetry is quantum mechanically corrected to $L^\hbar$, such that $L^\hbar$ is covariantly constant with respect to a connection $\hat \nabla^\hbar$ that has torsion
\begin{equation}\label{corrh}
H^{\hbar} = H - \frac{\hbar}{4\pi} Q^0_3(\omega, \Omega) + \mathcal{O}(\hbar^2)\, ,
\end{equation}
and $i_{L^\hbar} F^\hbar=0$. This is because the expression for the anomaly of the holonomy symmetry in \eqref{Lanom} can be viewed as a correction to the covariant constancy condition on $L$ required for the transformation $\delta_L$ to be a symmetry of the action. Clearly, the new transformation $\delta_{L^\hbar}$ will be a symmetry of the effective action and the anomaly will cancel. Such an anomaly cancellation mechanism is compatible with both the Green-Schwarz anomaly cancellation mechanism for gravitational anomalies \cite{mgjs} and the heterotic supergravity effective action \cite{roo}.

For the cancellation of holonomy symmetry anomalies of sigma models on supersymmetric backgrounds with non-compact holonomy groups, the second cancellation mechanism is the most appropriate with the expectation that $L^\hbar$ remains null along $K$. The latter assertion is justified as it is not expected for the Killing spinors to be corrected up to two loops in sigma model perturbation theory, after choosing an appropriate co-frame on the spacetime. Alternatively, the classification of the geometry of heterotic supersymmetric backgrounds remains the same, up to two loop level, for backgrounds with either a closed or non-closed 3-form field strength $H$. In the former case, the condition $dH=0$ is imposed at the end. Note though that the anomaly of the symmetry generated by $K$ can be removed with the addition of a finite local counterterm in the effective action. So $K$ need not to be corrected.

Furthermore, the above cancellation of the holonomy anomalies is consistent. The commutator of two holonomy symmetries for spacetimes with a non-compact holonomy group either vanishes or it closes to a type III transformation in the terminology of \cite{lggpepb}. In either case, the consistency condition is
\begin{equation}\label{concona}
	P(\omega, \Omega)_{\mu\nu[\rho|\sigma|} L^\mu{}_L M^\nu{}_{M]} = 0\, ,
\end{equation}
This is satisfied provided that the spacetime frame connection $\omega$ is chosen such that $i_KP(\omega, \Omega)=0$. There always exists such a connection in sigma model perturbation theory.

\chapter{Conclusions}

In chapter \ref{TCFH-General-Theory-Backgrounds}, we presented the TCFH of type IIA supergravity and demonstrated that the form bilinears satisfy a generalisation of the CKY equation with respect to the minimal TCFH connection in agreement with the general theorem in \cite{gptcfh}. Then prompted by the well-known result that KY forms generate (hidden) symmetries in spinning particle actions, we explored the question of whether the form bilinears of some known supergravity backgrounds, which include all type IIA branes, generate symmetries for various particle and string probes propagating on these backgrounds.

We have also explored the complete integrability of geodesic flow on all type II brane backgrounds. We demonstrated that if the harmonic function that the solutions depend on has at most one centre, i.e. they are spherically symmetric, then the geodesic flow is completely integrable. We have explicitly given all independent conserved charges in involution. We also presented the KS, KY and CCKY tensors of these brane backgrounds associated with their integrability structure.

Returning to the symmetries generated by the TCFH, supersymmetric type II common sector backgrounds admit form bilinears which are covariantly constant with respect to a connection with skew-symmetric torsion given by the NS-NS 3-form field strength. All these bilinears generate (hidden) symmetries for string and particle probe actions with 3-form couplings. The type II fundamental string and NS5-brane background form bilinears have explicitly been given. Common sector backgrounds admit additional form bilinears which satisfy a TCFH but they are not covariantly constant with respect to a connection with skew-symmetric torsion. Although these forms are part of the geometric structure of common sector backgrounds, their geometric interpretation is less straightforward.

Moreover we found that there are Killing spinors in all type IIA Dp-brane backgrounds such that the associated bilinears are KY forms and so generate (hidden) symmetries for spinning particle probes. All these form bilinears have components only along the worldvolume directions of the Dp-branes. 

It is fruitful to compare the KY forms we have obtained from the TCFH with those that are needed to investigate the integrability of the geodesic flow in type IIA brane backgrounds. TCFH KY forms exist for any choice of the harmonic function that the brane solutions depend on. Moreover, as we have mentioned, these KY forms have non-vanishing components only along the worldvolume directions of D-branes. It is clear from this that although they generate symmetries for particle probes propagating on D-brane backgrounds these symmetries are not necessarily connected to the integrability properties of such dynamical systems. This is because it is not expected, for example, that the geodesic flow of brane solutions which depend on a multi-centred harmonic function to be completely integrable. Indeed the KS and KY tensors we have found that are responsible for the integrability of the geodesic flow on spherically symmetric branes also have components along the transverse directions of these solutions. As the brane metrics have a non-trivial dependence on the transverse coordinates, this is essential for proving the integrability of the geodesic flow. Therefore one concludes that although the form bilinears of supersymmetric backgrounds can generate symmetries in string and particle probes propagating in these backgrounds, they are not sufficient to prove the complete integrability of probe dynamics. Nevertheless the TCFH KY tensors, when they exist, are associated with symmetries of probes propagating on brane backgrounds which are not necessarily spherically symmetric.

To find TCFH KY tensors, we have imposed a rather stringent set of conditions on the form bilinears. In particular in several D-brane backgrounds, we set all terms of the minimal TCFH connection that depend on a form field strength to zero. It is likely that such a restriction can be lifted and the only condition necessary for the invariance of a probe action will be that the terms in the TCFH which contain explicitly the metric should vanish. For this a new set of probe actions should be found that have couplings which depend on the form field strengths of the supergravity theories and generalise \eqref{1part} which exhibits only a 3-form coupling. We hope to report on such a development in the future.

In chapter \ref{TCFH-AdS-Backgrounds}, we presented all the TCFHs of massive IIA warped AdS backgrounds. In particular we have shown that the form bilinears of supersymmetric AdS backgrounds satisfy a generalisation of CKY equation with respect to the TCFH connection. In addition we have explored some of the properties of the minimal TCFH connection like its reduced holonomy. Furthermore we have investigated the question of whether the TCFHs give rise to hidden symmetries for probes propagating on the internal space of AdS backgrounds. For this we presented some examples of AdS backgrounds, namely those arising as near horizon geometries of intersecting IIA branes, and demonstrated that some of their form bilinears are KY forms. As a result they generate symmetries for spinning particles propagating on the internal space of such backgrounds. This work, together with those in \cite{epbgp, lggp}, completes the investigation of TCFHs of all warped AdS backgrounds of type II theories in 10 and 11 dimensions.

The extent of the interplay between TCFHs and symmetries of probes propagating on supersymmetric background remains open. There are certainly many examples of backgrounds that the TCFH conditions coincide with those required for the invariance of probe actions under transformations generated by the form bilinears. For example in the heterotic and common sector cases, all form bilinears generate symmetries for certain string and particle probes. However for generic type II theories, the relation between TCFH and probe symmetries can only be revealed on a case by case basis after exploring separately the geometric properties of each background. The difficulties lie both in the lack of classification of supersymmetric backgrounds in type II theories and the plethora of probes \cite{colesgp} that one can consider. A more systematic investigation will require developments both in the understanding of the supersymmetric backgrounds of type II theories as well as a better handle on probe actions and their symmetries.

Finally, in chapter \ref{W-symmetries-heterotic-backgrounds} we presented the W-algebra of holonomy symmetries of chiral sigma models propagating on supersymmetries heterotic backgrounds with non-compact holonomy group $G\ltimes \bR^8$. We demonstrated that the W-algebra, which depends on $G\ltimes \bR^8$, is completely determined by a Lie algebra structure on the space of $\hat\nabla$-covariantly constant forms, $\Omega^*_{\hat\nabla}(M)$, which are null along a null vector field $K$. The Lie algebra bracket on $\Omega^*_{\hat\nabla}(M)$ is a generalisation of that of the inner derivations on forms. Moreover, we determined the Lie algebra structure of $\Omega^*_{\hat\nabla}(M)$ for each holonomy group $G\ltimes \bR^8$ with $G$ given in \eqref{hol}.

In addition, we gave the anomalies of the holonomy symmetries up to and including 2-loops in sigma model perturbation theory. We demonstrate that they satisfy the Wess-Zumino consistency conditions, provided the frame connection is appropriately chosen on the spacetime. These anomalies are cancelled, provided that the forms that generate the symmetries are corrected such that they remain null along $K$ and are covariantly constant with respect to a connection with torsion. The torsion includes the difference of the Chern-Simons forms of the spacetime and gauge sector connections of the sigma model.

It is remarkable that there is such an extensive Lie algebraic structure underpinning the W-algebra of holonomy symmetries associated with supersymmetric heterotic backgrounds with a non-compact holonomy group. It is clear that this can be extended to all null forms along $K$, $\Omega^*_K(M)$, including those of even degree. The bracket will again be given by $\bar\curlywedge$. Of course $\Omega^*_K(M)$ is an infinite dimensional superalgebra where the odd degree forms span the Grassmannian even generators while the even degree forms span the Grassmannian odd generators of the superalgebra. This superalgebra structure can be restricted on $\Omega^*_{\hat\nabla}(M)$. This is relevant in the investigation of holonomy symmetries of sigma models on heterotic supersymmetric backgrounds with holonomy $\bR^8$, as $\Omega^*_{\hat\nabla}(M)$ includes forms of both even and odd degree. Notice that $\Omega^*_{\hat\nabla}(M)$ contains a sub-superalgebra $\Omega^1_{\hat\nabla}(M)\oplus \Omega^2_{\hat\nabla}(M)\oplus \Omega^3_{\hat\nabla}(M)$, which can be seen as an extension of $\mathfrak{so}(8)=\Omega^3_{\hat\nabla}(M)$ with new Grassmannian odd generators $\bR^8= \Omega^2_{\hat\nabla}(M)$ and central generator $\bR\langle K\rangle=\Omega^1_{\hat\nabla}(M)$. These structures can be further extended to all forms on Euclidean and Lorentzian manifolds, where now the Lie (super)bracket is given by the standard inner derivation $\bar\wedge$.

The Lie algebraic structure that we have uncovered that underpins the W-algebra of holonomy symmetries for backgrounds with non-compact holonomy groups does not naturally extend to include that of backgrounds with compact holonomy groups investigated in \cite{lggpepb}, see also \cite{sven, phgp2}. This is because in the latter case the W-algebra can close to symmetries which are not generated by $\hat\nabla$-covariantly constant forms. For example, the W-algebra can close to worldsheet translations generated by the sigma model energy-momentum tensor. Nevertheless, it may be possible to extend the Lie (super)algebra structure on $\Omega^*_{\hat\nabla}(M)$ with additional generators that are not forms in such a way that the new algebra specifies the associated W-algebra of symmetries of sigma models. It would be of interest to explore such a possibility in the future.


\appendix

\chapter{Notation and Convetions} \label{Conventions}

Let $M$ be an $n$-dimensional manifold with a (local) co-frame $\e^A$ and coordiantes $x^M$. A $p$-form, $\omega \in \Omega^p(M)$ is defined as:
\begin{equation}
	\omega = \frac{1}{p!} \omega_{M_1 \dots M_p} \dd x^{M_1} \wedge \dots \wedge \dd x^{M_p} = \frac{1}{p!} \omega_{A_1 \dots A_p} \e^{A_1} \wedge \dots \wedge \e^{A_p} \, .
\end{equation}
We take the Hodge duality operation to be:
\begin{equation}
	\hodge{\omega}_{M_1 \dots M_{n - p}} = \frac{1}{p!}\omega_{P_1\dots P_p}\epsilon^{P_1\dots P_p}{}_{M_1\dots M_{n - p}}
\end{equation}
where the orientation is taken to be $\epsilon_{012\dots (n - 1)} = -1$ for Lorentzian signature and $\epsilon_{12\dots n} = 1$ for Riemannian signature manifolds.

The exterior derivative is given by 
\begin{equation}
	\dd \omega \defeq \frac{1}{p!} \partial_{M_1} \omega_{M_2 \ldots M_{p+1}} \dd x^{M_1} \wedge \dots \wedge \dd x^{M_{p+1}} \, .
\end{equation}
The adjoint of the exterior derivative (co-derivative) is
\begin{equation}
	\delta \omega = (-1)^p \epsilon_{p-1} \star \dd \star \omega \, ,
\end{equation}
where $\epsilon_{p-1} \defeq (-1)^{p(n - p)} \frac{\det g}{|\det g|}$. Therefore, in local coordinates, one has
\begin{equation}
	(\dd \omega)_{M_1 \dots M_{p+1}}=(p+1) \partial_{\left[M_1\right.} \omega_{\left.M_2 \dots M_{p+1}\right]} \, , \quad (\delta \omega)_{M_2 \ldots M_p}=-\pd^M \omega_{M M_2 \ldots M_p} \, .
\end{equation}
The inner derivation $i_X$ of a $p$-form $\omega$ with respect to a vector field $X$ is
\begin{equation}
	i_X \omega \defeq \frac{1}{(p-1)!} X^B \omega_{B A_1 \dots A_{p-1}} \e^{A_1} \wedge \cdots \wedge \e^{A_{p-1}} \, .
\end{equation}
In general, the inner derivation of a $l$-form $\Phi$ with respect to a vector $(k-1)$-form $\xi$ is 
\begin{equation}
	i_\xi \Phi \equiv \xi\bar\wedge \Phi \defeq \frac{1}{(k-1)! (l-1)!} \xi^N{}_{L_1\dots L_{k-1}} \Phi_{N L_{k}\dots L_{k+l-2}} \dd x^{L_1} \wedge \dots \wedge \dd x^{L_{k+l-2}}.
\end{equation}
We shall use both notations $i_\xi$ and $\xi\bar\wedge$ to denote the inner derivation at convenience.

Given the form $\omega$ we can define a Clifford algebra element, $\slashed{\omega}$
\begin{equation}
	\slashed{\omega} = \omega_{N_1\dots N_p} \Gamma^{N_1\dots N_p} \, ,
\end{equation}
where $\Gamma^M$ are the Dirac gamma matrices. One can also define
\begin{equation}
	\slashed \omega_M = \omega_{M N_1\dots N_{p-1}} \Gamma^{N_1\dots N_{p-1}} \, .
\end{equation}

\chapter{Type IIA Warped AdS Backgrounds} \label{sec:Warped-AdS-Backgrounds}

In chapter \ref{TCFH-AdS-Backgrounds} (massive) IIA warped product AdS backgrounds are briefly reviewed, however here we aim to provide a more complete description of the geometry, fields and Killing spinors of these backgrounds. Many of these results were first presented in \cite{ggkpiia, bgpiia} and we refer the reader to these for an even more detailed account.

\section{Warped AdS\texorpdfstring{\textsubscript{2}}{2} Backgrounds}

The geometry of warped AdS$_2 \times_w M^8$ (massive) IIA backgrounds can be described by the near horizon geometry of extremal black holes. In this description the bosonic sector takes the form:
\begin{equation}
	\begin{gathered}
		g = 2\, \e^+ \e^- + g(M^8) \\
		G = \e^+ \wedge \e^- \wedge X + Y, \quad H = \e^+ \wedge \e^- \wedge W + Z \\
		F = N \, \e^+ \wedge \e^- + P, \quad S = S, \quad \Phi = \Phi,
	\end{gathered}
\end{equation}
where $g$ is the metric, $\Phi \in C^\infty(M^8)$ is the dilaton field and $N \in C^\infty(M^8)$, $W \in \Omega^1(M^8)$, $X, P \in \Omega^2(M^8)$, $Z \in \Omega^3(M^8)$ and $Y \in \Omega^4(M^8)$ are fluxes localised on the internal space. $S\in C^\infty(M^8)$ is the scalar field of massive IIA supergravity with $S = e^\Phi m$, where $m$ is a constant that is non-zero in massive IIA and vanishes in standard IIA supergravity. Further,
\begin{equation}
		\e^+ = \dd u, \quad \e^- = \dd r + r h - \frac{1}{2} r^2 \Delta \dd u, \quad \e^i = e^i{}_J \dd y^J
\end{equation}
is a null pseudo-orthonormal frame of AdS$_2 \times_w M^8$, where $h = \Delta^{-1} \dd \Delta$ with $\Delta = \ell^{-2} A^{-2}$. Here $(u, r)$ are the coordinates of AdS\textsubscript{2} and $y$ are the coordinates of $M^8$. $A$ is the warp factor and is a function of only the coordinates of $M^8$ and $\ell$ is the radius of AdS\textsubscript{2}. Moreover, 
\begin{equation}
	g(M^8) = \delta_{ij} \e^i \e^j
\end{equation}
is the metric on $M^8$, which is transverse to the lightcone directions given by $r = u = 0$. 

The solutions to the Killing spinor equations of massive IIA supergravity on warped AdS\textsubscript{2} backgrounds are a special case of the supersymmetric IIA horizons described in \cite{ggkpiia} and are presented in \cite{bgpiia}. What follows is a brief overview of these solutions.

Recall that the Killing spinor equations of type IIA supergravity are the conditions that the supersymmetry variations of the gravitino and dilatino fields vanish, evaluated on the locus of vanishing fermions. The gravitino Killing spinor equation is a parallel transport equation for the supercovariant connection, $\mathcal{D}$, whereas the dilatino Killing spinor equation is an algebraic condition on the supersymmetry parameter, $\epsilon$.

To construct the solution one first decomposes $\epsilon$ along the lightcone directions as:
\begin{equation}
	\epsilon = \epsilon_+ + \epsilon_-,
\end{equation}
subject to the lightcone projection conditions:
\begin{equation}
	\Gamma_\pm \epsilon_\pm = 0.
\end{equation}
The Killing spinors of type IIA supergravity are sections of the spin bundle associated with the 32 dimensional Majorana representation of $\mathfrak{spin}(9,1)$. The lightcone projection condition above halves the number of independent components of $\epsilon_\pm$ and as such these spinors each carry 16 degrees of freedom when counted over the reals.

After integrating along the AdS\textsubscript{2} directions, it can be shown that $\epsilon_\pm$ are constructed from a linear combination of spinors of the internal space $M^8$:
\begin{equation} \label{lightcone_spinors}
	\begin{aligned}
		\epsilon_+ &= \eta_+ + u \Gamma_+ \Theta_- \eta_- \\
		\epsilon_- &= \eta_- + r \Gamma_- \Theta_+ (\eta_+ + u \Gamma_+ \Theta_- \eta_-),
	\end{aligned}
\end{equation}
where $\eta_\pm$ depend only on the coordinates of $M^8$ and $\Theta_\pm$ is a Clifford algebra element constructed from the fluxes of the background. Where $\Theta_\pm$ is given by
\begin{equation}
	\Theta_{ \pm}=-\frac{1}{2} A^{-1} \slashed{\partial} A \mp \Gamma_{11} \slashed{W}-\frac{1}{16} \Gamma_{11}( \pm 2 N+ \slashed{P})-\frac{1}{8 \cdot 4!}( \pm 12 \slashed{X} + \slashed{Y})-\frac{1}{8} S \, .
\end{equation}
As spinors on $M^8$, $\eta_\pm$ are sections of the $\mathfrak{spin}(8)$ bundle on $M^8$ associated with the 16 dimensional Majorana representation.

The above summarises the solution along the AdS\textsubscript{2} directions; what remains are the independent equations along the $M^8$ directions. It can be shown that the gravitino Killing spinor equations of massive IIA supergravity on warped AdS$_2 \times_w M^8$ along the $M^8$ directions are:
\begin{equation} \label{AdS2_KSE}
	\mathcal{D}_m^{(\pm)} \eta_\pm = 0.
\end{equation}
This can be seen as a parallel transport equation for the supercovariant connection $\mathcal{D}^{(\pm)}$ on $M^8$. The supercovariant connection is given by:
\begin{equation} 
	\begin{split}
			\mathcal{D}_m^{(\pm)}\eta_\pm = \nabla_m \eta_\pm\, \pm & \frac{1}{2} A^{-1} \partial_{m} A \,\eta_\pm \mp \frac{1}{16} \slashed{X} \Gamma_{m} \eta_\pm + \frac{1}{8 \cdot 4 !} \slashed{Y} \Gamma_{m} \eta_\pm + \frac{1}{8} S \Gamma_{m} \eta_\pm \\
		&+\Gamma_{11}\left(\mp \frac{1}{4} W_{m} \eta_\pm + \frac{1}{8} \slashed{Z}_{m} \eta_\pm \pm \frac{1}{8} N \Gamma_{m} \eta_\pm - \frac{1}{16} \slashed{P} \Gamma_{m} \eta_\pm \right),
	\end{split}
\end{equation}
where $\nabla$ is the Levi-Civita connection induced on the spinor bundle of $M^8$. 

The Killing spinors $\eta_\pm$ satisfy the gravitino Killing spinor equation on $M^8$, \eqref{AdS2_KSE}, alongside an additional algebraic Killing spinor equation associated with the dilatino.

\section{Warped AdS\texorpdfstring{\textsubscript{3}}{3} Backgrounds}

As before, the geometry of warped AdS$_3 \times_w M^7$ (massive) IIA backgrounds can be described by the near horizon geometry of extremal black holes. In this description the bosonic sector takes the form:
\begin{equation} \label{AdS3background}
	\begin{gathered}
		g = 2\, \e^+ \e^- + A^2 \dd z^2 + g(M^7) \\
		G = A\, \e^+ \wedge \e^- \wedge \dd z \wedge X + Y, \quad H = A W\, \e^+ \wedge \e^- \wedge \dd z + Z \\
		F = F, \quad S = S, \quad \Phi = \Phi
	\end{gathered}
\end{equation}
where $g$ is the metric, $\Phi \in C^\infty(M^7)$ is the dilaton field and $W \in C^\infty(M^7)$, $X \in \Omega^1(M^7)$, $F \in \Omega^2(M^7)$, $Z \in \Omega^3(M^7)$ and $Y \in \Omega^4(M^7)$ are fluxes localised on the internal space. $S\in C^\infty(M^7)$ is the scalar field of massive IIA supergravity with $S = e^\Phi m$, where $m$ is a constant that is non-zero in massive IIA and vanishes in standard IIA supergravity. Further,
\begin{equation}
	\begin{gathered}
		\e^+ = \dd u, \quad \e^- = \dd r + r h, \quad \e^i = e^i{}_J \dd y^J \\
		\Delta = 0, \quad h = - \frac{2}{\ell} \dd z - 2 A^{-1} \dd A
	\end{gathered}
\end{equation}
is a null pseudo-orthonormal frame of AdS$_3 \times_w M^7$. Here $(u, r, z)$ are the coordinates of AdS\textsubscript{3} and $y$ are the coordinates of $M^7$. $A$ is the warp factor and is a function of only the coordinates of $M^7$ and $\ell$ is the radius of AdS\textsubscript{3}. Moreover, 
\begin{equation}
	g(M^7) = \delta_{ij} \e^i \e^j
\end{equation}
is the metric on $M^7$.

Analysis of the Bianchi identities reveals that either $S = 0$ or $W = 0$. Therefore, there are two distinct IIA AdS\textsubscript{3} supergravity backgrounds to consider. The first is a standard IIA supergravity background where the $H$ flux has a non-vanishing component on AdS\textsubscript{3} and the second is a massive IIA supergravity background where the $H$ flux is localised on $M^7$. 

The solutions to the Killing spinor equations of massive IIA supergravity on warped AdS\textsubscript{3} are described in \cite{bgpiia} and are constructed similarly to the AdS\textsubscript{2} case. Here we provide a brief overview of these solutions.

As before, one first decomposes $\epsilon$ along the lightcone directions as:
\begin{equation}
	\epsilon = \epsilon_+ + \epsilon_-,
\end{equation}
subject to the lightcone projection conditions:
\begin{equation}
	\Gamma_\pm \epsilon_\pm = 0 .
\end{equation}
As in the AdS\textsubscript{2} case, after integrating along the $u, r$ directions one finds that the Killing spinors can be expressed as in \eqref{lightcone_spinors}. It can be shown that after integrating along the remaining AdS\textsubscript{3} direction, $z$, one finds:
\begin{equation} \label{AdS_KS}
	\eta_\pm = \sigma_\pm + e^{\mp \frac{z}{\ell}} \tau_\pm,
\end{equation}
where the spinors $\sigma_\pm, \tau_\pm$ are subject to an additional algebraic constraint (separate from the dilatino Killing spinor equations):
\begin{equation} \label{AdS3_Add_KSE}
	\Xi_\pm \sigma_\pm = 0 , \qquad \Xi_\pm \tau_\pm = \mp \ell^{-1} \tau_\pm,
\end{equation}
where
\begin{equation}
	\Xi_{\pm} = \mp \frac{1}{2 \ell} + \frac{1}{2} \slashed{\partial} A \Gamma_{z} \pm \frac{1}{4} A W \Gamma_{11} - \frac{1}{8} A S \Gamma_{z} - \frac{1}{16} A \slashed{F} \Gamma_{z} \Gamma_{11} - \frac{1}{192} A \slashed{Y} \Gamma_{z} \mp \frac{1}{8} A \slashed{X}.
\end{equation}

As spinors on $M^7$, $\sigma_\pm$ and $\tau_\pm$ are sections of the $\mathfrak{spin}(7)$ bundle on $M^7$ associated with the (reducible) 16 dimensional Majorana representation. Bringing this together we find the Killing spinors along the AdS\textsubscript{3} directions:
\begin{equation}
	\epsilon = \sigma_{+} + e^{-\frac{z}{\ell}} \tau_{+} + \sigma_{-}+e^{\frac{z}{\ell}} \tau_{-} - \ell^{-1} u A^{-1} \Gamma_{+z} \sigma_{-} - \ell^{-1} r A^{-1} e^{-\frac{z}{\ell}} \Gamma_{-z} \tau_{+} .
\end{equation}
The above summarises the solution along the AdS\textsubscript{3} directions; what remains are the independent equations along the $M^7$ directions. It can be shown that the gravitino Killing spinor equations of massive IIA supergravity on warped AdS$_3 \times_w M^7$ along the $M^7$ directions are:
\begin{equation} \label{AdS3_KSE}
	\mathcal{D}_m^{(\pm)} \chi_\pm = 0, 
\end{equation}
where $\chi_\pm \in \{\sigma_\pm, \tau_\pm\}$. As before, this can be seen as a parallel transport equation for the supercovariant connection $\mathcal{D}^{(\pm)}$ on $M^7$. The supercovariant connection is given by:
\begin{equation} 
	\begin{split}
			\mathcal{D}_m^{(\pm)}\chi_\pm = \nabla_m \chi_\pm\, \pm & \frac{1}{2} A^{-1} \partial_{m} A \,\chi_\pm + \frac{1}{8} \slashed{Z}_{m} \Gamma_{11} \chi_\pm + \frac{1}{8} S \Gamma_{m} \chi_\pm \\&+ \frac{1}{16} \slashed{F}\Gamma_{m}\Gamma_{11} \chi_\pm + \frac{1}{192} \slashed{Y} \Gamma_{m} \chi_\pm \pm \frac{1}{8} \slashed{X} \Gamma_{z m}\chi_\pm,
	\end{split}
\end{equation}
where $\nabla$ is the Levi-Civita connection induced on the spinor bundle of $M^7$.

\section{Warped AdS\texorpdfstring{\textsubscript{4}}{4} Backgrounds}

As in the previous cases, the bosonic sector of warped AdS$_4 \times_w M^6$ (massive) IIA backgrounds can be expressed in the near horizon geometry as follows:
\begin{equation}
	\begin{gathered}
		g = 2\, \e^+ \e^- + A^2 ( \dd z^2 + e^{2z/\ell} \dd x^2 ) + g(M^6) \\
		G = e^{z/\ell} A^2 X \, \e^+ \wedge \e^- \wedge \dd z \wedge \dd x + Y \\
		H = H, \quad F = F, \quad S = S, \quad \Phi = \Phi
	\end{gathered}
\end{equation} 
where $\Phi \in C^\infty(M^6)$ is the dilaton field and $X \in C^\infty(M^6)$, $F \in \Omega^2(M^6)$, $H \in \Omega^3(M^6)$ and $Y \in \Omega^4(M^6)$ are fluxes localised on the internal space. $S\in C^\infty(M^6)$ is the scalar field of massive IIA supergravity with $S = e^\Phi m$, where $m$ is a constant that is non-zero in massive IIA and vanishes in standard IIA supergravity. Further,
\begin{equation}
	\begin{gathered}
		\e^+ = \dd u, \quad \e^- = \dd r + r h, \quad \e^i = e^i{}_J \dd y^J \\
		\Delta = 0, \quad h = - \frac{2}{\ell} \dd z - 2 A^{-1} \dd A
	\end{gathered}
\end{equation}
is a null pseudo-orthonormal frame of AdS$_4 \times_w M^6$. Here $(u, r, z, x)$ are the coordinates of AdS\textsubscript{4} and $y$ are the coordinates of $M^6$. $A$ is the warp factor and is a function of only the coordinates of $M^6$ and $\ell$ is the radius of AdS\textsubscript{4}. Moreover, 
\begin{equation}
	g(M^6) = \delta_{ij} \e^i \e^j
\end{equation}
is the metric on $M^6$.

As in the previous cases, the Killing spinor equations can be integrated along the AdS directions; after integrating along the $u, r$ directions one finds that the Killing spinors can be expressed as in \eqref{lightcone_spinors}. Integrating along $z$ results in \eqref{AdS_KS}. A further integration along $x$ yields:
\begin{equation}
	\begin{aligned}
	\epsilon =\, \sigma_{+}-\ell^{-1} x \Gamma_{x z} \tau_{+}&+e^{-\frac{z}{\ell}} \tau_{+}+\sigma_{-}+e^{\frac{z}{\ell}}\left(\tau_{-}-\ell^{-1} x \Gamma_{x z} \sigma_{-}\right) \\
	&-\ell^{-1} u A^{-1} \Gamma_{+z} \sigma_{-}-\ell^{-1} r A^{-1} e^{-\frac{z}{\ell}} \Gamma_{-z} \tau_{+},
	\end{aligned}
\end{equation}
where the spinors $\sigma_\pm, \tau_\pm$ are subject to an additional algebraic constraint:
\begin{equation}
	\Xi_\pm \sigma_\pm = 0 , \qquad \Xi_\pm \tau_\pm = \mp \ell^{-1} \tau_\pm,
\end{equation}
where
\begin{equation}
	\Xi_{\pm}=\mp \frac{1}{2 \ell}+\frac{1}{2} \slashed \partial A \Gamma_{z}-\frac{1}{8} A S \Gamma_{z}-\frac{1}{16} A \slashed F\Gamma_{z} \Gamma_{11}-\frac{1}{192} A \slashed Y \Gamma_{z} \mp \frac{1}{8} A X \Gamma_{x} .
\end{equation}

As spinors on $M^6$, $\sigma_\pm$ and $\tau_\pm$ are sections of the $\mathfrak{spin}(6)$ bundle on $M^6$ associated with the (reducible) 16 dimensional Majorana representation.

The above summarises the solution along the AdS\textsubscript{4} directions; what remains are the independent equations along the $M^6$ directions. It can be shown that the gravitino Killing spinor equations of massive IIA supergravity on warped AdS$_4 \times_w M^6$ along the $M^6$ directions are:
\begin{equation}
	\mathcal{D}_m^{(\pm)} \chi_\pm = 0
\end{equation}
where $\chi_\pm \in \{\sigma_\pm, \tau_\pm\}$. As before, this can be seen as a parallel transport equation for the supercovariant connection $\mathcal{D}^{(\pm)}$ on $M^6$. The supercovariant connection is given by:
\begin{equation}
	\begin{split}
			\mathcal{D}_m^{(\pm)}\chi_\pm = \nabla_m \chi_\pm\, \pm & \frac{1}{2} A^{-1} \partial_{m} A \,\chi_\pm +\frac{1}{8} \slashed{H}_{m}\Gamma_{11}\chi_\pm + \frac{1}{8} S \Gamma_{m}\chi_\pm \\
			+& \frac{1}{16} \slashed{F} \Gamma_{m} \Gamma_{11}\chi_\pm + \frac{1}{192} \slashed{Y} \Gamma_{m}\chi_\pm \mp \frac{1}{8} X \Gamma_{z x m}\chi_\pm,
	\end{split}
\end{equation}
where $\nabla_m$ is the Levi-Civita (spin) connection on $M^6$. 

\section{Warped AdS\texorpdfstring{\textsubscript{n}, $n \geq 5$}{n, n >= 5} Backgrounds}

As for all previous cases, the bosonic sector of warped AdS$_n \times_w M^{10 - n}$, $n \geq 5$ (massive) IIA backgrounds can be expressed in the near horizon geometry as follows:
\begin{equation}
	\begin{gathered}
		g = 2\, \e^+ \e^- + A^2 \left( \dd z^2 + e^{2z/\ell} \sum\limits_{a = 1}^{n - 3} (\dd x^a)^2 \right) + g(M^{10 - n}) \\
		G = G, \quad H = H, \quad F = F, \quad S = S, \quad \Phi = \Phi
	\end{gathered}
\end{equation} 
where $g$ is the metric, $\Phi \in C^\infty(M^{10 - n})$ is the dilaton field and $F \in \Omega^2(M^{10 - n})$, $H \in \Omega^3(M^{10 - n})$ and $G \in \Omega^4(M^{10 - n})$ are fluxes localised on the internal space. $S\in C^\infty(M^{10 - n})$ is the scalar field of massive IIA supergravity with $S = e^\Phi m$, where $m$ is a constant that is non-zero in massive IIA and vanishes in standard IIA supergravity. For sufficiently large $n$, some of the fluxes vanish; for example $G$ vanishes for $n \geq 7$. Further,
\begin{equation}
	\begin{gathered}
		\e^+ = \dd u, \quad \e^- = \dd r + r h, \quad \e^i = e^i{}_J \dd y^J \\
		\Delta = 0, \quad h = - \frac{2}{\ell} \dd z - 2 A^{-1} \dd A
	\end{gathered}
\end{equation}
is a null pseudo-orthonormal frame of AdS$_n \times_w M^{10 - n}$. Here $(u, r, z, x^a)$ are the coordinates of AdS\textsubscript{n} and $y$ are the coordinates of $M^{10 - n}$. $A$ is the warp factor and is a function of only the coordinates of $M^{10 - n}$ and $\ell$ is the radius of AdS\textsubscript{n}. Moreover, 
\begin{equation}
	g(M^{10 - n}) = \delta_{ij} \e^i \e^j
\end{equation}
is the metric on $M^{10 - n}$.

As in the previous cases, the Killing spinor equations can be integrated along the AdS directions; after integrating along the $u, r$ directions one finds that the Killing spinors can be expressed as in \eqref{lightcone_spinors}. Integrating along $z$ results in \eqref{AdS_KS}. A further integration along $x^a$ yields:
\begin{equation}
	\begin{aligned}
	\epsilon= \, \sigma_{+} &-\ell^{-1} \sum_{a=1}^{n-3} x^{a} \Gamma_{a z} \tau_{+}+e^{-\frac{z}{\ell}} \tau_{+}+\sigma_{-}+e^{\frac{z}{\ell}}\left(\tau_{-}-\ell^{-1} \sum_{a=1}^{n-3} x^{a} \Gamma_{a z} \sigma_{-}\right) \\
	&-\ell^{-1} u A^{-1} \Gamma_{+z} \sigma_{-}-\ell^{-1} r A^{-1} e^{-\frac{z}{\ell}} \Gamma_{-z} \tau_{+}
	\end{aligned}
\end{equation}
where the spinors $\sigma_\pm, \tau_\pm$ are subject to an additional algebraic constraint:
\begin{equation}
	\Xi_\pm \sigma_\pm = 0 , \qquad \Xi_\pm \tau_\pm = \mp \ell^{-1} \tau_\pm,
\end{equation}
where
\begin{equation}
	\Xi_{\pm}=\mp \frac{1}{2 \ell}+\frac{1}{2} \slashed \partial A \Gamma_{z}-\frac{1}{8} A S \Gamma_{z}-\frac{1}{16} A \slashed F \Gamma_{z} \Gamma_{11}-\frac{1}{192} A \slashed G \Gamma_{z} .
\end{equation}

As spinors on $M^{10 - n}$, $\sigma_\pm$ and $\tau_\pm$ are sections of the $\mathfrak{spin}(10 - n)$ bundle on $M^{10 - n}$ associated with the (reducible) 16 dimensional Majorana representation.

The above summarises the solution along the AdS\textsubscript{n} directions; what remains are the independent equations along the $M^{10 - n}$ directions. It can be shown that the gravitino Killing spinor equations of massive IIA supergravity on warped AdS$_n \times_w M^{10 - n}$ along the $M^{10 - n}$ directions are:
\begin{equation}
	\mathcal{D}_m^{(\pm)} \chi_\pm = 0
\end{equation}
where $\chi_\pm \in \{\sigma_\pm, \tau_\pm\}$. As before, this can be seen as a parallel transport equation for the supercovariant connection $\mathcal{D}^{(\pm)}$ on $M^{10 - n}$. The supercovariant connection is given by:
\begin{equation}
	\begin{split}
			\mathcal{D}_m^{(\pm)}\chi_\pm = \nabla_m \chi_\pm\, \pm & \frac{1}{2} A^{-1} \partial_{m} A \,\chi_\pm +\frac{1}{8} \slashed{H}_{m}\Gamma_{11}\chi_\pm + \frac{1}{8} S \Gamma_{m}\chi_\pm \\
			&+ \frac{1}{16} \slashed{F} \Gamma_{m} \Gamma_{11}\chi_\pm + \frac{1}{192} \slashed{G} \Gamma_{m}\chi_\pm ,
	\end{split}
\end{equation}
where $\nabla_m$ is the Levi-Civita (spin) connection on $M^{10 - n}$.

\chapter{Killing Spinor Form Bilinears of Brane Backgrounds} \label{IIA-bilinears}

\section{Common Sector Brane Form Bilinears}\label{common-sector-bilinears}

\subsection{Fundamental String}

A direct computation using \eqref{fsol} reveals that the form bilinears of IIA fundamental string are
\begin{equation}
	\begin{split}
		&\sigma^{rs}= h^{-\frac{1}{2}}\big(-\Herm{\eta^r}{\lambda^s}+ \Herm{\lambda^r}{\eta^s}\big)~ , \quad \\
		&k^{rs}= h^{-\frac{1}{2}}\Herm{\eta^r}{\eta^s} (\e^0-\e^5)+ h^{-\frac{1}{2}} \Herm{\lambda^r}{\lambda^s} (\e^0 + \e^5)~, \\
		&\omega^{rs} = h^{-\frac{1}{2}}\big(\Herm{\eta^r}{\lambda^s}+ \Herm{\lambda^r}{\eta^s}\big) \e^0\wedge \e^5 + \frac{1}{2} h^{-\frac{1}{2}}
		\big(-\Herm{\eta^r}{\Gamma_{ij}\lambda^s}+ \Herm{\lambda^r}{\Gamma_{ij}\eta^s}\big) \e^i\wedge \e^j~, \\
		&\pi^{rs} = \frac{1}{2}h^{-\frac{1}{2}} \Herm{\eta^r}{\Gamma_{ij}\eta^s} (\e^0-\e^5)\wedge \e^i\wedge \e^j
		+\frac{1}{2} h^{-\frac{1}{2}}\Herm{\lambda^r}{\Gamma_{ij} \lambda^s} (\e^0 + \e^5)\wedge \e^i\wedge \e^j~, \\
		&\zeta^{rs} = \frac{1}{2}h^{-\frac{1}{2}}\big(\Herm{\eta^r}{\Gamma_{ij}\lambda^s}+ \Herm{\lambda^r}{\Gamma_{ij}\eta^s}\big) \e^0\wedge \e^5\wedge \e^i\wedge \e^j \\
		&\qquad\qquad +\frac{1}{4!}h^{-\frac{1}{2}}\big(-\Herm{\eta^r}{\Gamma_{ijk\ell}\lambda^s}+ \Herm{\lambda^r}{\Gamma_{ijk\ell}\eta^s}\big) \e^i\wedge \e^j\wedge \e^k\wedge \e^\ell ~, \quad \\
		&\tau^{rs} = \frac{1}{4!}h^{-\frac{1}{2}}\Herm{\eta^r}{\Gamma_{ijk\ell} \eta^s} (\e^0-\e^5)\wedge \e^i\wedge \e^j\wedge \e^k\wedge \e^\ell \\
		&\qquad\qquad+ \frac{1}{4!} h^{-\frac{1}{2}} \Herm{\lambda^r}{\Gamma_{ijk\ell} \lambda^s} (\e^0 + \e^5)\wedge \e^i\wedge \e^j\wedge \e^k\wedge \e^\ell~,
	\end{split}
\end{equation}
where $i,j,k,\ell = 1, 2, 3, 4, 6, 7, 8, 9$ are the transverse directions of the string and $(\e^0, \e^5, \e^i)$ is a pseudo-orthonormal frame of the fundamental string metric \eqref{fstring}, i.e $g = -(\e^0)^2 + (\e^5)^2+ \sum_i (\e^i)^2$. The remaining form bilinears $\tilde\sigma$, $\tilde k$, $\tilde \omega$, $\tilde \pi$, $\tilde \zeta$ and $\tilde \tau$ can be obtained from the expressions above upon setting $\lambda^s$ to $-\lambda^s$.

\subsection{NS5-brane}

A direct computation using \eqref{ns5s} reveals that the form bilinears of NS5-brane are
\begin{equation}
	k^{rs} = 2 \big(\mathrm{Re}\Dirac{\eta^{1r}}{\Gamma_a\eta^{1s}} + \mathrm{Re}\Dirac{\eta^{2r}}{\Gamma_a\eta^{2s}}\big)~\e^a~,
\end{equation}
\begin{equation}
	\begin{split}
		\omega^{rs} &= 2 \big(\mathrm{Re}\Dirac{\eta^{1r}}{\Gamma_{a}\eta^{2s}} + \mathrm{Re}\Dirac{\eta^{2r}}{\Gamma_{a}\eta^{1s}}\big)~\e^a\wedge \e^3 \\
		&+ 2 \big(\mathrm{Re}\Dirac{\eta^{1r}}{\Gamma_{a}\lambda^{2s}}- \mathrm{Re}\Dirac{\eta^{2r}}{\Gamma_{a}\lambda^{1s}}\big)~\e^a\wedge \e^4 \\
		&+ 2 \big(\mathrm{Im}\Dirac{\eta^{1r}}{\Gamma_{a}\eta^{2s}}-\mathrm{Im}\Dirac{\eta^{2r}}{\Gamma_{a}\eta^{1s}}\big)~\e^a\wedge \e^8 \\
		&+ 2 \big(\mathrm{Im}\Dirac{\eta^{1r}}{\Gamma_{a}\lambda^{2s}}- \mathrm{Im}\Dirac{\eta^{2r}}{\Gamma_{a}\lambda^{1s}}\big)~\e^a\wedge \e^9~,
	\end{split}
\end{equation}
\begin{equation}
	\begin{split}
		\pi^{rs} &= \frac{1}{3}\big(\mathrm{Re}\Dirac{\eta^{1r}}{\Gamma_{abc} \eta^{1s}} + \mathrm{Re}\Dirac{\eta^{2r}}{\Gamma_{abc} \eta^{2r}} \big) \e^a\wedge \e^b\wedge \e^c \\
		&- 2\mathrm{Re}\Dirac{\eta^{1r}}{\Gamma_{a} \lambda^{1s}} (\e^3\wedge \e^4 -\e^8\wedge \e^9)\wedge \e^a \\
		&+ 2\mathrm{Re}\Dirac{\eta^{2r}}{\Gamma_{a} \lambda^{2s}} (\e^3\wedge \e^4 +\e^8\wedge \e^9)\wedge \e^a \\
		&- 2\mathrm{Im}\Dirac{\eta^{1r}}{\Gamma_{a} \eta^{1s}} (\e^3\wedge \e^8 +\e^4\wedge \e^9)\wedge \e^a \\
		&+ 2\mathrm{Im}\Dirac{\eta^{2r}}{\Gamma_{a} \eta^{2s}} (\e^3\wedge \e^8 -\e^4\wedge \e^9)\wedge \e^a \\
		&- 2\mathrm{Im}\Dirac{\eta^{1r}}{\Gamma_{a} \lambda^{1s}} (\e^3\wedge \e^9 -\e^4\wedge \e^8)\wedge \e^a \\
		&+ 2\mathrm{Im}\Dirac{\eta^{2r}}{\Gamma_{a} \lambda^{2s}} (\e^3\wedge \e^9 +\e^4\wedge \e^8)\wedge \e^a~,
	\end{split}
\end{equation}
\begin{equation}
	\begin{split}
		\zeta^{rs} &= \frac{1}{6} \tilde \omega^{rs}_{a\ell} \epsilon^\ell{}_{ijk} \e^a\wedge \e^i\wedge \e^j\wedge \e^k \\
		&+ \frac{1}{3} \big(\mathrm{Re}\Dirac{\eta^{1r}}{\Gamma_{abc}\eta^{2s}} + \mathrm{Re}\Dirac{\eta^{2r}}{\Gamma_{a}\eta^{1s}}\big)~\e^a\wedge \e^b\wedge \e^c\wedge \e^3 \\
		&+ \frac{1}{3} \big(\mathrm{Re}\Dirac{\eta^{1r}}{\Gamma_{abc}\lambda^{2s}}- \mathrm{Re}\Dirac{\eta^{2r}}{\Gamma_{abc}\lambda^{1s}}\big)~\e^a\wedge \e^b\wedge \e^c\wedge \e^4 \\
		&+ \frac{1}{3} \big(\mathrm{Im}\Dirac{\eta^{1r}}{\Gamma_{abc}\eta^{2s}}-\mathrm{Im}\Dirac{\eta^{2r}}{\Gamma_{abc}\eta^{1s}}\big)~\e^a\wedge \e^b\wedge \e^c\wedge \e^8 \\
		&+ \frac{1}{3} \big(\mathrm{Im}\Dirac{\eta^{1r}}{\Gamma_{abc}\lambda^{2s}}- \mathrm{Im}\Dirac{\eta^{2r}}{\Gamma_{abc}\lambda^{1s}}\big)~\e^a\wedge \e^b\wedge \e^c\wedge \e^9~,
	\end{split}
\end{equation}
\begin{equation}
	\begin{split}
		\tau^{rs} &= \tilde k^{rs}\wedge \e^3\wedge \e^4\wedge \e^8\wedge \e^9 \\
		&- \frac{1}{3}\mathrm{Re}\Dirac{\eta^{1r}}{\Gamma_{abc} \lambda^{1s}} (\e^3\wedge \e^4 -\e^8\wedge \e^9)\wedge \e^a\wedge \e^b\wedge \e^c \\
		&+ \frac{1}{3}\mathrm{Re}\Dirac{\eta^{2r}}{\Gamma_{abc} \lambda^{2s}} (\e^3\wedge \e^4 +\e^8\wedge \e^9)\wedge \e^a\wedge \e^b\wedge \e^c \\
		&- \frac{1}{3}\mathrm{Im}\Dirac{\eta^{1r}}{\Gamma_{abc} \eta^{1s}} (\e^3\wedge \e^8 +\e^4\wedge \e^9)\wedge \e^a\wedge \e^b\wedge \e^c \\
		&+ \frac{1}{3}\mathrm{Im}\Dirac{\eta^{2r}}{\Gamma_{abc} \eta^{2s}} (\e^3\wedge \e^8 -\e^4\wedge \e^9)\wedge \e^a\wedge \e^b\wedge \e^c \\
		&- \frac{1}{3}\mathrm{Im}\Dirac{\eta^{1r}}{\Gamma_{abc} \lambda^{1s}} (\e^3\wedge \e^9 -\e^4\wedge \e^8)\wedge \e^a\wedge \e^b\wedge \e^c \\
		&+ \frac{1}{3}\mathrm{Im}\Dirac{\eta^{2r}}{\Gamma_{abc} \lambda^{2s}} (\e^3\wedge \e^9 +\e^4\wedge \e^8)\wedge \e^a\wedge \e^b\wedge \e^c \\
		&+ \frac{2}{5!} \big(\mathrm{Re}\Dirac{\eta^{1r}}{\Gamma_{a_1\dots a_5}\eta^{1s}} + \mathrm{Re}\Dirac{\eta^{2r}}{\Gamma_{a_1\dots a_5}\eta^{2s}}\big)~\e^{a_1}\wedge\dots\wedge \e^{a_5}~,
	\end{split}
\end{equation}
where $a,b,c = 0,1,2,5,6,7$ are the worldvolume directions, $\epsilon_{3489} = 1$ and $(\e^a, \e^3, \e^4, \e^8, \e^9)$ is a pseudo-orthonormal frame for the NS5-brane metric \eqref{ns5}. The remaining form bilinears $\tilde \sigma$, $\tilde k$, $\tilde \omega$, $\tilde \pi$, $\tilde \zeta$ and $\tilde \tau$ bilinears can be constructed from those above upon replacing both
$\eta^{2s}$ and $\lambda^{2s}$ with $-\eta^{2s}$ and $-\lambda^{2s}$, respectively.

\section{Form Bilinears of IIA D-branes} \label{Dbrane-bilinears}

\subsection{D0-brane}

Using the expression for the Killing spinors of the D0-brane \eqref{d0sol}, one finds that the non-vanishing from bilinears of the solution are
\begin{equation}
	\tilde \sigma^{rs} = -2 h^{-\frac{1}{4}}\, \Herm{\eta^r}{\eta^s}~, ~~~ k^{rs} = 2 h^{-\frac{1}{4}}\, \Herm{\eta^r}{\eta^s}\, \e^0~,
\end{equation}
\begin{equation}
	\tilde k^{rs} = -2 h^{-\frac{1}{4}} \Herm{\eta^r}{\Gamma_{11}\eta^s}\, \e^5+ 2 h^{-\frac{1}{4}} \Herm{\eta^r}{\Gamma_i\eta^s}\, \e^i~,
\end{equation}
\begin{equation}
	\omega^{rs} = -2 h^{-\frac{1}{4}}\, \Herm{\eta^r}{\Gamma_{11} \eta^s}\, \e^0\wedge \e^5+ 2 h^{-\frac{1}{4}} \Herm{\eta^r}{\Gamma_i\eta^s} \e^0\wedge \e^i~,
\end{equation}
\begin{equation}
	\tilde\omega^{rs} = -2 h^{-\frac{1}{4}}\, \Herm{\eta^r}{\Gamma_i \Gamma_{11} \eta^s}\, \e^5\wedge \e^i- h^{-\frac{1}{4}} \Herm{\eta^r}{\Gamma_{ij}\eta^s} \e^i\wedge \e^j~,
\end{equation}
\begin{equation}
	\pi^{rs} = 2 h^{-\frac{1}{4}}\, \Herm{\eta^r}{\Gamma_i \Gamma_{11} \eta^s}\, \e^0\wedge \e^5\wedge \e^i+ h^{-\frac{1}{4}} \Herm{\eta^r}{\Gamma_{ij}\eta^s} \e^0\wedge \e^i\wedge \e^j~,
\end{equation}
\begin{equation}
	\tilde\pi^{rs} = - h^{-\frac{1}{4}} \Herm{\eta^r}{\Gamma_{ij}\Gamma_{11}\eta^s}\, \e^5\wedge \e^i\wedge \e^j + \frac{1}{3} h^{-\frac{1}{4}} \Herm{\eta^r}{\Gamma_{ijk}\eta^s} \e^i\wedge \e^j\wedge \e^k~,
\end{equation}
\begin{equation}
	\zeta^{rs} = -h^{-\frac{1}{4}}\, \Herm{\eta^r}{\Gamma_{ij} \Gamma_{11} \eta^s}\, \e^0\wedge \e^5\wedge \e^i\wedge \e^j + \frac{1}{3} h^{-\frac{1}{4}} \Herm{\eta^r}{\Gamma_{ijk}\eta^s}\, \e^0\wedge \e^i\wedge \e^j \wedge \e^k~,
\end{equation}
\begin{equation}
	\tilde \zeta^{rs} = -\frac{1}{3} h^{-\frac{1}{4}}\, \Herm{\eta^r}{\Gamma_{ijk} \Gamma_{11} \eta^s}\, \e^5\wedge \e^i\wedge \e^j\wedge \e^k - \frac{2}{4!} h^{-\frac{1}{4}}\,\Herm{\eta^r}{\Gamma_{i_1\dots i_4}\eta^s} \e^{i_1}\wedge \dots \wedge \e^{i_4}~,
\end{equation}
\begin{equation}
	\begin{split}
	\tau^{rs} &= \frac{1}{3} h^{-\frac{1}{4}}\, \Herm{\eta^r}{\Gamma_{ijk} \Gamma_{11} \eta^s}\, \e^0\wedge \e^5\wedge \e^i\wedge \e^j\wedge \e^k \\
	&+ \frac{2}{4!} h^{-\frac{1}{4}} \Herm{\eta^r}{\Gamma_{i_1\dots i_4}\eta^s}\, \e^0\wedge \e^{i_1}\wedge\dots \wedge \e^{i_4}~,
	\end{split}
\end{equation}
where $i,j,k = 1,2,3,4,6,7,8,9$ and $(\e^0, \e^5, \e^i)$ is a pseudo-orthonormal frame of the D0-brane metric \eqref{dbrane} for $p = 0$.

\subsection{D6-brane}

Using the expression for the Killing spinors in \eqref{d6sol}, one can easily compute the non-vanishing form bilinears of D6-brane as follows
\begin{equation}
	\sigma^{rs} = 2 h^{-\frac{1}{4}} \mathrm{Re}{\Dirac{\eta^r}{\eta^s}}~, \quad 
	k^{rs} = 2 h^{-\frac{1}{4}} \mathrm{Re}{\Dirac{\eta^r}{\Gamma_a \eta^s}}\, \e^a~,
\end{equation}
\begin{equation}
	\tilde{k}^{rs} = -2 h^{-\frac{1}{4}} \mathrm{Re}{\Dirac{\eta^r}{\Gamma_{11} \lambda^s}}\, \e^4 + 2 h^{-\frac{1}{4}} \mathrm{Im} \Dirac{\eta^r}{\eta^s}\, \e^5 - 2 h^{-\frac{1}{4}} \mathrm{Im}{\Dirac{\eta^r}{\Gamma_{11} \lambda^s}}\, \e^9~,
\end{equation}
\begin{equation}
	\begin{split}
		\omega^{rs} = -&2 h^{-\frac{1}{4}} \mathrm{Re}{\Dirac{\eta^r}{\Gamma_5 \lambda^s}} \e^4 \wedge \e^5 - 2 h^{-\frac{1}{4}} \mathrm{Im}{\Dirac{\eta^r}{\eta^s}} \e^4 \wedge \e^9 \\
		+& 2 h^{-\frac{1}{4}} \mathrm{Im}{\Dirac{\eta^r}{\Gamma_5 \lambda^s}} \e^5 \wedge \e^9 + h^{-\frac{1}{4}} \mathrm{Re}{\Dirac{\eta^r}{\Gamma_{ab}\eta^s}} \e^a \wedge \e^b~,
	\end{split}
\end{equation}
\begin{equation}
	\begin{split}
		\tilde\omega^{rs} = -&2 h^{-\frac{1}{4}} \mathrm{Re}{\Dirac{\eta^r}{\Gamma_a \Gamma_{11} \lambda^s}} \e^a \wedge \e^4 + 2 h^{-\frac{1}{4}} \mathrm{Im} {\Dirac{\eta^r}{\Gamma_a \eta^s}} \e^a \wedge \e^5 \\
		-& 2 h^{-\frac{1}{4}} \mathrm{Im}{\Dirac{\eta^r}{\Gamma_a \Gamma_{11} \lambda^s}} \e^a \wedge \e^9~,
	\end{split}
\end{equation}
\begin{equation}
	\begin{split}
		\pi^{rs} = -& 2 h^{-\frac{1}{4}} \mathrm{Re}{\Dirac{\eta^r}{\Gamma_a \Gamma_{5} \lambda^s}} \e^a \wedge \e^4 \wedge \e^5 - 2 h^{-\frac{1}{4}} \mathrm{Im}{\Dirac{\eta^r}{\Gamma_a \eta^s}} \e^a \wedge \e^4 \wedge \e^9 \\
		+& 2 h^{-\frac{1}{4}} \mathrm{Im}{\Dirac{\eta^r}{\Gamma_a \Gamma_{5} \lambda^s}} \e^a \wedge \e^5 \wedge \e^9 + \frac{1}{3} h^{-\frac{1}{4}} \mathrm{Re}{\Dirac{\eta^r}{\Gamma_{abc} \eta^s}} \e^a \wedge \e^b \wedge \e^c~,
	\end{split}
\end{equation}
\begin{equation}
	\begin{split}
		\tilde{\pi}^{rs} = -& 2 h^{-\frac{1}{4}} \mathrm{Re}{\Dirac{\eta^r}{\eta^s}} \e^4 \wedge \e^5 \wedge \e^9 - h^{-\frac{1}{4}} \mathrm{Re}{\Dirac{\eta^r}{\Gamma_{ab} \Gamma_{11} \lambda^s}} \e^a \wedge \e^b \wedge \e^4 \\
		+& h^{-\frac{1}{4}} \mathrm{Im}{\Dirac{\eta^r}{\Gamma_{ab} \eta^s}} \e^a \wedge \e^b \wedge \e^5 - h^{-\frac{1}{4}} \mathrm{Im}{\Dirac{\eta^r}{\Gamma_{ab} \Gamma_{11} \lambda^s}} \e^a \wedge \e^b \wedge \e^9~,
	\end{split}
\end{equation}
\begin{equation}
	\begin{split}
		&\zeta^{rs} = - h^{-\frac{1}{4}} \mathrm{Re}{\Dirac{\eta^r}{\Gamma_{ab} \Gamma_5 \lambda^s}} \e^a \wedge \e^b \wedge \e^4 \wedge \e^5 - h^{-\frac{1}{4}} \mathrm{Im}{\Dirac{\eta^r}{\Gamma_{ab} \eta^s}} \e^a \wedge \e^b \wedge \e^4 \wedge \e^9 \\
		&+ h^{-\frac{1}{4}} \mathrm{Im}{\Dirac{\eta^r}{\Gamma_{ab} \Gamma_5 \lambda^s}} \e^a \wedge \e^b \wedge \e^5 \wedge \e^9 + \frac{1}{12} h^{-\frac{1}{4}} \mathrm{Re}{\Dirac{\eta^r}{\Gamma_{abcd}\eta^s}} \e^a \wedge \e^b \wedge \e^c \wedge \e^d~,
	\end{split}
\end{equation}
\begin{equation}
	\begin{split}
		&\tilde{\zeta}^{rs} = -2 h^{-\frac{1}{4}} \mathrm{Re}{\Dirac{\eta^r}{\Gamma_a \eta^s}} \e^a \wedge \e^4 \wedge \e^5 \wedge \e^9 - \frac{1}{3} h^{-\frac{1}{4}} \mathrm{Re}{\Dirac{\eta^r}{\Gamma_{abc} \Gamma_{11} \lambda^s}} \e^a \wedge \e^b \wedge \e^c \wedge \e^4 \\
		+& \frac{1}{3} h^{-\frac{1}{4}} \mathrm{Im}{\Dirac{\eta^r}{\Gamma_{abc} \eta^s}} \e^a \wedge \e^b \wedge \e^c \wedge \e^5 - \frac{1}{3} h^{-\frac{1}{4}} \mathrm{Im}{\Dirac{\eta^r}{\Gamma_{abc} \Gamma_{11} \lambda^s}} \e^a \wedge \e^b \wedge \e^c \wedge \e^9~,
	\end{split}
\end{equation}
\begin{equation}
	\begin{split}
		\tau^{rs} = &- \frac{1}{3} h^{-\frac{1}{4}} \mathrm{Re}{\Dirac{\eta^r}{\Gamma_{abc} \Gamma_5 \lambda^s}} \e^a \wedge \e^b \wedge \e^c \wedge \e^4 \wedge \e^5 \\
		&- \frac{1}{3} h^{-\frac{1}{4}} \mathrm{Im}{\Dirac{\eta^r}{\Gamma_{abc} \eta^s}} \e^a \wedge \e^b \wedge \e^c \wedge \e^4 \wedge \e^9 \\
		&+ \frac{1}{3} h^{-\frac{1}{4}} \mathrm{Im}{\Dirac{\eta^r}{\Gamma_{abc} \Gamma_5 \lambda^s}} \e^a \wedge \e^b \wedge \e^c \wedge \e^5 \wedge \e^9 \\
		&+ \frac{2}{5!} h^{-\frac{1}{4}} \mathrm{Re}{\Dirac{\eta^r}{\Gamma_{a_1 \dots a_5} \eta^s}} \e^{a_1} \wedge \dots \wedge \e^{a_5}~,
	\end{split}
\end{equation}
where $a,b,c = 0, 1, 2, 3, 6, 7, 8$ and $(\e^a, \e^5, \e^4, \e^9)$ is a pseudo-orthonormal frame of the metric \eqref{dbrane} with $p = 6$.

\subsection{D2-brane}

Using the Killing spinors \eqref{d2solscon}, one finds that the non-vanishing form bilinears of the D2-brane solution are as follows
\begin{equation}
	\sigma^{rs} = h^{-\frac{1}{4}}(-\Herm{\eta^r}{\lambda^s} + \Herm{\lambda^r}{\eta^s})~,~~\tilde k^{rs} = h^{-\frac{1}{4}}(-\Herm{\eta^r}{\Gamma_i \Gamma_{11}\lambda^s}+ \Herm{\lambda^r}{\Gamma_i \Gamma_{11} \eta^s})\, \e^i~,
\end{equation}
\begin{equation}
	\begin{split}
		k^{rs} &= h^{-\frac{1}{4}}(\Herm{\eta^r}{\eta^s}+ \Herm{\lambda^r}{\lambda^s})\, \e^0 + h^{-\frac{1}{4}}(-\Herm{\eta^r}{\eta^s}+ \Herm{\lambda^r}{\lambda^s})\, \e^5 \\
		&+ h^{-\frac{1}{4}}(\Herm{\eta^r}{\lambda^s} + \Herm{\lambda^r}{\eta^s})\, \e^1~,
	\end{split}
\end{equation}
\begin{equation}
	\begin{split}
		\omega^{rs} &= \frac{h^{-\frac{1}{4}}}{2} (-\Herm{\eta^r}{\Gamma_{ij} \lambda^s} + \Herm{\lambda^r}{\Gamma_{ij} \eta^s})\, \e^i\wedge \e^j + h^{-\frac{1}{4}}(\Herm{\eta^r}{\lambda^s}+ \Herm{\lambda^r}{\eta^s})\, \e^0\wedge \e^5 \\
		&+ h^{-\frac{1}{4}}(\Herm{\eta^r}{\eta^s}- \Herm{\lambda^r}{\lambda^s})\, \e^0\wedge \e^1 + h^{-\frac{1}{4}} (\Herm{\eta^r}{\eta^s}+ \Herm{\lambda^r}{\lambda^s})\, \e^1\wedge \e^5~,
	\end{split}
\end{equation}
\begin{equation}
	\begin{split}
		\tilde \omega^{rs} = &-h^{-\frac{1}{4}}(\Herm{\eta^r}{\Gamma_{i} \Gamma_{11}\eta^s} + \Herm{\lambda^r}{\Gamma_{i} \Gamma_{11}\lambda^s})\,\e^i\wedge \e^0
		\\
		&+ h^{-\frac{1}{4}} (\Herm{\eta^r}{\Gamma_{i} \Gamma_{11}\eta^s}-\Herm{\lambda^r}{\Gamma_{i} \Gamma_{11}\lambda^s})\, \e^i\wedge \e^5 \\
		&- h^{-\frac{1}{4}} (\Herm{\eta^r}{\Gamma_i \Gamma_{11} \lambda^s}+ \Herm{\lambda^r}{\Gamma_i \Gamma_{11}\eta^s})\, \e^i\wedge \e^1~,
	\end{split}
\end{equation}
\begin{equation}
	\begin{split}
		\pi^{rs} &=h^{-\frac{1}{4}}(-\Herm{\eta^r}{\lambda^s}+ \Herm{\lambda^r}{\eta^s})\,\e^0\wedge \e^5\wedge \e^1 \\
		& + \frac{h^{-\frac{1}{4}}}{2}\Big((\Herm{\eta^r}{\Gamma_{ij} \eta^s}+ \Herm{\lambda^r}{\Gamma_{ij}\lambda^s})\, \e^0 + (-\Herm{\eta^r}{\Gamma_{ij}\eta^s}+ \Herm{\lambda^r}{\Gamma_{ij} \lambda^s})\, \e^5 \\
		& + (\Herm{\eta^r}{\Gamma_{ij}\lambda^s}+ \Herm{\lambda^r}{\Gamma_{ij}\eta^s})\, \e^1 \Big)\,\wedge \e^i\wedge \e^j~,
	\end{split}
\end{equation}
\begin{equation}
	\begin{split}
		\tilde \pi^{rs} &=h^{-\frac{1}{4}}(\Herm{\lambda^r}{\Gamma_i \Gamma_{11} \eta^s}+ \Herm{\eta^r}{\Gamma_i \Gamma_{11}\lambda^s})\, \e^0\wedge \e^5\wedge \e^i \\
		& + h^{-\frac{1}{4}}
		(\Herm{\eta^r}{\Gamma_i \Gamma_{11} \eta^s}- \Herm{\lambda^r}{\Gamma_i \Gamma_{11}\lambda^s})\, \e^0\wedge \e^1\wedge \e^i \\
		& - h^{-\frac{1}{4}}(\Herm{\eta^r}{\Gamma_i \Gamma_{11} \eta^s}+ \Herm{\lambda^r}{\Gamma_i \Gamma_{11}\lambda^s})\, \e^5\wedge \e^1\wedge \e^i \\
		& + \frac{h^{-\frac{1}{4}}}{3!}(-\Herm{\eta^r}{\Gamma_{ijk} \Gamma_{11}\lambda^s}+ \Herm{\lambda^r}{\Gamma_{ijk} \Gamma_{11} \eta^s})\, \e^i\wedge \e^j\wedge \e^k~,
	\end{split}
\end{equation}
\begin{equation}
	\begin{split}
		\zeta^{rs} &= \frac{h^{-\frac{1}{4}}}{4!}(-\Herm{\eta^r}{\Gamma_{i_1\dots i_4} \lambda^s} + \Herm{\lambda^r}{\Gamma_{i_1\dots i_4}\eta^s})\,
		\e^{i_1}\wedge\dots \wedge \e^{i_4} \\
		&+ \frac{h^{-\frac{1}{4}}}{2} \Big((\Herm{\eta^r}{\Gamma_{ij}\lambda^s}+ \Herm{\lambda^r}{\Gamma_{ij} \eta^s})\, \e^0\wedge \e^5
		+(\Herm{\eta^r}{\Gamma_{ij}\eta^s}- \Herm{\lambda^r}{\Gamma_{ij}\lambda^s})\, \e^0\wedge \e^1 \\
		&+ (\Herm{\eta^r}{\Gamma_{ij}\eta^s}+ \Herm{\lambda^r}{\Gamma_{ij}\lambda^s})\, \e^1\wedge \e^5\Big)\wedge \e^i\wedge \e^j~,
	\end{split}
\end{equation}
\begin{equation}
	\begin{split}
		\tilde \zeta^{rs} &= \frac{h^{-\frac{1}{4}}}{3!} \Big(-(\Herm{\eta^r}{\Gamma_{ijk} \Gamma_{11}\eta^s} + \Herm{\lambda^r}{\Gamma_{ijk} \Gamma_{11}\lambda^s})\,\e^i\wedge \e^j\wedge \e^k\wedge \e^0 \\
		&+ (\Herm{\eta^r}{\Gamma_{ijk} \Gamma_{11}\eta^s}-\Herm{\lambda^r}{\Gamma_{ijk} \Gamma_{11}\lambda^s})\, \e^i\wedge \e^j\wedge \e^k\wedge \e^5 \\
		&- (\Herm{\eta^r}{\Gamma_{ijk} \Gamma_{11} \lambda^s}+ \Herm{\lambda^r}{\Gamma_{ijk} \Gamma_{11}\eta^s})\Big)\, \e^i\wedge \e^j\wedge \e^k\wedge \e^1 \\
		&+ h^{-\frac{1}{4}}(-\Herm{\eta^r}{\Gamma_i \Gamma_{11} \lambda^s}+ \Herm{\lambda^r}{\Gamma_i \Gamma_{11}\eta^s})\, \e^0\wedge \e^5\wedge \e^1\wedge \e^i~,
	\end{split}
\end{equation}
\begin{equation}
	\begin{split}
		\tau^{rs} &= \frac{h^{-\frac{1}{4}}}{2}(-\Herm{\eta^r}{\Gamma_{ij} \lambda^s}+ \Herm{\lambda^r}{\Gamma_{ij}\eta^s})\,\e^0\wedge \e^5\wedge \e^1\wedge \e^i\wedge \e^j \\
		&+ \frac{h^{-\frac{1}{4}}}{4!}\Big((\Herm{\eta^r}{\Gamma_{i_1\dots i_4} \eta^s}
		+ \Herm{\lambda^r}{\Gamma_{i_1\dots i_4}\lambda^s})\, \e^0 \\
		&+ (-\Herm{\eta^r}{\Gamma_{i_1\dots i_4}\eta^s}+ \Herm{\lambda^r}{\Gamma_{i_1\dots i_4} \lambda^s})\, \e^5 \\
		&+ (\Herm{\eta^r}{\Gamma_{i_1\dots i_4}\lambda^s}+ \Herm{\lambda^r}{\Gamma_{i_1\dots i_4}\eta^s})\, \e^1\Big)\wedge \e^{i_1}\wedge \dots \wedge \e^{i_4}~,
	\end{split}
\end{equation}
where $i,j,k = 2,3,4,6,7,8,9 $ and $(\e^0, \e^5, \e^1, \e^i)$ is a pseudo-orthonormal frame of the metric \eqref{dbrane} with $p = 2$.

\subsection{D4-brane}

Using the Killing spinors \eqref{d4sol}, one finds that the non-vanishing form bilinears of the D4-brane solution are as follows
\begin{equation}
	\tilde\sigma^{rs} = 2h^{-\frac{1}{4}} \mathrm{Re} \Dirac{\eta^{1r}}{\Gamma_{11} \eta^{1s}} + 2h^{-\frac{1}{4}} \mathrm{Re} \Dirac{\lambda^{1r}}{\Gamma_{11} \lambda^{1s}}~,~~
\end{equation}
\begin{equation}
	k^{rs} = 2 h^{-\frac{1}{4}} ( \mathrm{Re} \Dirac{\eta^{1r}}{\Gamma_{a} \eta^{1s}}+ \mathrm{Re} \Dirac{\lambda^{1r}}{\Gamma_{a} \lambda^{1s}}) \, \e^a~,
\end{equation}
\begin{equation}
	\begin{split}
		\tilde k^{rs} &=2 h^{-\frac{1}{4}}(\mathrm{Re} \Dirac{\eta^{1r}}{\Gamma_{11} \eta^{1s}}- \mathrm{Re} \Dirac{\lambda^{1r}}{\Gamma_{11} \lambda^{1s}})\, \e^2 \\
		&+ 2h^{-\frac{1}{4}} (-\mathrm{Re} \Dirac{\eta^{1r}}{\Gamma_{11} \lambda^{2s}}+ \mathrm{Re} \Dirac{\lambda^{1r}}{\Gamma_{11} \eta^{2s}})\, \e^4 \\
		&- 2 h^{-\frac{1}{4}}(\mathrm{Re} \Dirac{\eta^{1r}}{\Gamma_{11} \lambda^{1s}}+ \mathrm{Re} \Dirac{\lambda^{1r}}{\Gamma_{11} \eta^{1s}})\, \e^3 \\
		&+ 2 h^{-\frac{1}{4}}(-\mathrm{Im} \Dirac{\eta^{1r}}{\Gamma_{11} \lambda^{1s}}+ \mathrm{Im} \Dirac{\lambda^{1r}}{\Gamma_{11} \eta^{1s}})\, \e^8 \\
		&+ 2h^{-\frac{1}{4}} (-\mathrm{Im} \Dirac{\eta^{1r}}{\Gamma_{11} \lambda^{2s}}+ \mathrm{Im} \Dirac{\lambda^{1r}}{\Gamma_{11} \eta^{2s}})\, \e^9~,
	\end{split}
\end{equation}
\begin{equation}
	\begin{split}
		\omega^{rs} &= 2 h^{-\frac{1}{4}}(-\mathrm{Re} \Dirac{\eta^{1r}}{\Gamma_a \eta^{1s}}+ \mathrm{Re} \Dirac{\lambda^{1r}}{\Gamma_a \lambda^{1s}})\, \e^a\wedge \e^2 \\
		&+ 2 h^{-\frac{1}{4}}(\mathrm{Re} \Dirac{\eta^{1r}}{\Gamma_a \lambda^{1s}}+ \mathrm{Re} \Dirac{\lambda^{1r}}{\Gamma_a \eta^{1s}})\,\e^a\wedge \e^3 \\
		&+ 2h^{-\frac{1}{4}} (\mathrm{Re} \Dirac{\eta^{1r}}{\Gamma_a \lambda^{2s}}- \mathrm{Re} \Dirac{\lambda^{1r}}{\Gamma_a \eta^{2s}})\, \e^a\wedge \e^4 \\
		&+ 2h^{-\frac{1}{4}} (\mathrm{Im} \Dirac{\eta^{1r}}{\Gamma_a \lambda^{1s}}- \mathrm{Im} \Dirac{\lambda^{1r}}{\Gamma_a \eta^{1s}})\, \e^a\wedge \e^8 \\
		&+ 2h^{-\frac{1}{4}} (\mathrm{Im} \Dirac{\eta^{1r}}{\Gamma_a \lambda^{2s}}- \mathrm{Im} \Dirac{\lambda^{1r}}{\Gamma_a \eta^{2s}})\, \e^a\wedge \e^9~,
	\end{split}
\end{equation}
\begin{equation}
	\begin{split}
		\tilde \omega^{rs} &= h^{-\frac{1}{4}}(\mathrm{Re} \Dirac{\eta^{1r}}{\Gamma_{ab} \Gamma_{11} \eta^{1s}}+ \mathrm{Re} \Dirac{\lambda^{1r}}{\Gamma_{ab} \Gamma_{11} \lambda^{1s}})\, \e^a\wedge \e^b \\
		&-2h^{-\frac{1}{4}}(\mathrm{Re} \Dirac{\eta^{1r}}{\Gamma_{11} \lambda^{1s}}- \mathrm{Re} \Dirac{\lambda^{1r}}{\Gamma_{11} \eta^{1s}})\, \e^2\wedge \e^3 \\
		&+2h^{-\frac{1}{4}}(-\mathrm{Re} \Dirac{\eta^{1r}}{\Gamma_{11} \lambda^{2s}}- \mathrm{Re} \Dirac{\lambda^{1r}}{\Gamma_{11} \eta^{2s}})\, \e^2\wedge \e^4 \\
		&-2h^{-\frac{1}{4}}(\mathrm{Im} \Dirac{\eta^{1r}}{\Gamma_{11} \lambda^{1s}}+ \mathrm{Im} \Dirac{\lambda^{1r}}{\Gamma_{11} \eta^{1s}})\, \e^2\wedge \e^8 \\
		&-2h^{-\frac{1}{4}}(\mathrm{Im} \Dirac{\eta^{1r}}{\Gamma_{11} \lambda^{2s}}+ \mathrm{Im} \Dirac{\lambda^{1r}}{\Gamma_{11} \eta^{2s}})\, \e^2\wedge \e^9 \\
		&+2h^{-\frac{1}{4}}(-\mathrm{Re} \Dirac{\eta^{1r}}{\Gamma_{11} \eta^{2s}}+ \mathrm{Re} \Dirac{\lambda^{1r}}{\Gamma_{11} \lambda^{2s}})\, \e^3\wedge \e^4 \\
		&+2h^{-\frac{1}{4}}(-\mathrm{Im} \Dirac{\eta^{1r}}{\Gamma_{11} \eta^{1s}}+ \mathrm{Im} \Dirac{\lambda^{1r}}{\Gamma_{11} \lambda^{1s}})\, \e^3\wedge \e^8 \\
		&+2h^{-\frac{1}{4}}(-\mathrm{Im} \Dirac{\eta^{1r}}{\Gamma_{11} \eta^{2s}}+ \mathrm{Im} \Dirac{\lambda^{1r}}{\Gamma_{11} \lambda^{2s}})\, \e^3\wedge \e^9 \\
		&+2h^{-\frac{1}{4}}(\mathrm{Im} \Dirac{\eta^{1r}}{\Gamma_{11} \eta^{2s}}+ \mathrm{Im} \Dirac{\lambda^{1r}}{\Gamma_{11} \lambda^{2s}})\, \e^4\wedge \e^8 \\
		&-2h^{-\frac{1}{4}}(\mathrm{Im} \Dirac{\eta^{1r}}{\Gamma_{11} \eta^{1s}}+ \mathrm{Im} \Dirac{\lambda^{1r}}{\Gamma_{11} \lambda^{1s}})\, \e^4\wedge \e^9 \\
		&+2h^{-\frac{1}{4}}(\mathrm{Re} \Dirac{\eta^{1r}}{\Gamma_{11} \eta^{2s}}+ \mathrm{Re} \Dirac{\lambda^{1r}}{\Gamma_{11} \lambda^{2s}})\, \e^8\wedge \e^9~,
	\end{split}
\end{equation}
\begin{equation}
	\begin{split}
		\pi^{rs}&=\frac{h^{-\frac{1}{4}}}{3} (\mathrm{Re} \Dirac{\eta^{1r}}{\Gamma_{abc} \eta^{1s}}+ \mathrm{Re} \Dirac{\lambda^{1r}}{\Gamma_{abc} \lambda^{1s}})\, \e^a\wedge \e^b\wedge \e^c \\
		&-2h^{-\frac{1}{4}}(\mathrm{Re} \Dirac{\eta^{1r}}{\Gamma_a \lambda^{1s}}- \mathrm{Re} \Dirac{\lambda^{1r}}{\Gamma_a\eta^{1s}})\, \e^2\wedge \e^3\wedge \e^a \\
		&+2h^{-\frac{1}{4}}(-\mathrm{Re} \Dirac{\eta^{1r}}{\Gamma_a \lambda^{2s}}- \mathrm{Re} \Dirac{\lambda^{1r}}{\Gamma_a \eta^{2s}})\, \e^2\wedge \e^4\wedge \e^a \\
		&-2h^{-\frac{1}{4}}(\mathrm{Im} \Dirac{\eta^{1r}}{\Gamma_a \lambda^{1s}}+ \mathrm{Im} \Dirac{\lambda^{1r}}{\Gamma_a \eta^{1s}})\, \e^2\wedge \e^8\wedge \e^a \\
		&-2h^{-\frac{1}{4}}(\mathrm{Im} \Dirac{\eta^{1r}}{\Gamma_a \lambda^{2s}}+ \mathrm{Im} \Dirac{\lambda^{1r}}{\Gamma_a \eta^{2s}})\, \e^2\wedge \e^9\wedge \e^a \\
		&+2h^{-\frac{1}{4}}(-\mathrm{Re} \Dirac{\eta^{1r}}{\Gamma_a \eta^{2s}}+ \mathrm{Re} \Dirac{\lambda^{1r}}{\Gamma_a \lambda^{2s}})\, \e^3\wedge \e^4\wedge \e^a \\
		&+2h^{-\frac{1}{4}}(-\mathrm{Im} \Dirac{\eta^{1r}}{\Gamma_a \eta^{1s}}+ \mathrm{Im} \Dirac{\lambda^{1r}}{\Gamma_a \lambda^{1s}})\, \e^3\wedge \e^8\wedge \e^a \\
		&+2h^{-\frac{1}{4}}(-\mathrm{Im} \Dirac{\eta^{1r}}{\Gamma_a \eta^{2s}}+ \mathrm{Im} \Dirac{\lambda^{1r}}{\Gamma_a \lambda^{2s}})\, \e^3\wedge \e^9\wedge \e^a \\
		&+2h^{-\frac{1}{4}}(\mathrm{Im} \Dirac{\eta^{1r}}{\Gamma_a \eta^{2s}}+ \mathrm{Im} \Dirac{\lambda^{1r}}{\Gamma_a \lambda^{2s}})\, \e^4\wedge \e^8\wedge \e^a \\
		&-2h^{-\frac{1}{4}}(\mathrm{Im} \Dirac{\eta^{1r}}{\Gamma_a \eta^{1s}}+ \mathrm{Im} \Dirac{\lambda^{1r}}{\Gamma_a \lambda^{1s}})\, \e^4\wedge \e^9\wedge \e^a \\
		&+2h^{-\frac{1}{4}}(\mathrm{Re} \Dirac{\eta^{1r}}{\Gamma_a \eta^{2s}}+ \mathrm{Re} \Dirac{\lambda^{1r}}{\Gamma_a \lambda^{2s}})\, \e^8\wedge \e^9\wedge \e^a
	\end{split}
\end{equation}
\begin{equation}
	\begin{split}
		\tilde\pi^{rs} &= h^{-\frac{1}{4}}\Big( (\mathrm{Re} \Dirac{\eta^{1r}}{\Gamma_{ab} \Gamma_{11} \eta^{1s}}- \mathrm{Re} \Dirac{\lambda^{1r}}{\Gamma_{ab}\Gamma_{11} \lambda^{1s}})\, \e^2 \\
		& + (-\mathrm{Re} \Dirac{\eta^{1r}}{\Gamma_{ab} \Gamma_{11} \lambda^{2s}}+ \mathrm{Re} \Dirac{\lambda^{1r}}{\Gamma_{ab}\Gamma_{11} \eta^{2s}})\, \e^4 \\
		& - (\mathrm{Re} \Dirac{\eta^{1r}}{\Gamma_{ab}\Gamma_{11} \lambda^{1s}}+ \mathrm{Re} \Dirac{\lambda^{1r}}{\Gamma_{ab}\Gamma_{11} \eta^{1s}})\, \e^3 \\
		& + (-\mathrm{Im} \Dirac{\eta^{1r}}{\Gamma_{ab}\Gamma_{11} \lambda^{1s}}+ \mathrm{Im} \Dirac{\lambda^{1r}}{\Gamma_{ab} \Gamma_{11} \eta^{1s}})\, \e^8 \\
		& + (-\mathrm{Im} \Dirac{\eta^{1r}}{\Gamma_{ab}\Gamma_{11} \lambda^{2s}}+ \mathrm{Im} \Dirac{\lambda^{1r}}{\Gamma_{ab}\Gamma_{11} \eta^{2s}})\, \e^9\Big) \wedge \e^a\wedge \e^b \\
		& +\frac{1}{2\cdot 3!} \epsilon_{ijk}{}^{mn} \tilde \omega^{rs}_{mn}\, \e^i\wedge \e^i\wedge \e^k~,
	\end{split}
\end{equation}
\begin{equation}
	\begin{split}
		\zeta^{rs} &= \frac{h^{-\frac{1}{4}}}{3} (-\mathrm{Re} \Dirac{\eta^{1r}}{\Gamma_{abc} \eta^{1s}}+ \mathrm{Re} \Dirac{\lambda^{1r}}{\Gamma_{abc} \lambda^{1s}})\, \e^a\wedge \e^b\wedge \e^c\wedge \e^2 \\
		& +\frac{h^{-\frac{1}{4}}}{3} (\mathrm{Re} \Dirac{\eta^{1r}}{\Gamma_{abc} \lambda^{1s}}+ \mathrm{Re} \Dirac{\lambda^{1r}}{\Gamma_{abc} \eta^{1s}})\,\e^a\wedge \e^b\wedge \e^c\wedge \e^3 \\
		& +\frac{h^{-\frac{1}{4}}}{3} (\mathrm{Re} \Dirac{\eta^{1r}}{\Gamma_{abc} \lambda^{2s}}- \mathrm{Re} \Dirac{\lambda^{1r}}{\Gamma_{abc} \eta^{2s}})\, \e^a\wedge \e^b\wedge \e^c\wedge \e^4 \\
		& +\frac{h^{-\frac{1}{4}}}{3} (\mathrm{Im} \Dirac{\eta^{1r}}{\Gamma_{abc} \lambda^{1s}}- \mathrm{Im} \Dirac{\lambda^{1r}}{\Gamma_{abc} \eta^{1s}})\, \e^a\wedge \e^b\wedge \e^c\wedge \e^8 \\
		& +\frac{h^{-\frac{1}{4}}}{3} (\mathrm{Im} \Dirac{\eta^{1r}}{\Gamma_{abc} \lambda^{2s}}- \mathrm{Im} \Dirac{\lambda^{1r}}{\Gamma_{abc} \eta^{2s}})\, \e^a\wedge \e^b\wedge \e^c\wedge \e^9 \\
		& -\frac{1}{12} \pi^{rs}_{amn} \epsilon_{ijk}{}^{mn} \e^a\wedge \e^i\wedge \e^j\wedge \e^k~,
	\end{split}
\end{equation}
\begin{equation}
	\begin{split}
		\tilde\zeta^{rs} &= \frac{2 h^{-\frac{1}{4}}}{4!} (\mathrm{Re} \Dirac{\eta^{1r}}{\Gamma_{a_1\dots a_4} \Gamma_{11} \eta^{1s}}+ \mathrm{Re} \Dirac{\lambda^{1r}}{\Gamma_{a_1\dots a_4} \Gamma_{11} \lambda^{1s}}) \, \e^{a_1}\wedge \dots\wedge \e^{a_4} \\
		& -h^{-\frac{1}{4}}(\mathrm{Re} \Dirac{\eta^{1r}}{\Gamma_{ab} \Gamma_{11} \lambda^{1s}}- \mathrm{Re} \Dirac{\lambda^{1r}}{\Gamma_{ab} \Gamma_{11} \eta^{1s}})\, \e^a\wedge \e^b\wedge \e^2\wedge \e^3 \\
		& +h^{-\frac{1}{4}}(-\mathrm{Re} \Dirac{\eta^{1r}}{\Gamma_{ab} \Gamma_{11} \lambda^{2s}}- \mathrm{Re} \Dirac{\lambda^{1r}}{\Gamma_{ab} \Gamma_{11} \eta^{2s}})\, \e^a\wedge \e^b\wedge \e^2\wedge \e^4 \\
		& -h^{-\frac{1}{4}}(\mathrm{Im} \Dirac{\eta^{1r}}{\Gamma_{ab} \Gamma_{11} \lambda^{1s}}+ \mathrm{Im} \Dirac{\lambda^{1r}}{\Gamma_{ab} \Gamma_{11} \eta^{1s}})\, \e^a\wedge \e^b\wedge \e^2\wedge \e^8 \\
		& -h^{-\frac{1}{4}}(\mathrm{Im} \Dirac{\eta^{1r}}{\Gamma_{ab} \Gamma_{11} \lambda^{2s}}+ \mathrm{Im} \Dirac{\lambda^{1r}}{\Gamma_{ab} \Gamma_{11} \eta^{2s}})\, \e^a\wedge \e^b\wedge \e^2\wedge \e^9 \\
		& +h^{-\frac{1}{4}}(-\mathrm{Re} \Dirac{\eta^{1r}}{\Gamma_{ab} \Gamma_{11} \eta^{2s}}+ \mathrm{Re} \Dirac{\lambda^{1r}}{\Gamma_{ab} \Gamma_{11} \lambda^{2s}})\, \e^a\wedge \e^b\wedge \e^3\wedge \e^4 \\
		& +h^{-\frac{1}{4}}(-\mathrm{Im} \Dirac{\eta^{1r}}{\Gamma_{ab} \Gamma_{11} \eta^{1s}}+ \mathrm{Im} \Dirac{\lambda^{1r}}{\Gamma_{ab} \Gamma_{11} \lambda^{1s}})\, \e^a\wedge \e^b\wedge \e^3\wedge \e^8 \\
		& +h^{-\frac{1}{4}}(-\mathrm{Im} \Dirac{\eta^{1r}}{\Gamma_{ab} \Gamma_{11} \eta^{2s}}+ \mathrm{Im} \Dirac{\lambda^{1r}}{\Gamma_{ab} \Gamma_{11} \lambda^{2s}})\, \e^a\wedge \e^b\wedge \e^3\wedge \e^9 \\
		& +h^{-\frac{1}{4}}(\mathrm{Im} \Dirac{\eta^{1r}}{\Gamma_{ab} \Gamma_{11} \eta^{2s}}+ \mathrm{Im} \Dirac{\lambda^{1r}}{\Gamma_{ab} \Gamma_{11} \lambda^{2s}})\, \e^a\wedge \e^b\wedge \e^4\wedge \e^8 \\
		& -h^{-\frac{1}{4}}(\mathrm{Im} \Dirac{\eta^{1r}}{\Gamma_{ab} \Gamma_{11} \eta^{1s}}+ \mathrm{Im} \Dirac{\lambda^{1r}}{\Gamma_{ab} \Gamma_{11} \lambda^{1s}})\, \e^a\wedge \e^b\wedge \e^4\wedge \e^9 \\
		& +h^{-\frac{1}{4}}(\mathrm{Re} \Dirac{\eta^{1r}}{\Gamma_{ab} \Gamma_{11} \eta^{2s}}+ \mathrm{Re} \Dirac{\lambda^{1r}}{\Gamma_{ab} \Gamma_{11} \lambda^{2s}})\, \e^a\wedge \e^b\wedge \e^8\wedge \e^9 \\
		& -\frac{1}{4!} \epsilon_{i_1\dots i_4}{}^j \tilde k^{rs}_j\, \e^{i_1}\wedge\dots\wedge \e^{i_4}~,
	\end{split}
\end{equation}
\begin{equation}
	\begin{split}
		\tau^{rs} &= \frac{2h^{-\frac{1}{4}}}{5!}( \mathrm{Re} \Dirac{\eta^{1r}}{\Gamma_{a_1\dots a_5} \eta^{1s}}+ \mathrm{Re} \Dirac{\lambda^{1r}}{\Gamma_{a_1\dots a_5} \lambda^{1s}}) \, \e^{a_1}\wedge\dots\wedge \e^{a_5} \\
		& -\frac{h^{-\frac{1}{4}}}{3}(\mathrm{Re} \Dirac{\eta^{1r}}{\Gamma_{abc} \lambda^{1s}}- \mathrm{Re} \Dirac{\lambda^{1r}}{\Gamma_{abc}\eta^{1s}})\, \e^2\wedge \e^3\wedge \e^a \wedge \e^b\wedge \e^c \\
		& +\frac{h^{-\frac{1}{4}}}{3}(-\mathrm{Re} \Dirac{\eta^{1r}}{\Gamma_{abc} \lambda^{2s}}- \mathrm{Re} \Dirac{\lambda^{1r}}{\Gamma_{abc}\eta^{2s}})\, \e^2\wedge \e^4\wedge \e^a\wedge \e^b\wedge \e^c \\
		& -\frac{h^{-\frac{1}{4}}}{3}(\mathrm{Im} \Dirac{\eta^{1r}}{\Gamma_{abc} \lambda^{1s}}+ \mathrm{Im} \Dirac{\lambda^{1r}}{\Gamma_{abc} \eta^{1s}})\, \e^2\wedge \e^8\wedge \e^a\wedge \e^b\wedge \e^c \\
		& -\frac{h^{-\frac{1}{4}}}{3}(\mathrm{Im} \Dirac{\eta^{1r}}{\Gamma_{abc} \lambda^{2s}}+ \mathrm{Im} \Dirac{\lambda^{1r}}{\Gamma_{abc} \eta^{2s}})\, \e^2\wedge \e^9\wedge \e^a\wedge \e^b\wedge \e^c \\
		& +\frac{h^{-\frac{1}{4}}}{3}(-\mathrm{Re} \Dirac{\eta^{1r}}{\Gamma_{abc} \eta^{2s}}+ \mathrm{Re} \Dirac{\lambda^{1r}}{\Gamma_{abc} \lambda^{2s}})\, \e^3\wedge \e^4\wedge \e^a\wedge \e^b\wedge \e^c \\
		& +\frac{h^{-\frac{1}{4}}}{3}(-\mathrm{Im} \Dirac{\eta^{1r}}{\Gamma_{abc} \eta^{1s}}+ \mathrm{Im} \Dirac{\lambda^{1r}}{\Gamma_{abc} \lambda^{1s}})\, \e^3\wedge \e^8\wedge \e^a\wedge \e^b\wedge \e^c \\
		& +\frac{h^{-\frac{1}{4}}}{3}(-\mathrm{Im} \Dirac{\eta^{1r}}{\Gamma_{abc} \eta^{2s}}+ \mathrm{Im} \Dirac{\lambda^{1r}}{\Gamma_{abc} \lambda^{2s}})\, \e^3\wedge \e^9\wedge \e^a\wedge \e^b\wedge \e^c \\
		& +\frac{h^{-\frac{1}{4}}}{3}(\mathrm{Im} \Dirac{\eta^{1r}}{\Gamma_{abc} \eta^{2s}}+ \mathrm{Im} \Dirac{\lambda^{1r}}{\Gamma_{abc} \lambda^{2s}})\, \e^4\wedge \e^8\wedge \e^a\wedge \e^b\wedge \e^c \\
		& -\frac{h^{-\frac{1}{4}}}{3}(\mathrm{Im} \Dirac{\eta^{1r}}{\Gamma_{abc} \eta^{1s}}+ \mathrm{Im} \Dirac{\lambda^{1r}}{\Gamma_{abc} \lambda^{1s}})\, \e^4\wedge \e^9\wedge \e^a\wedge \e^b\wedge \e^c \\
		& +\frac{h^{-\frac{1}{4}}}{3}(\mathrm{Re} \Dirac{\eta^{1r}}{\Gamma_{abc} \eta^{2s}}+ \mathrm{Re} \Dirac{\lambda^{1r}}{\Gamma_{abc} \lambda^{2s}})\, \e^8\wedge \e^9\wedge \e^a\wedge \e^b\wedge \e^c \\
		& +\frac{1}{4!} \epsilon_{i_1\dots i_4}{}^j \omega_{aj} \e^a\wedge \e^{i_1}\wedge \dots\wedge \e^{i_4}~,
	\end{split}
\end{equation}
where $\epsilon_{23849} = 1$, $a,b,c = 0, 5, 1, 6, 7$, $i,j,k = 2,3,4,8,9$ and $(\e^a, \e^i)$ is a pseudo-orthonormal frame of the metric \eqref{dbrane} for $p = 4$.

\subsection{D8-brane}

Using the Killing spinors \eqref{d8sol}, the non-vanishing form bilinears of D8-brane are as follows
\begin{equation}
	\begin{split}
		&\tilde \sigma^{rs}= 2 h^{-\frac{1}{4}} \Herm{\eta^r}{\Gamma_{11} \eta^s}~, \quad k^{rs} = 2 h^{-\frac{1}{4}} \Herm{\eta^r}{\eta^s}\, \e^0 + 2 h^{-\frac{1}{4}} \Herm{\eta^r}{\Gamma_{a'} \eta^s}\, \e^{a'}~, \quad \\
		&\tilde k^{rs} =-2h^{-\frac{1}{4}} \Herm{\eta^r}{\Gamma_{11} \eta^s}\, \e^5~, \quad \omega^{rs} = 2 h^{-\frac{1}{4}}\Herm{\eta^r}{\eta^s}~\e^0\wedge \e^5 + 2 h^{-\frac{1}{4}}\Herm{\eta^r}{\Gamma_{a'} \eta^s}~\e^{a'}\wedge \e^5~, \quad \\
		&\tilde \omega^{rs} = 2 h^{-\frac{1}{4}} \Herm{\eta^r}{\Gamma_{a'} \Gamma_{11} \eta^s}~\e^0\wedge \e^{a'} + h^{-\frac{1}{4}} \Herm{\eta^r}{\Gamma_{a'b'} \Gamma_{11} \eta^s}~\e^{a'}\wedge \e^{b'}~, \\
		&\pi^{rs} = h^{-\frac{1}{4}} \Herm{\eta^r}{\Gamma_{b'c'} \eta^s}~\e^0\wedge \e^{b'}\wedge \e^{c'} + \frac{h^{-\frac{1}{4}}}{3} \Herm{\eta^r}{\Gamma_{a'b'c'} \eta^s}~\e^{a'}\wedge \e^{b'}\wedge \e^{c'}~, \quad \\
		&\tilde \pi^{rs} = -2h^{-\frac{1}{4}}\Herm{\eta^r}{\Gamma_{a'}\Gamma_{11} \eta^s}~\e^{0}\wedge \e^{a'}\wedge \e^5-h^{-\frac{1}{4}}\Herm{\eta^r}{\Gamma_{ab}\Gamma_{11} \eta^s}~\e^a\wedge \e^b\wedge \e^5~, \\
		&\zeta^{rs} = {h^{-\frac{1}{4}}} \Herm{\eta^r}{\Gamma_{b'c'} \eta^s}~\e^0\wedge \e^{b'}\wedge \e^{c'}\wedge \e^5 + \frac{h^{-\frac{1}{4}}}{3} \Herm{\eta^r}{\Gamma_{a'b'c'} \eta^s}~\e^{a'}\wedge \e^{b'}\wedge \e^{c'}\wedge \e^5~, \quad \\
		&\tilde \zeta^{rs} = \frac{h^{-\frac{1}{4}}}{3} \Herm{\eta^r}{\Gamma_{a'b'c'} \Gamma_{11} \eta^s}~\e^{0}\wedge \e^{a'}\wedge \e^{b'} \wedge \e^{c'} + \frac{h^{-\frac{1}{4}}}{12} \Herm{\eta^r}{\Gamma_{a'_1\dots a'_4} \Gamma_{11} \eta^s}~\e^{a'_1}\wedge \dots \wedge \e^{a'_4}~, \\
		&\tau^{rs} = \frac{h^{-\frac{1}{4}}}{12} \Herm{\eta^r}{\Gamma_{a'_1\dots a'_4} \eta^s}~\e^0\wedge \e^{a'_1}\wedge \dots\wedge \e^{a'_4} + \frac{2h^{-\frac{1}{4}}}{5!} \Herm{\eta^r}{\Gamma_{a'_1\dots a'_5} \eta^s}~\e^{a'_1}\wedge \dots\wedge \e^{a'_5}~,
	\end{split}
\end{equation}
where $a',b',c' = 1,6,2,7,3,8,4,9$ and $(\e^a, \e^5)$ is a pseudo-orthonormal frame of the D8-brane metric \eqref{dbrane} for $p = 8$.


\bibliographystyle{jhep} 
\bibliography{references} 


\end{document}